\documentclass[traditabstract]{aa} 
\usepackage{graphicx,natbib,longtable,txfonts,lscape,lineno}
\newcommand{\FR}{FRI{\sl{CAT}}}
\newcommand{\FRII}{FRII{\sl{CAT}}}

\newcommand{\ergs}{\>{\rm erg}\,{\rm s}^{-1}}
\newcommand{\kms}{$\rm{\,km \,s}^{-1}$}

\begin{document}

   \title{FRII{\sl{CAT}}: A FIRST catalog of FR~II radio galaxies.}

   \subtitle{}

   \author{A. Capetti\inst{1}
          \and
          F. Massaro\inst{2}
          \and
          R.D. Baldi\inst{3}
          }

   \institute{INAF-Osservatorio Astrofisico di Torino, via Osservatorio 20,
     10025 Pino Torinese, Italy, 
\and
Dipartimento di Fisica, Universit\`a degli
     Studi di Torino, via Pietro Giuria 1, 10125 Torino, Italy, 
\and
Department of
     Physics and Astronomy, University of Southampton, Highfield, SO17 1BJ, UK}
   \date{}

   \abstract{We built a catalog of 122 FR~II radio galaxies, called
     FRII{\sl{CAT}}, selected from a published sample obtained by
     combining observations from the NVSS, FIRST, and SDSS
     surveys. The catalog includes sources with redshift $\leq 0.15$,
     an edge-brightened radio morphology, and those with at least one
     of the emission peaks located at radius $r$ larger than 30 kpc
     from the center of the host.

     The radio luminosity at 1.4 GHz of the \FRII\ sources covers the
     range $L_{1.4} \sim 10^{39.5} - 10^{42.5}$ $\ergs$. The \FRII\
     catalog has 90\% of low and 10\% of high excitation galaxies
     (LEGs and HEGs), respectively. The properties of these two
     classes are significantly different.  The FRII{\sl{CAT}} LEGs are
     mostly luminous ($-20 \gtrsim M_r \gtrsim -24$), red early-type
     galaxies with black hole masses in the range
     $10^8 \lesssim M_{\rm BH} \lesssim 10^9 M_\odot$; they are
     essentially indistinguishable from the FR~Is belonging to the
     FRI{\sl{CAT}}. The HEG FR~IIs are associated with optically bluer
     and mid-IR redder hosts than the LEG FR~IIs and to galaxies and
     black holes that are smaller, on average, by a factor $\sim$2.

FR~IIs have a factor $\sim$ 3 higher average radio luminosity than
FR~Is. Nonetheless, most ($\sim 90$ \%) of the selected FR~IIs have a radio
power that is lower, by as much as a factor of $\sim$100, than the transition
value between FR~Is and FR~IIs found in the 3C sample. The correspondence
between the morphological classification of FR~I and FR~II and the separation
in radio power disappears when including sources selected at low radio flux
thresholds, which is in line with previous results. In conclusion, a radio source
  produced by a low power jet can be edge brightened or edge darkened, and the
  outcome is not related to differences in the optical properties of the
  host galaxy.}  \maketitle

\section{Introduction}

\citet{fanaroff74} introduced the first classification scheme for
extragalactic radio sources with large-scale structures (i.e., greater than
$\sim$15-20 kpc in size) based on the ratio $R_{FR}$ of the distance between
the regions of highest surface brightness on opposite sides of the central
host galaxy to the total extent of the source up to the lowest brightness
contour in the radio images. Radio sources with $R_{FR}<$0.5 were placed in
Class I (i.e., the edge-darkened FR~Is) and sources with $R_{FR}>$0.5 in Class
II (i.e., the edge-brightened FR~IIs). This morphology-based classification
scheme was found to be linked to their intrinsic power; Fanaroff and Riley
found that all sources in their sample with luminosity at 178 MHz smaller than
2$\times$10$^{25}$ W Hz$^{-1}$ sr$^{-1}$ (for a Hubble constant of 50
\kms\ Mpc$^{-1}$) were classified as FR~I while the brighter sources all were
FR~II. The luminosity distinction between FR classes is fairly sharp at 178
MHz; their separation is even cleaner in an optical-radio luminosity plane,
implying that the FR~I/FR~II dichotomy depends on both optical and radio
luminosity \citep{ledlow96}.

The two Fanaroff-Riley classes do not instead correspond to a division from
the point of view of the optical spectroscopic properties of their hosts.
\citet{laing94} defined low and high excitation galaxies (LEGs and HEGs) based
on the ratios of the diagnostic optical emission lines in a scheme similar to
that adopted to distinguish LINERs and Seyferts in radio quiet AGN
(\citealt{kewley06}). While all FR~Is for which a reliable classification can
be obtained are LEGs, both LEGs and HEGs are found among the FR~IIs (e.g.,
\citealt{buttiglione10}). Buttiglione et al. find that LEGs and HEGs also differ
from other points of view. While LEGs do not show prominent broad lines,
they are observed in $\sim 30\%$ of HEGs; narrow emission lines are a factor
$\sim$ 10 brighter in HEGs than in LEGs at the same radio luminosity.  Also, HEGs
show bluer colors than LEGs \citep{smolcic09}. \citet{baldi08} identify
compact knots in the UV images of 3C HEG radio galaxies, a morphological
evidence of recent star formation extending over 5$-$20 kpc; conversely, LEGs
hosts are usually red, passive galaxies. These results suggest that the radio
galaxies belonging to the two spectroscopic classes correspond to different
manifestation of the radio loud AGN phenomenon
\citep{hardcastle07,buttiglione10,best12}.

The recent multiwavelength large-area surveys are a unique tool to further
explore the connection between the morphological and spectroscopic classes of
radio galaxies, providing us with large samples of radio emitting AGN
extending to lower luminosities than in previous studies.

In \citet{capetti16} we created a catalog of 219 edge-darkened FR~I radio
galaxies called \FR. We found that the \FR\ hosts are remarkably homogeneous,
as they are all luminous red early-type galaxies (ETGs) with large black hole
masses and spectroscopically classified as LEGs. All these properties are
shared by the hosts of more powerful FR~Is in the 3C sample.  They do not show
significant differences from the point of view of their colors with respect to
the general population of massive ETGs. The presence of an active nucleus (and
its level of activity) does not appear to affect the hosts of FR~Is.

We now extend this study to the population of edge-brightened FR~II radio
galaxies with the main aim of comparing the properties of FR~Is and FR~IIs
by also considering their spectroscopic classification.

This paper is organized as follows. In Sect.\ 2 we present the selection
criteria of the sample of FR~IIs.  The radio and optical properties of the
selected sources are presented in Sect.\ 3. Sect.\ 4 is devoted to results
and conclusions.

Throughout the paper we adopt the same cosmology parameters used in
\citet{capetti16}, i.e., $H_0=67.8 \, \rm km \, s^{-1} \, Mpc^{-1}$,
$\Omega_{\rm M}=0.308$, and $\Omega_\Lambda=0.692$ \citep{ade16}.

For our numerical results, we use c.g.s. units unless stated
otherwise. Spectral indices $\alpha$ are defined in the usual convention on
the flux density, $S_{\nu}\propto\,\nu^{-\alpha}$. The SDSS magnitudes are in
the AB system and are corrected for the Galactic extinction; {\em WISE}
magnitudes are instead in the Vega system and are not corrected for extinction
since, as shown by, for example, \citet{dabrusco14}, such correction affects
mostly the magnitude at 3.4 $\mu$m of sources lying at low Galactic latitudes
(and by less than $\sim$3\%).

\section{Sample selection}
\label{sample}

We searched for FR~II radio galaxies in the sample of 18,286 radio sources
built by Best \& Heckman (2012; hereafter the BH12 sample) by limiting our
search to the subsample of objects in which, according to these authors, the
radio emission is produced by an active nucleus.  They cross-matched the
optical spectroscopic catalogs produced by the group from the Max Planck
Institute for Astrophysics and Johns Hopkins University \citep{bri04,tre04}
based on data from the data release 7 of the Sloan Digital Sky Survey
(DR7/SDSS; \citealt{abazajian09}),\footnote{Available at {\tt
    http://www.mpa-garching.mpg.de/SDSS/}.} with the National Radio Astronomy
Observatory Very Large Array Sky Survey (NVSS; \citealt{condon98}) and the
Faint Images of the Radio Sky at Twenty centimeters survey (FIRST;
\citealt{becker95,helfand15}) adopting a radio flux density limit of 5 mJy in
the NVSS.  We focused on the sources with redshift $z < 0.15$.

We adopted a purely morphological classification based on the radio
structure shown by the FIRST images. We visually inspected the FIRST
images of each source and preserved those with an edge brightened
morphology in which at least one of the emission peaks lies at a
distance of at least 30 kpc from the center of the optical host. The
30 kpc radius corresponds to 11$\farcs$4 for the farthest objects; the
$z<0.15$ redshift limit ensures that all the selected sources are well
resolved with the 5$\arcsec$ resolution of the FIRST images. The three
authors performed this analysis independently and we included only the
sources for which a FR~II classification is proposed by at least two
of us. We allowed for the presence of diffuse emission leading to X-,
Z-, or C-shaped morphologies, but not extending at larger distances
with respect to the emission peaks, thus excluding wide angle tail
sources \citep{owen76}. Most of these sources are double, i.e., they
do not show nuclear radio emission; the lack of this precise position
reference requires a further check of the original optical
identifications. We discarded three objects in which the
identification of the host is not secure.

The resulting sample, to which we refer as \FRII, is formed by 122 FR~IIs
whose FIRST images are presented in the Appendix. Their main properties
are presented in Table \ref{tab}, where we report the SDSS name, redshift, and
NVSS 1.4 GHz flux density (from BH12).  The [O~III] line flux, the r-band SDSS
AB magnitude, $m_r$, the Dn(4000) index (see Section 3 for its definition),
and the stellar velocity dispersion $\sigma_*$ are instead from the MPA-JHU
DR7 release of spectrum measurements. The concentration index $C_r$ was
obtained for each source directly from the SDSS database. For sake of clarity,
uncertainties are not shown in the table; we estimated a median uncertainty of
0.09 on $C_r$, of 0.03 on Dn(4000), of 0.005 magnitudes on $m_r$, and of 10
\kms\ on $\sigma_*$. We also list the resulting radio and line luminosity and
the black hole masses estimated from the stellar velocity dispersion and the
relation $\sigma_* - M_{\rm BH}$ of \citet{tremaine02}. The uncertainty in the
$M_{\rm BH}$ value is dominated by the spread of the relation used (rather
than by the errors in the measurements of $\sigma_*$) resulting in an
uncertainty of a factor $\sim$ 2. Finally, we give the classification (from
BH12) into LEGs and HEGs based on the optical emission line ratios in their
SDSS spectra.

\section{\FRII\ hosts and radio properties}
\label{hosts}

\subsection{Hosts properties}

The majority (107) of the selected FR~IIs are classified as LEG, but there
are also 14 HEG and just one source that cannot be classified spectroscopically
because of the lack of emission lines, namely J1446+2142.

\begin{figure*}
\includegraphics[width=9.5cm]{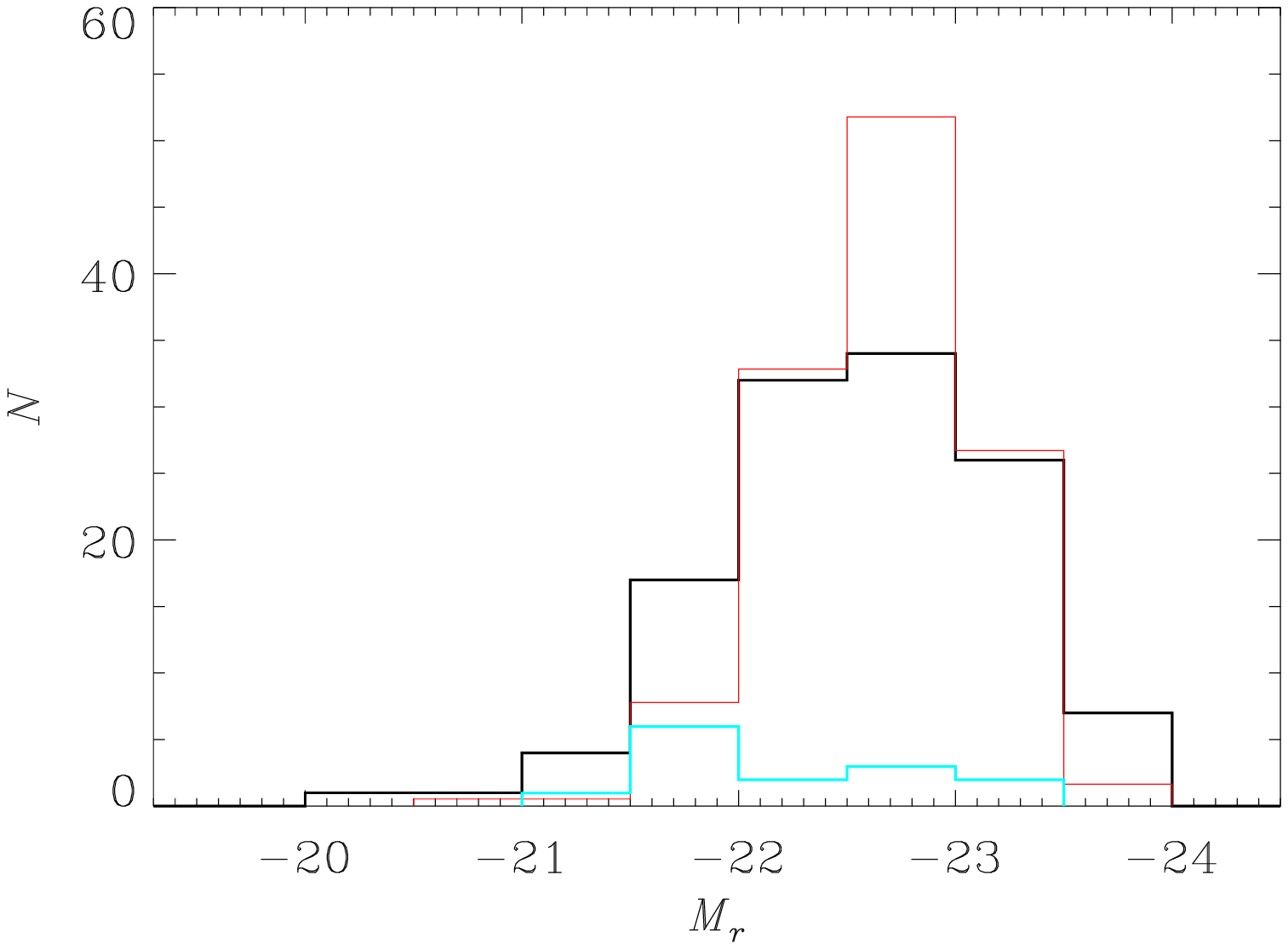}
\includegraphics[width=9.5cm]{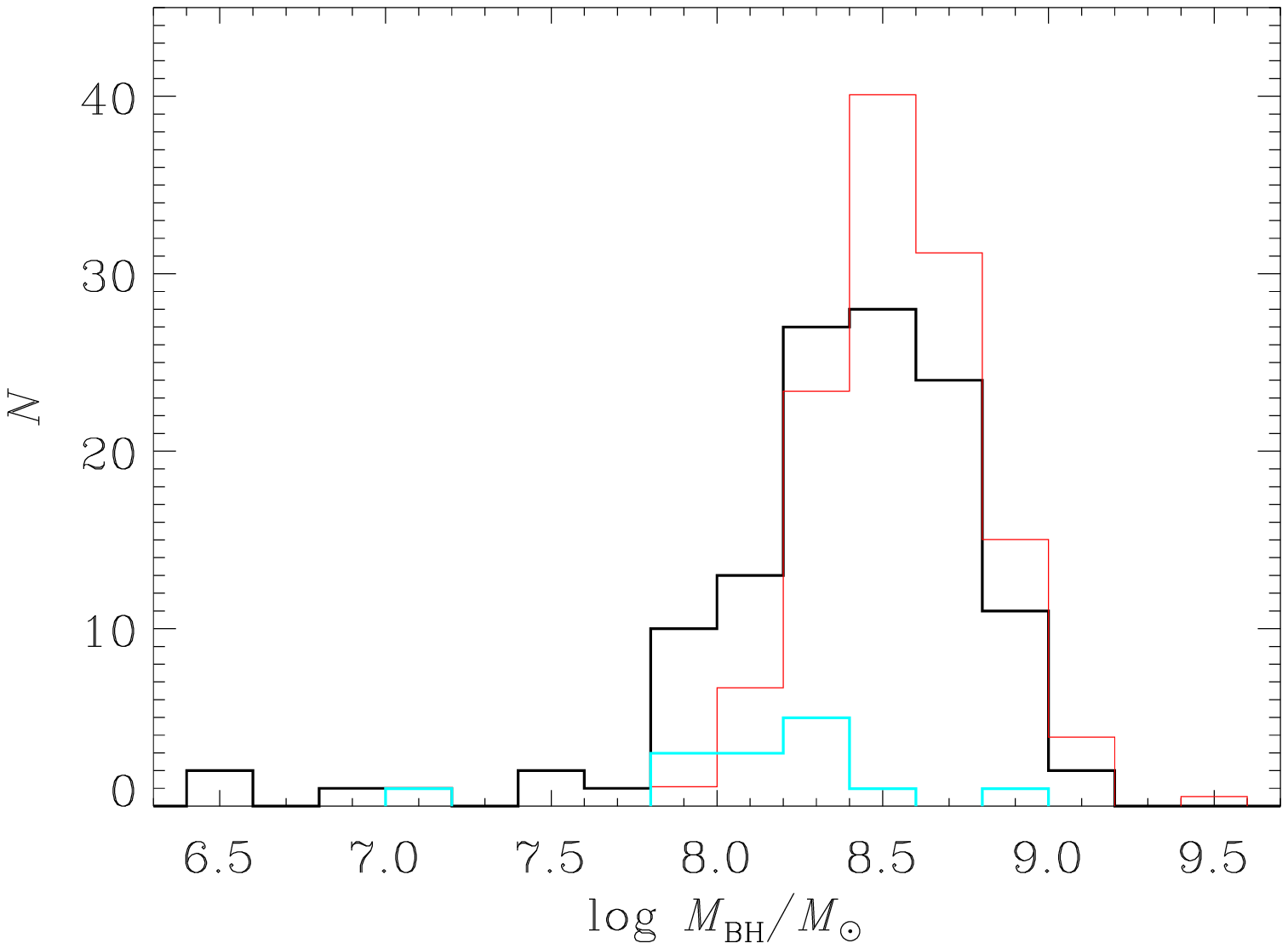}
\caption{Distributions of the $r$-band absolute magnitude (left) and black
  hole masses (right). The black histograms are for \FRII\ (cyan for the HEG
  \FRII\ subsample), the red histograms for the \FR. The \FR\ histograms are all scaled by the relative number of FR~I and FR~II, i.e., by 122/219.}
\label{mhist}
\end{figure*}

The distribution of absolute magnitude of the \FRII\ hosts covers the range
$-20 \gtrsim M_r \gtrsim -24$ (see Fig. \ref{mhist}, left panel). The
distribution of black hole masses (Fig. \ref{mhist}, right panel) is rather
broad. Most sources have $8.0 \lesssim \log M_{\rm BH} \lesssim 9.0 M_\odot$,
but a tail toward smaller values, down to $M_{\rm BH} \sim10^{6.5} M_\odot$,
that includes $\sim 15\%$ (13 LEGs and 4 HEGs) of the sample.

 The FR~II HEGs are less luminous, overall, and harbor less massive black holes
with respect to the FR~II LEGs; the medians of their distributions are
$<M_r({\rm HEG})> = -21.97 \pm 0.17$, $<M_r({\rm LEG})> = -22.62 \pm 0.06$,
$<\log M_{\rm BH}({\rm HEG})> = 8.21 \pm 0.11$, and $<\log M_{\rm BH}(LEG)> =
8.46 \pm 0.04$, respectively. The comparison between the \FRII\ LEG sources
and the \FR\ hosts ($< M_r({\rm FR~I}) > = -22.69 \pm 0.03$ and $<\log M_{\rm
  BH}({\rm FR~I})> = 8.55 \pm 0.02$) indicates that only marginal differences
(and not statistically significant) are present between the medians of these
distributions.

The Dn(4000) spectroscopic index, defined according to
\citet{balogh99} as the ratio between the flux density measured on the
red side and blue side of the Ca~II break, is an indicator of the
presence of young stars or of nonstellar emission. Low redshift
($z < 0.1$) red galaxies have Dn(4000)$= 1.98 \pm 0.05$, which is a
value that decreases to $= 1.95 \pm 0.05$ for $0.1 < z < 0.15$
galaxies \citep{capetti15}.

The concentration index $C_r$, which is defined as the ratio of the
radii including 90\% and 50\% of the light in the $r$ band, can be
used for a morphological classification of galaxies, in which
early-type galaxies have higher values of $C_r$ than late-type
galaxies. Two thresholds have been suggested to define ETGs: a more
conservative value at $C_r \gtrsim$ 2.86 \citep{nakamura03,shen03} and
a more relaxed selection at $C_r \gtrsim$ 2.6
\citep{strateva01,kauffmann03b,bell03}. \citet{bernardi10} found that
the second threshold of the concentration index corresponds to a mix
of E+S0+Sa types, while the first mainly selects ellipticals galaxies,
removing the majority of Sas, but also some Es and S0s.

\begin{figure*}
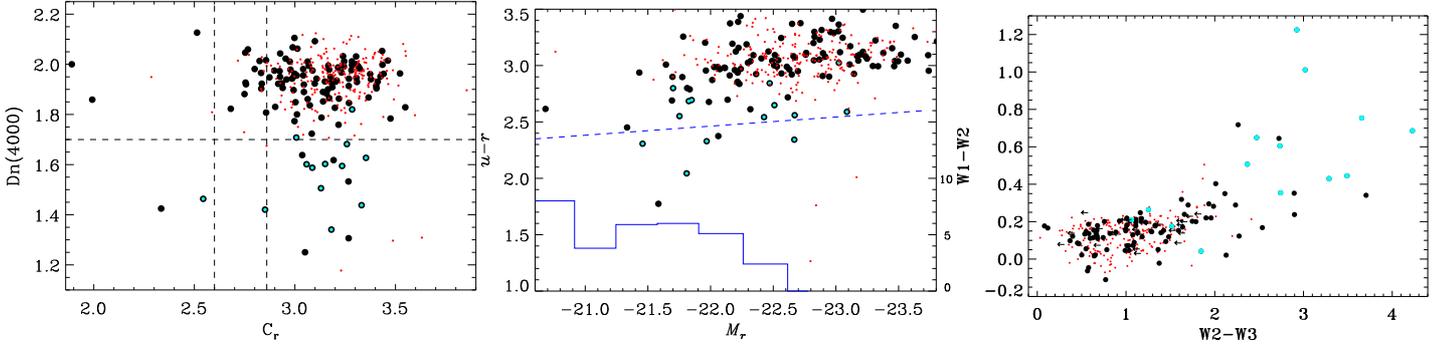

\includegraphics[width=6.2cm]{crdn.epsi}
\includegraphics[width=6.2cm]{ur.epsi} 
\includegraphics[width=6.2cm]{wise.epsi} 
\caption{Left: concentration index $C_r$ vs. Dn(4000) index for the \FRII\ and
  \FR\ samples indicated by black and red dots, respectively. The HEG FR~II are
  represented by cyan circles. Center: absolute $r$-band magnitude, $M_r$,
  vs. $u-r$ color. Right: {\em{WISE}} mid-IR colors.}
\label{crdn}
\end{figure*}

In Fig. \ref{crdn} we show the concentration index $C_r$ versus the Dn(4000)
index (left panel) for the \FRII\ sources. More than $\sim$90\% of the LEG
FR~II hosts lie in the region of high $C_r$ and Dn(4000) values, indicating
that they are red ETGs. The HEG FR~II are still ETGs, but they show generally
lower values of Dn(4000).

We also consider the $u-r$ color of the galaxies, obtained from SDSS
  imaging and thus referred to the whole galaxy rather than just the
3\arcsec\ circular region covered by the SDSS spectroscopic aperture. In
Fig.~\ref{crdn} (central panel) we show the $u-r$ color versus the absolute
r-band magnitude $M_r$ of the hosts. As already found by considering the
Dn(4000) index, the FR~II HEGs show bluer color than the FR~II LEGs; in this
latter class, only two sources have a $u-r$ color that is smaller than the threshold
separating red and blue ETGs \citep{schawinski09}.

In Fig. \ref{crdn}, right panel, we show the comparison of the
{\em{WISE}} mid-IR colors of \FRII\ and \FR\ sources; the associations
with the {\em{WISE}} catalog are computed by adopting a 3\farcs3
search radius \citep{dabrusco13}. All but one of the (J1121+5344)
\FRII\ sources have a {\em{WISE}} counterpart, but 25 of these sources
are undetected in the $W3$ band. The LEG FR~IIs have mid-IR colors
similar to those of the \FR\ sources; their mid-IR emission is
dominated by their host galaxies since they fall in the same region of
elliptical galaxies \citep{wright10}. Only five LEGs have
$W2-W3 > 2.5$, exceeding the highest value measured for FR~Is.

Conversely, HEGs reach mid-IR colors as high as $W2-W3$=4.3, colors similar to
those of Seyfert and starburst galaxies (e.g., \citealt{stern05}). Their red
colors are likely due to a combination of star-forming regions and/or emission
from hot dust within a circumnuclear dusty torus.

Overall, we found 10 LEG FR~IIs whose properties do not conform with
the general behavior of their class, for example, showing blue colors
or being associated with small black holes. In some cases, this is due
to relatively large errors particularly in the measurement of
$\sigma_*$, a possible uncertain identification of their spectroscopic
class, or a substantial contribution from a bright nonthermal
nucleus. However, there are three objects (namely J0755+5204,
J1158+3006, and J1226+2538) for which we obtain estimates of the black
hole mass of $\log M_{\rm BH} \sim 6.5 - 6.8$; based on their $C_r$
value these objects are late-type galaxies and two of them also show
blue optical colors (and red mid-IR colors). These properties are all
typical of radio quiet AGN. This contrasts with the observed radio
power ($\log \nu L_r \sim 40.5 - 40.9$) and morphology.

\subsection{Radio properties}

\begin{figure*}
\includegraphics[width=9.5cm]{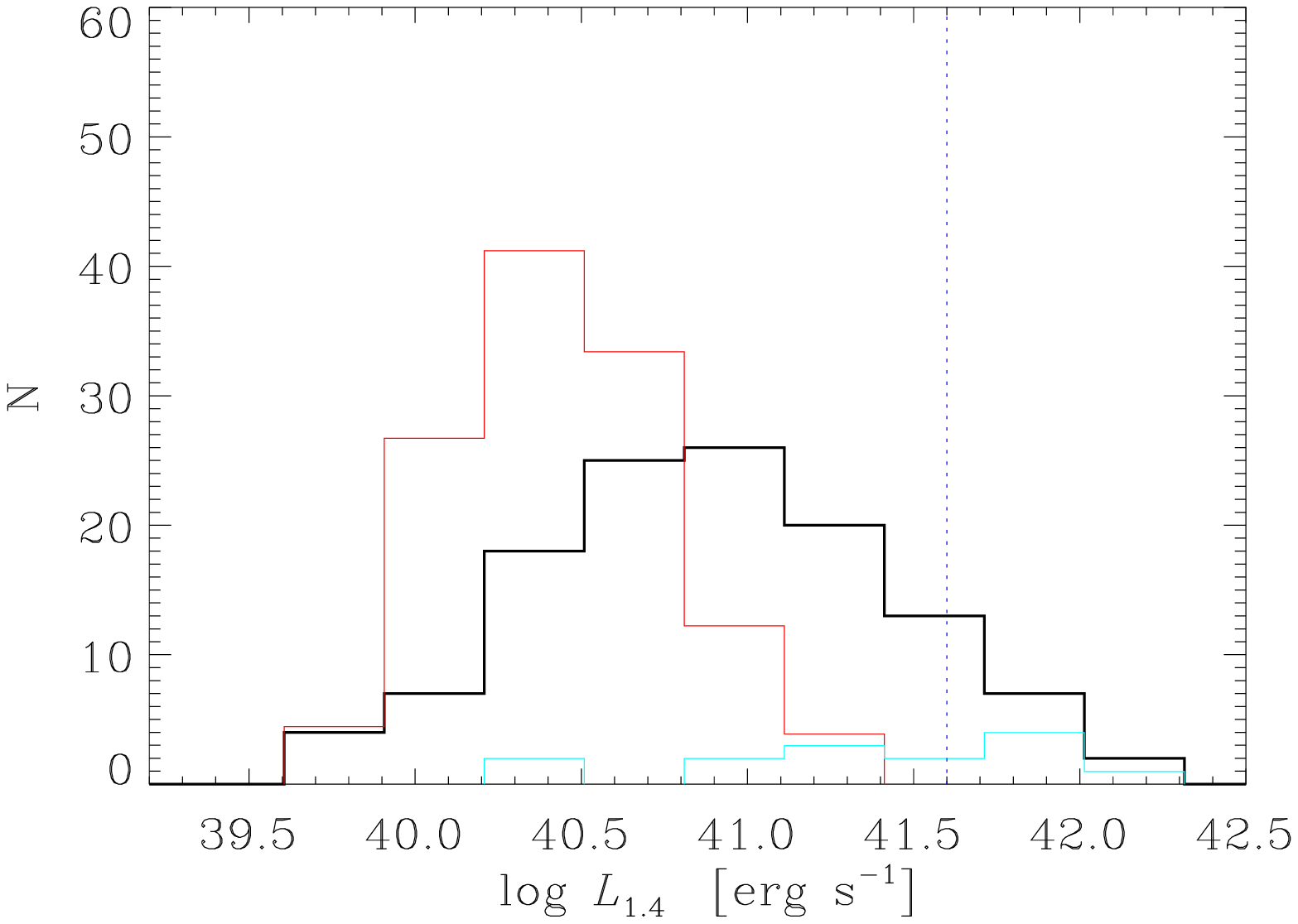}
\includegraphics[width=9.5cm]{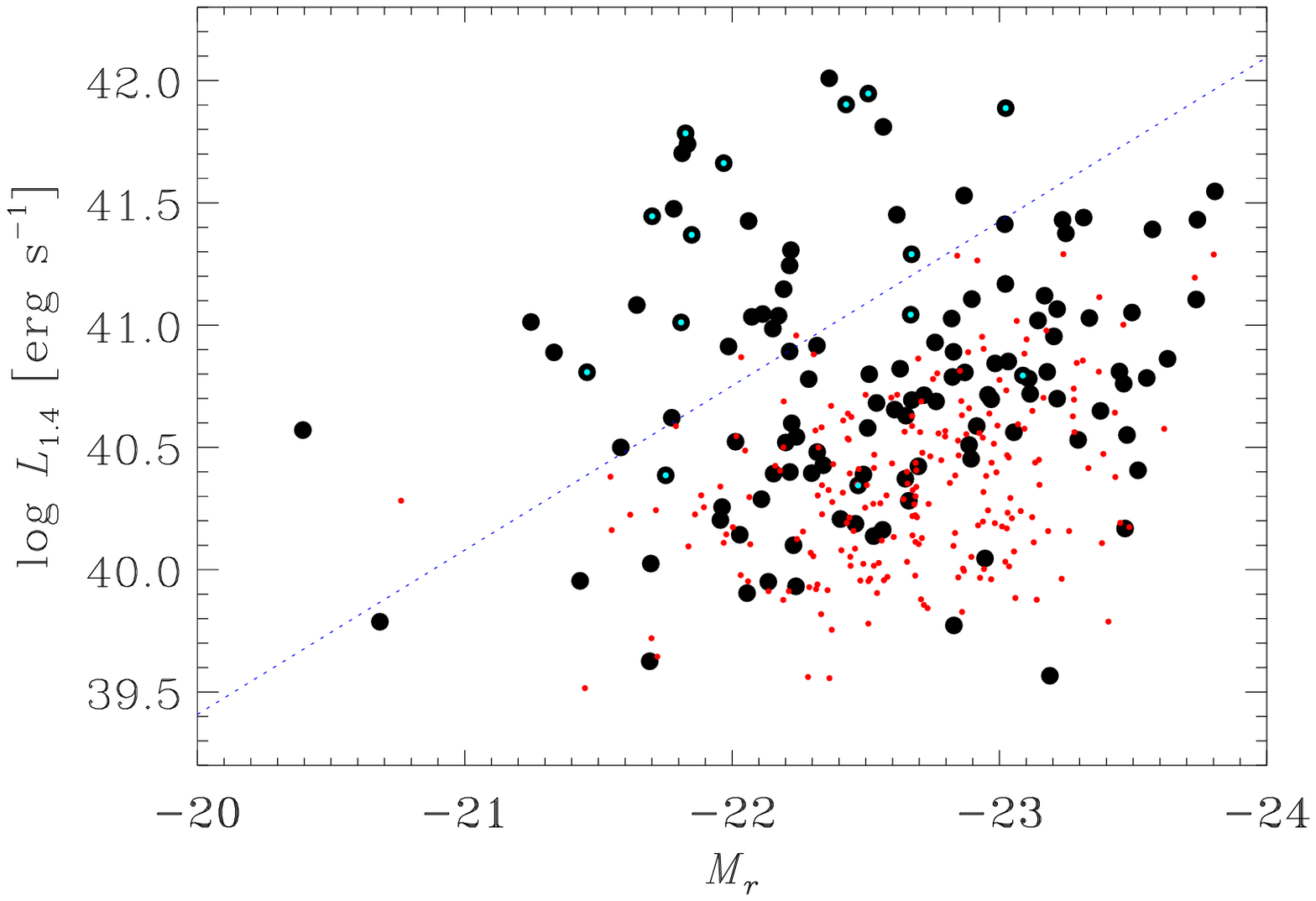}
\caption{Left panel: radio luminosity distribution of the \FRII\ (black, cyan
  for the HEGs) and \FR\ sources (red). The dotted vertical line indicates the
  transition power between FR~I and FR~II reported by \citet{fanaroff74}.
  Right panel: radio luminosity (NVSS) vs. host absolute magnitude, $M_r$, for
  \FRII\ and \FR\ (black and red, respectively; cyan for the HEG FR~II). The
  dotted line shows the separation between FR~I and FR~II reported by
  \citet{ledlow96} to which we applied a correction of 0.34 mag to account for
  the different magnitude definition and the color transformation between the
  SDSS and Cousin systems.}
\label{lr}
\end{figure*}

The radio luminosity at 1.4 GHz of the \FRII\ covers the range $L_{1.4}$ =
$\nu_{\rm r} l_{\rm r}$ $\sim 10^{39.5} - 10^{42.5}$ $\ergs$ (Fig. \ref{lr},
left panel), reaching a radio power almost two orders of magnitude lower than
the FR~IIs in the 3C sample. The HEGs are brighter than LEGs (with median of $\log
L_{1.4} = 41.37$ and 40.76, respectively) and LEGs are brighter than the
\FR\ sources by a factor $\sim$3; 90\% of the \FRII\ fall below the
separation between FR~Is and FR~IIs originally reported by \citet{fanaroff74}
which translates, with our adopted cosmology and by assuming a spectral index
of 0.7 between 178 MHz and 1.4 GHz, into $L_{1.4} \sim 10^{41.6}$ $\ergs$.
Similarly, we find that $\sim$75\% of the \FRII\ sources (and including also
four HEGs) are located {\sl below} the dividing line in the optical-radio
luminosity plane defined by \citet{ledlow96};\footnote{We shifted the dividing
  line to the right of the diagram to include a correction of 0.12 mag to
  scale our total host magnitude to the M$_{24.5}$ used by these authors, and
  an additional 0.22 mag to convert the Cousin system into the SDSS system
  \citep{fukugita96}.}  see Fig. 3, right panel.

\begin{figure}
\includegraphics[width=9.5cm]{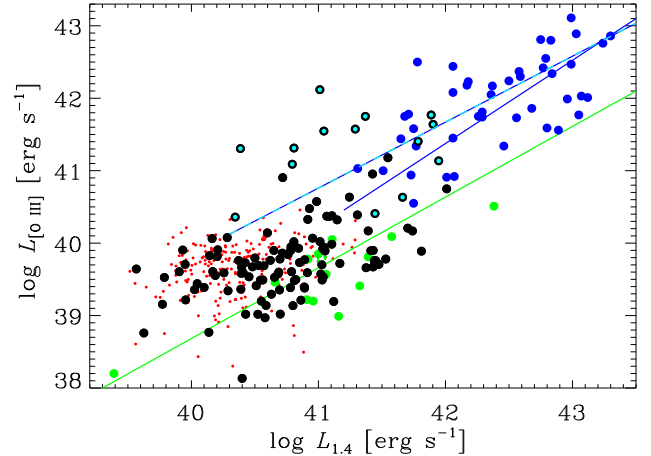}
\caption{Radio (NVSS) vs. [O~III] line luminosity of the \FRII\ (black),
  \FR\ (red), 3C-FR~I (green), and 3C-FR~II samples (blue) The green line
  (blue) shows the linear correlation between these two quantities derived
  from the FR~Is (FR~IIs) of the 3C sample from \citep{buttiglione10}.
    The dashed blue-cyan line is instead the linear fit on both the 3C 
    and the \FRII\ HEGs.}
\label{lrlo3}
\end{figure}

FR~IIs show a large spread in both radio and [O~III] line luminosities
(see Fig. \ref{lrlo3}). In this plane, the FR~II LEGs cover
essentially the same region of the FR~Is with just a tail toward
higher power both in line and in radio; no correlation is seen between
these two quantities. The FR~II HEGs generally have higher ratios
between $L_{\rm[O~III]}$ and $L_{1.4}$ than LEGs, which is an effect
already found in the 3C sample \citep{buttiglione10}. The HEGs in
\FRII\ are mostly located above the correlation defined by the 3C
HEGs. A linear fit including both samples is indeed shallower (with a
slope of 0.91) than that obtained from the 3C sources alone (whose
slope is 1.15).

\subsection{Comparison with previous works}

As discussed in Sect. 2, we decided to maintain the traditional
  morphological visual classification into edge-brightened FR~IIs and
  edge-darkened FR~Is rather than adopting quantitative methods such as those
  used by \citet{lin10}. The comparison of our classification with that
  proposed by these authors indicates that, among the 96 sources in common,
  $\sim 80\%$ of the \FRII\ sources are classified as class {\sl a} in their
  nomenclature, i.e., sources with two hot spots on either side of the
  galaxy. Most of the remaining objects fall in class {\sl b}, in which the
  emission peak is coincident with the host galaxy; however, the inspection of
  these FR~II radio sources does not show any clear distinguishing feature
  from those in the main class, other than having a relatively brighter
  central source in addition to the two lobes.

  The main drawback of our scheme, based on the subjective visual
  inspection of radio images, is the relatively high fraction of
  sources of uncertain classification. The rather strict criteria
  adopted for a positive classification as FR~I in \citet{capetti16}
  and here for the FR~II enabled us to select only 219 FR~Is and 122
  FR~II; more than half of the 714 radio galaxies extended more than
  30 kpc cannot be allocated to any FR class. On the other hand, this
  strategy allows us to select samples that are very uniform from a
  morphological point of view and that are optimally suited for our
  main purpose, i.e., the comparison of the properties of the two
  classes.
 
More recently, \citet{miraghaei17} performed an analysis on the same initial
sample, with an apparently similar selection strategy based on visual
inspection. However, the resulting sample of both FR~I and FR~II differ
significantly from those we obtained with only $\sim$ 25\% of objects in
common for both classes, even restricting the comparison to the same range of
redshift, $0.03 < z < 0.15$. This mismatch is likely due to the different
requirements (based, e.g., on linear instead of angular sizes and different
radio flux limits); most importantly they considered only sources
corresponding to multicomponents objects in FIRST and this rejects most of
the edge darkened sources we included in \FR. Overall, their results do not
strongly differ from ours, probably because (leaving aside the HEGs) the
properties of low z radio AGN are very homogeneous regardless of their radio
morphology. However, for example, we do not find significative differences in
the $C_r$ values between FRI and FRII hosts.

\section{Discussion and conclusions}

The properties of the \FRII\ sources differ between those
spectroscopically identified as LEGs or HEGs. The HEGs have lower
optical luminosities, smaller black hole masses, and higher radio
luminosities with respect to LEGs, although a substantial overlap
between the two classes exist for all these quantities. The clearest
differences are related to the ratio between line and radio
luminosities and to their colors; HEGs are bluer in the optical and
redder in the mid-IR. These results confirm the conclusions of
previous studies (e.g.,
\citealt{baldi08,buttiglione10,baldi10b,best12}).

The population of the LEG FR~IIs included in the \FRII\ is remarkably
uniform. They are all luminous red ETGs with large black hole masses ($M_{\rm
  BH} \gtrsim 10^8 M_\odot$); only $\sim$ 10\% of the LEG\ FR\ IIs depart from
this general description. All these properties are shared with the hosts of
the \FR\ sources. The distributions of $M_{\rm BH}$ and $M_{r}$ differ with a
statistical significance higher than 95\% according to the Kolmogoroff-Smirnov
test; this is due to the presence of a tail of low $M_{\rm BH}$ sources,
reaching value as low as 3$\times 10^6 M_\odot$.  However, the median of
$M_{\rm BH}$ and $M_{r}$ differ only marginally by less than 0.1 dex. Even
the median radio luminosity of LEG FR~II is just a factor $\sim$3 higher than
that measured in FR~I. Apparently, the difference in radio morphology between
edge-brightened and edge-darkened radio sources does not translate into a
clear separation between the nuclear and host properties, while the
spectroscopic classes, LEG and HEG, do.

The \FRII\ sample unveils a population of FR~IIs of much lower radio
power with respect to those obtained at high radio flux thresholds,
extending it downward by two orders of magnitude. The correspondence
of the morphological classification of FR~I and FR~II with a
separation in radio power that is observed, for example, in the 3C
sample, disappears.  This conclusion is in line with previous results
\citep{best09,lin10,wing11}. A radio source produced by a low power
jet can be edge brightened or edge darkened and the outcome is not due
to differences in the optical properties of the host galaxy.

Nonetheless, \citet{capetti16} find that the connection between radio
morphology and host properties is preserved in FR~Is; there is a
well-defined threshold of radio power above which an edge darkened
radio source does not form and this limit has a strong positive
dependence on the host luminosity. This effect was originally seen by
\citet{ledlow96} but partly lost in subsequent studies; we believe
that we recover it because of the stricter criteria we adopted for the
selection of FR~Is.

It can be envisaged that brighter galaxies are associated with denser and more
extended hot coronae that are able to disrupt more powerful jets.  But the
large population of low power FR~IIs indicates that the situation is more
complex; there is a large overlap of radio power between FR~Is and FR~IIs and
radio power is believed to be a robust proxy for the jet power (e.g.,
\citealt{willott99,birzan04}). Apparently, jets of the same power expanding in
similar galaxies can form both FR~I and FR~II. This indicates that the optical
properties of the host and radio luminosity are not the only parameters
driving the evolution of low power radio sources. Further studies of, for example,
the X-rays properties and the larger scale environment are needed to clarify
this issue.

\bibliographystyle{aa} 

\begin{acknowledgements}

F.M. gratefully acknowledges the financial support of the Programma Giovani
Ricercatori -- Rita Levi Montalcini -- Rientro dei Cervelli (2012) awarded by
the Italian Ministry of Education, Universities and Research (MIUR).

Part of this work is based on the NVSS (NRAO VLA Sky Survey): The National
Radio Astronomy Observatory is operated by Associated Universities, Inc.,
under contract with the National Science Foundation.

This publication makes use of data products from the Wide-field Infrared
Survey Explorer, which is a joint project of the University of California, Los
Angeles, and the Jet Propulsion Laboratory/California Institute of Technology,
funded by the National Aeronautics and Space Administration.

This research made use of the NASA/ IPAC Infrared Science Archive and
Extragalactic Database (NED), which are operated by the Jet Propulsion
Laboratory, California Institute of Technology, under contract with the
National Aeronautics and Space Administration.

We acknowledge the usage of the HyperLeda database (http://leda.univ-lyon1.fr).

Funding for SDSS-III has been provided by the Alfred P. Sloan Foundation, the
Participating Institutions, the National Science Foundation, and the
U.S. Department of Energy Office of Science. The SDSS-III website is
http://www.sdss3.org/.  SDSS-III is managed by the Astrophysical Research
Consortium for the Participating Institutions of the SDSS-III Collaboration,
including the University of Arizona, the Brazilian Participation Group,
Brookhaven National Laboratory, University of Cambridge, Carnegie Mellon
University, University of Florida, the French Participation Group, the German
Participation Group, Harvard University, the Instituto de Astrofisica de
Canarias, the Michigan State/Notre Dame/JINA Participation Group, Johns
Hopkins University, Lawrence Berkeley National Laboratory, Max Planck
Institute for Astrophysics, Max Planck Institute for Extraterrestrial Physics,
New Mexico State University, New York University, Ohio State University,
Pennsylvania State University, University of Portsmouth, Princeton University,
the Spanish Participation Group, University of Tokyo, University of Utah,
Vanderbilt University, University of Virginia, University of Washington, and
Yale University.

\end{acknowledgements}

\newpage
\clearpage
\onecolumn
\begin{center}
\begin{longtable}{l l r r l l r l l c r r r}

\caption[Properties of the \FRII\ sources.]{Properties of the \FRII\ sources.} 
\label{tab} \\

\hline \hline 
 &  \,\,\,\,z & NVSS & [O~III] & \,\,\,\,\,m$_{\rm r}$ & \,\,Dn & \,\,\,$\sigma_*$ & \,\,\,C$_{\rm r}$ & \,\,\,\, $\nu L_r$ &  $L_{\rm[O~III]}$ & $M_{\rm BH}$ & Class \\
\hline  
\endfirsthead

\multicolumn{3}{c}{{\tablename} \thetable{} -- Continued} \\[0.5ex]
\hline \hline 
& \,\,\,\,\,z & NVSS & [O~III] & \,\,\,\,\,m$_{\rm r}$ & \,\,Dn & \,\,\,$\sigma_*$ & \,\,\,C$_{\rm r}$ & \,\,\,\, $\nu L_r$ &  $L_{\rm[O~III]}$ & $M_{\rm BH}$ & Class \\
\hline
\endhead

\hline
  \multicolumn{10}{c}{{Continued on Next Page}} \\
\endfoot

  \\[-1.8ex] 
\endlastfoot

SDSS~J001247.57+004715.8 & 0.148 &   58.6 &   11.5 & 16.352 &   2.07 & 263 &   2.99 &  40.72 &  39.86 &   8.6 & LEG  \\
SDSS~J002107.62$-$005531.4 & 0.108 &111.9 &    3.3 & 15.351 &   1.99 & 250 &   3.17 &  40.70 &  39.02 &   8.5 & LEG  \\
SDSS~J004326.80$-$105421.8 & 0.127 & 82.3 &  176.6 & 15.834 &   1.92 & 247 &   2.92 &  40.72 &  40.90 &   8.5 & LEG  \\
SDSS~J004404.66+010152.9 & 0.112 &  366.0 &  126.1 & 16.426 &   1.78 & 219 &   3.48 &  41.24 &  40.64 &   8.3 & LEG  \\
SDSS~J011830.65$-$104356.4 & 0.126 & 56.3 &    7.0 & 16.681 &   1.83 & 224 &   2.90 &  40.54 &  39.49 &   8.3 & LEG  \\
SDSS~J024558.54$-$064900.6 & 0.139 &102.0 &    4.4 & 16.318 &   1.86 & 261 &   3.06 &  40.89 &  39.38 &   8.6 & LEG  \\
SDSS~J075221.83+333348.9 & 0.140 &   66.0 &    --- & 16.458 &   1.95 & 269 &   3.31 &  40.71 &    --- &   8.6 & LEG  \\
SDSS~J075529.95+520450.6 & 0.140 &   99.4 &    4.5 & 17.835 &   1.42 &  81 &   2.33 &  40.89 &  39.39 &   6.6 & LEG  \\
SDSS~J075628.78+501716.3 & 0.134 &  165.0 &   46.8 & 15.844 &   1.93 & 296 &   2.75 &  41.07 &  40.37 &   8.8 & LEG  \\
SDSS~J080107.19+175849.7 & 0.140 &  847.0 &    9.7 & 17.417 &   1.57 & 161 &   2.85 &  41.82 &  39.73 &   7.7 & LEG  \\
SDSS~J080158.09+175152.6 & 0.147 &   16.9 &   18.8 & 16.715 &   1.99 & 258 &   3.24 &  40.16 &  40.06 &   8.6 & LEG  \\
SDSS~J080535.00+240950.3 & 0.060 & 4781.0 & 2791.1 & 15.376 &   1.68 & 205 &   3.26 &  41.78 &  41.40 &   8.2 & HEG  \\
SDSS~J081343.61+525738.2 & 0.138 &  194.6 &    9.7 & 16.116 &   1.98 & 250 &   2.80 &  41.17 &  39.72 &   8.5 & LEG  \\
SDSS~J081512.33+384045.4 & 0.125 &  487.8 &   11.6 & 17.127 &   1.88 & 333 &   2.75 &  41.48 &  39.71 &   9.0 & LEG  \\
SDSS~J081734.46+445850.8 & 0.142 &  135.6 &    6.0 & 17.032 &   1.98 & 234 &   3.28 &  41.04 &  39.54 &   8.4 & LEG  \\
SDSS~J082247.63+041750.0 & 0.095 &  132.0 &   13.7 & 15.666 &   1.94 & 248 &   3.27 &  40.65 &  39.53 &   8.5 & LEG  \\
SDSS~J083109.54+414738.1 & 0.132 &  155.0 &   10.2 & 16.211 &   2.01 & 236 &   3.28 &  41.03 &  39.70 &   8.4 & LEG  \\
SDSS~J083112.15+434158.4 & 0.111 &  121.0 &   20.4 & 15.170 &   1.84 & 275 &   3.07 &  40.76 &  39.84 &   8.7 & LEG  \\
SDSS~J084529.04+421952.9 & 0.133 &  201.0 &   42.0 & 16.857 &   1.79 & 230 &   3.13 &  41.15 &  40.32 &   8.4 & LEG  \\
SDSS~J084759.04+314708.3 & 0.067 & 1647.0 &   67.7 & 13.735 &   2.02 & 338 &   2.95 &  41.43 &  39.90 &   9.0 & LEG  \\
SDSS~J090150.32+555527.4 & 0.141 &  638.0 &   28.5 & 17.372 &   1.85 & 191 &   3.00 &  41.70 &  40.21 &   8.1 & LEG  \\
SDSS~J091153.61+372413.3 & 0.104 &  596.4 &   50.0 & 14.908 &   1.95 & 301 &   2.87 &  41.39 &  40.17 &   8.8 & LEG  \\
SDSS~J091225.08+534139.1 & 0.101 &  204.1 &   21.6 & 16.181 &   1.77 & 183 &   3.00 &  40.89 &  39.77 &   8.0 & LEG  \\
SDSS~J091445.53+413714.2 & 0.140 &  449.0 &  269.1 & 15.371 &   1.82 & 300 &   2.68 &  41.55 &  41.18 &   8.8 & LEG  \\
SDSS~J091948.45+575055.9 & 0.137 &   11.5 &   15.1 & 16.879 &   1.94 & 261 &   3.21 &  39.93 &  39.90 &   8.6 & LEG  \\
SDSS~J092611.26+170357.9 & 0.116 &  163.3 &   80.6 & 15.972 &   1.93 & 233 &   3.44 &  40.93 &  40.48 &   8.4 & LEG  \\
SDSS~J093215.91+180419.9 & 0.147 &   57.0 &    9.7 & 16.324 &   2.00 & 271 &   3.41 &  40.70 &  39.78 &   8.7 & LEG  \\
SDSS~J093641.99+111350.9 & 0.120 &   48.3 &    2.6 & 16.457 &   1.96 & 298 &   3.44 &  40.43 &  39.02 &   8.8 & LEG  \\
SDSS~J094124.02+394441.8 & 0.108 & 1811.0 & 1382.7 & 16.125 &   1.60 & 209 &   3.06 &  41.90 &  41.64 &   8.2 & HEG  \\
SDSS~J094201.76+084736.8 & 0.134 &   97.6 &   16.6 & 16.092 &   1.93 & 292 &   3.15 &  40.84 &  39.93 &   8.8 & LEG  \\
SDSS~J094703.01+231614.2 & 0.084 &   30.8 &   21.7 & 15.923 &   1.93 & 234 &   3.18 &  39.90 &  39.61 &   8.4 & LEG  \\
SDSS~J095640.77$-$000124.0 & 0.139 &166.0 &   43.4 & 16.259 &   2.04 & 265 &   2.93 &  41.11 &  40.38 &   8.6 & LEG  \\
SDSS~J101558.24+404647.2 & 0.128 & 1009.0 &   16.9 & 16.392 &   1.91 & 236 &   3.40 &  41.81 &  39.89 &   8.4 & LEG  \\
SDSS~J101954.48+393022.8 & 0.112 &   52.0 &    0.4 & 16.433 &   1.86 & 227 &   3.22 &  40.40 &  38.13 &   8.4 & LEG  \\
SDSS~J102150.38+080833.8 & 0.103 &  585.7 &   15.8 & 15.211 &   1.90 & 253 &   2.99 &  41.38 &  39.66 &   8.5 & LEG  \\
SDSS~J102156.67+144331.4 & 0.111 &  412.0 & 1110.3 & 15.956 &   1.60 & 193 &   3.05 &  41.29 &  41.57 &   8.1 & HEG  \\
SDSS~J102237.40+383444.9 & 0.052 &  117.4 &   40.8 & 13.935 &   1.96 & 268 &   3.01 &  40.05 &  39.44 &   8.6 & LEG  \\
SDSS~J103128.22+084324.1 & 0.141 &  347.7 &   10.2 & 15.871 &   1.97 & 229 &   2.92 &  41.44 &  39.76 &   8.4 & LEG  \\
SDSS~J103443.26+053319.8 & 0.141 &  167.1 &    2.8 & 16.015 &   2.03 & 279 &   3.30 &  41.12 &  39.19 &   8.7 & LEG  \\
SDSS~J103602.94+525936.1 & 0.142 &   30.1 &  350.0 & 17.456 &   1.59 & 170 &   3.09 &  40.39 &  41.31 &   7.8 & HEG  \\
SDSS~J104502.74+471759.1 & 0.145 &   76.2 &   17.4 & 16.381 &   1.98 & 289 &   3.32 &  40.81 &  40.02 &   8.8 & LEG  \\
SDSS~J104632.21+543559.7 & 0.145 &  279.0 &  938.6 & 17.400 &   1.44 & 302 &   3.33 &  41.37 &  41.75 &   8.8 & HEG  \\
SDSS~J104742.83+434652.7 & 0.086 &  104.0 &   17.6 & 15.137 &   2.01 & 230 &   2.97 &  40.45 &  39.54 &   8.4 & LEG  \\
SDSS~J105052.47+400050.7 & 0.129 &  147.0 &   79.7 & 16.833 &   1.83 & 204 &   3.20 &  40.99 &  40.57 &   8.2 & LEG  \\
SDSS~J105439.55+060630.7 & 0.135 &   58.7 &    3.8 & 16.442 &   1.94 & 226 &   3.38 &  40.63 &  39.29 &   8.3 & LEG  \\
SDSS~J110035.08+253911.0 & 0.145 &  242.0 &   41.0 & 17.027 &   1.95 & 228 &   3.20 &  41.31 &  40.39 &   8.4 & LEG  \\
SDSS~J110151.89+164038.6 & 0.069 &  699.0 &  188.9 & 15.891 &   1.93 & 226 &   3.23 &  41.08 &  40.37 &   8.3 & LEG  \\
SDSS~J110214.15+242954.7 & 0.144 &  108.0 &   19.7 & 16.038 &   1.94 & 282 &   3.31 &  40.95 &  40.07 &   8.7 & LEG  \\
SDSS~J110214.91+234111.9 & 0.146 &   29.0 &    6.3 & 17.111 &   1.90 & 184 &   3.11 &  40.39 &  39.58 &   8.0 & LEG  \\
SDSS~J110215.68+290725.2 & 0.106 &  145.6 &  402.7 & 15.429 &   1.51 & 219 &   3.08 &  40.79 &  41.09 &   8.3 & HEG  \\
SDSS~J110253.95+125904.0 & 0.140 &   82.0 &   16.6 & 15.999 &   1.96 & 253 &   3.04 &  40.81 &  39.97 &   8.5 & LEG  \\
SDSS~J111317.91+412429.3 & 0.095 &  115.2 &    5.8 & 15.340 &   1.94 & 202 &   2.99 &  40.59 &  39.14 &   8.1 & LEG  \\
SDSS~J112126.44+534456.7 & 0.104 &   83.0 &   16.8 & 15.175 &   1.90 & 215 &   2.83 &  40.53 &  39.69 &   8.3 & LEG  \\
SDSS~J113133.57+604727.9 & 0.145 &   19.0 &   10.9 & 17.294 &   1.91 & 190 &   2.94 &  40.20 &  39.81 &   8.0 & LEG  \\
SDSS~J113626.35+501323.7 & 0.054 &   41.0 &    7.8 & 15.280 &   1.91 & 173 &   3.19 &  39.63 &  38.76 &   7.9 & LEG  \\
SDSS~J114427.19+370831.8 & 0.115 & 2012.0 &  154.7 & 16.341 &   1.87 & 225 &   3.52 &  42.01 &  40.75 &   8.3 & LEG  \\
SDSS~J114428.11+035815.7 & 0.127 &   75.6 &    8.3 & 16.408 &   1.85 & 248 &   3.14 &  40.68 &  39.57 &   8.5 & LEG  \\
SDSS~J114432.99+213217.0 & 0.109 &   35.3 &   24.8 & 16.183 &   1.97 & 263 &   3.21 &  40.21 &  39.91 &   8.6 & LEG  \\
SDSS~J114525.98$-$022332.9 & 0.128 & 46.8 &   10.2 & 16.648 &   1.92 & 248 &   2.94 &  40.48 &  39.67 &   8.5 & LEG  \\
SDSS~J114948.11+435412.6 & 0.071 &   58.0 &   17.5 & 15.898 &   1.96 & 229 &   3.37 &  40.02 &  39.36 &   8.4 & LEG  \\
SDSS~J115358.87+093929.8 & 0.103 &  672.0 &   16.4 & 15.211 &   1.99 & 267 &   3.07 &  41.43 &  39.67 &   8.6 & LEG  \\
SDSS~J115525.19+253222.4 & 0.137 &   26.1 &    4.1 & 17.009 &   1.99 & 263 &   3.15 &  40.29 &  39.34 &   8.6 & LEG  \\
SDSS~J115812.96+300625.9 & 0.139 &  106.7 &    7.1 & 17.165 &   1.72 &  96 &   1.99 &  40.91 &  39.59 &   6.8 & LEG  \\
SDSS~J115820.13+262112.0 & 0.112 & 1142.0 &   42.8 & 16.813 &   1.86 & 198 &   3.18 &  41.74 &  40.17 &   8.1 & LEG  \\
SDSS~J120732.92+335240.1 & 0.079 &  480.7 & 2148.6 & 15.175 &   1.34 & 192 &   2.75 &  41.04 &  41.55 &   8.1 & HEG  \\
SDSS~J122052.45+313308.4 & 0.104 &   60.3 &   15.8 & 16.179 &   1.96 & 241 &   3.18 &  40.39 &  39.67 &   8.5 & LEG  \\
SDSS~J122640.22+253855.5 & 0.134 &   44.3 &   12.2 & 17.490 &   1.25 &  76 &   1.89 &  40.50 &  39.80 &   6.5 & LEG  \\
SDSS~J124021.98+465636.3 & 0.141 &  135.5 &   14.5 & 17.121 &   1.85 & 175 &   3.27 &  41.03 &  39.92 &   7.9 & LEG  \\
SDSS~J125222.61+031554.0 & 0.099 &  919.0 &   22.9 & 15.489 &   1.98 & 246 &   3.04 &  41.53 &  39.78 &   8.5 & LEG  \\
SDSS~J125303.17+450044.8 & 0.078 &  472.0 &   23.7 & 14.659 &   1.97 & 253 &   2.90 &  41.02 &  39.57 &   8.5 & LEG  \\
SDSS~J131509.84+084053.3 & 0.093 &  869.0 &  111.7 & 16.504 &   1.82 & 219 &   3.28 &  41.45 &  40.41 &   8.3 & HEG  \\
SDSS~J131904.16+293834.8 & 0.073 & 1467.0 &   41.5 & 15.036 &   1.87 & 202 &   3.17 &  41.45 &  39.76 &   8.2 & LEG  \\
SDSS~J132117.81+423515.2 & 0.079 & 1985.0 &  259.8 & 15.883 &   1.46 & 176 &   2.54 &  41.66 &  40.63 &   7.9 & HEG  \\
SDSS~J132848.45+275227.8 & 0.091 &  207.0 &  927.0 & 16.711 &   1.42 & 177 &   3.37 &  40.81 &  41.31 &   7.9 & HEG  \\
SDSS~J133453.36$-$013238.5 & 0.087 & 22.0 &   16.9 & 17.369 &   1.81 & 132 &   2.86 &  39.79 &  39.53 &   7.4 & LEG  \\
SDSS~J133729.21+481820.5 & 0.119 &   69.0 &    2.4 & 16.285 &   1.89 & 234 &   3.04 &  40.58 &  38.97 &   8.4 & LEG  \\
SDSS~J133742.35+294223.2 & 0.115 &  219.0 &   21.2 & 15.222 &   2.05 & 269 &   3.01 &  41.05 &  39.89 &   8.6 & LEG  \\
SDSS~J134134.85+534443.7 & 0.141 & 1118.0 &  241.9 & 16.677 &   1.59 & 240 &   3.23 &  41.95 &  41.14 &   8.4 & HEG  \\
SDSS~J134503.02+094724.0 & 0.130 &   73.0 &    8.3 & 16.239 &   1.92 & 242 &   3.04 &  40.69 &  39.60 &   8.5 & LEG  \\
SDSS~J134532.76+054610.4 & 0.125 &   81.0 &    5.2 & 16.231 &   1.93 & 239 &   3.41 &  40.69 &  39.35 &   8.4 & LEG  \\
SDSS~J135117.87+640936.9 & 0.109 &   39.3 &   11.9 & 16.632 &   1.89 & 165 &   3.14 &  40.26 &  39.59 &   7.8 & LEG  \\
SDSS~J135124.55+085216.4 & 0.136 &   88.0 &    5.8 & 15.656 &   2.01 & 298 &   3.06 &  40.81 &  39.48 &   8.8 & LEG  \\
SDSS~J135432.96+281436.1 & 0.065 &   84.0 &   22.8 & 15.150 &   2.06 & 221 &   3.28 &  40.10 &  39.39 &   8.3 & LEG  \\
SDSS~J135526.20+352544.1 & 0.108 &   53.0 &   18.1 & 15.911 &   1.91 & 205 &   3.27 &  40.37 &  39.76 &   8.2 & LEG  \\
SDSS~J141231.76+140041.0 & 0.140 &   46.9 &    5.6 & 16.113 &   2.01 & 305 &   3.24 &  40.56 &  39.49 &   8.9 & LEG  \\
SDSS~J141247.44+045431.5 & 0.136 &   36.3 &   10.3 & 16.401 &   1.90 & 233 &   3.16 &  40.42 &  39.73 &   8.4 & LEG  \\
SDSS~J141527.22+172431.1 & 0.124 &    9.8 &    3.3 & 16.063 &   2.00 & 313 &   3.46 &  39.77 &  39.16 &   8.9 & LEG  \\
SDSS~J142341.44+250149.2 & 0.146 &   28.8 &    3.8 & 16.773 &   2.00 & 263 &   3.00 &  40.39 &  39.36 &   8.6 & LEG  \\
SDSS~J142557.05+392444.9 & 0.143 &   27.0 &   39.1 & 16.752 &   1.71 & 220 &   3.01 &  40.34 &  40.36 &   8.3 & HEG  \\
SDSS~J143010.53+304428.3 & 0.129 &   49.4 &    5.5 & 16.095 &   1.95 & 211 &   2.96 &  40.51 &  39.41 &   8.2 & LEG  \\
SDSS~J143715.00+244532.2 & 0.086 &  151.6 &   29.4 & 16.267 &   1.94 & 210 &   3.35 &  40.62 &  39.76 &   8.2 & LEG  \\
SDSS~J144625.13+214209.8 & 0.112 &    7.6 &   12.7 & 15.465 &   1.98 & 264 &   3.31 &  39.57 &  39.64 &   8.6 &      \\
SDSS~J144626.18+063359.1 & 0.063 &   64.1 &   16.4 & 15.877 &   1.95 & 168 &   2.93 &  39.95 &  39.22 &   7.8 & LEG  \\
SDSS~J144948.90+335126.8 & 0.088 &  116.0 &    5.1 & 16.074 &   1.83 & 223 &   3.55 &  40.52 &  39.02 &   8.3 & LEG  \\
SDSS~J145423.45+162119.0 & 0.045 & 1489.0 &   60.5 & 13.245 &   2.01 & 304 &   3.09 &  41.03 &  39.49 &   8.9 & LEG  \\
SDSS~J145752.76+111809.5 & 0.122 &  105.2 &    9.7 & 16.031 &   2.03 & 247 &   3.24 &  40.79 &  39.61 &   8.5 & LEG  \\
SDSS~J145753.81+283218.7 & 0.144 &  930.0 &  990.4 & 16.214 &   1.63 & 232 &   2.84 &  41.89 &  41.77 &   8.4 & HEG  \\
SDSS~J150031.48+362849.8 & 0.093 &   59.7 &   52.1 & 15.543 &   1.92 & 260 &   3.10 &  40.28 &  40.08 &   8.6 & LEG  \\
SDSS~J150146.57+243916.0 & 0.120 &  108.3 &   22.2 & 15.263 &   1.92 & 261 &   2.77 &  40.78 &  39.95 &   8.6 & LEG  \\
SDSS~J150827.31+541507.4 & 0.096 &   25.7 &   20.5 & 16.156 &   1.62 & 187 &   3.19 &  39.95 &  39.71 &   8.0 & LEG  \\
SDSS~J151247.16$-$014423.3 & 0.146 &120.0 & 2148.5 & 17.461 &   1.04 & 110 &   2.54 &  41.01 &  42.12 &   7.1 & HEG  \\
SDSS~J151414.68+232708.4 & 0.088 &  232.7 &   15.4 & 15.449 &   2.06 & 238 &   3.27 &  40.82 &  39.50 &   8.4 & LEG  \\
SDSS~J151423.26+011130.7 & 0.125 &   54.0 &    --- & 16.709 &   1.97 & 189 &   2.83 &  40.52 &    --- &   8.0 & LEG  \\
SDSS~J151454.07+420047.3 & 0.135 &  177.0 &   18.8 & 15.346 &   1.98 & 315 &   3.17 &  41.10 &  39.99 &   8.9 & LEG  \\
SDSS~J151639.48+024712.6 & 0.110 &  131.0 &   24.1 & 15.491 &   1.88 & 273 &   3.16 &  40.78 &  39.90 &   8.7 & LEG  \\
SDSS~J151640.21+001501.9 & 0.053 & 2722.0 & 1286.6 & 14.855 &   1.31 & 293 &   3.27 &  41.43 &  40.95 &   8.8 & LEG  \\
SDSS~J152245.61+194220.3 & 0.109 &   86.7 &   42.3 & 16.368 &   1.53 & 209 &   3.10 &  40.60 &  40.14 &   8.2 & LEG  \\
SDSS~J152257.38+025512.1 & 0.110 &   29.4 &    1.8 & 16.084 &   2.13 & 213 &   2.51 &  40.14 &  38.77 &   8.2 & LEG  \\
SDSS~J153007.96+231616.0 & 0.090 &  236.0 &   25.0 & 15.103 &   2.09 & 287 &   3.00 &  40.85 &  39.73 &   8.8 & LEG  \\
SDSS~J154042.39+422136.0 & 0.142 &   55.3 &   13.8 & 15.829 &   2.01 & 279 &   2.82 &  40.65 &  39.90 &   8.7 & LEG  \\
SDSS~J154118.78+514043.6 & 0.148 &   40.3 &    2.5 & 15.827 &   2.11 & 260 &   3.00 &  40.55 &  39.20 &   8.6 & LEG  \\
SDSS~J155700.17+413111.1 & 0.085 &  140.0 &   25.7 & 17.607 &   1.76 & 139 &   3.22 &  40.57 &  39.69 &   7.5 & LEG  \\
SDSS~J155852.66+262618.9 & 0.087 &   52.0 &   20.0 & 14.601 &   1.98 & 282 &   2.98 &  40.17 &  39.61 &   8.7 & LEG  \\
SDSS~J160909.44+162549.9 & 0.143 &   88.7 &    2.8 & 15.598 &   1.96 & 297 &   3.13 &  40.86 &  39.21 &   8.8 & LEG  \\
SDSS~J161057.11+032202.8 & 0.119 &  109.1 &    6.2 & 16.508 &   2.00 & 229 &   3.31 &  40.78 &  39.39 &   8.4 & LEG  \\
SDSS~J162046.64+430911.4 & 0.134 &   35.9 &    7.7 & 15.550 &   2.02 & 293 &   2.86 &  40.41 &  39.59 &   8.8 & LEG  \\
SDSS~J163604.21+271829.0 & 0.134 &  363.0 &   15.4 & 16.051 &   1.99 & 263 &   3.01 &  41.41 &  39.89 &   8.6 & LEG  \\
SDSS~J164857.35+260441.1 & 0.137 &  137.7 &   19.6 & 17.877 &   1.94 & 178 &   3.18 &  41.01 &  40.02 &   7.9 & LEG  \\
SDSS~J164924.01+263502.6 & 0.055 &  146.0 &   47.4 & 14.537 &   1.92 & 223 &   2.76 &  40.19 &  39.55 &   8.3 & LEG  \\
SDSS~J165847.14+625624.6 & 0.106 &  258.0 &   28.4 & 16.410 &   1.80 & 175 &   3.01 &  41.05 &  39.94 &   7.9 & LEG  \\
SDSS~J214801.27$-$082527.5 & 0.135 & 88.0 &    2.7 & 16.564 &   2.05 & 189 &   3.43 &  40.80 &  39.14 &   8.0 & LEG  \\
SDSS~J231542.10$-$002607.0 & 0.091 & 45.0 &   30.5 & 16.136 &   1.95 & 280 &   3.43 &  40.14 &  39.83 &   8.7 & LEG  \\
SDSS~J232710.69$-$004157.8 & 0.099 &223.0 &   80.4 & 16.043 &   1.64 & 223 &   3.04 &  40.92 &  40.33 &   8.3 & LEG  \\
\hline
\hline
\end{longtable}
\end{center}
Column description: (1) source name; (2)
redshift; (3) NVSS 1.4 GHz flux density [mJy]; (4) [O~III] flux [in
10$^{-17}$ erg cm$^{-2}$ s$^{-1}$ units]; (5) SDSS DR7 r band AB magnitude; (6)
concentration index $C_r$; (7) Dn(4000) index; (8) stellar velocity dispersion
[\kms]; (9) logarithm of the radio luminosity [erg s$^{-1}$]; (10) logarithm
of the [O~III] line luminosity [erg s$^{-1}$]; (11) logarithm of the black
hole mass [in solar units]; (12) spectroscopic class.
\twocolumn

\appendix

\section{FIRST images of the \FRII\ sources.}

\begin{figure*}
\includegraphics[width=6.3cm,height=6.3cm]{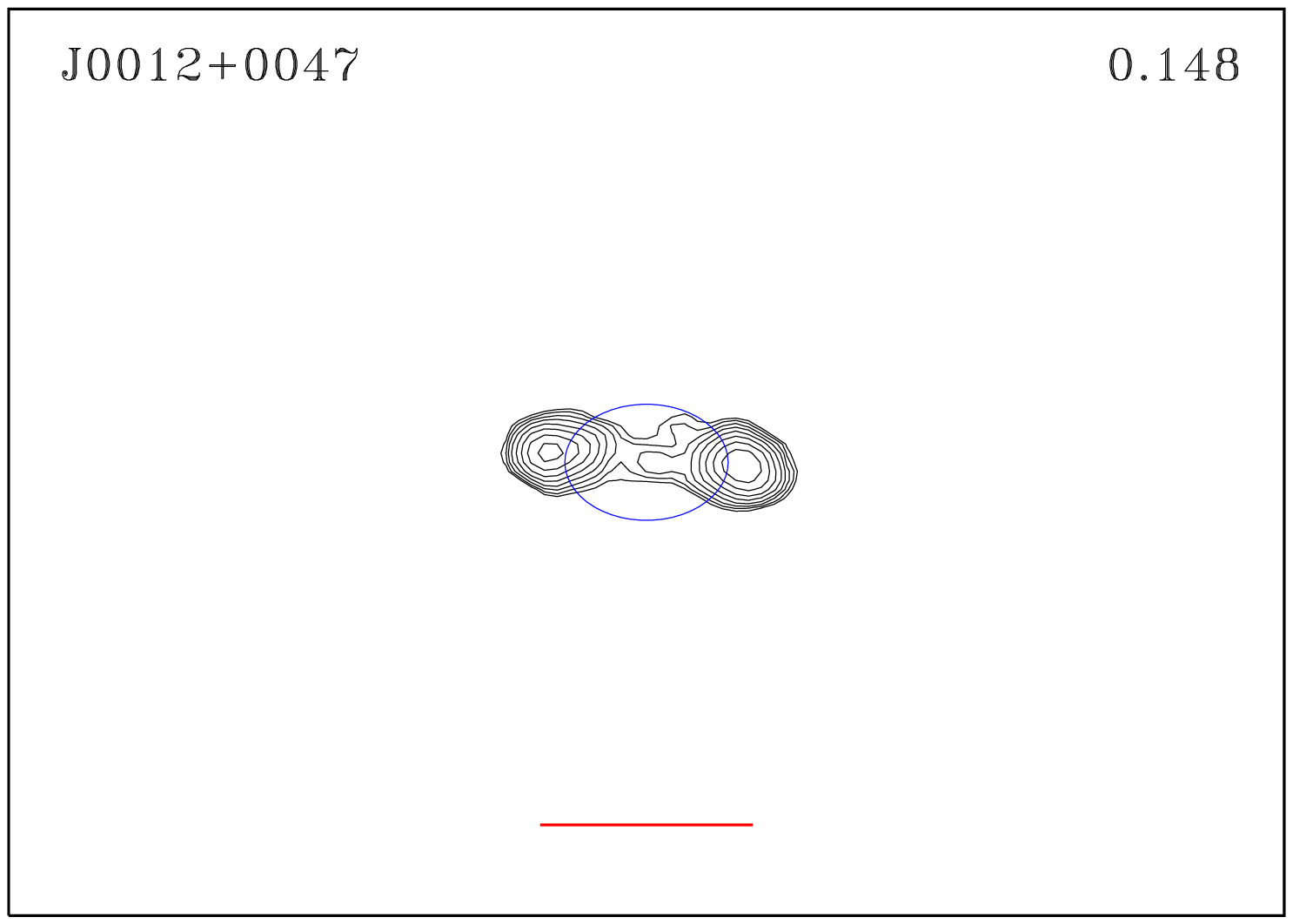}
\includegraphics[width=6.3cm,height=6.3cm]{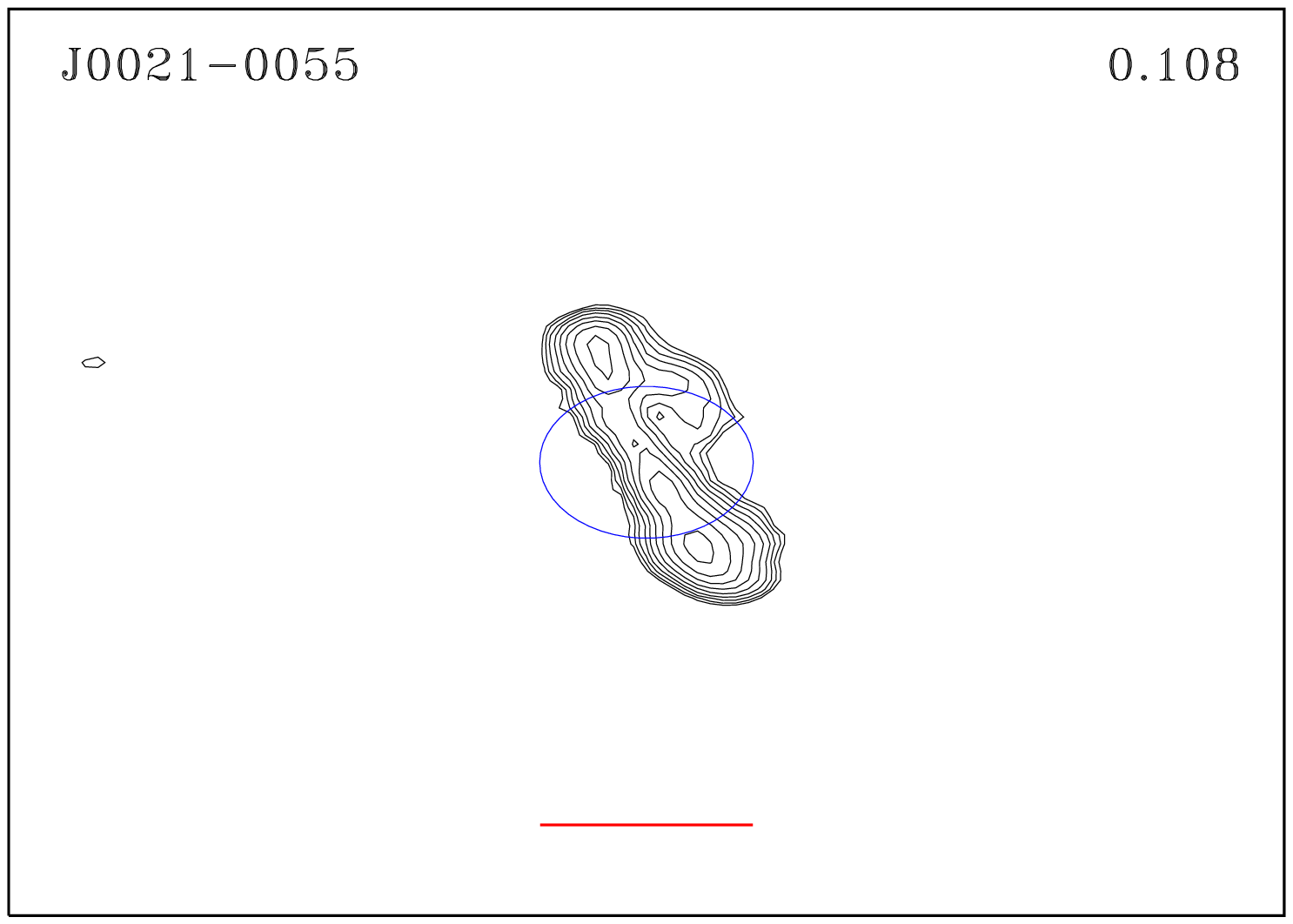}
\includegraphics[width=6.3cm,height=6.3cm]{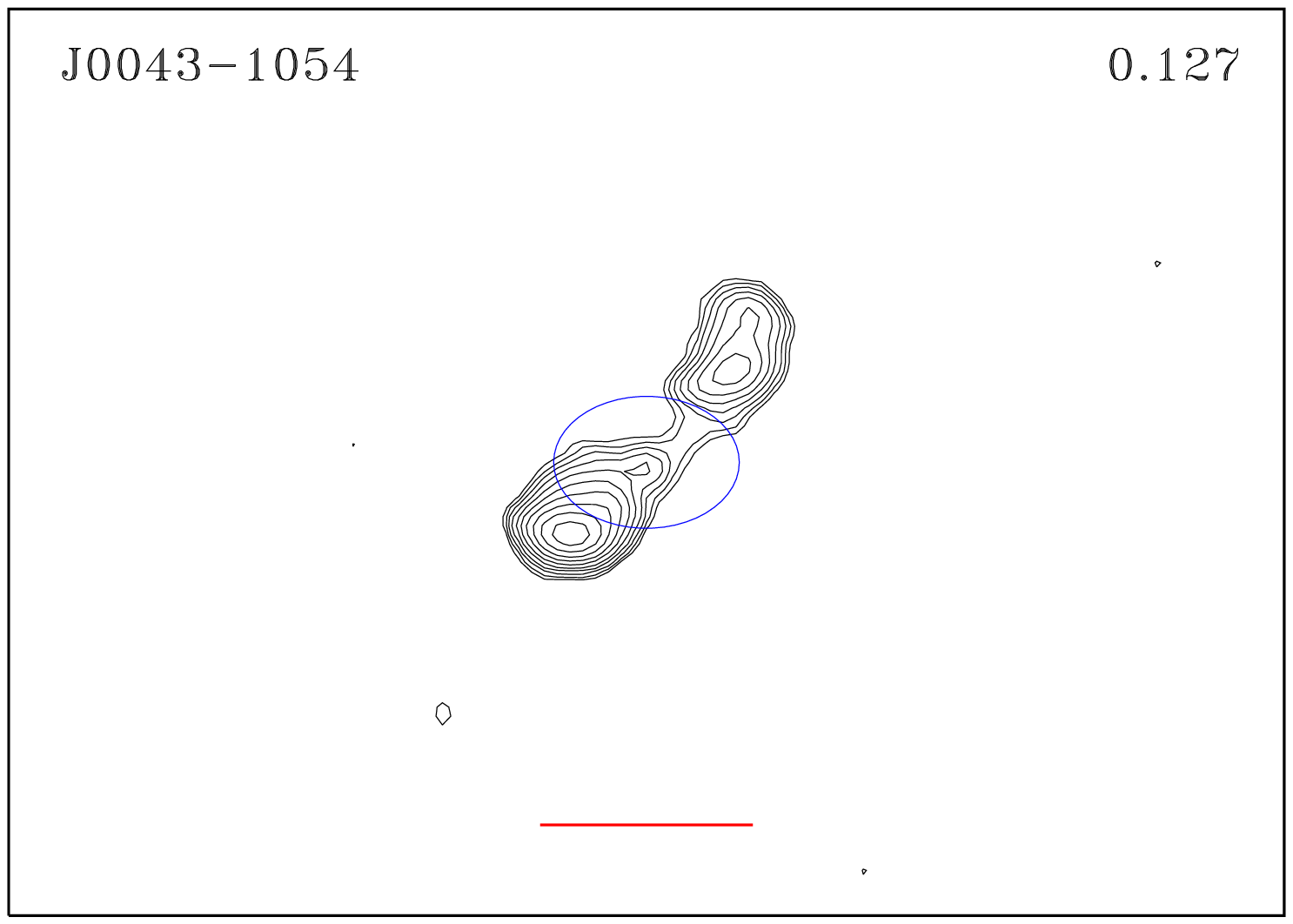}

\includegraphics[width=6.3cm,height=6.3cm]{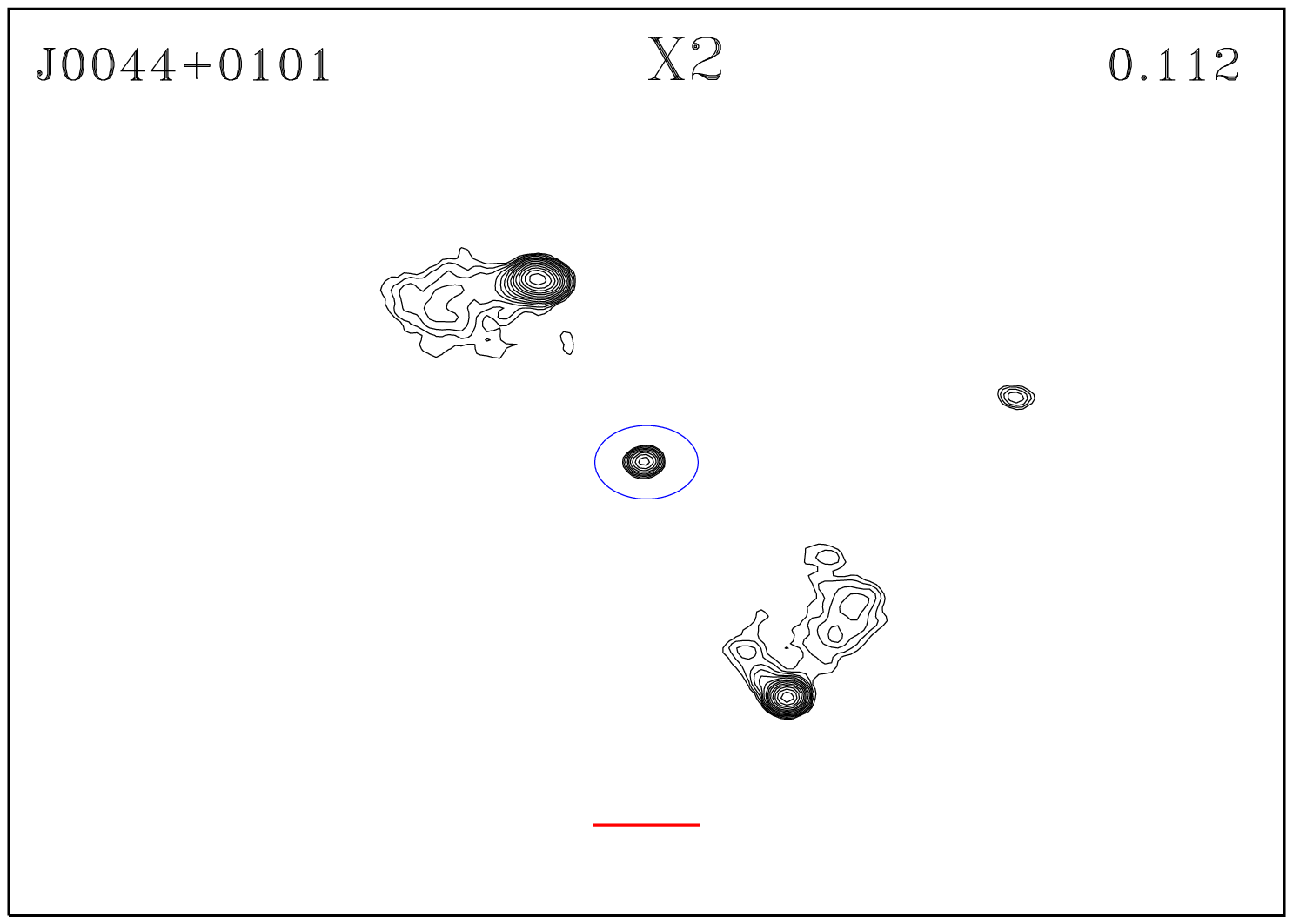}
\includegraphics[width=6.3cm,height=6.3cm]{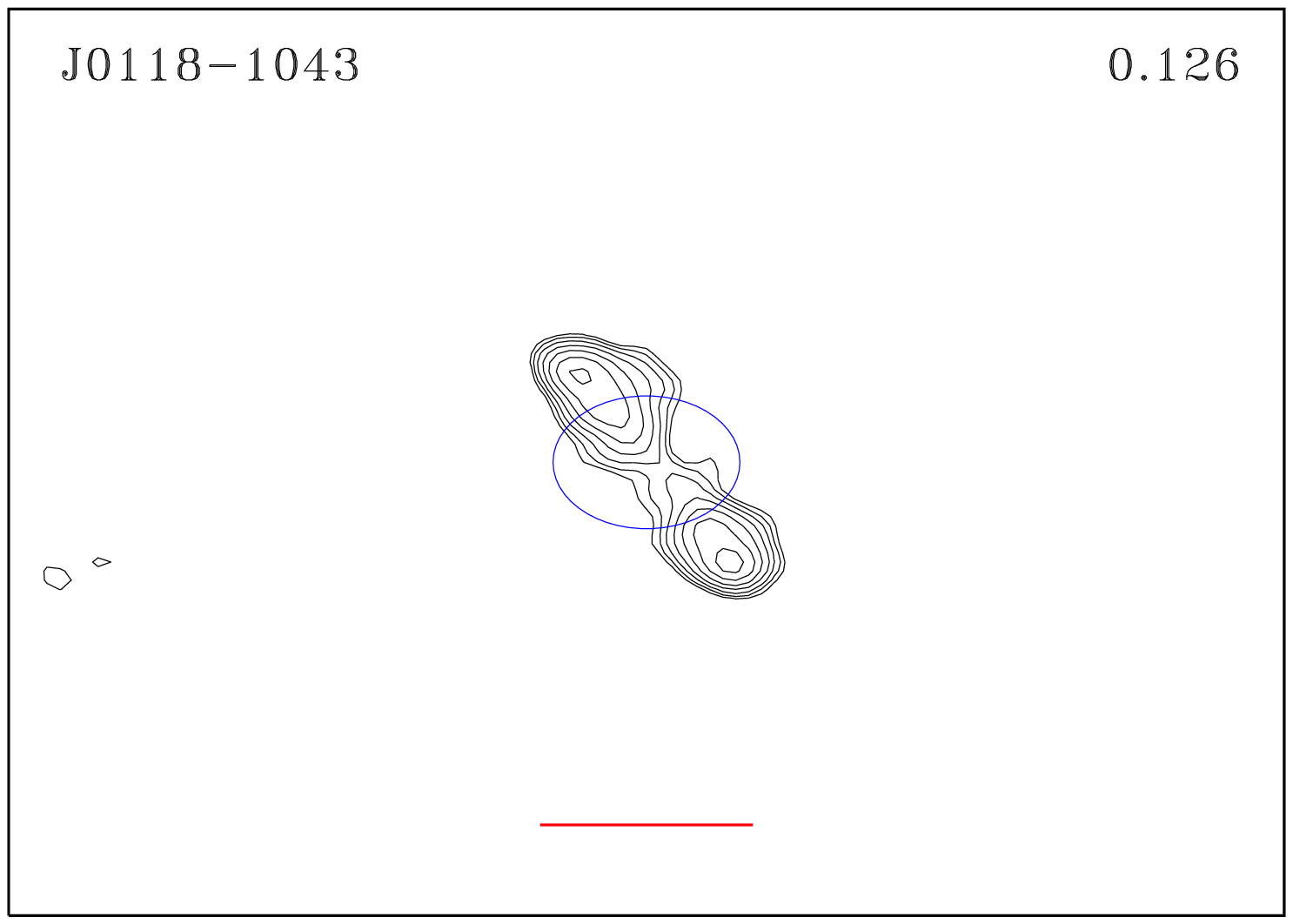}
\includegraphics[width=6.3cm,height=6.3cm]{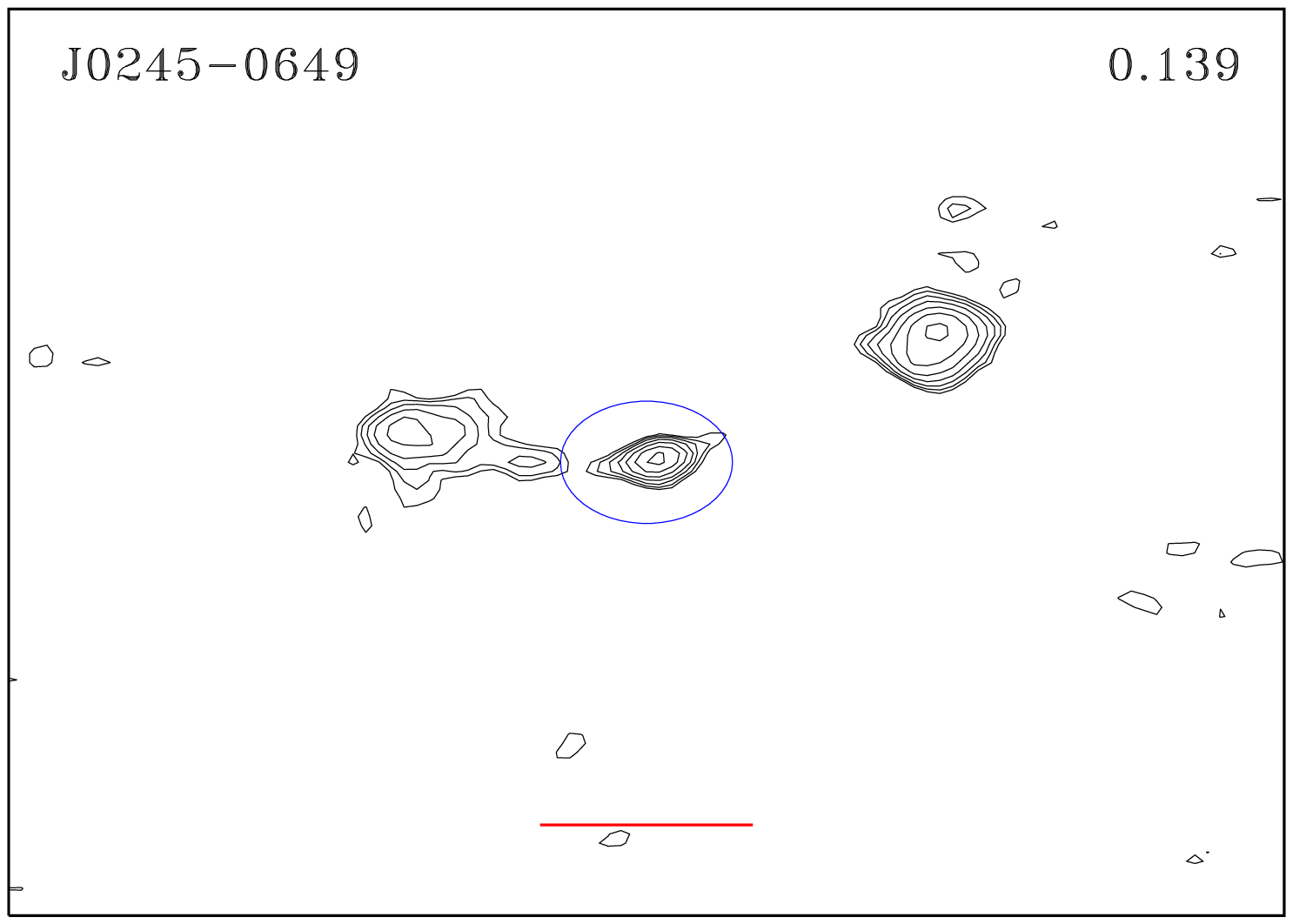}

\includegraphics[width=6.3cm,height=6.3cm]{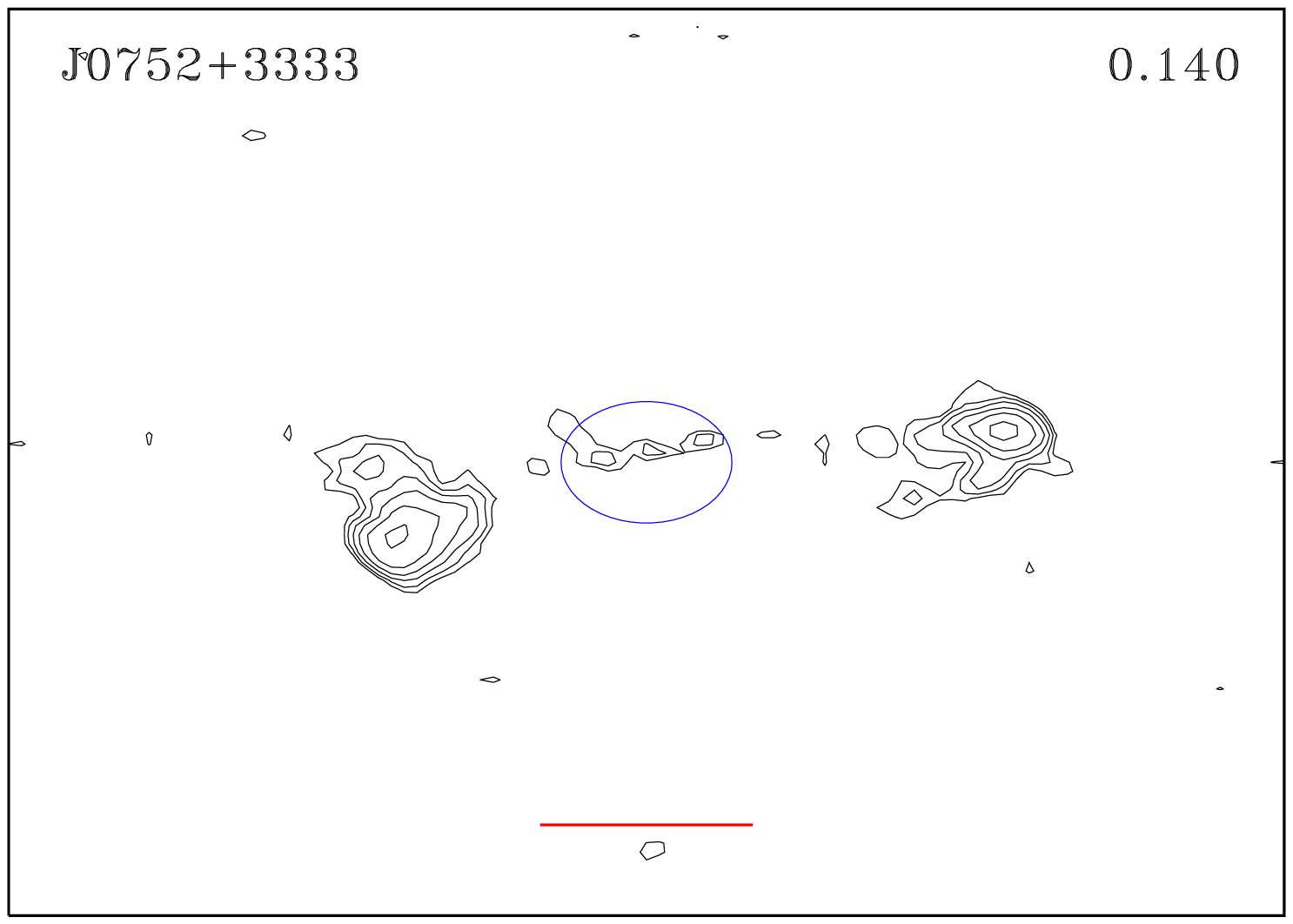}
\includegraphics[width=6.3cm,height=6.3cm]{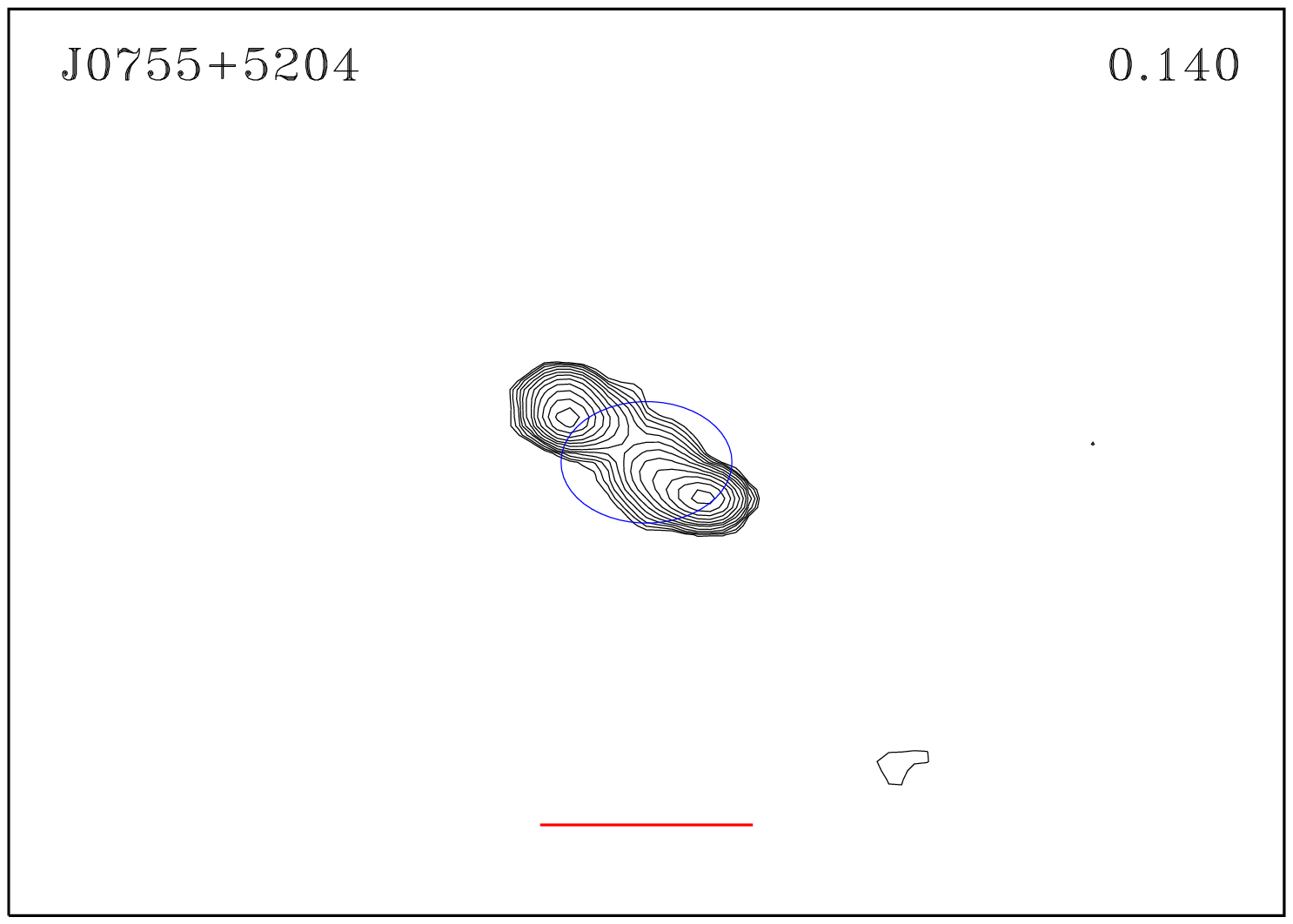}
\includegraphics[width=6.3cm,height=6.3cm]{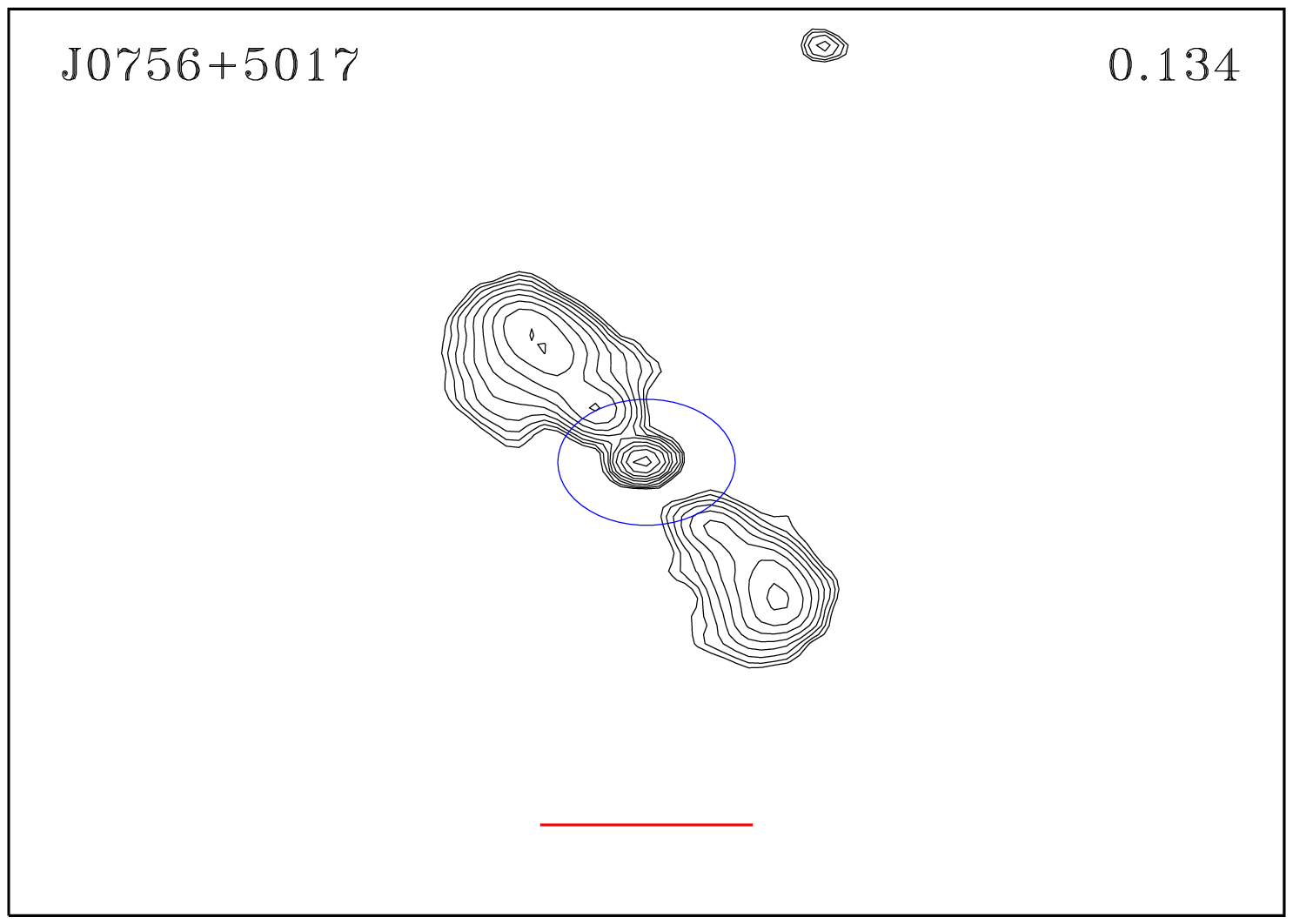}

\includegraphics[width=6.3cm,height=6.3cm]{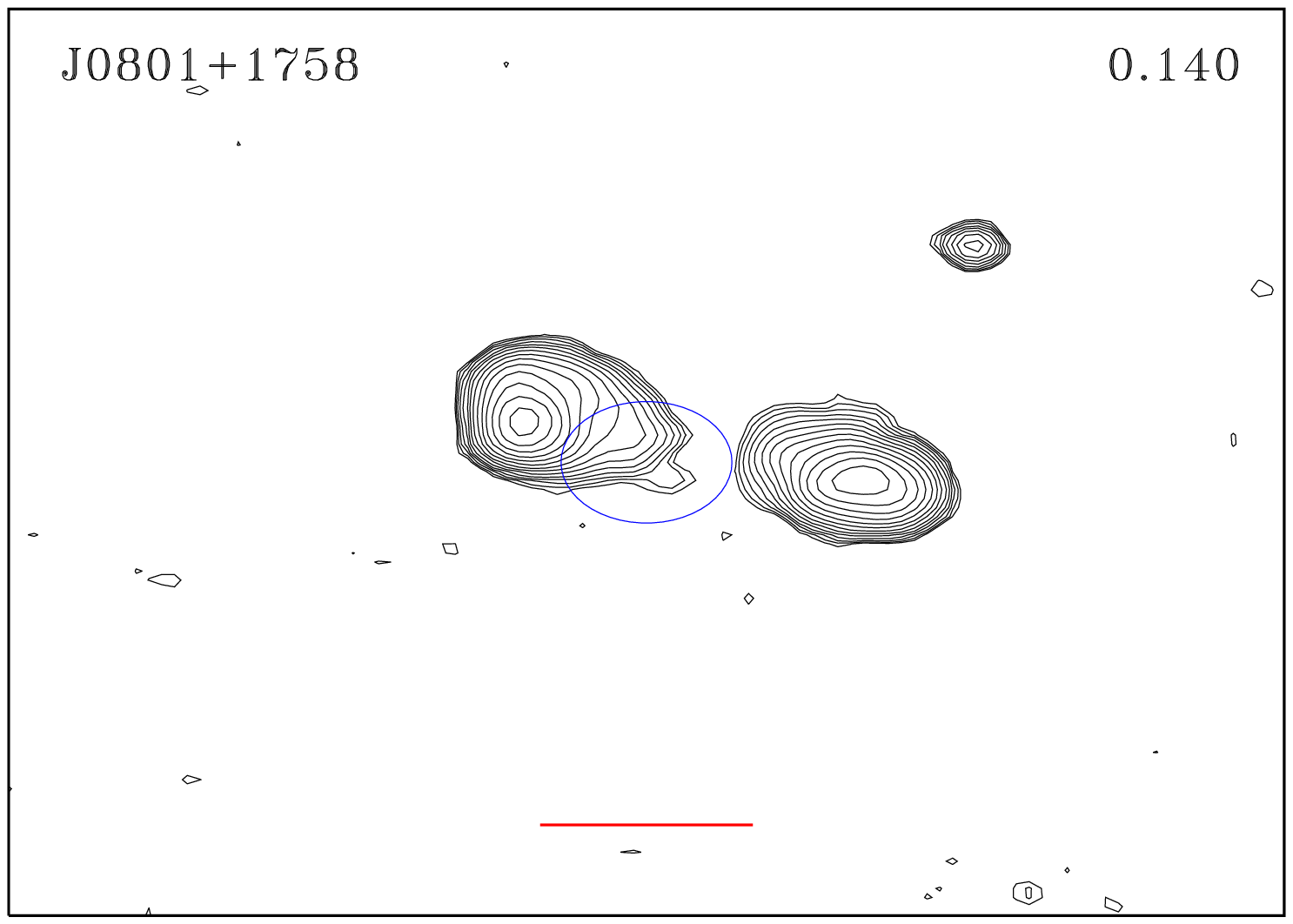}
\includegraphics[width=6.3cm,height=6.3cm]{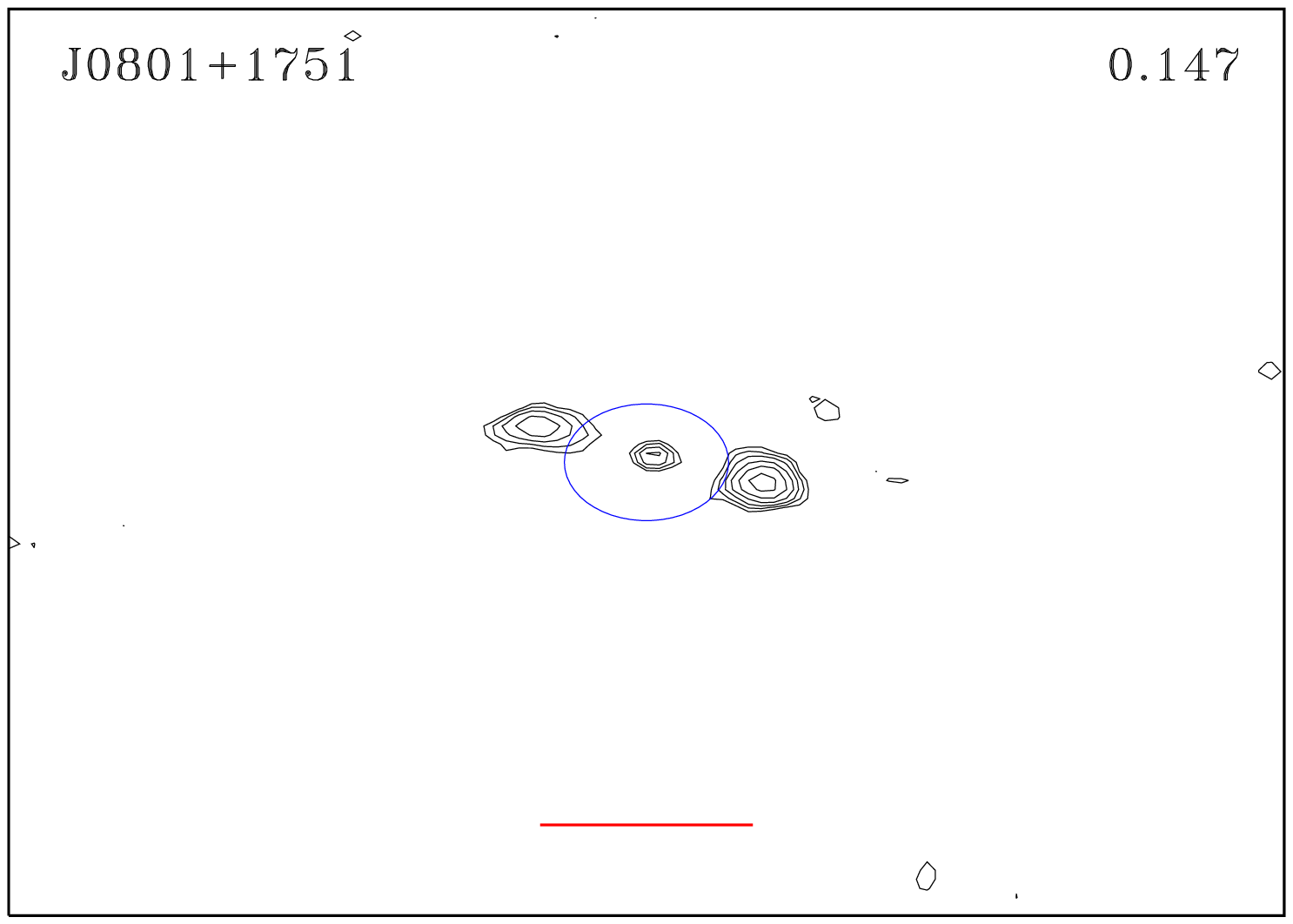}
\includegraphics[width=6.3cm,height=6.3cm]{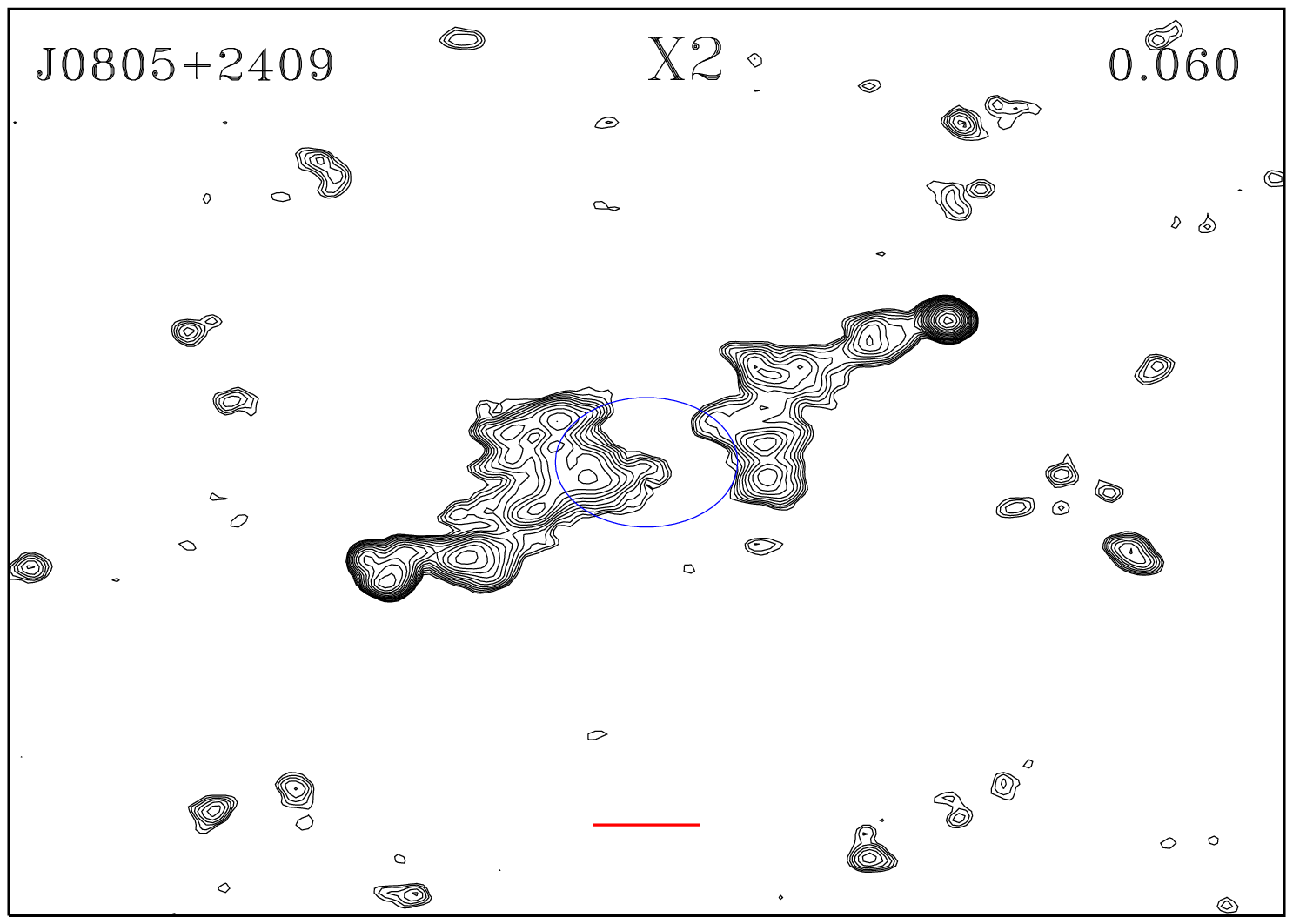}
\caption{Images of the FR~IIs selected. Contours are drawn starting from 0.45
  mJy/beam and increase with a geometric progression with a common ratio of
  $\sqrt2$. The field of view is 3'$\times$3' in most cases, except for those
  marked with a 'X2' at the top; the red tick at the bottom is 30$\arcsec$
  long. The blue circle is centered on the host galaxy and has a radius of 30
  kpc. The source ID and redshift are reported in the upper corners.}
\label{images3}
\end{figure*}

\begin{figure*}
\includegraphics[width=6.3cm,height=6.3cm]{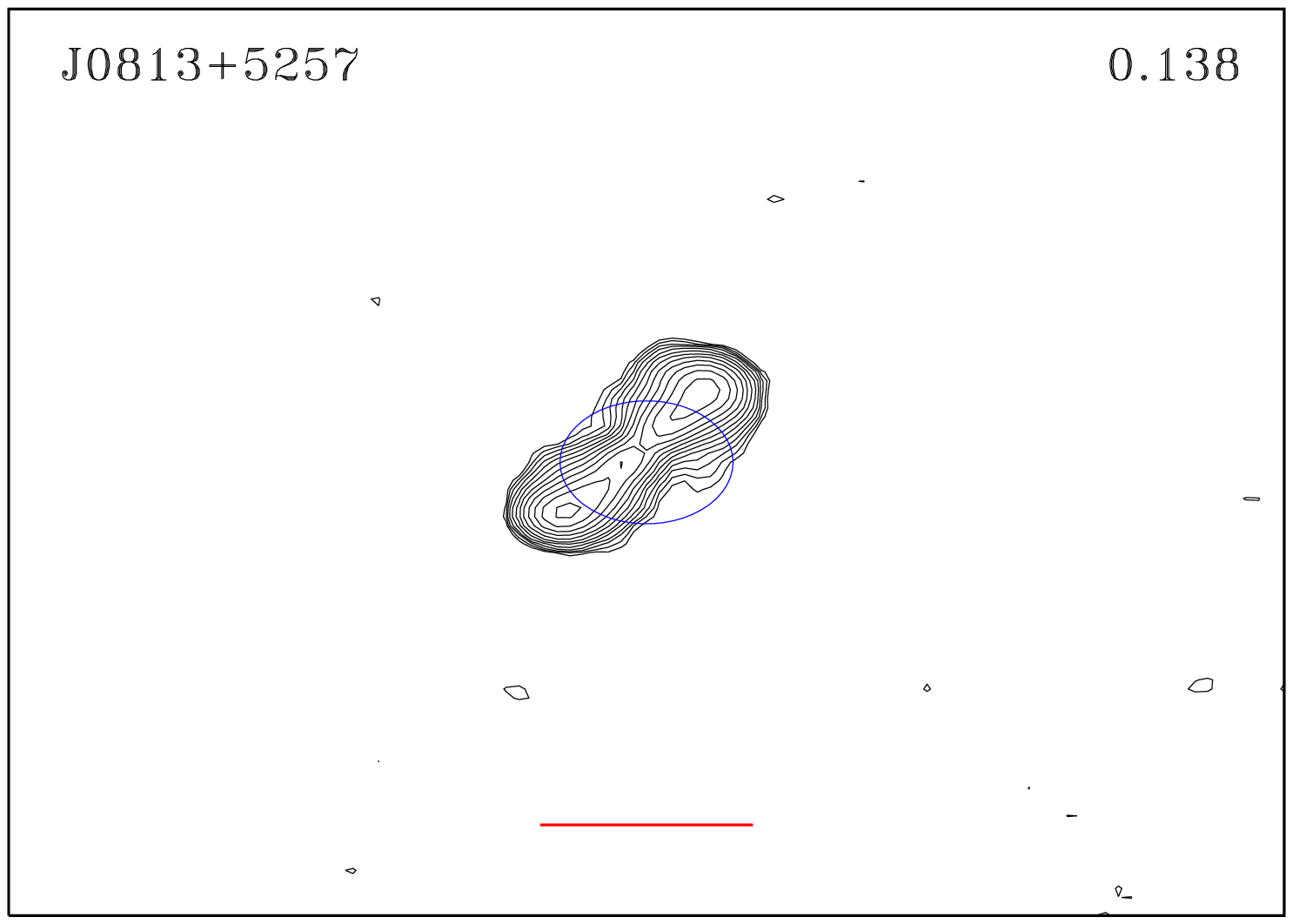}
\includegraphics[width=6.3cm,height=6.3cm]{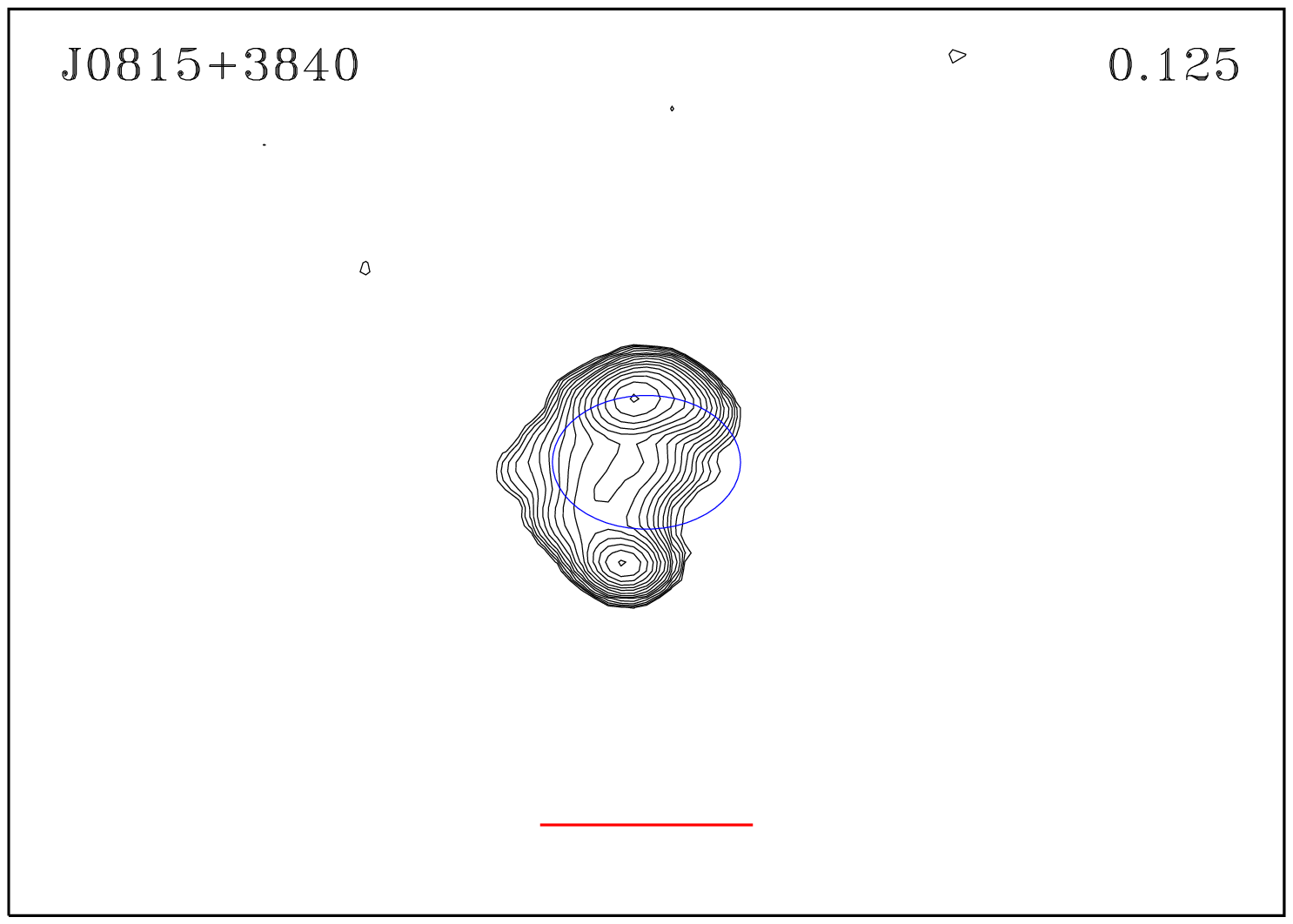}
\includegraphics[width=6.3cm,height=6.3cm]{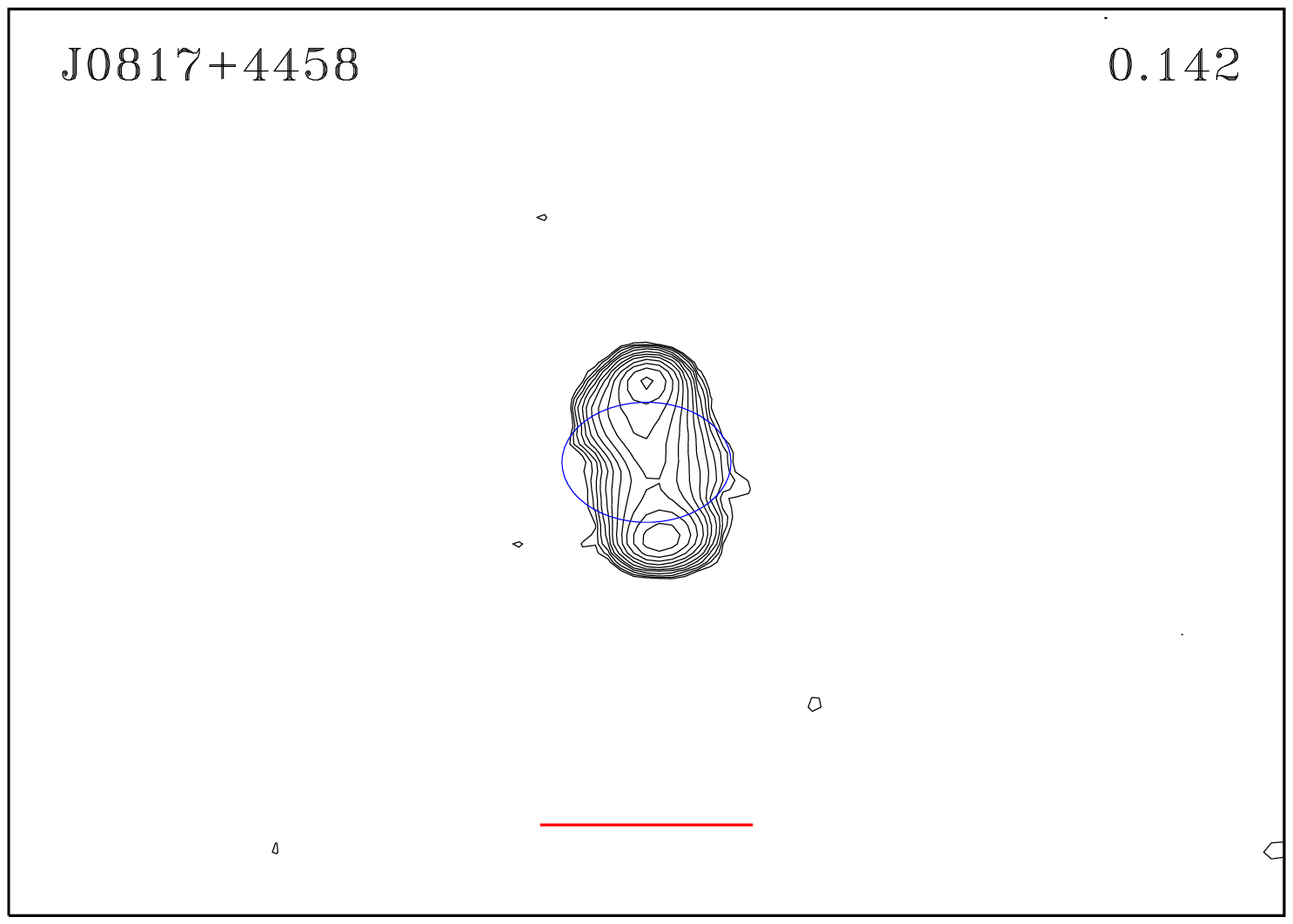}

\includegraphics[width=6.3cm,height=6.3cm]{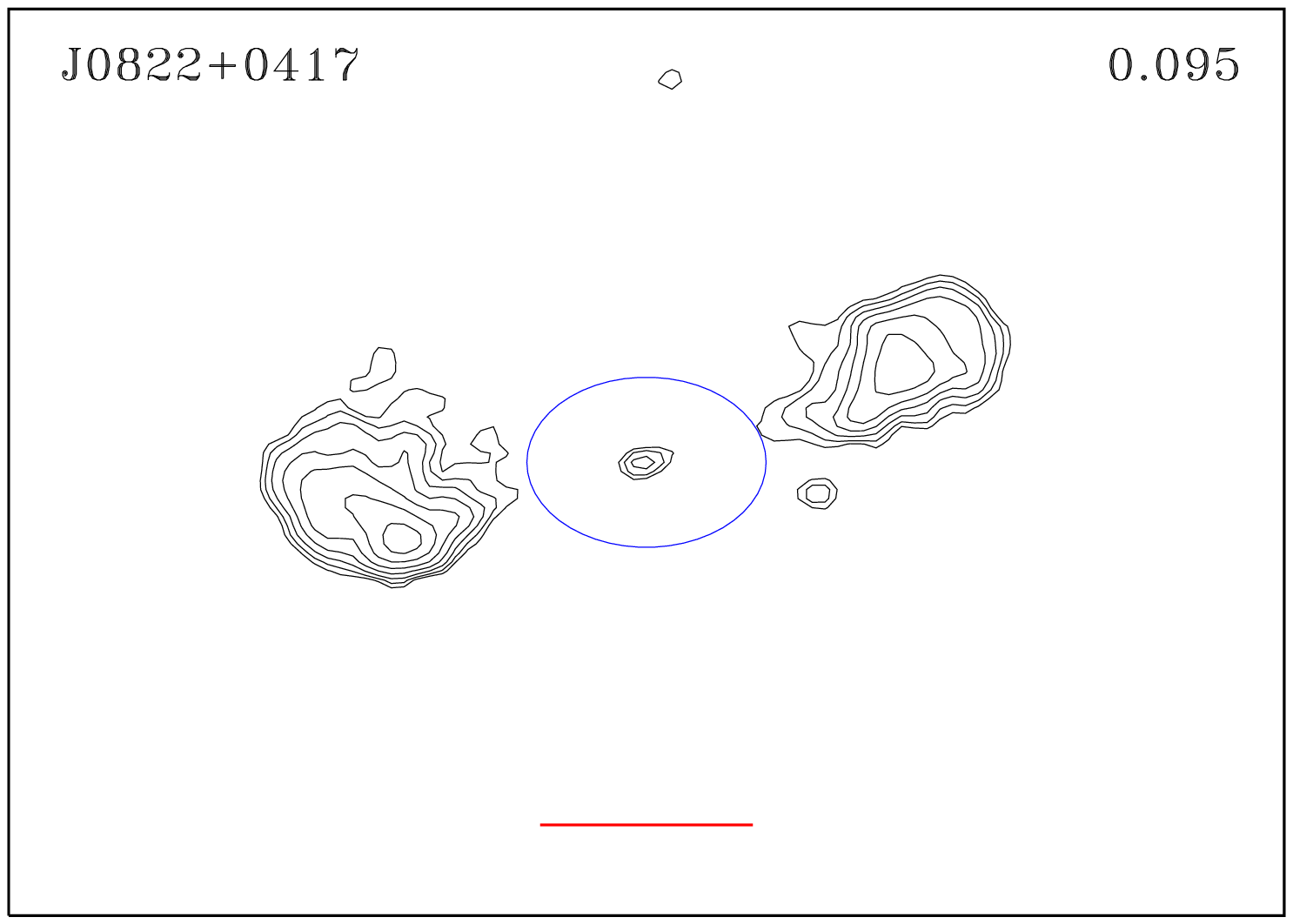}
\includegraphics[width=6.3cm,height=6.3cm]{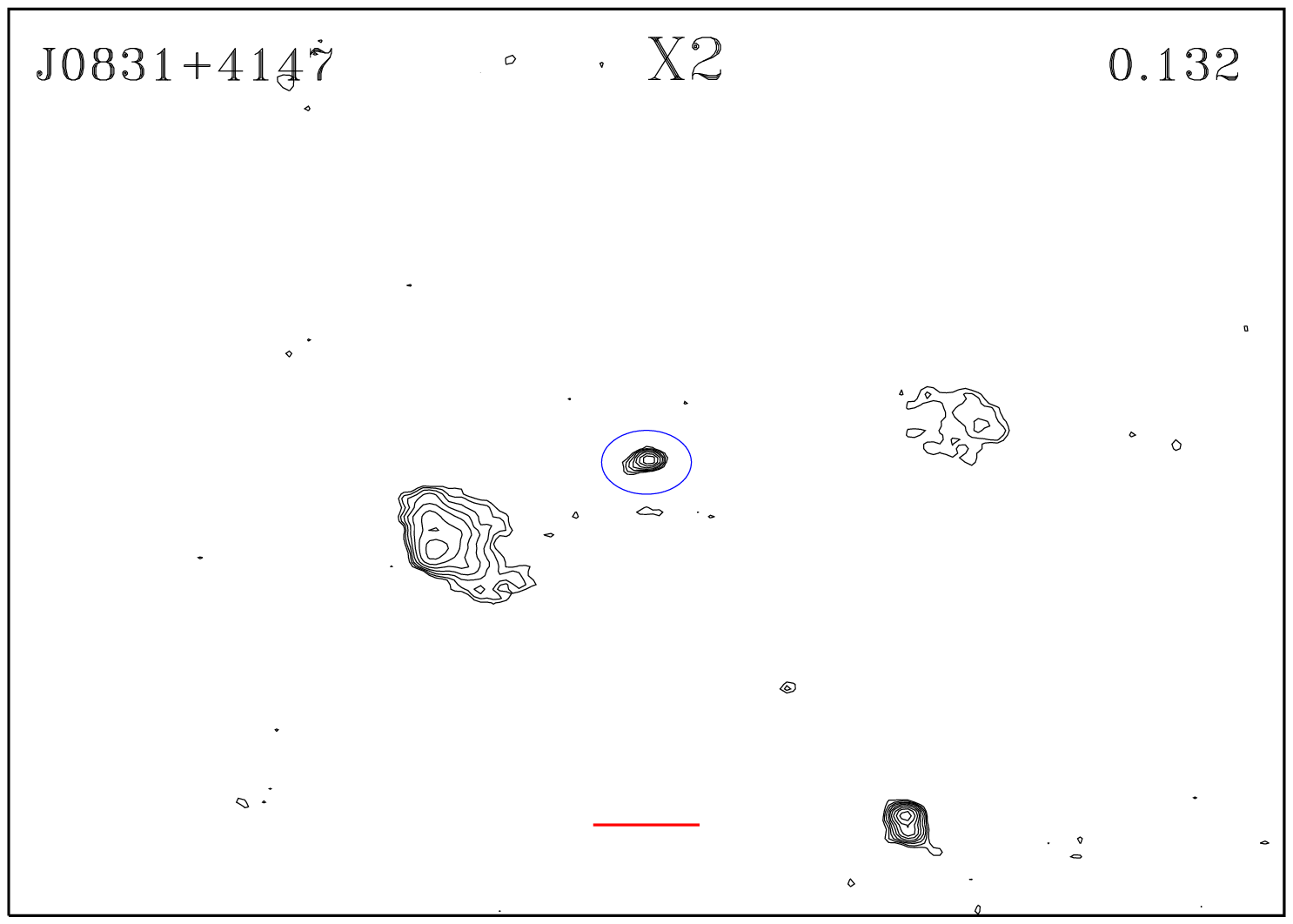}
\includegraphics[width=6.3cm,height=6.3cm]{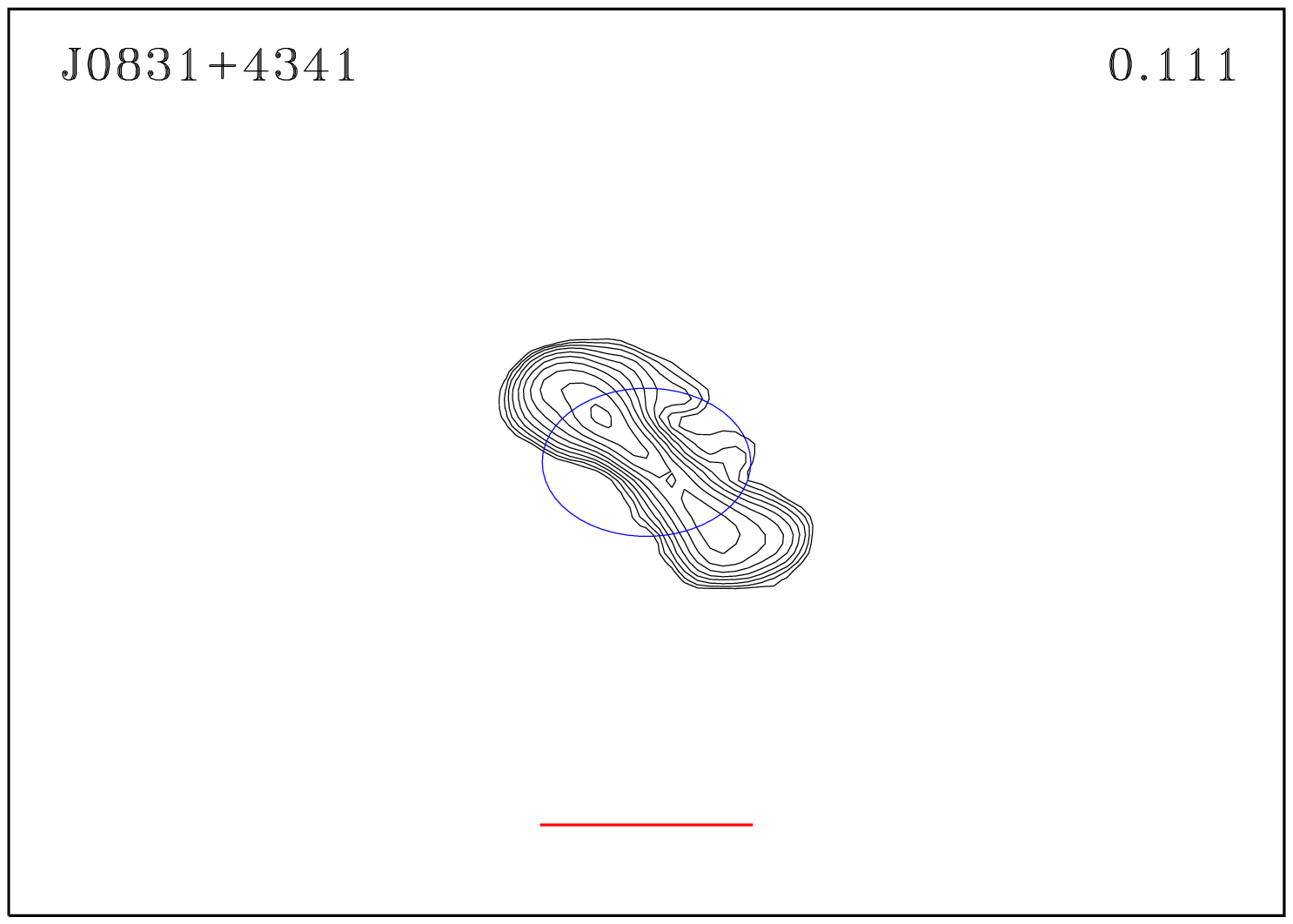}

\includegraphics[width=6.3cm,height=6.3cm]{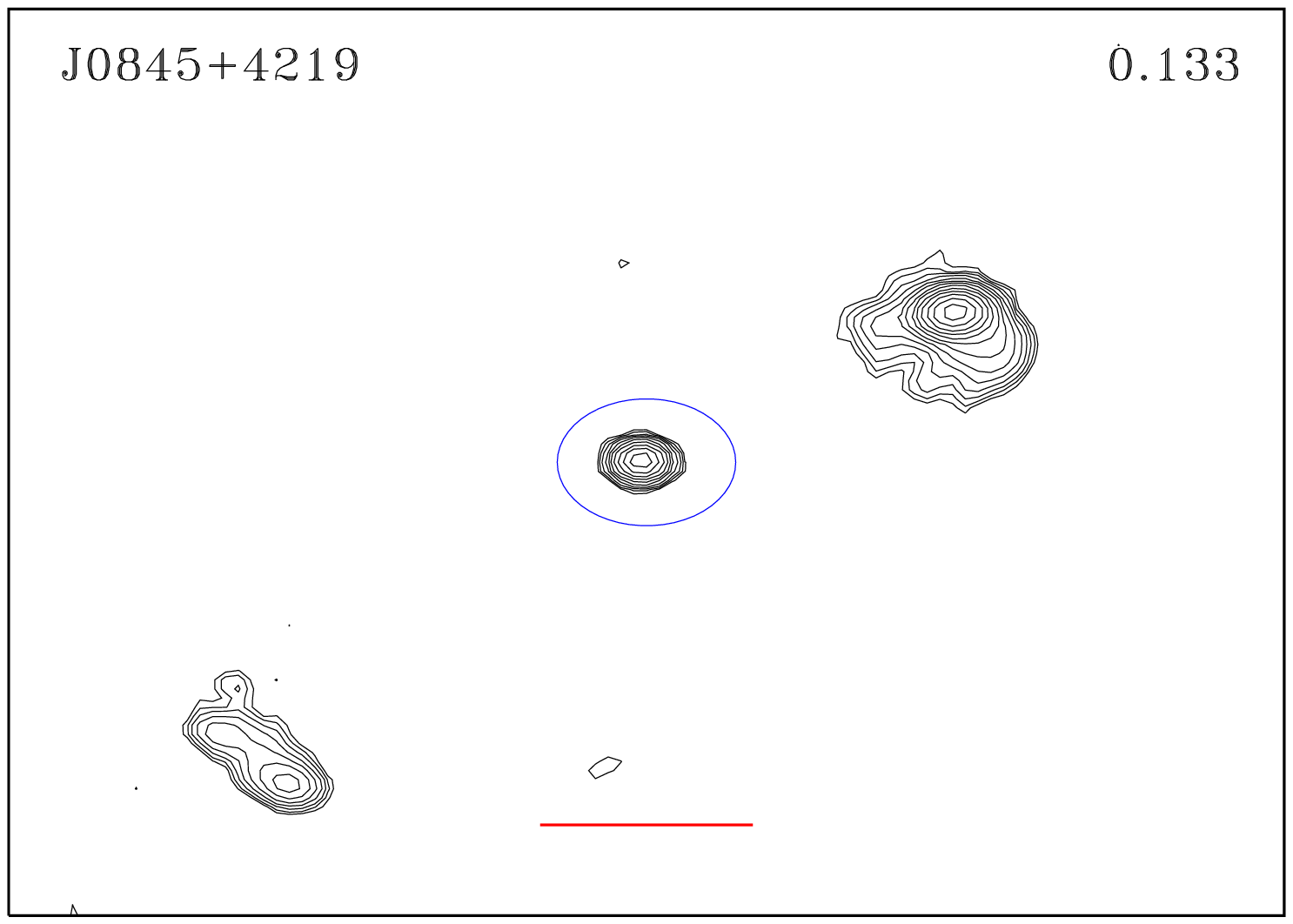}
\includegraphics[width=6.3cm,height=6.3cm]{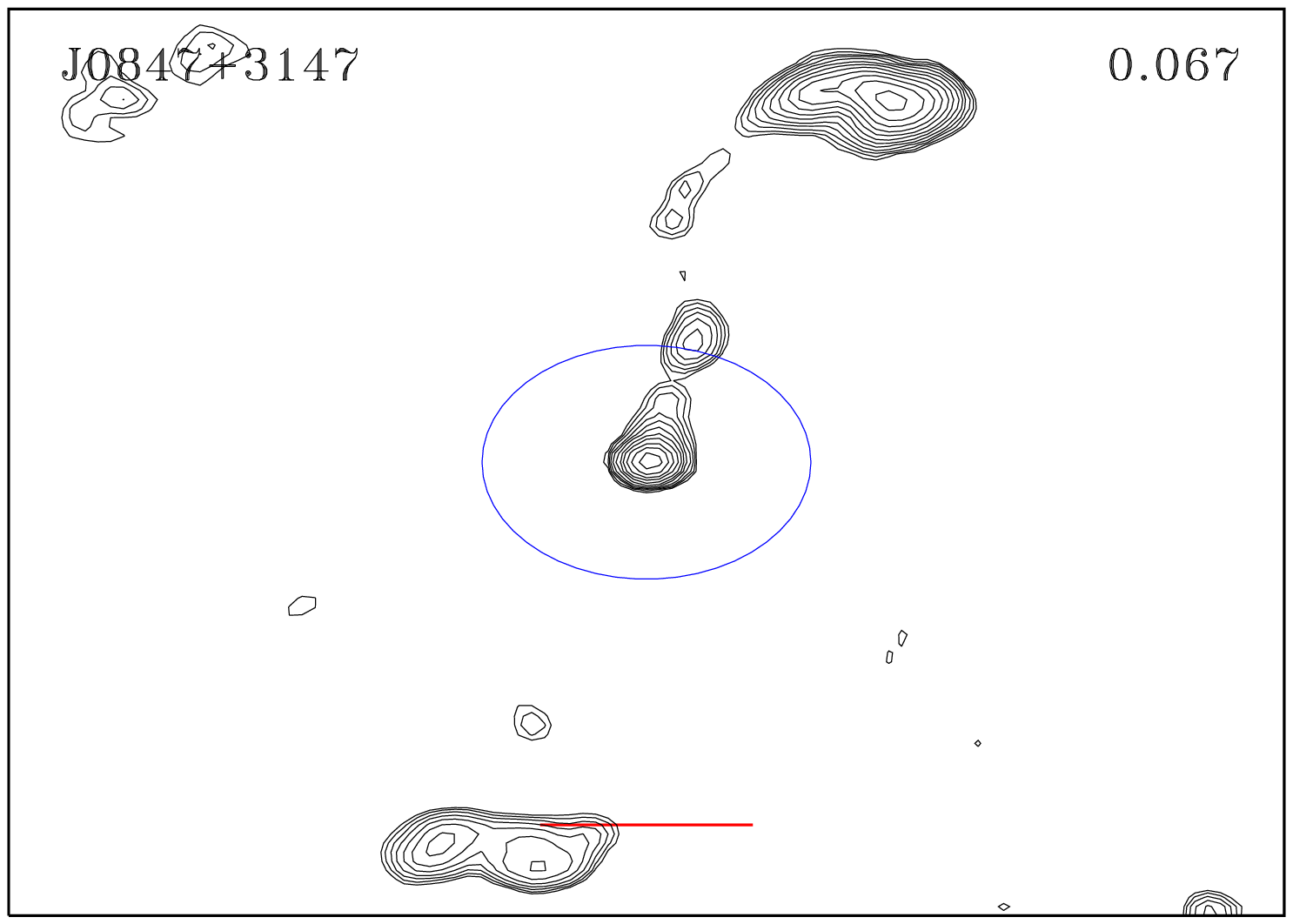}
\includegraphics[width=6.3cm,height=6.3cm]{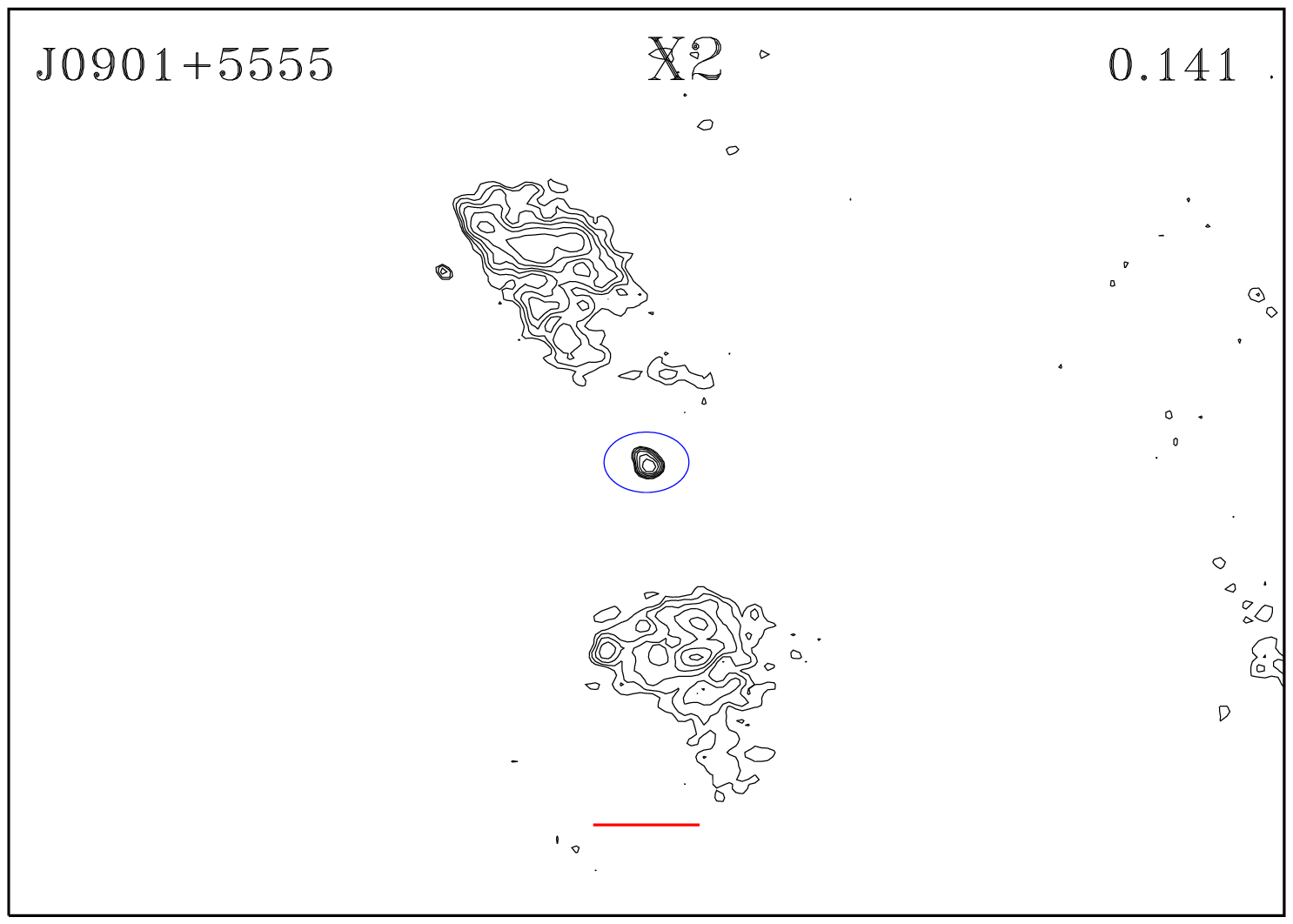}

\includegraphics[width=6.3cm,height=6.3cm]{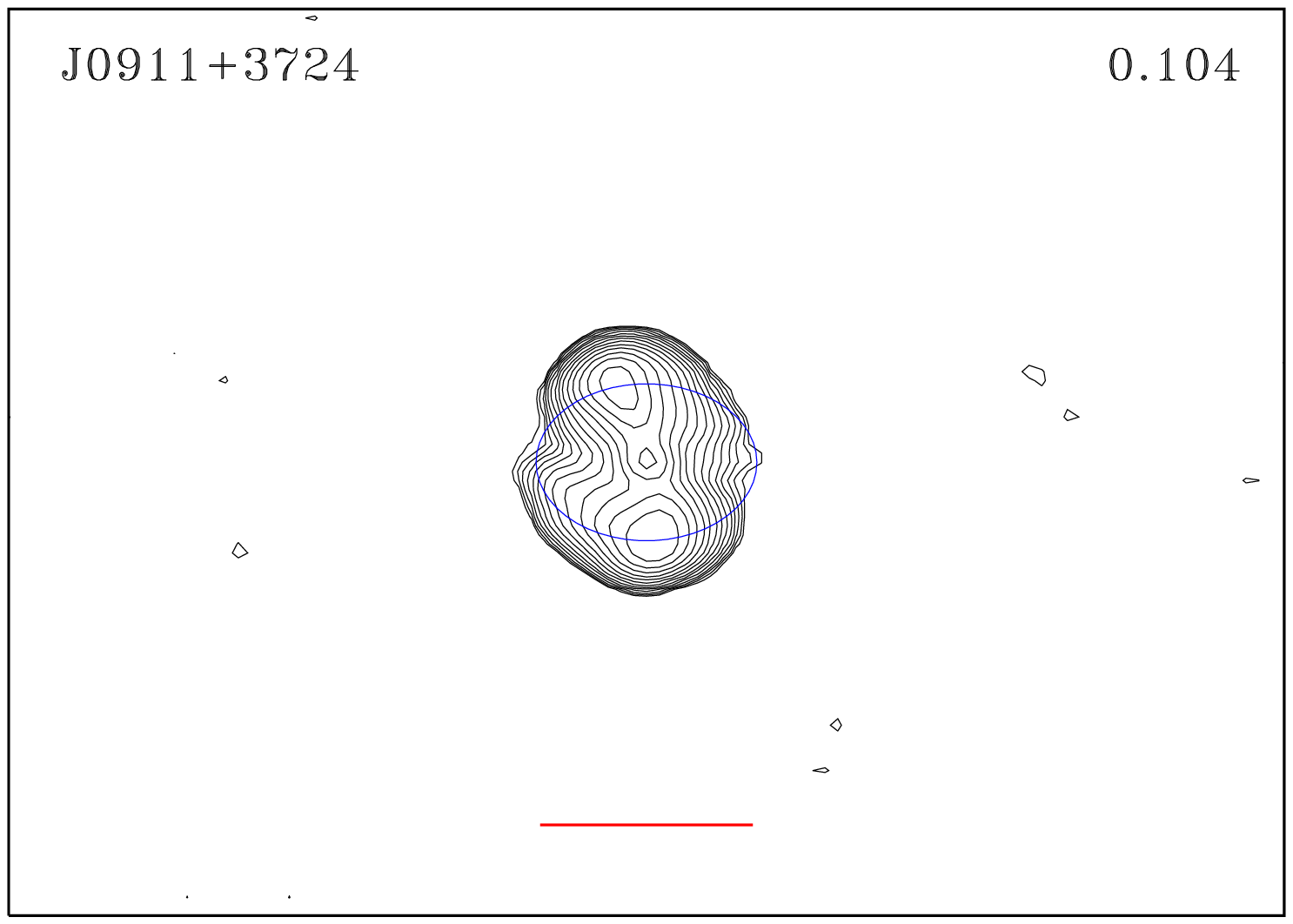}
\includegraphics[width=6.3cm,height=6.3cm]{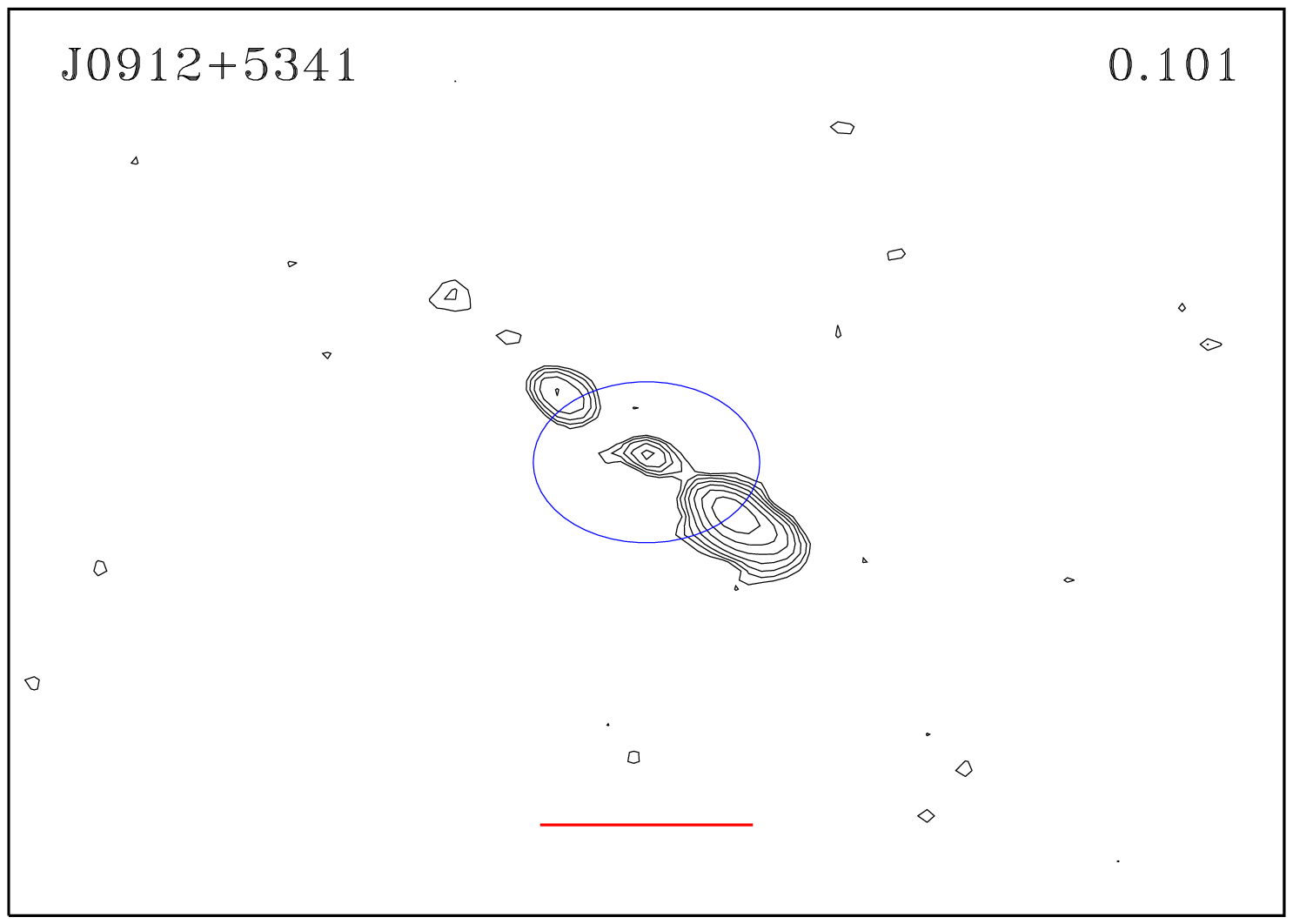}
\includegraphics[width=6.3cm,height=6.3cm]{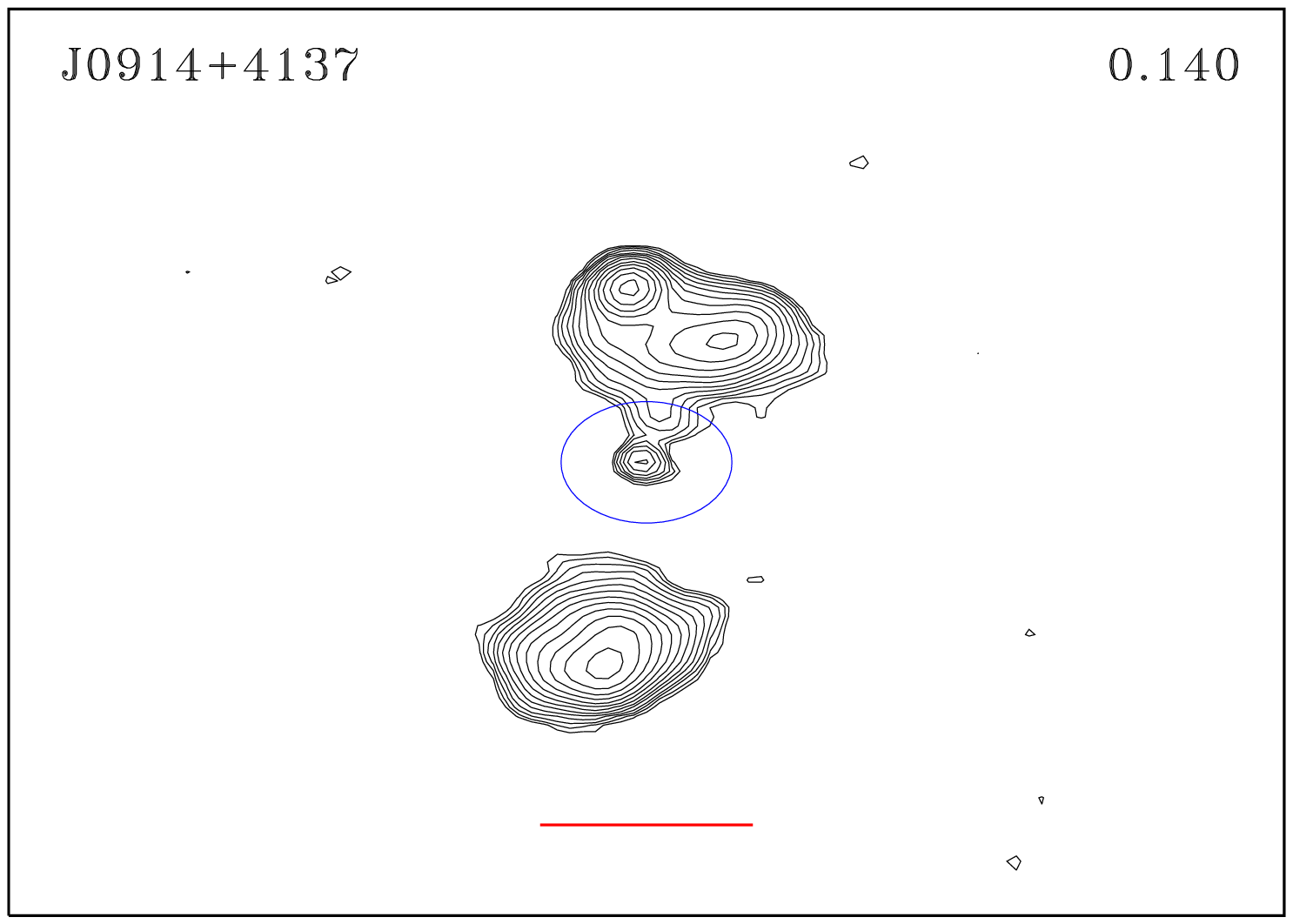}
\caption{(continued)}
\end{figure*}

\addtocounter{figure}{-1}
\begin{figure*}
\includegraphics[width=6.3cm,height=6.3cm]{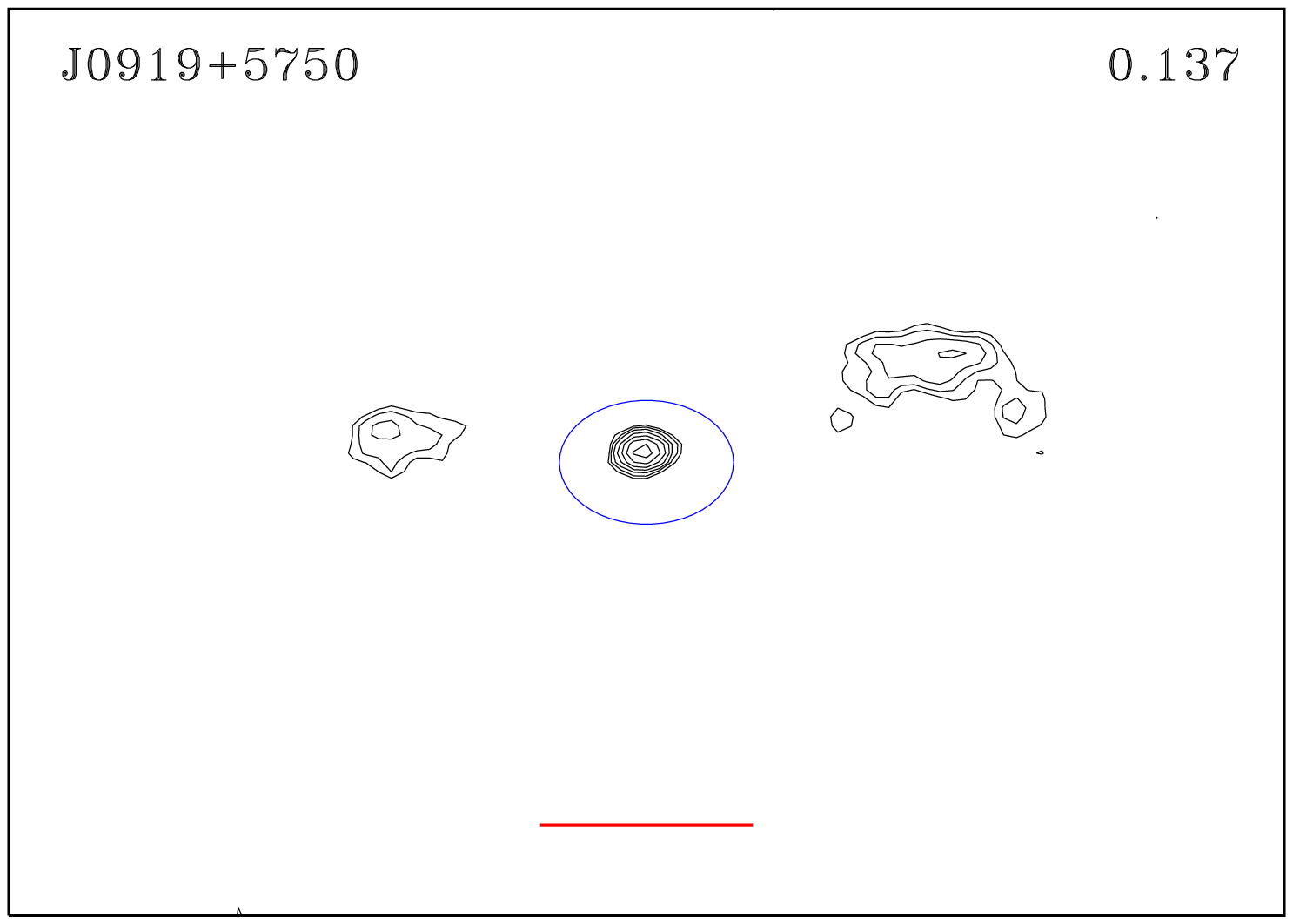}
\includegraphics[width=6.3cm,height=6.3cm]{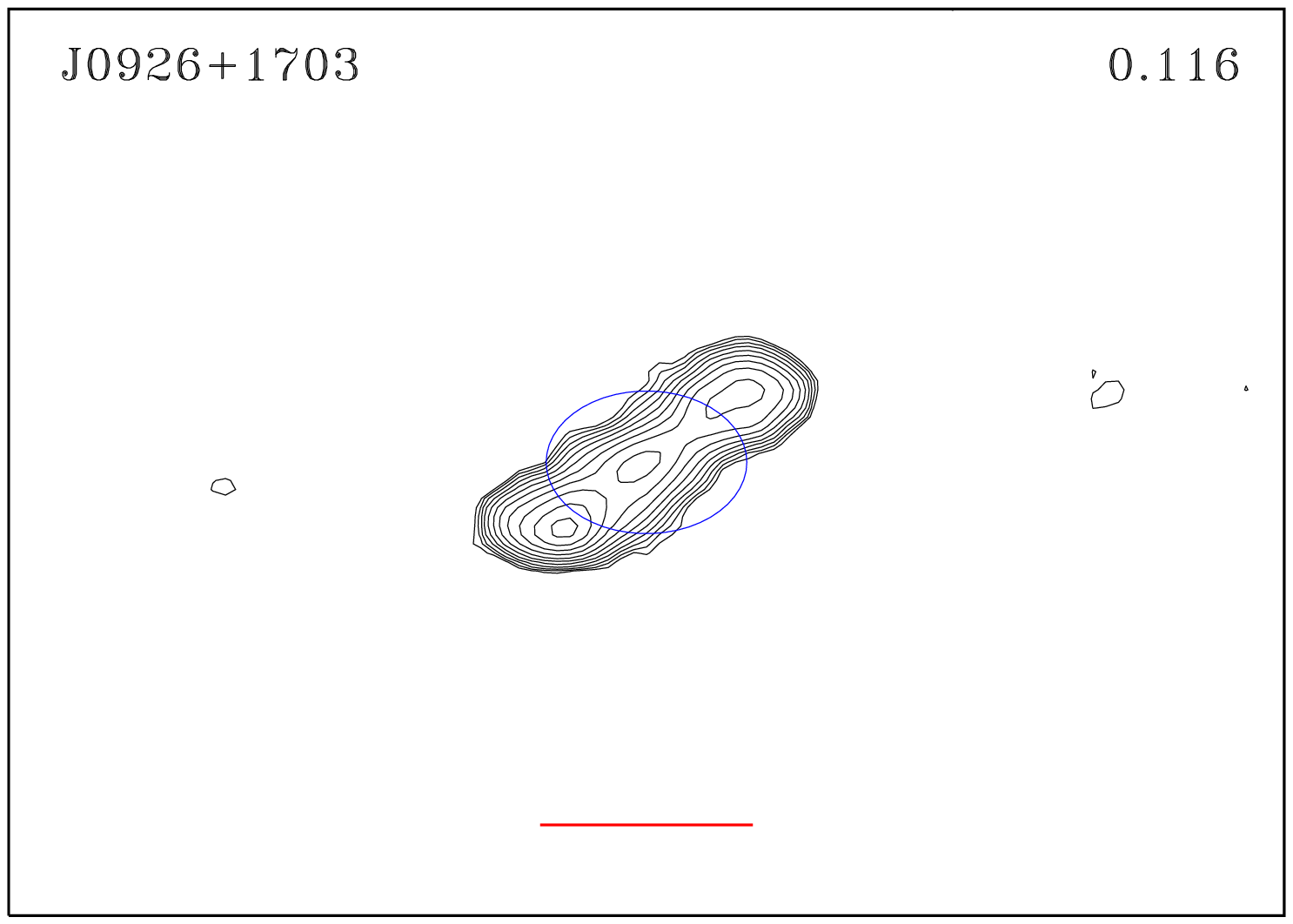}
\includegraphics[width=6.3cm,height=6.3cm]{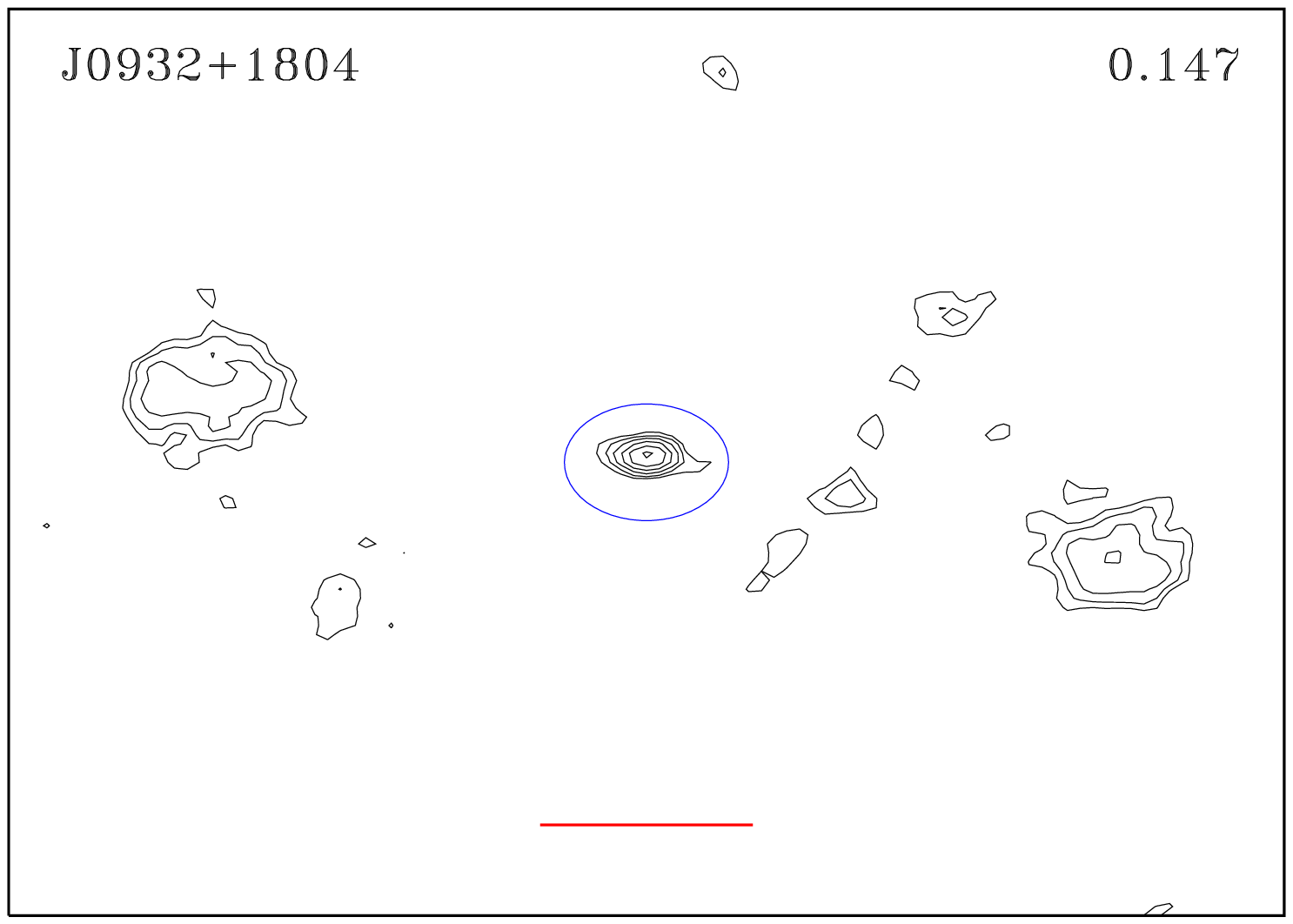}

\includegraphics[width=6.3cm,height=6.3cm]{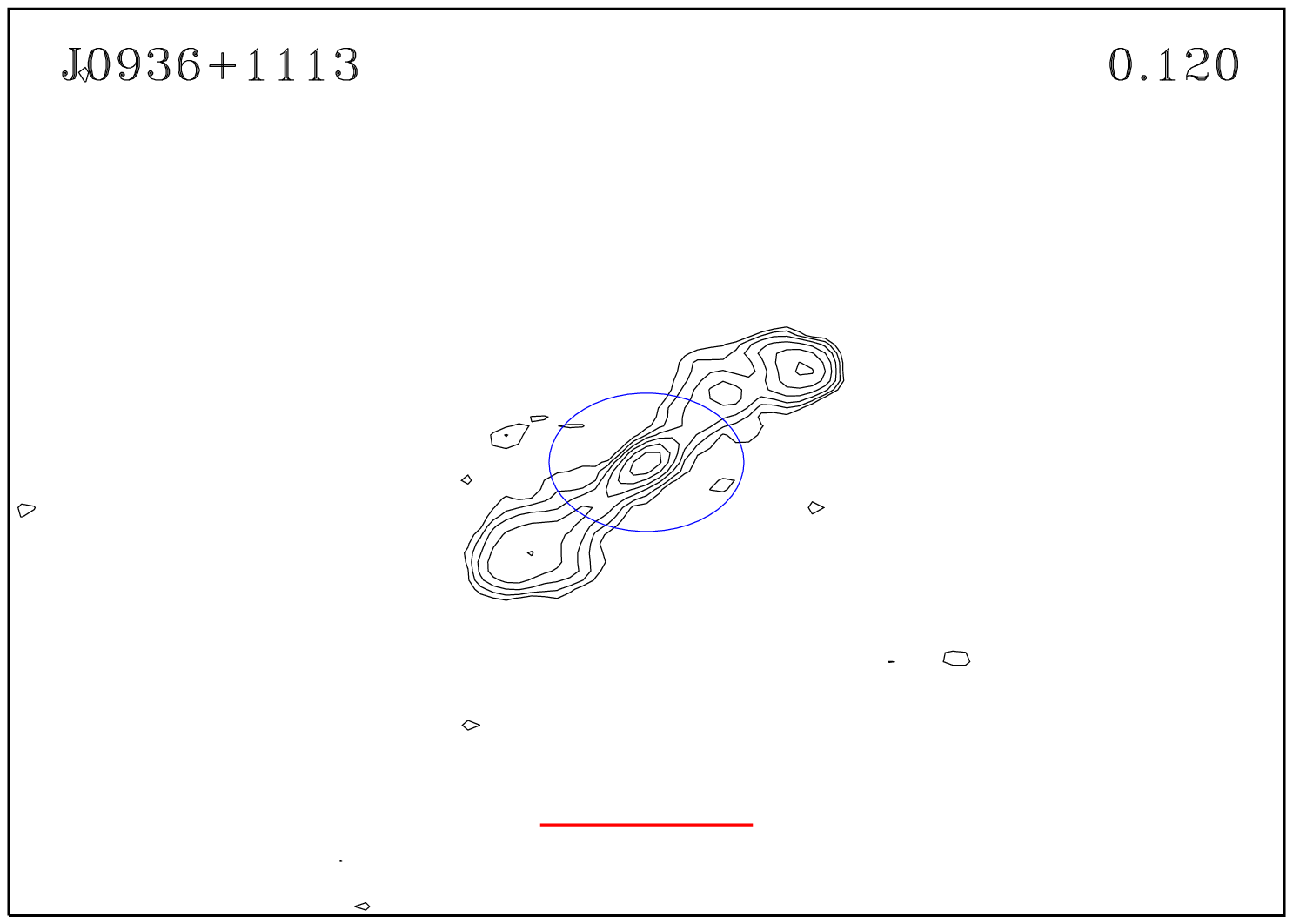}
\includegraphics[width=6.3cm,height=6.3cm]{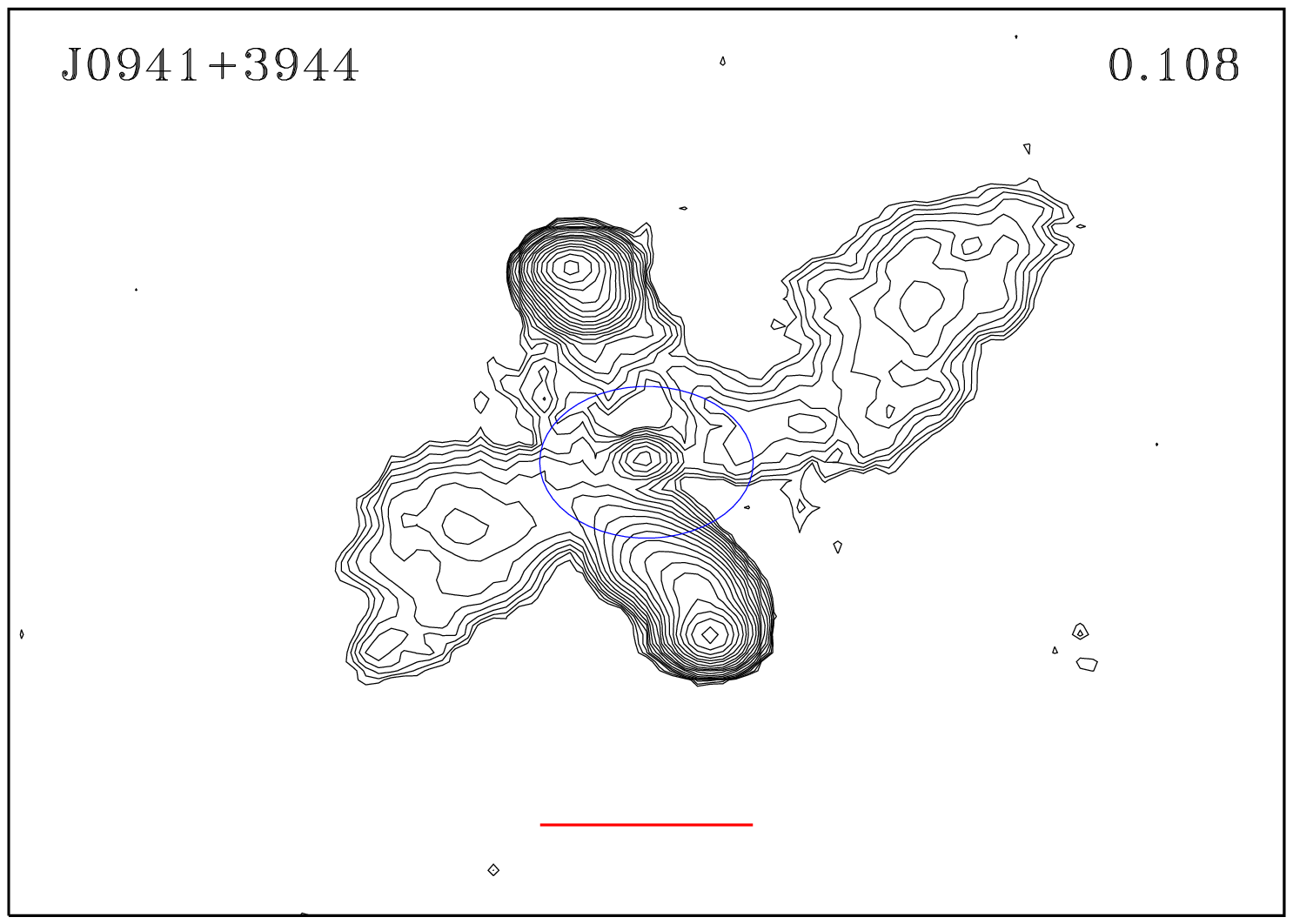}
\includegraphics[width=6.3cm,height=6.3cm]{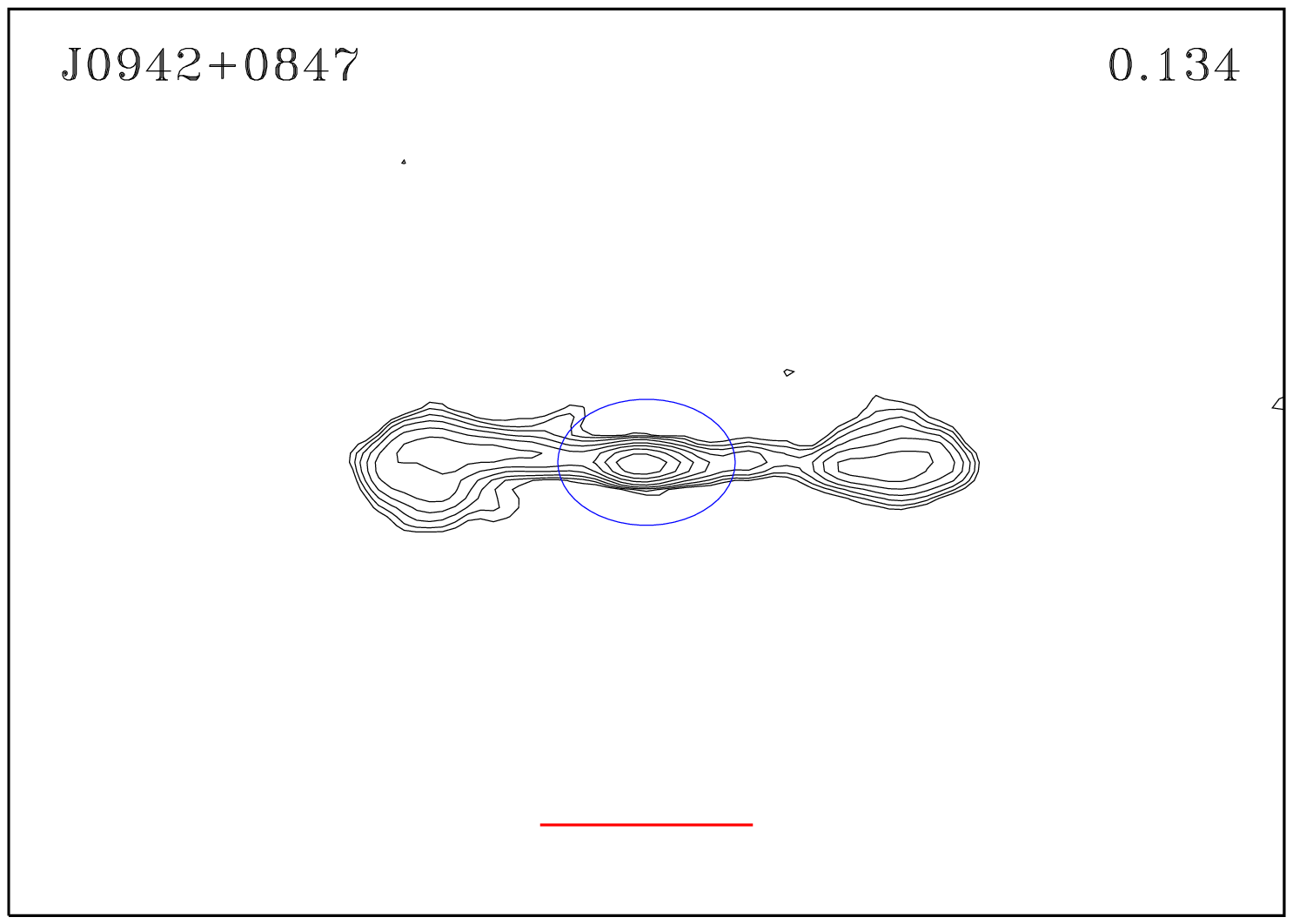}

\includegraphics[width=6.3cm,height=6.3cm]{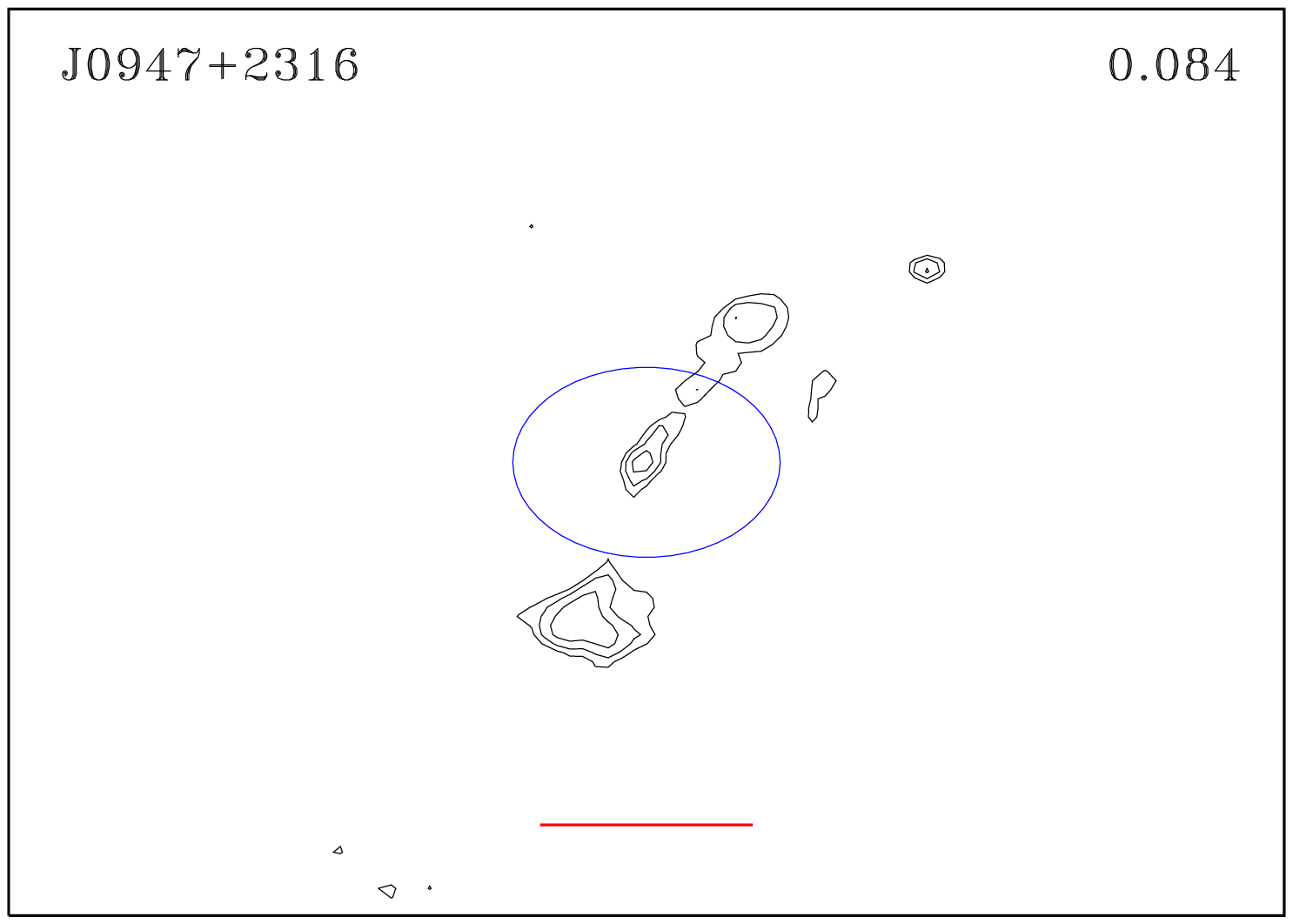}
\includegraphics[width=6.3cm,height=6.3cm]{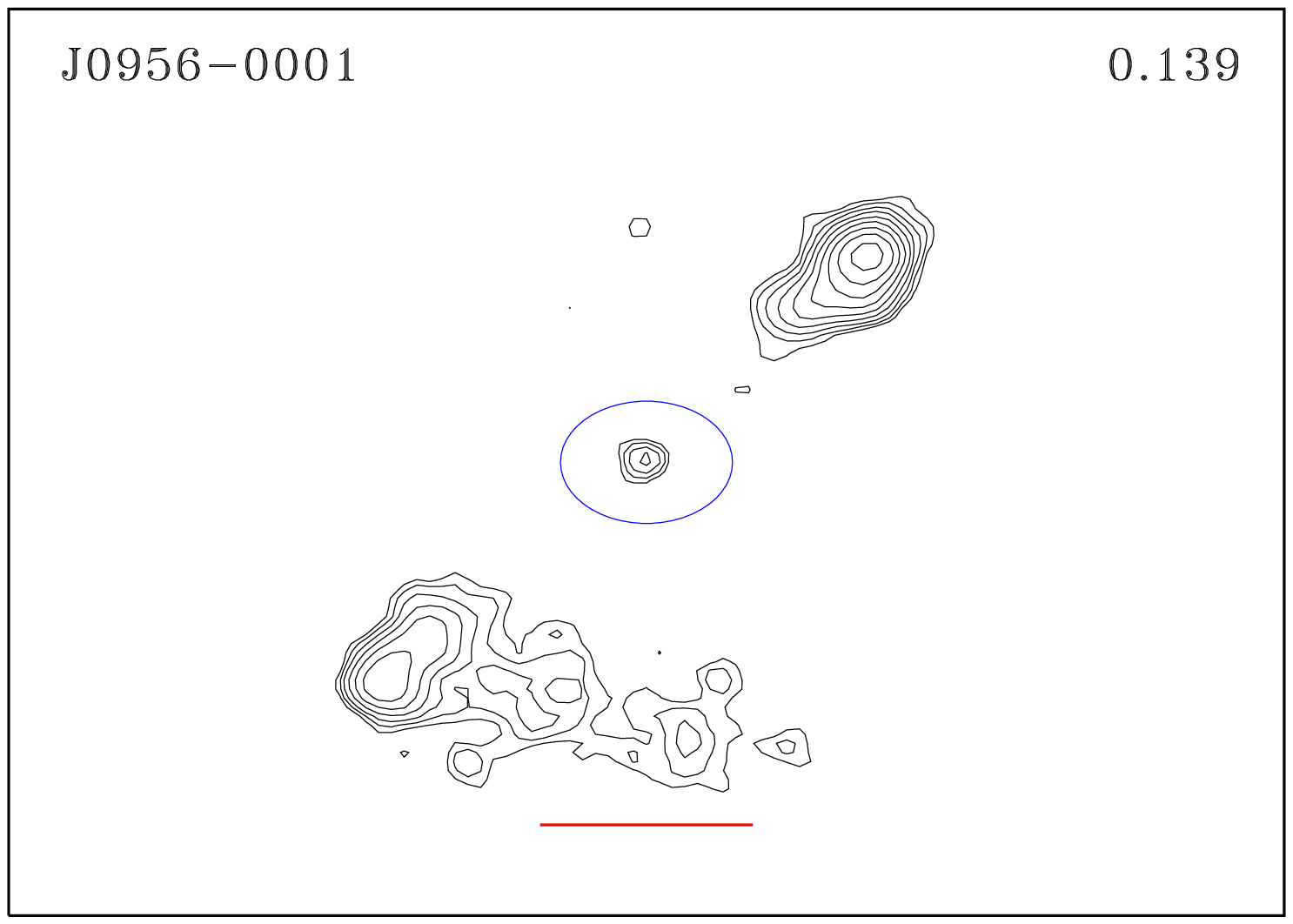}
\includegraphics[width=6.3cm,height=6.3cm]{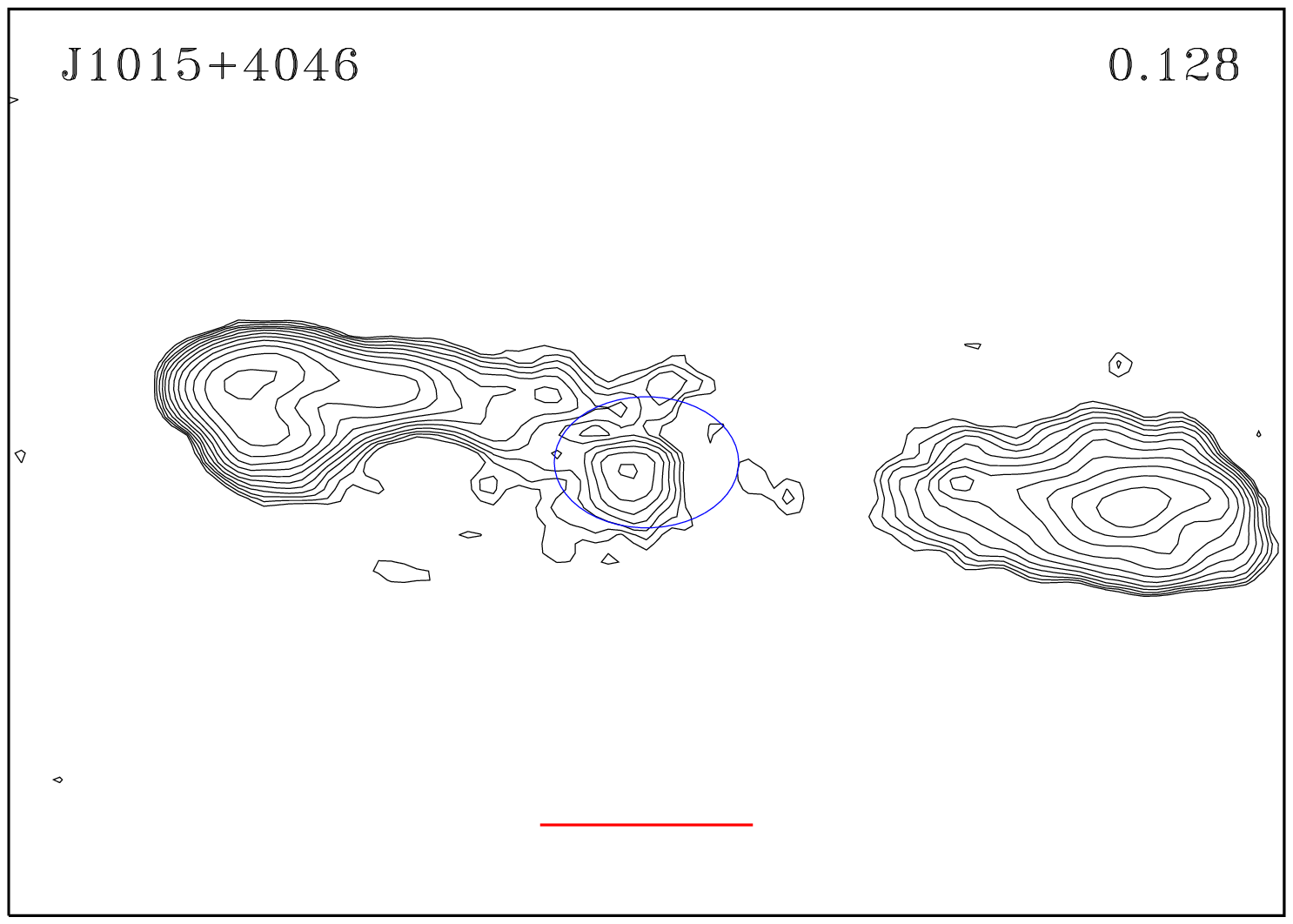}

\includegraphics[width=6.3cm,height=6.3cm]{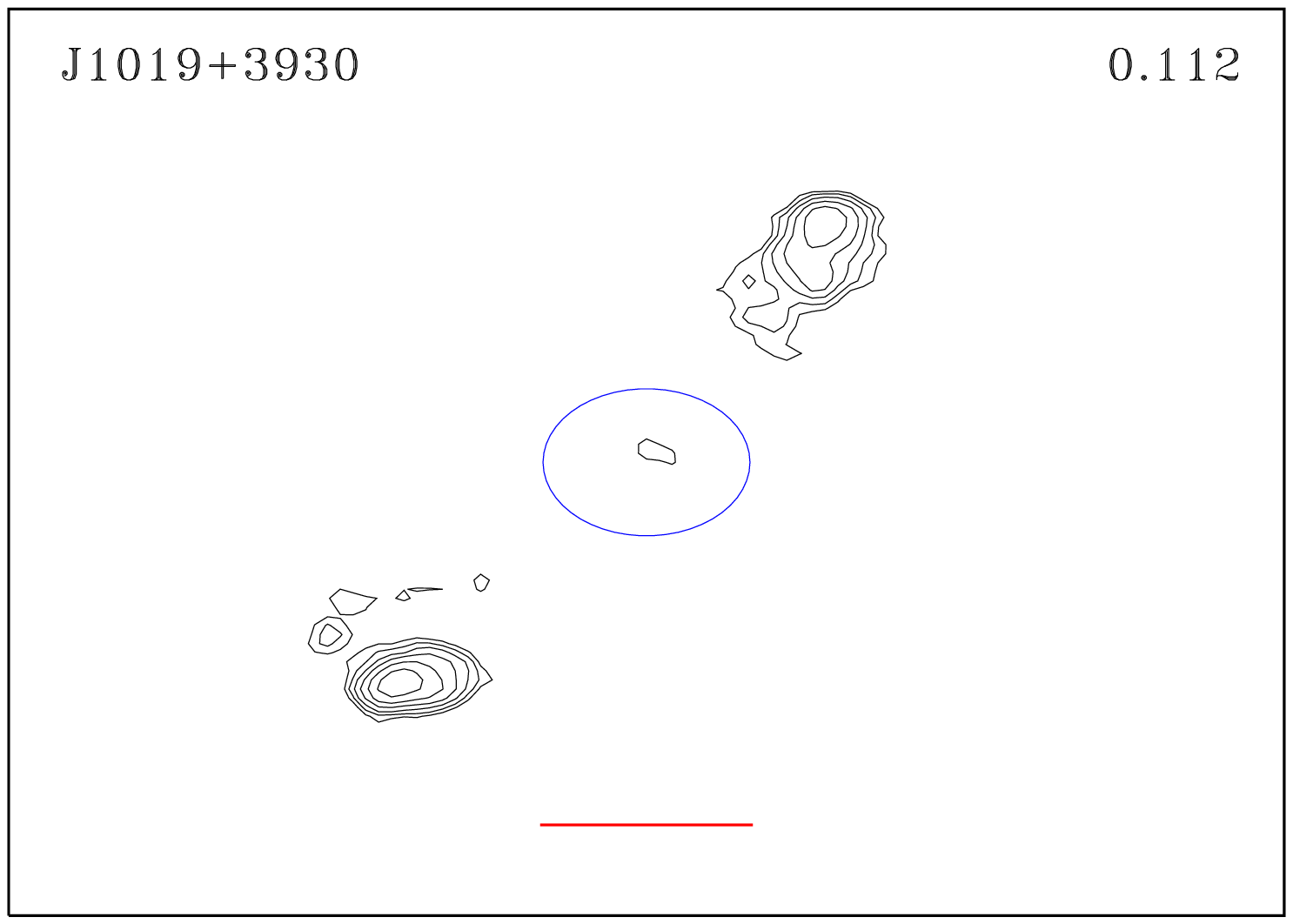}
\includegraphics[width=6.3cm,height=6.3cm]{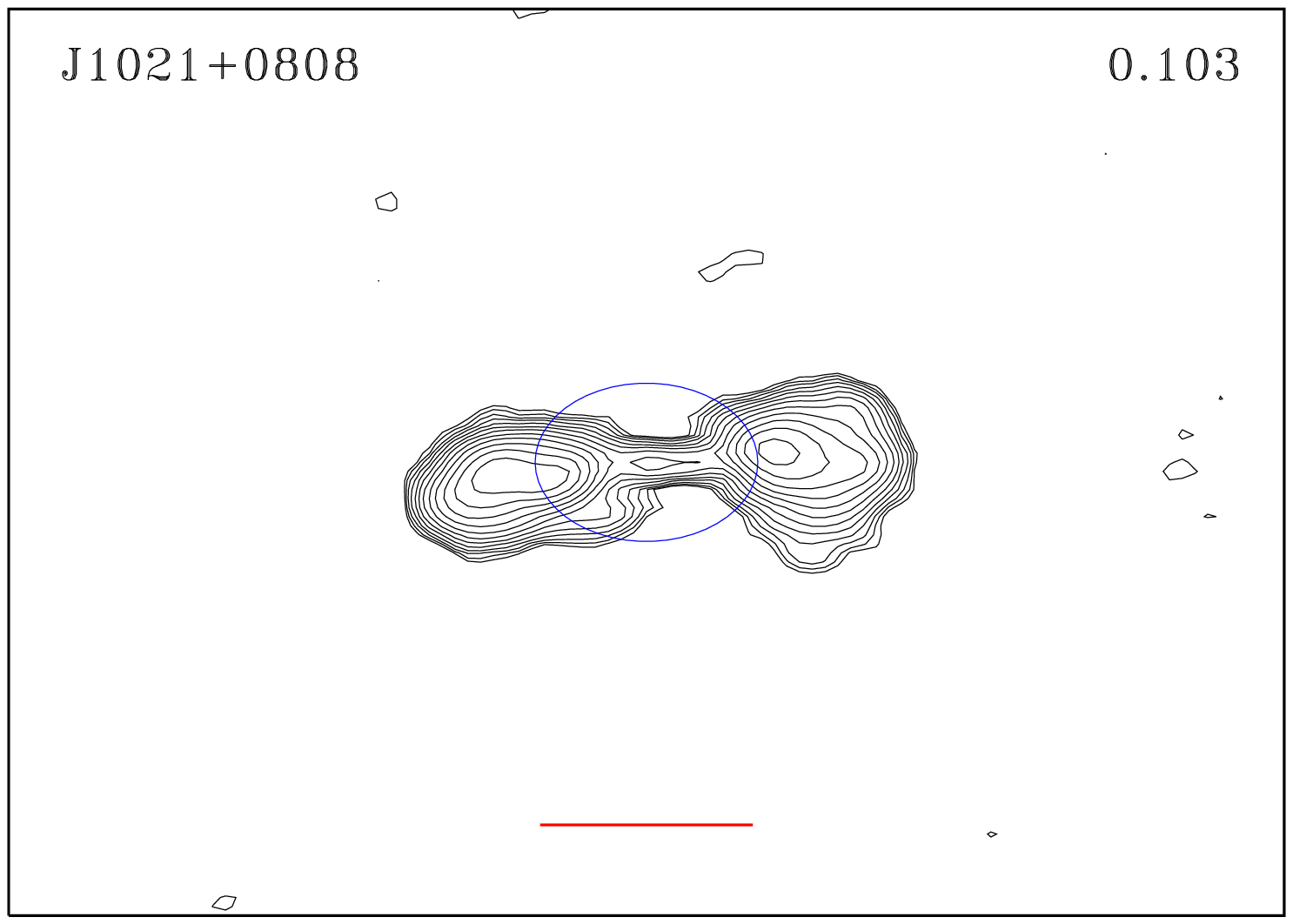}
\includegraphics[width=6.3cm,height=6.3cm]{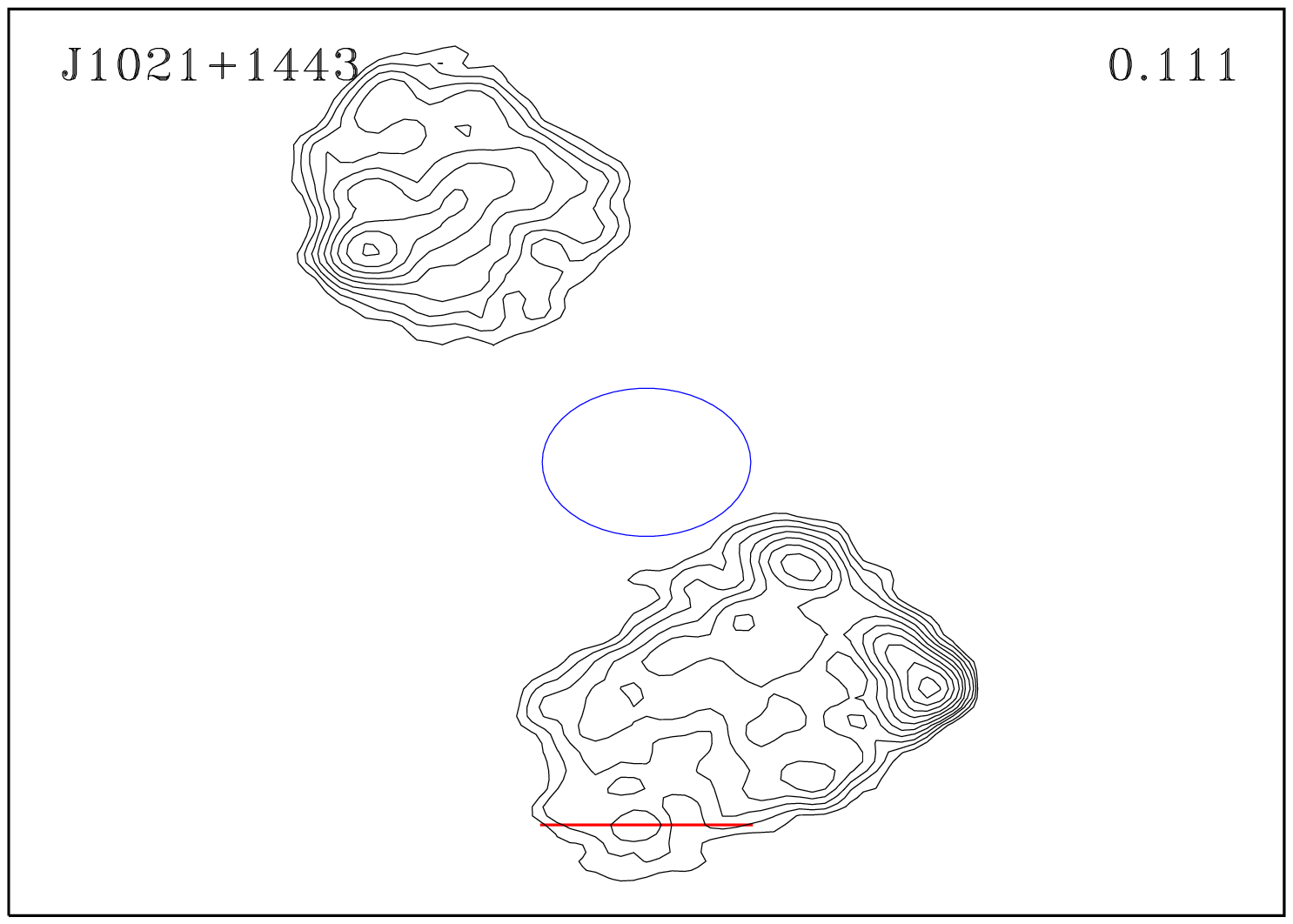}
\caption{(continued)}
\end{figure*}

\addtocounter{figure}{-1}
\begin{figure*}
\includegraphics[width=6.3cm,height=6.3cm]{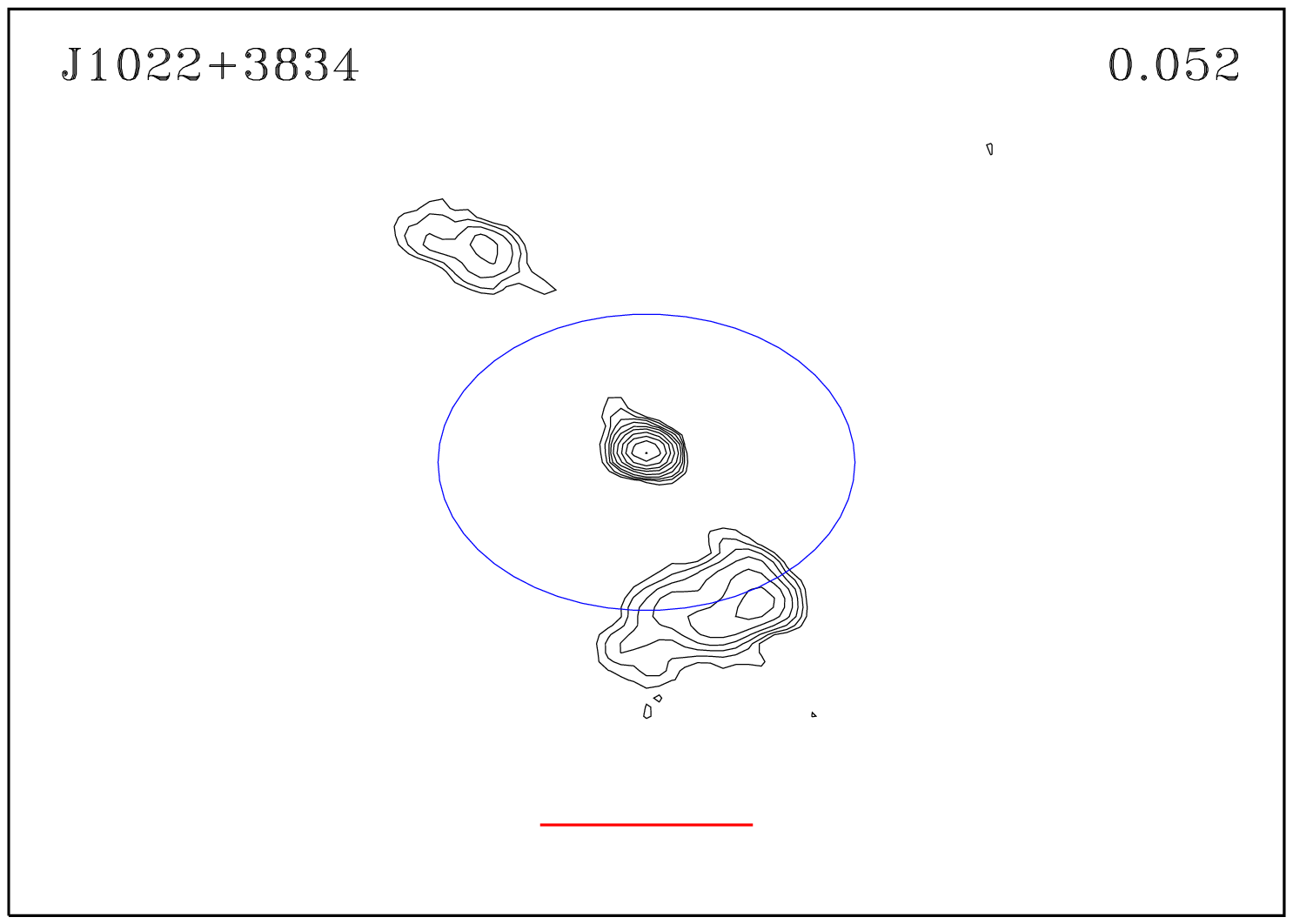}
\includegraphics[width=6.3cm,height=6.3cm]{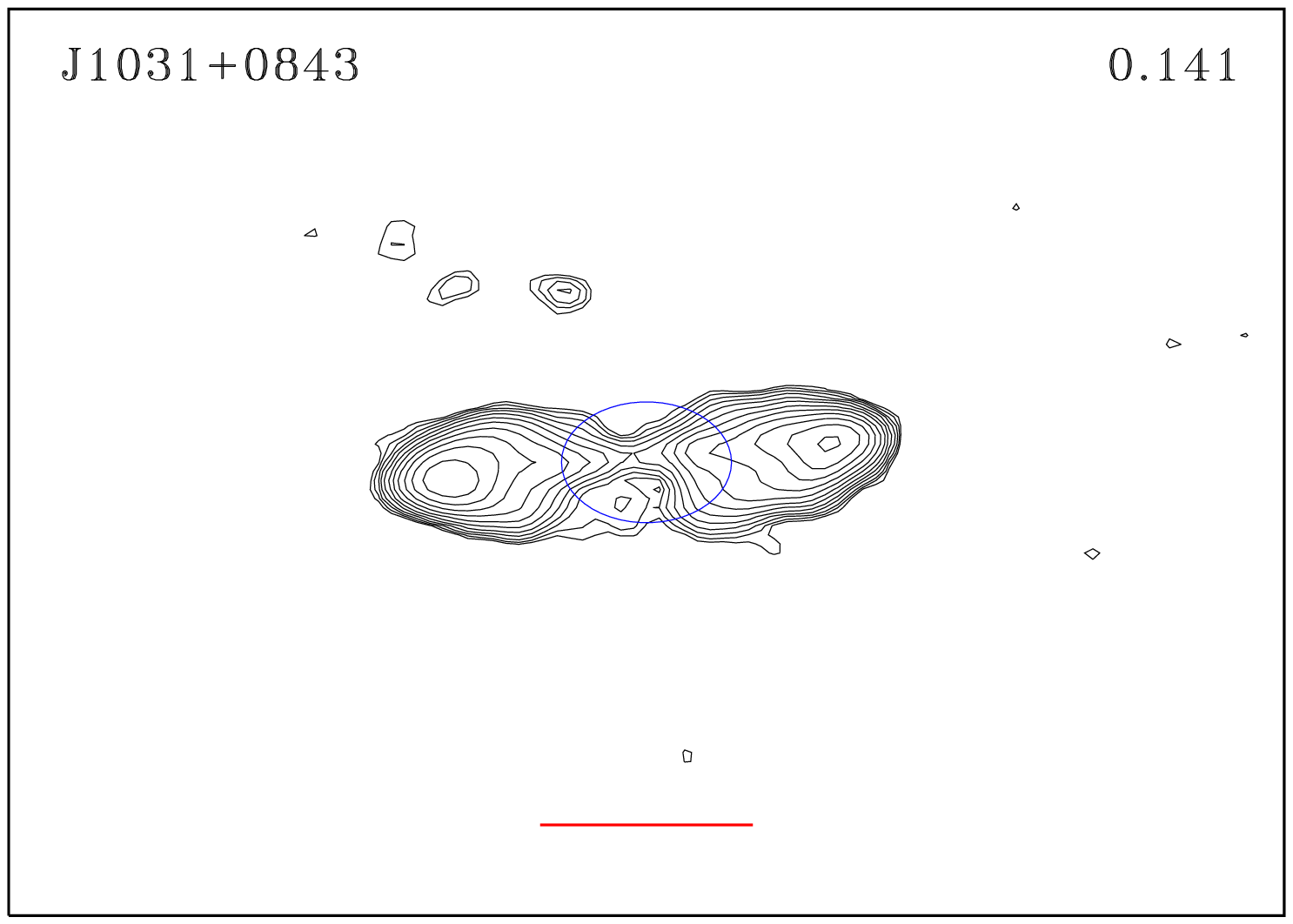}
\includegraphics[width=6.3cm,height=6.3cm]{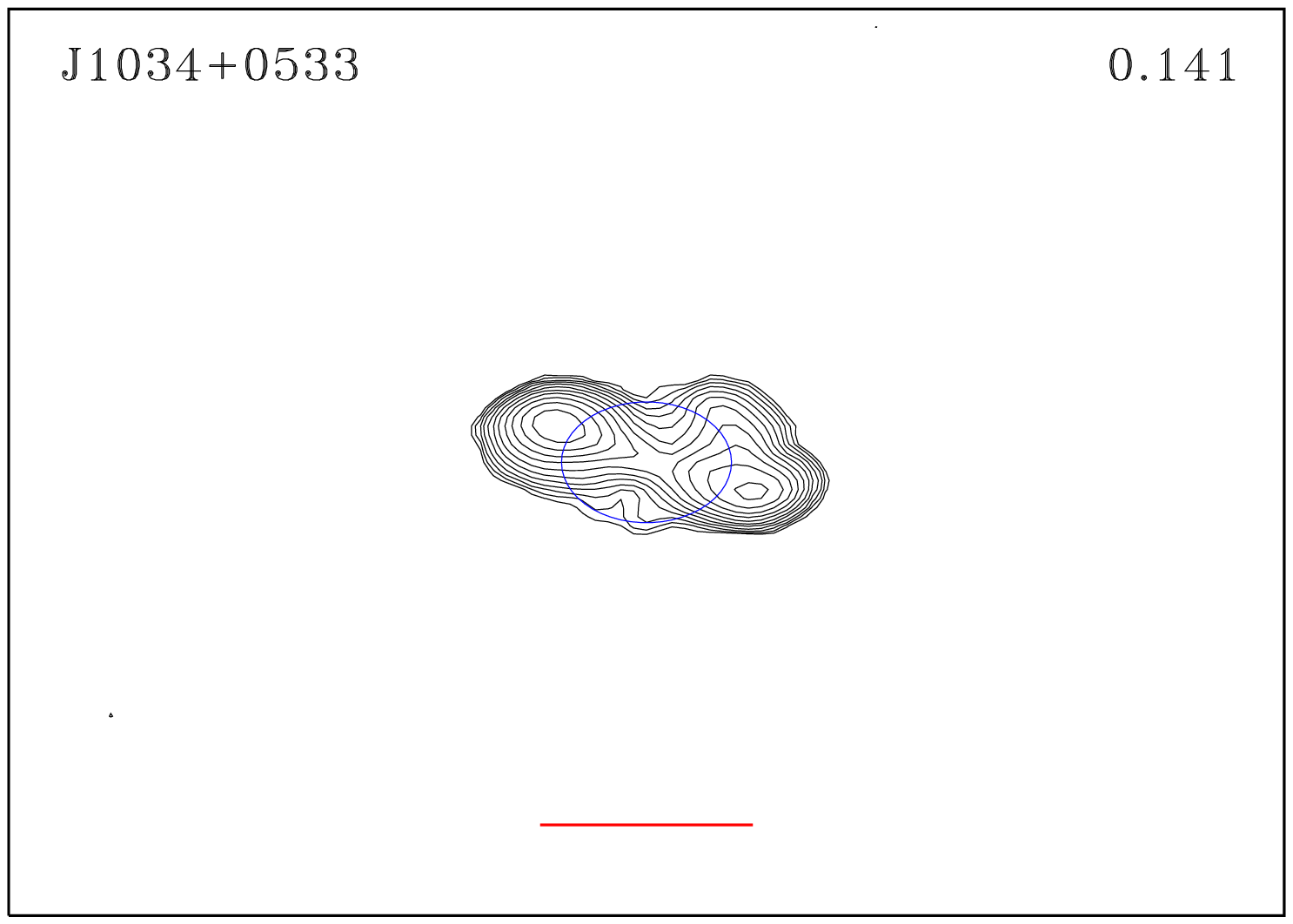}

\includegraphics[width=6.3cm,height=6.3cm]{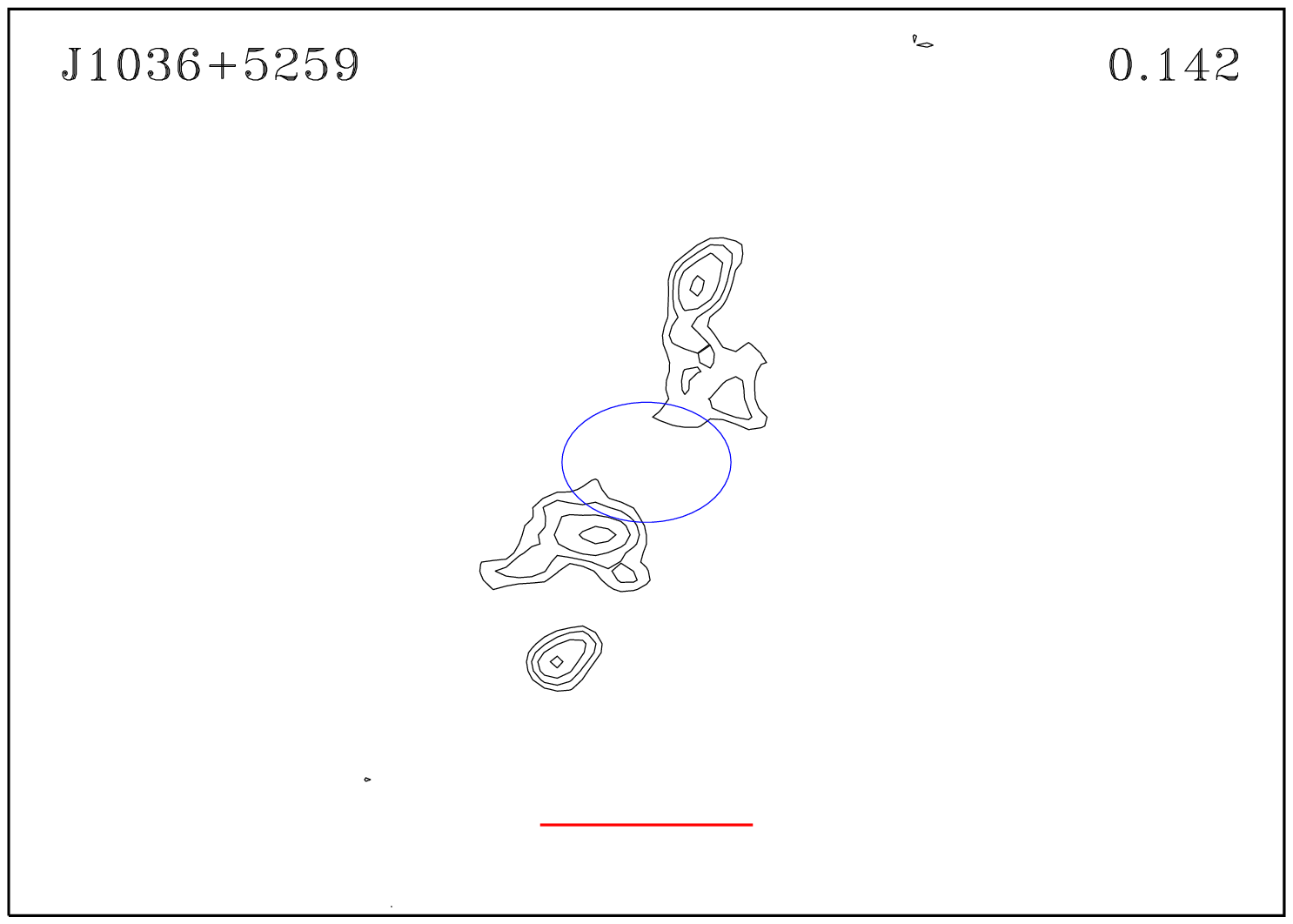}
\includegraphics[width=6.3cm,height=6.3cm]{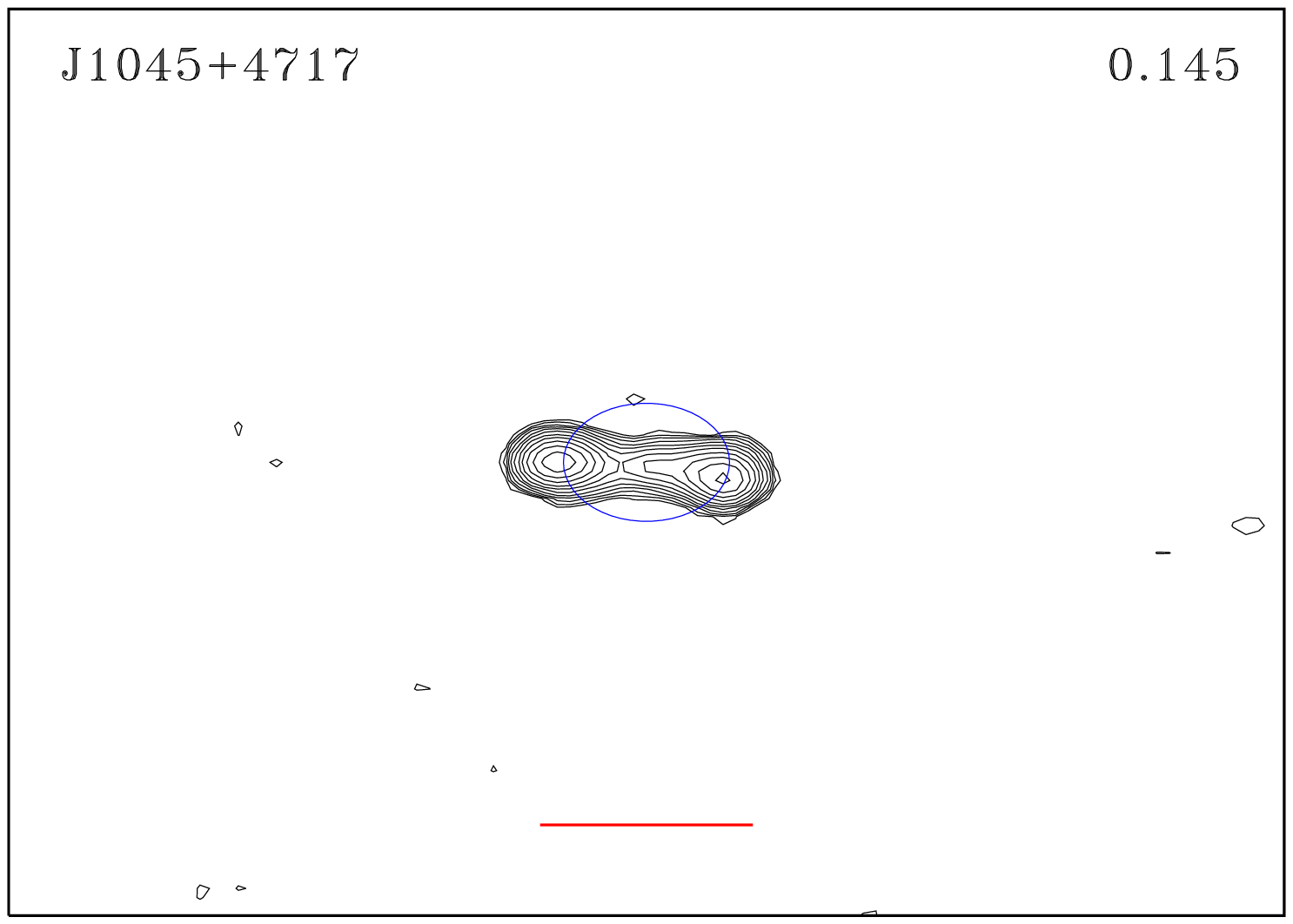}
\includegraphics[width=6.3cm,height=6.3cm]{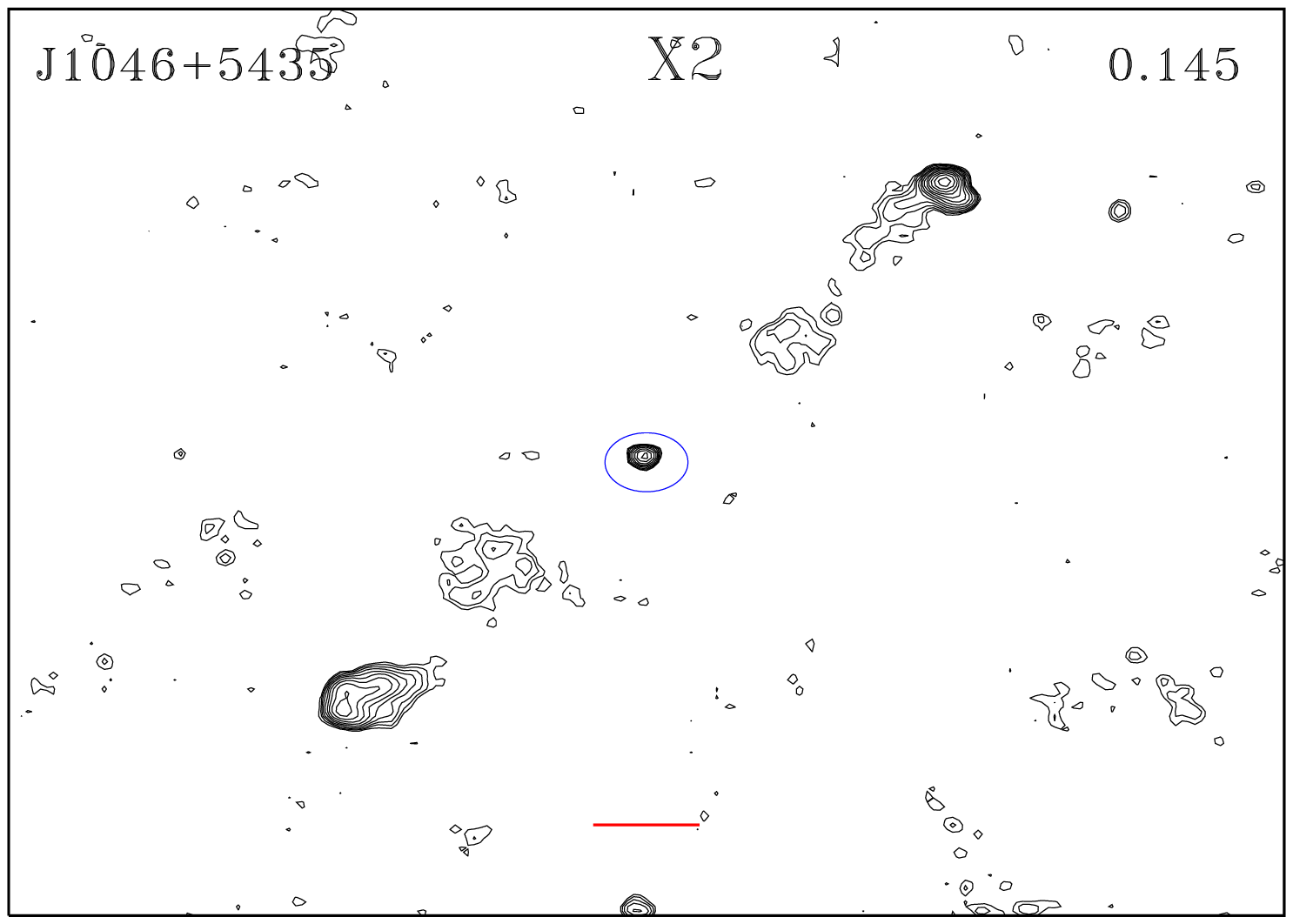}

\includegraphics[width=6.3cm,height=6.3cm]{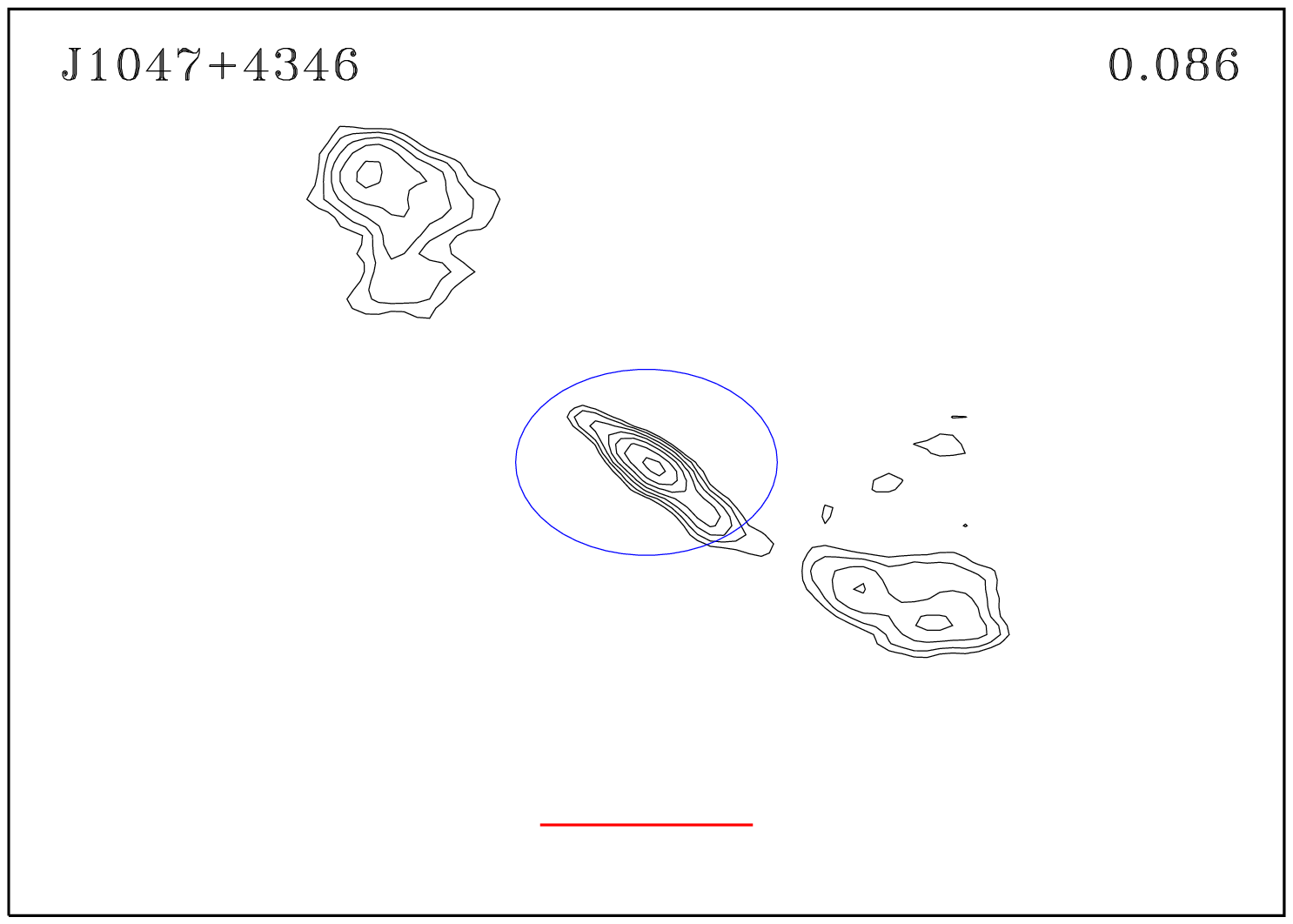}
\includegraphics[width=6.3cm,height=6.3cm]{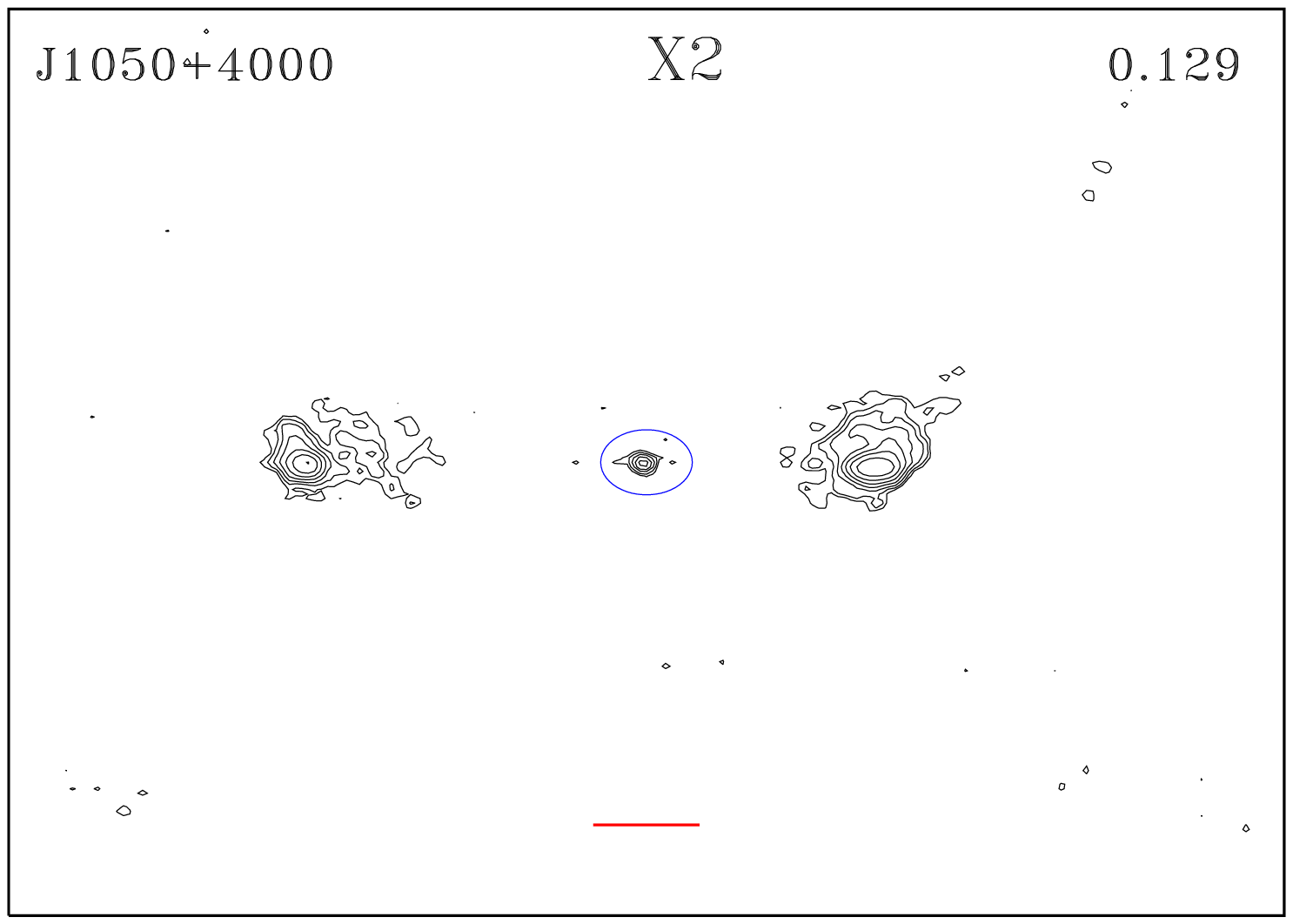}
\includegraphics[width=6.3cm,height=6.3cm]{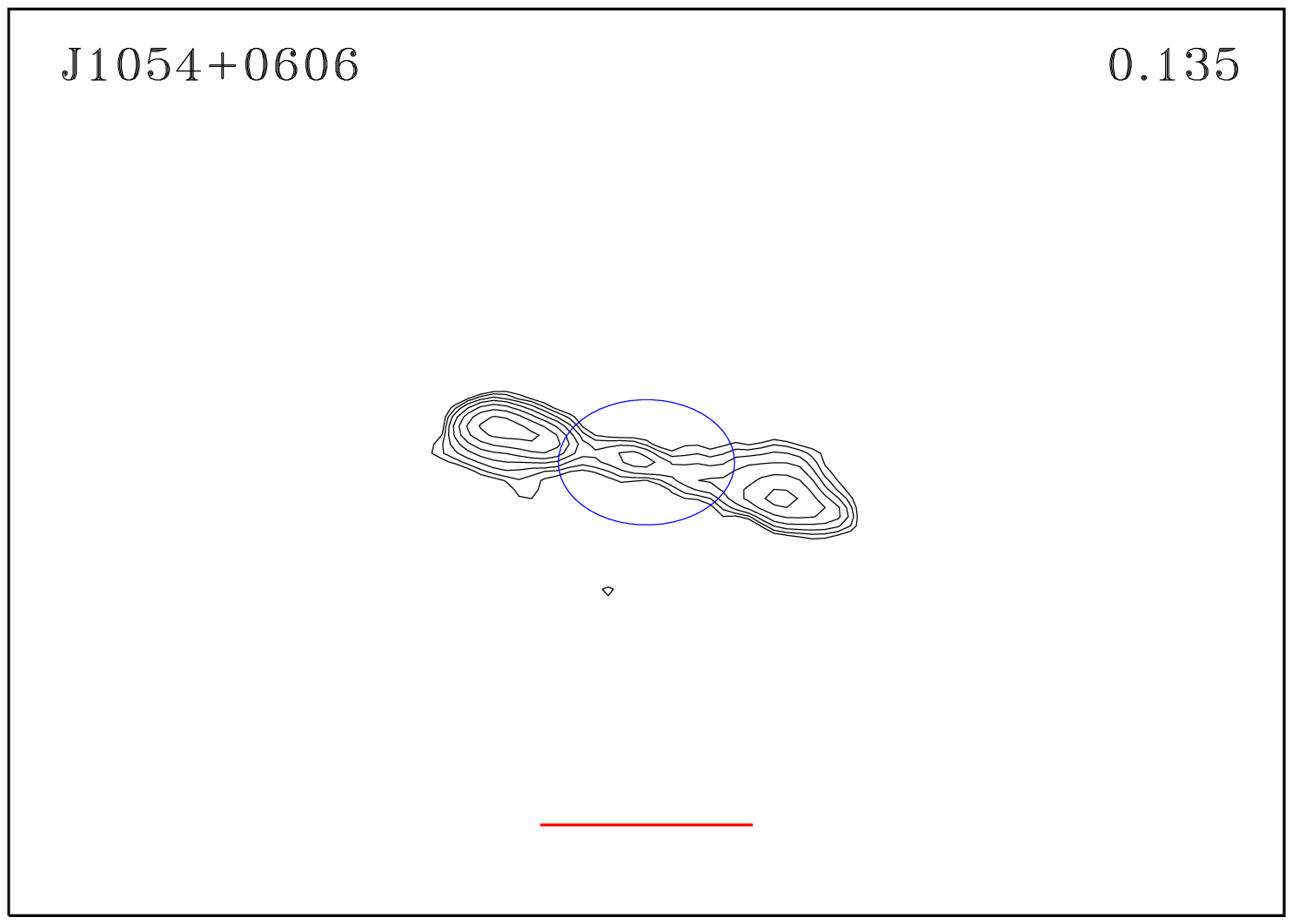}

\includegraphics[width=6.3cm,height=6.3cm]{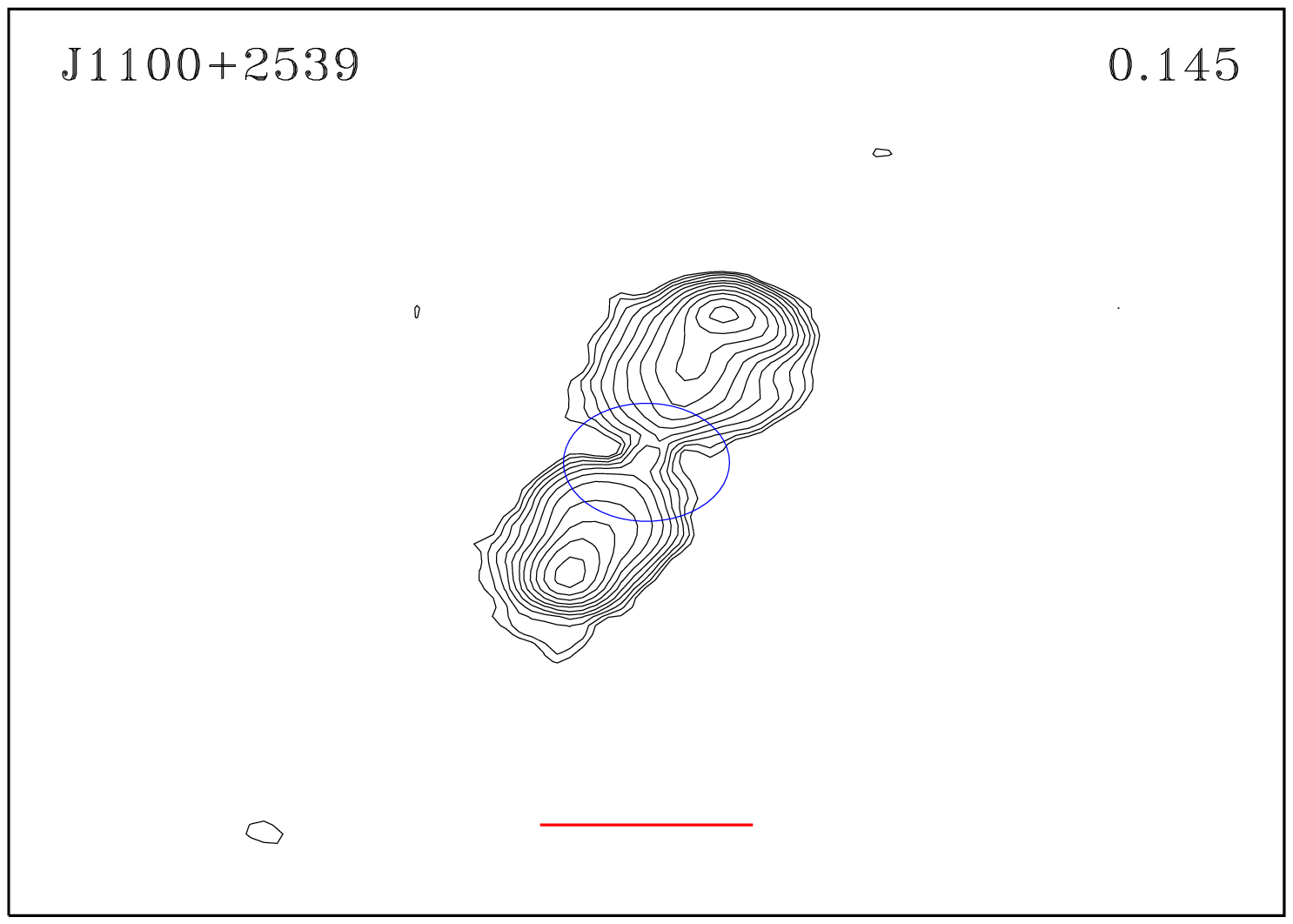}
\includegraphics[width=6.3cm,height=6.3cm]{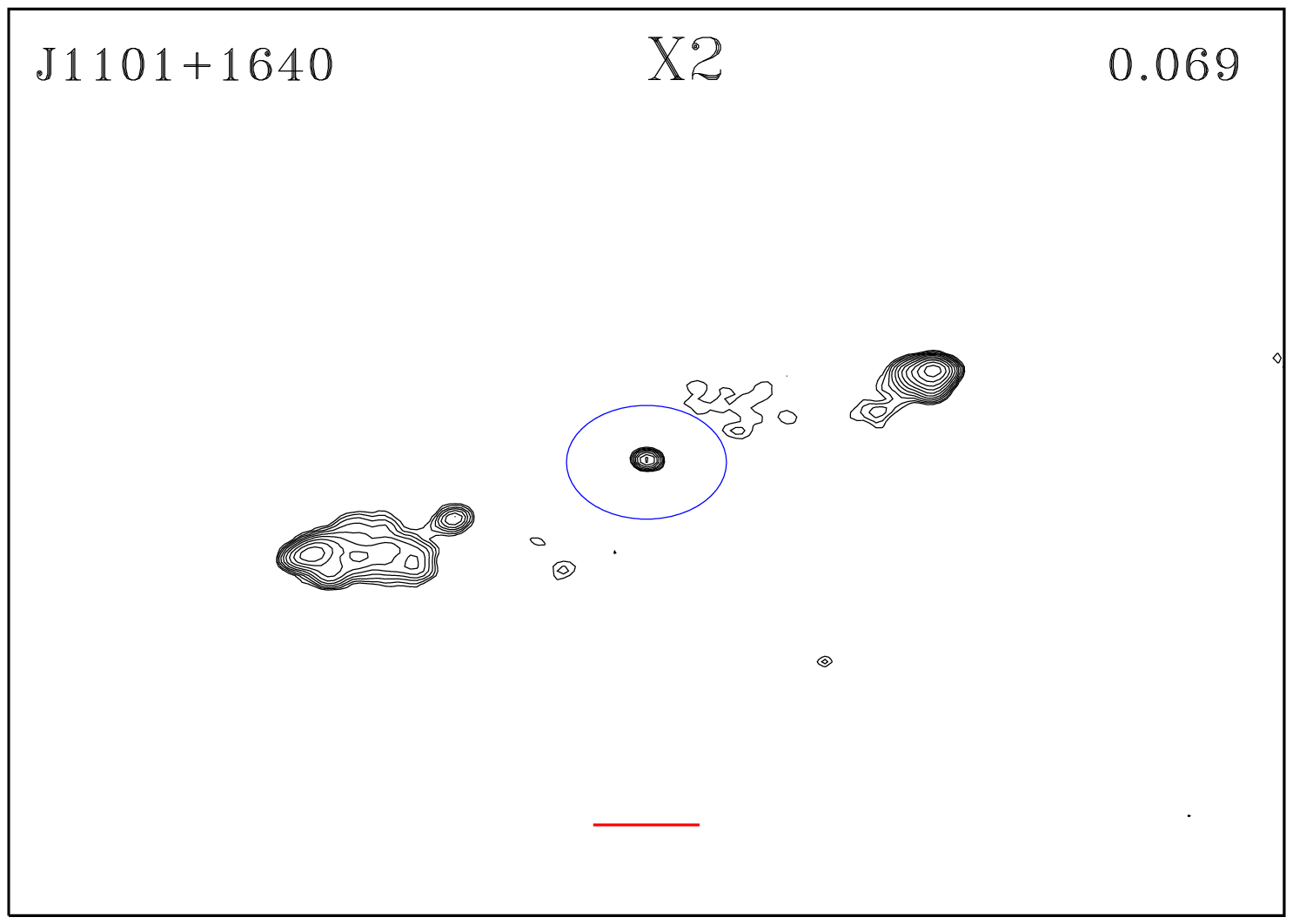}
\includegraphics[width=6.3cm,height=6.3cm]{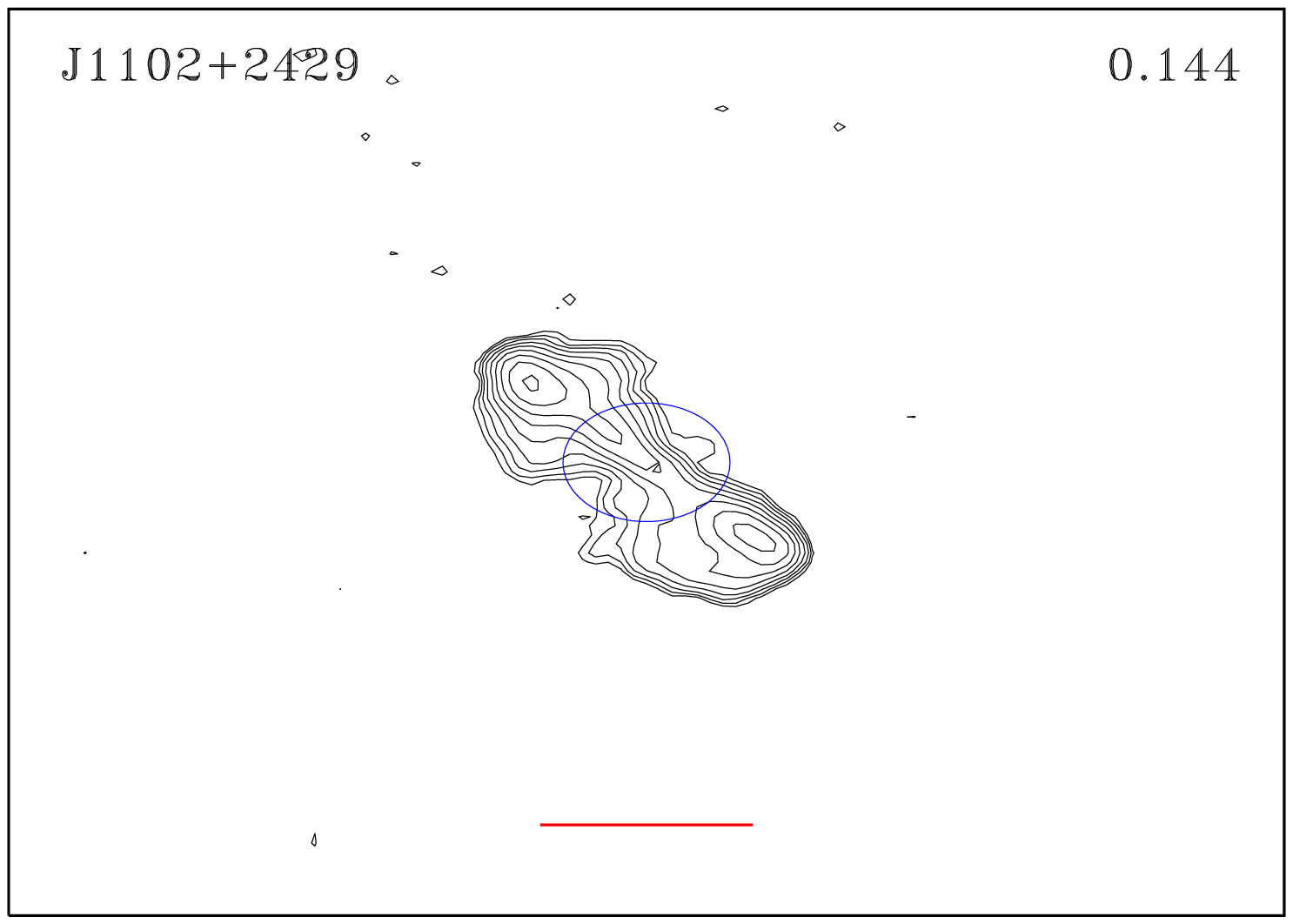}
\caption{(continued)}
\end{figure*}

\addtocounter{figure}{-1}
\begin{figure*}
\includegraphics[width=6.3cm,height=6.3cm]{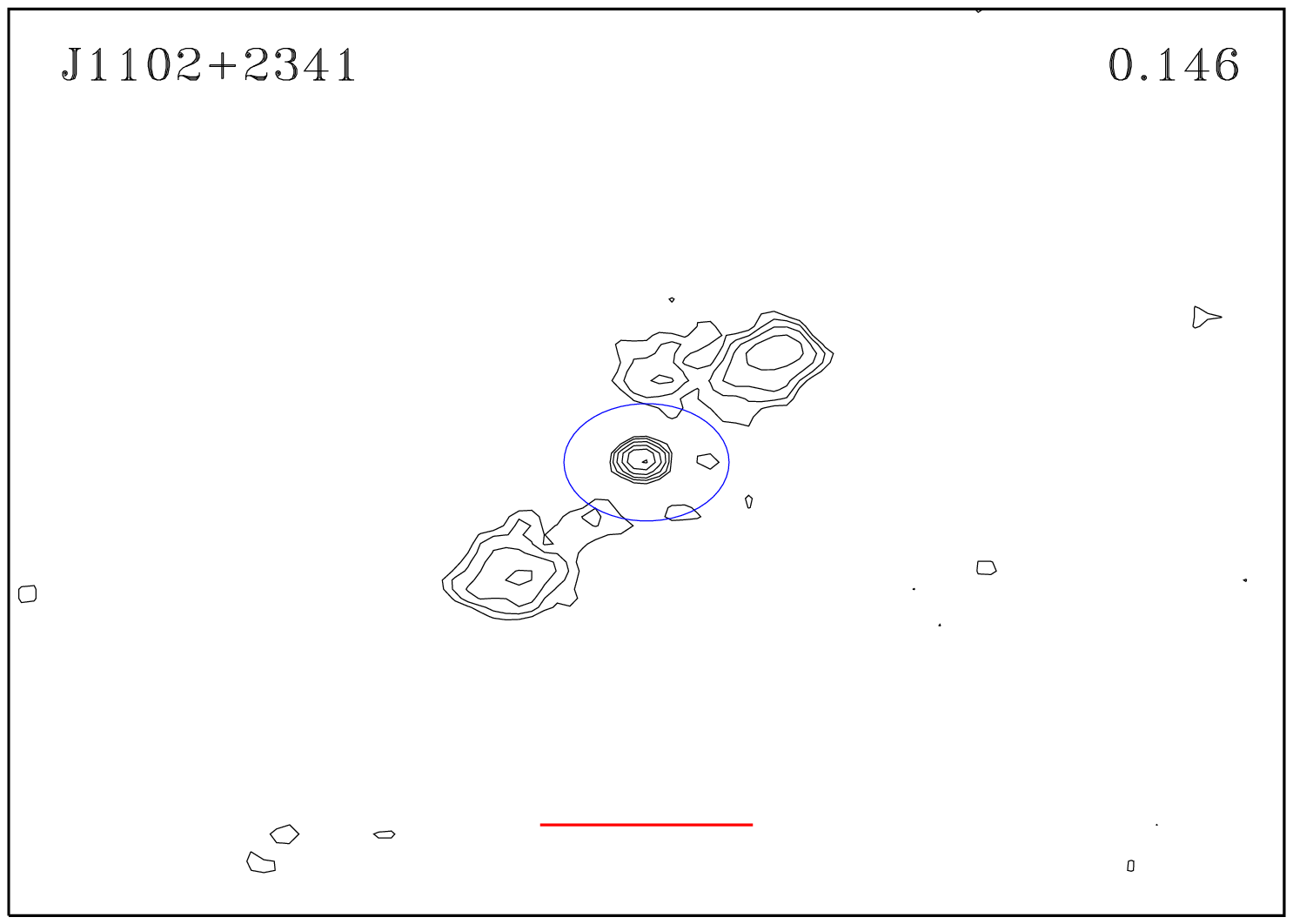}
\includegraphics[width=6.3cm,height=6.3cm]{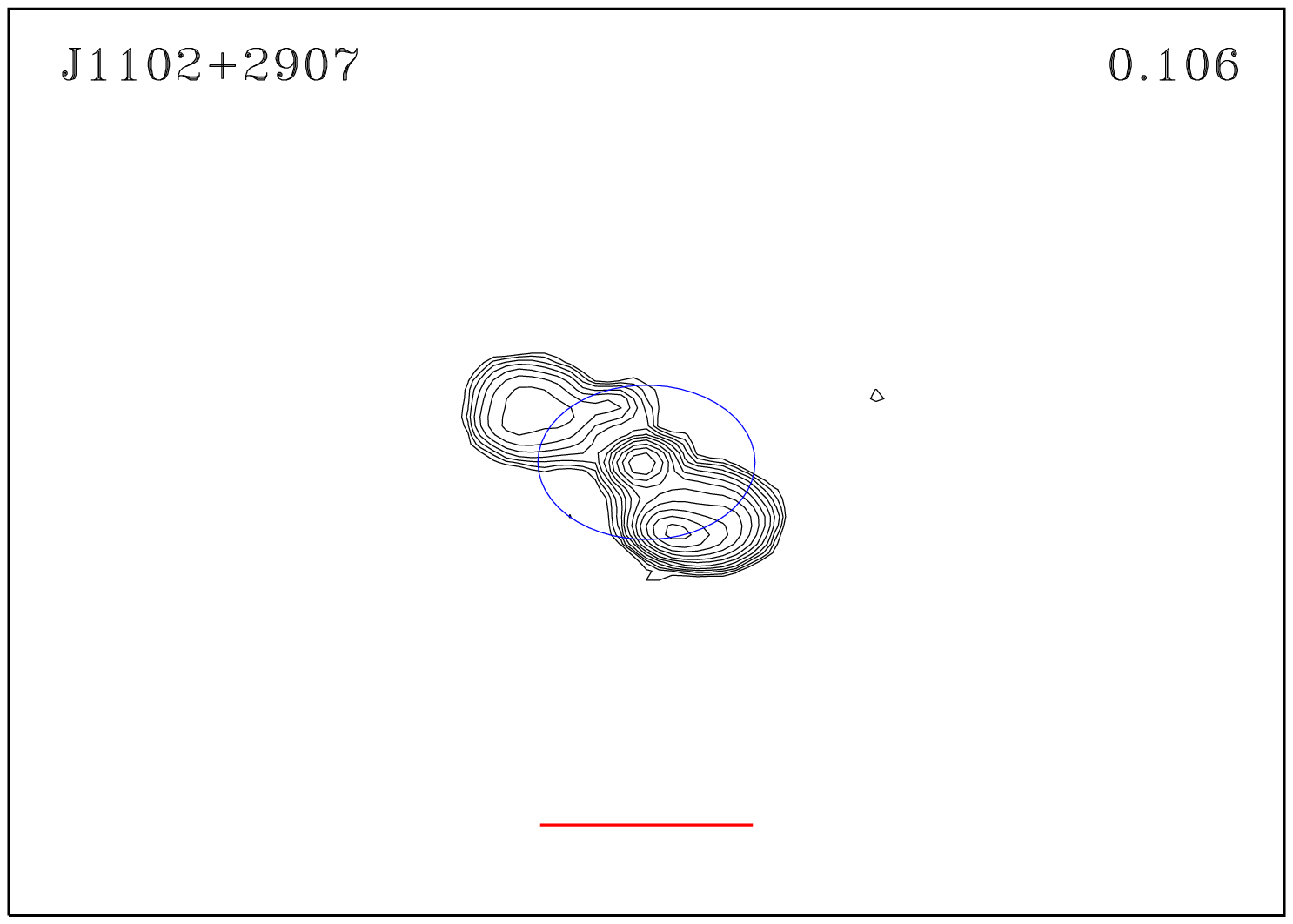}
\includegraphics[width=6.3cm,height=6.3cm]{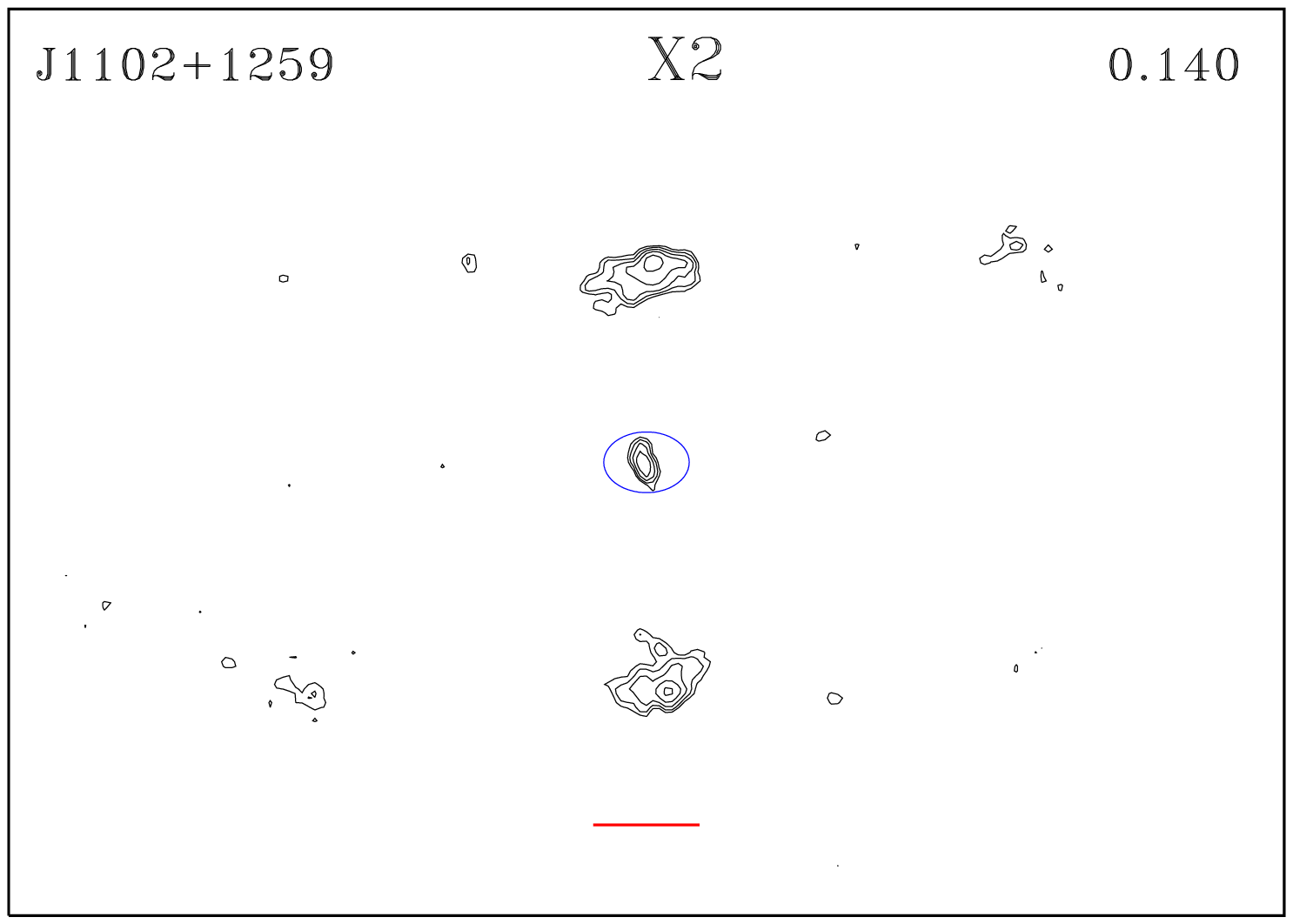}

\includegraphics[width=6.3cm,height=6.3cm]{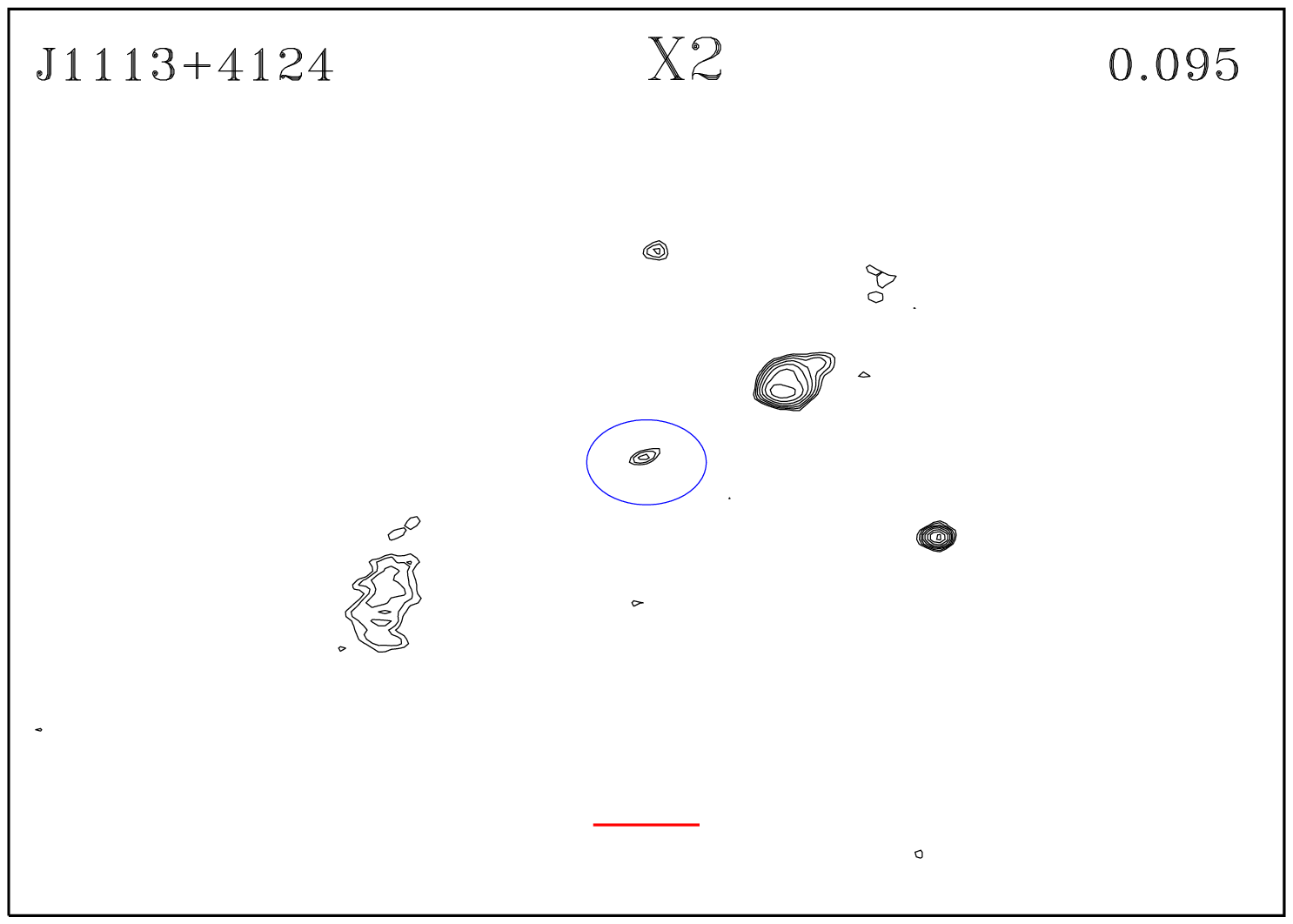}
\includegraphics[width=6.3cm,height=6.3cm]{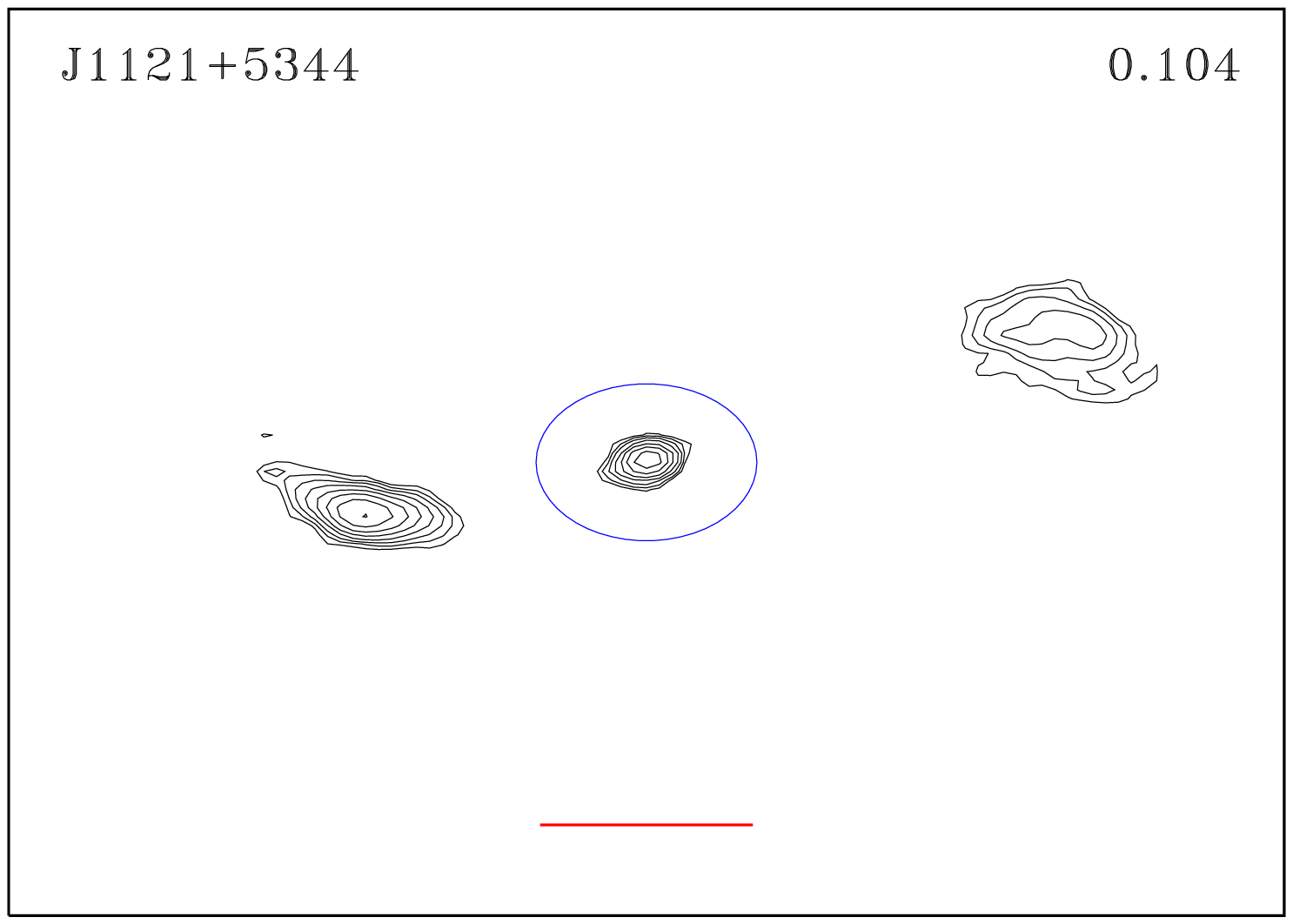}
\includegraphics[width=6.3cm,height=6.3cm]{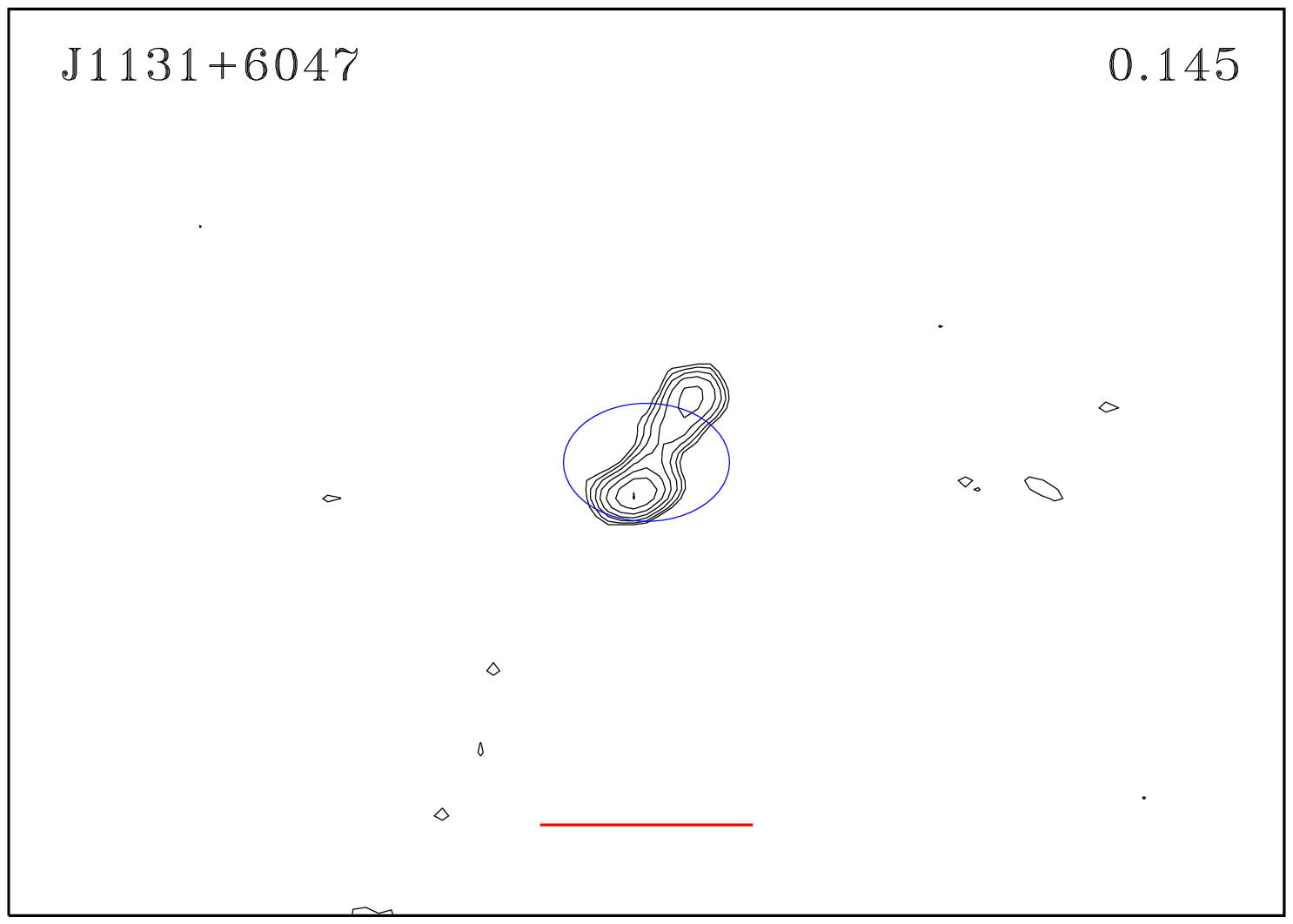}

\includegraphics[width=6.3cm,height=6.3cm]{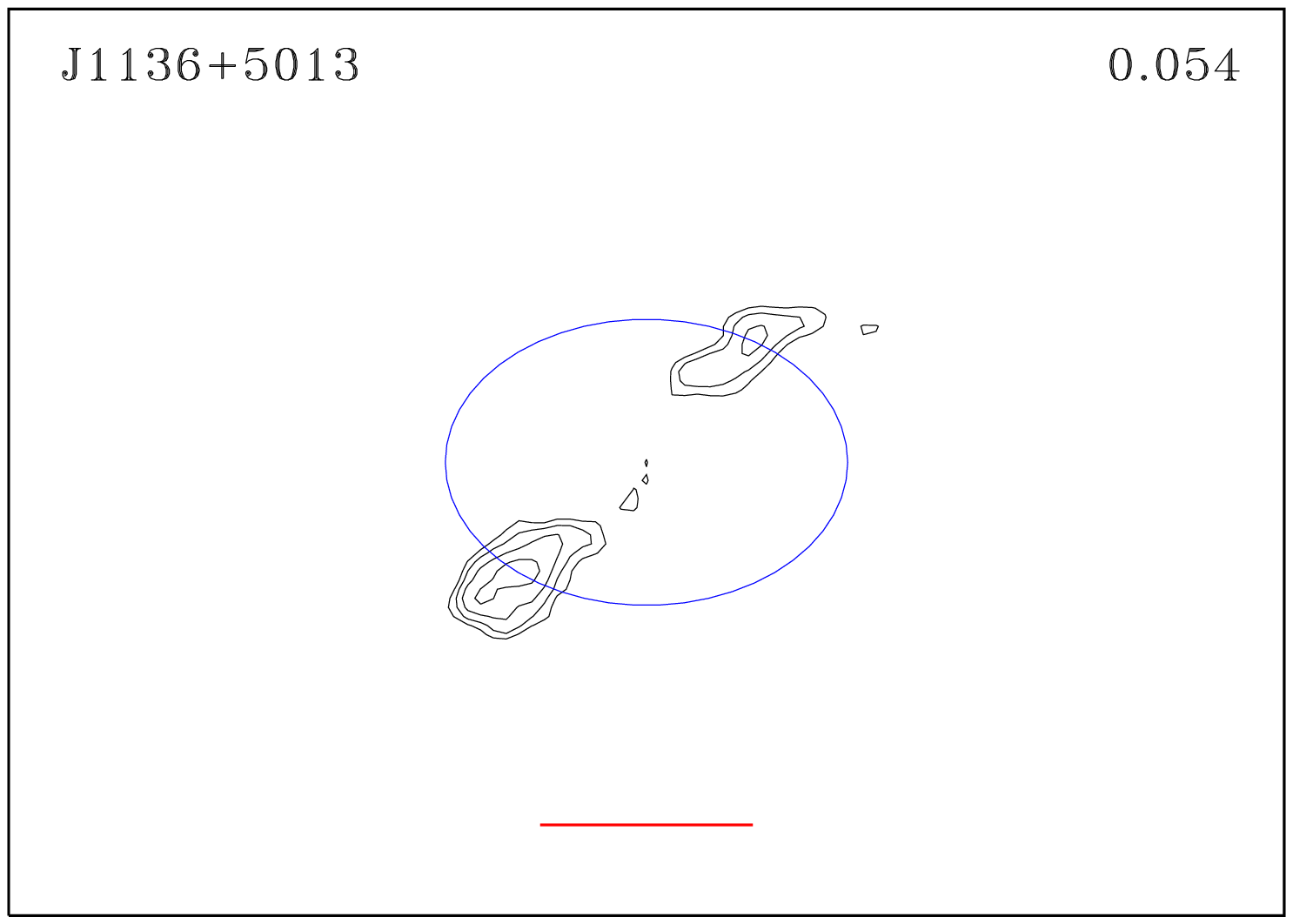}
\includegraphics[width=6.3cm,height=6.3cm]{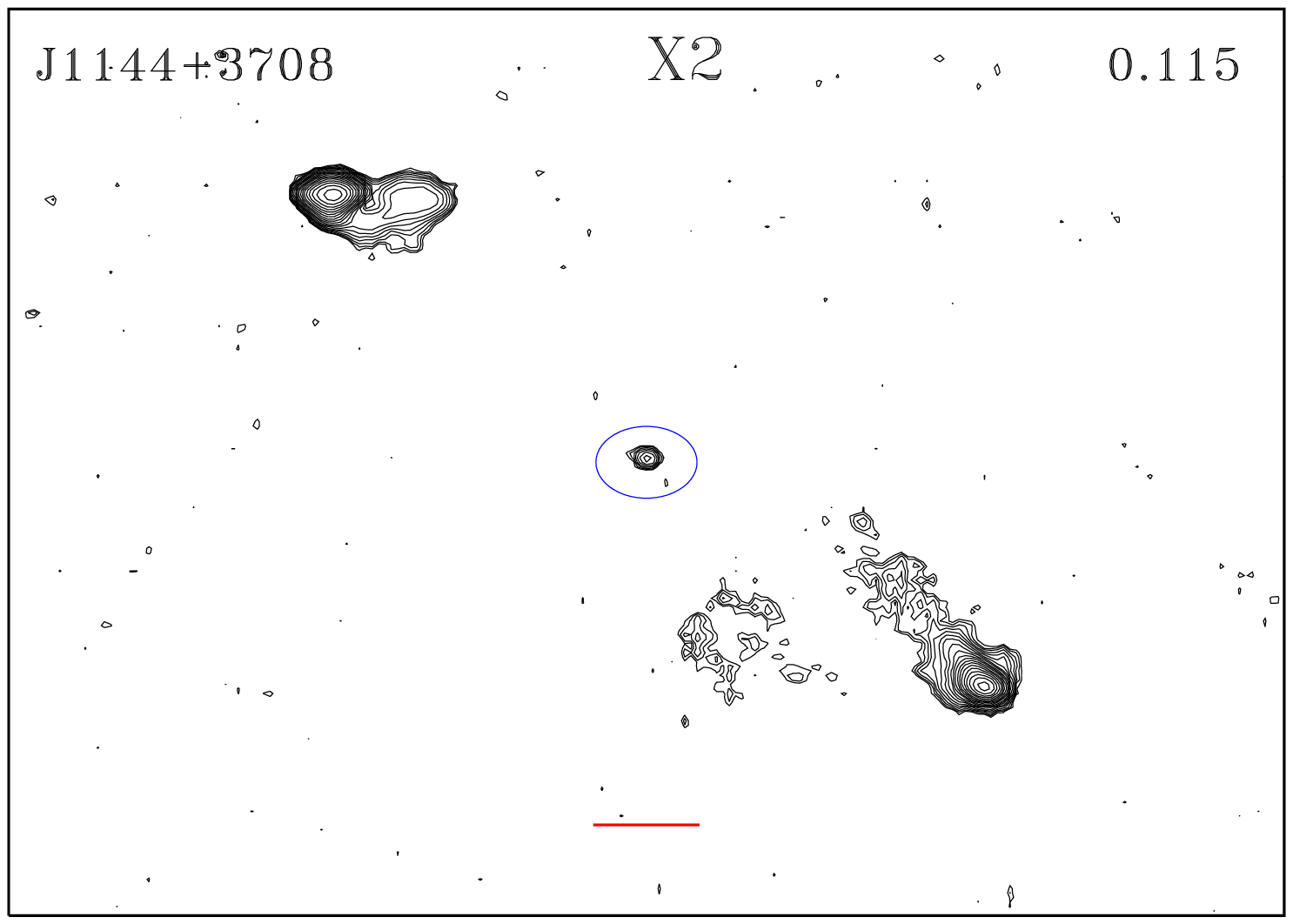}
\includegraphics[width=6.3cm,height=6.3cm]{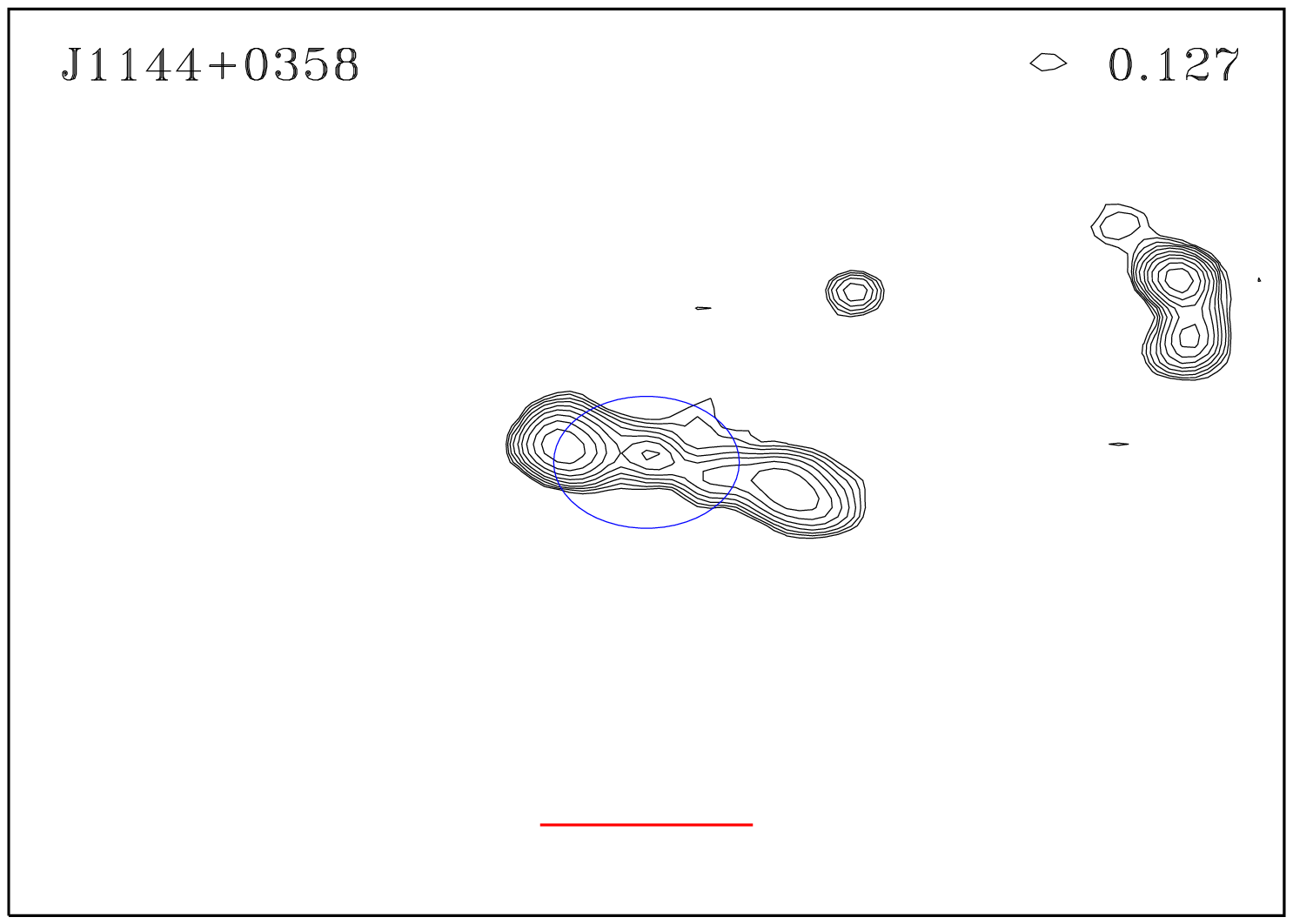}

\includegraphics[width=6.3cm,height=6.3cm]{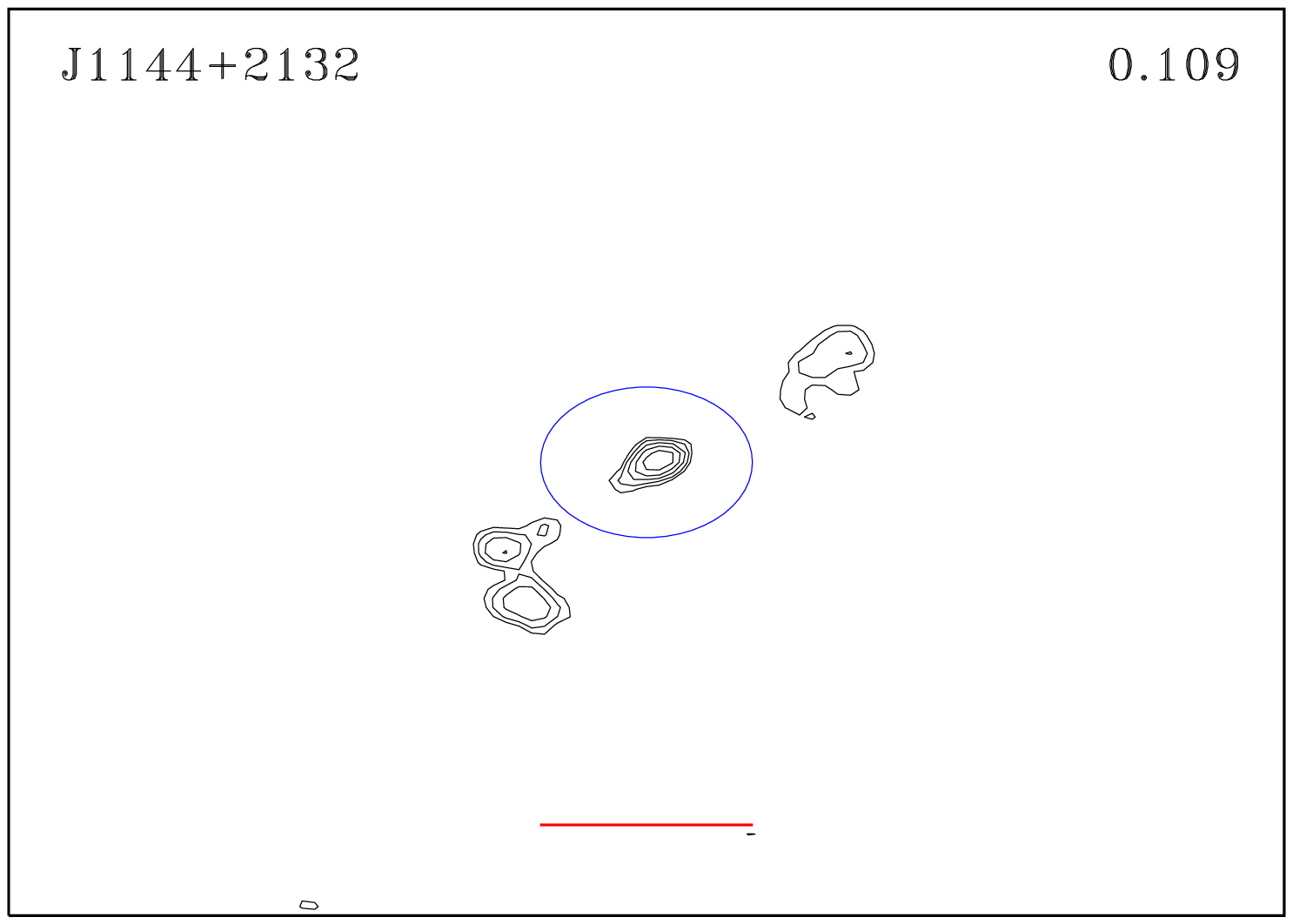}
\includegraphics[width=6.3cm,height=6.3cm]{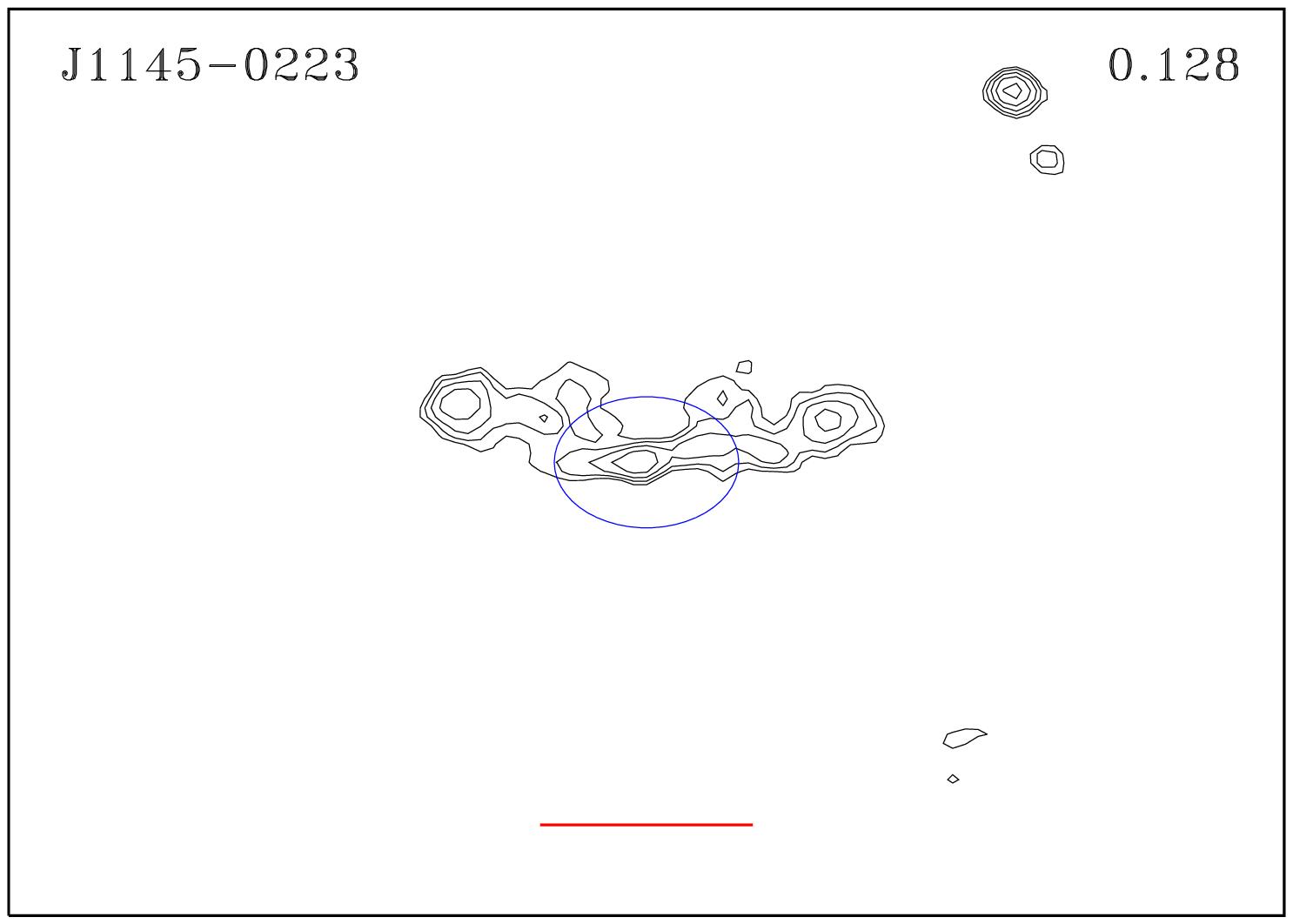}
\includegraphics[width=6.3cm,height=6.3cm]{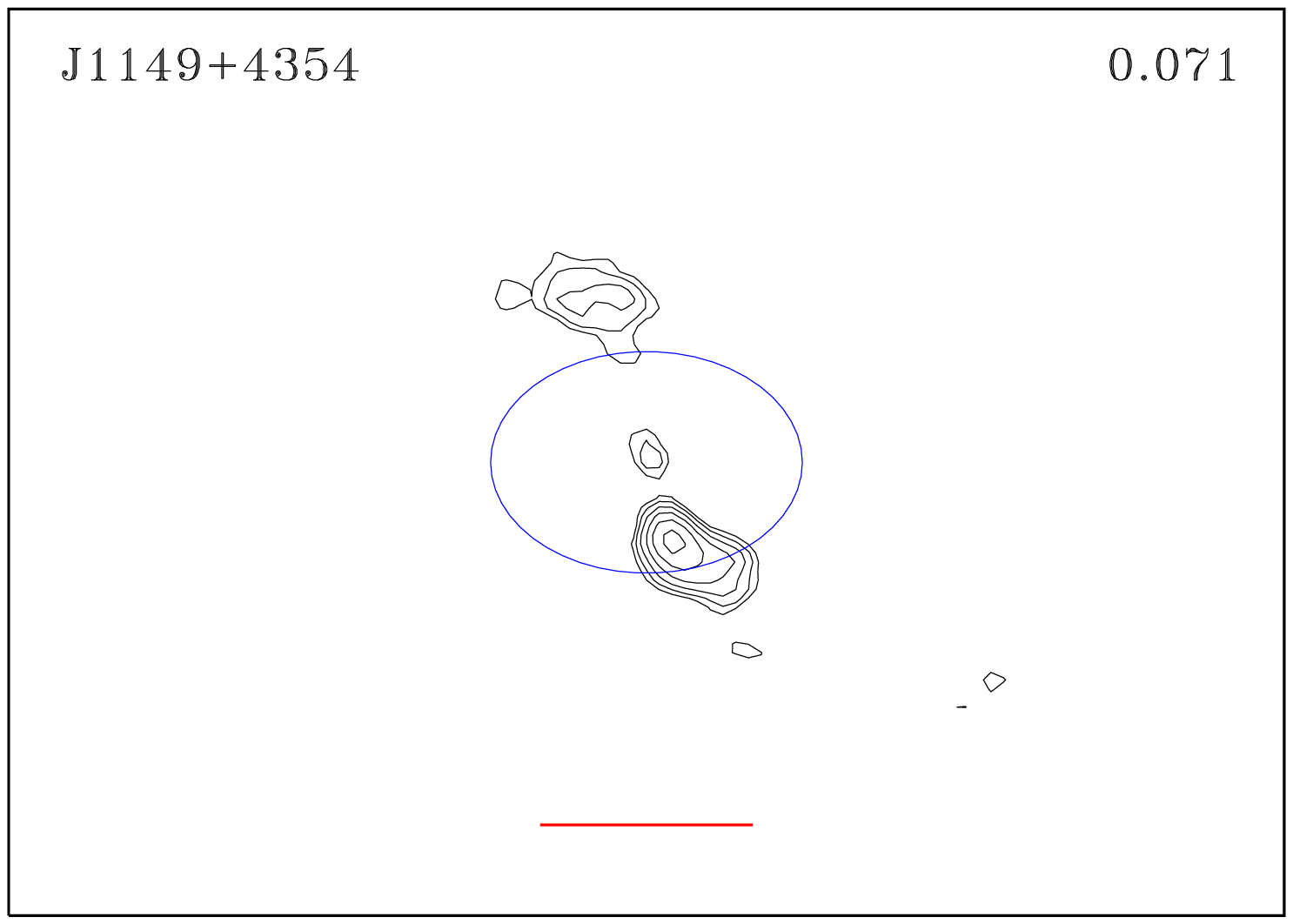}
\caption{(continued)}
\end{figure*}

\addtocounter{figure}{-1}
\begin{figure*}
\includegraphics[width=6.3cm,height=6.3cm]{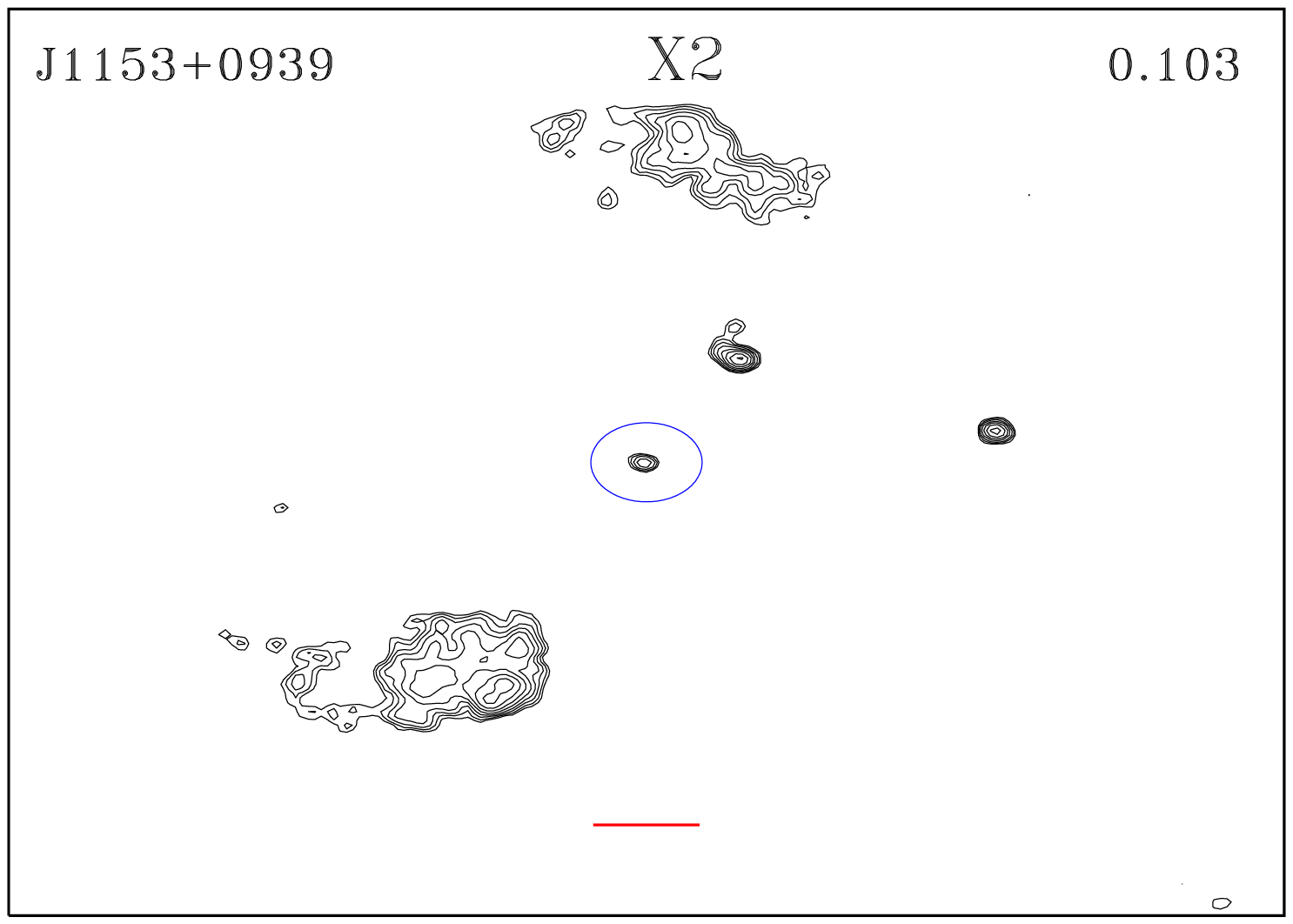}
\includegraphics[width=6.3cm,height=6.3cm]{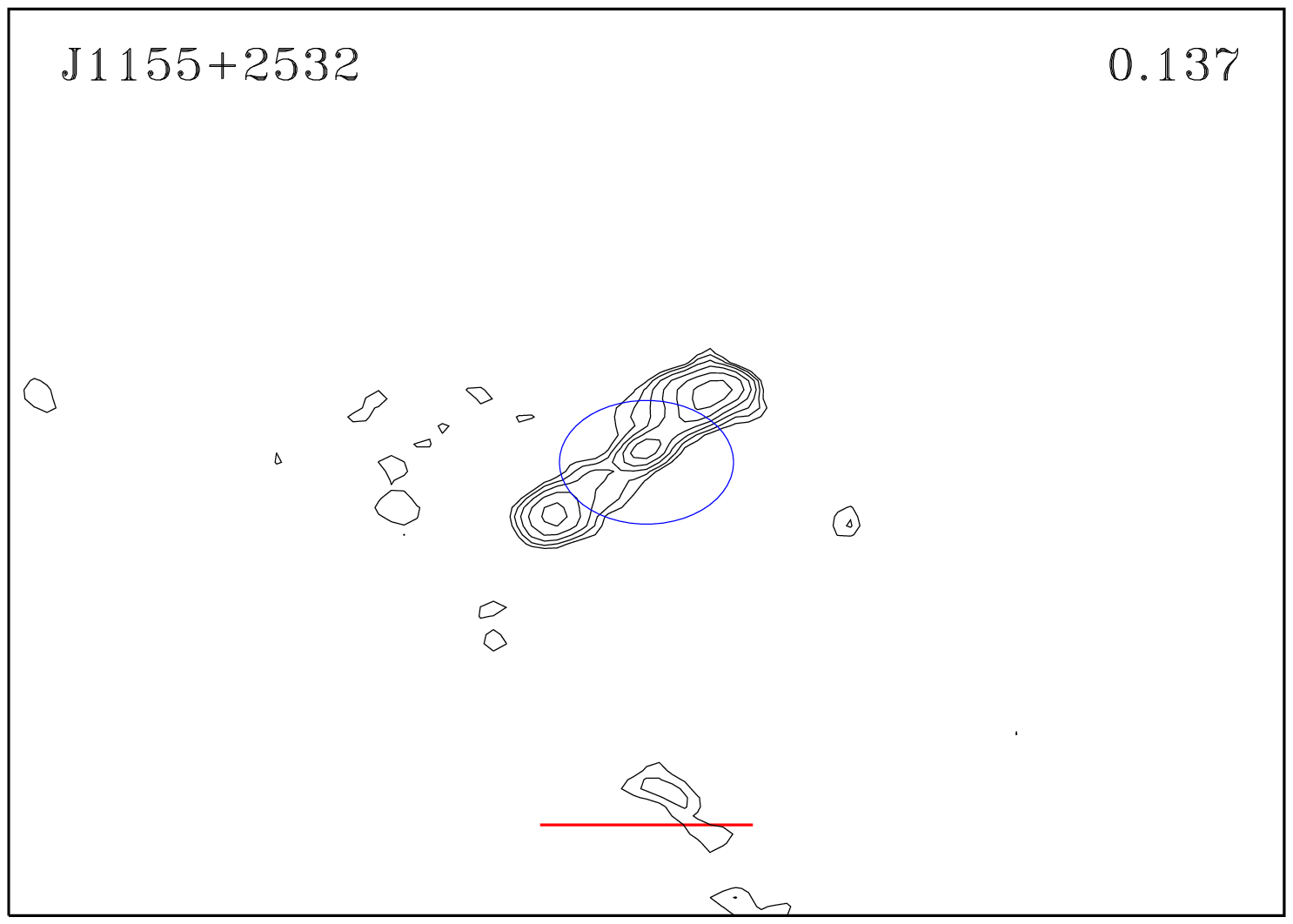}
\includegraphics[width=6.3cm,height=6.3cm]{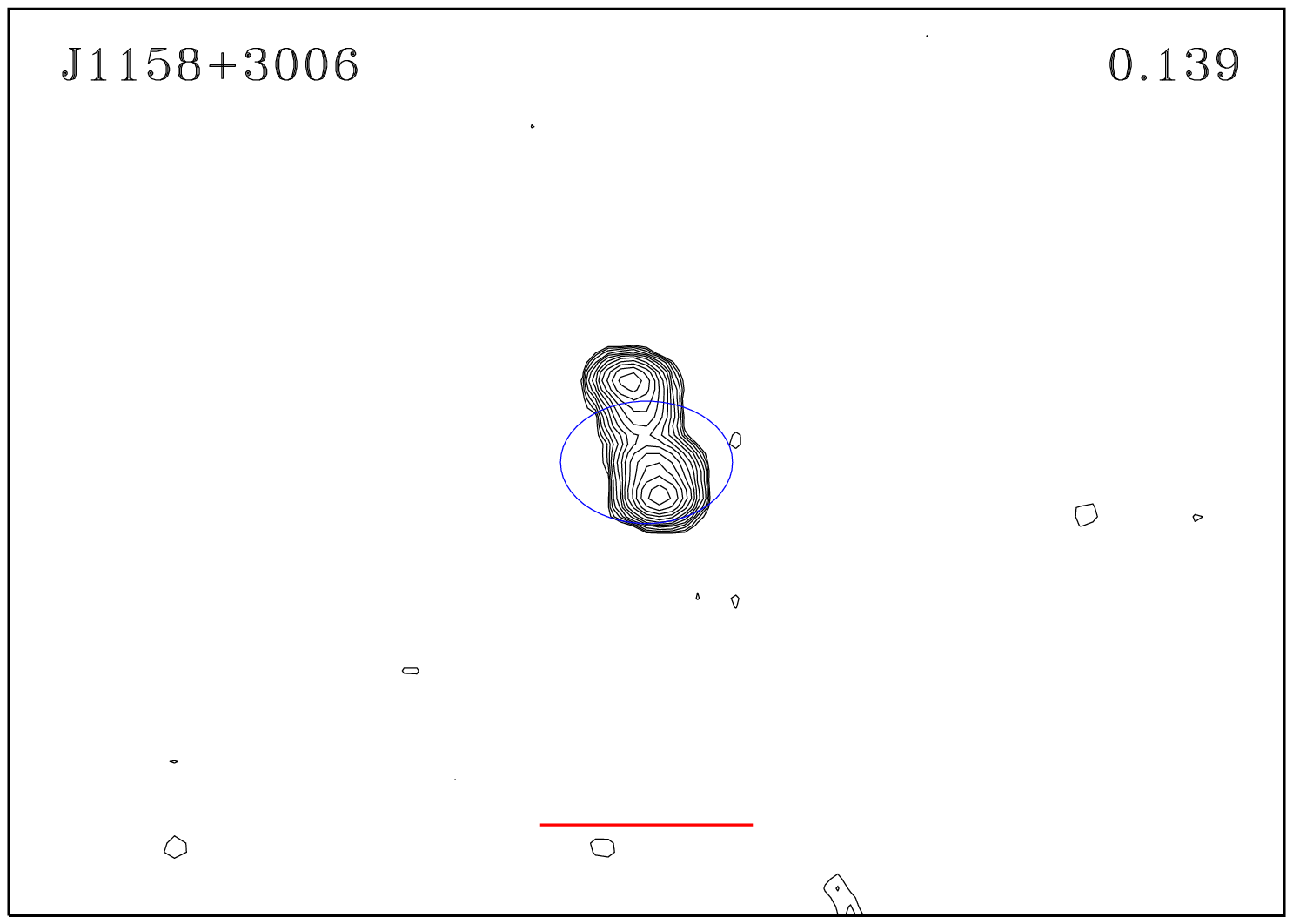}

\includegraphics[width=6.3cm,height=6.3cm]{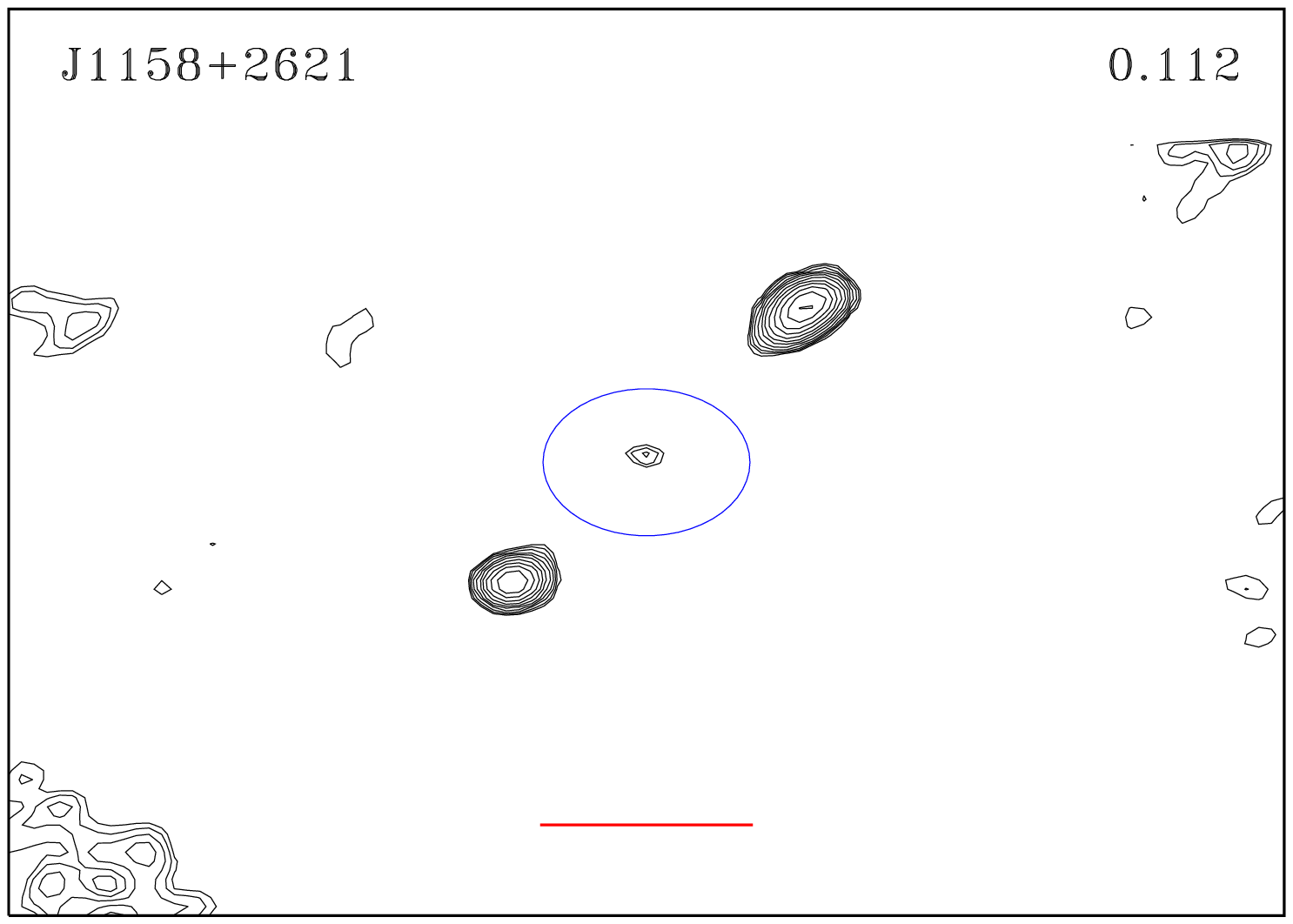}
\includegraphics[width=6.3cm,height=6.3cm]{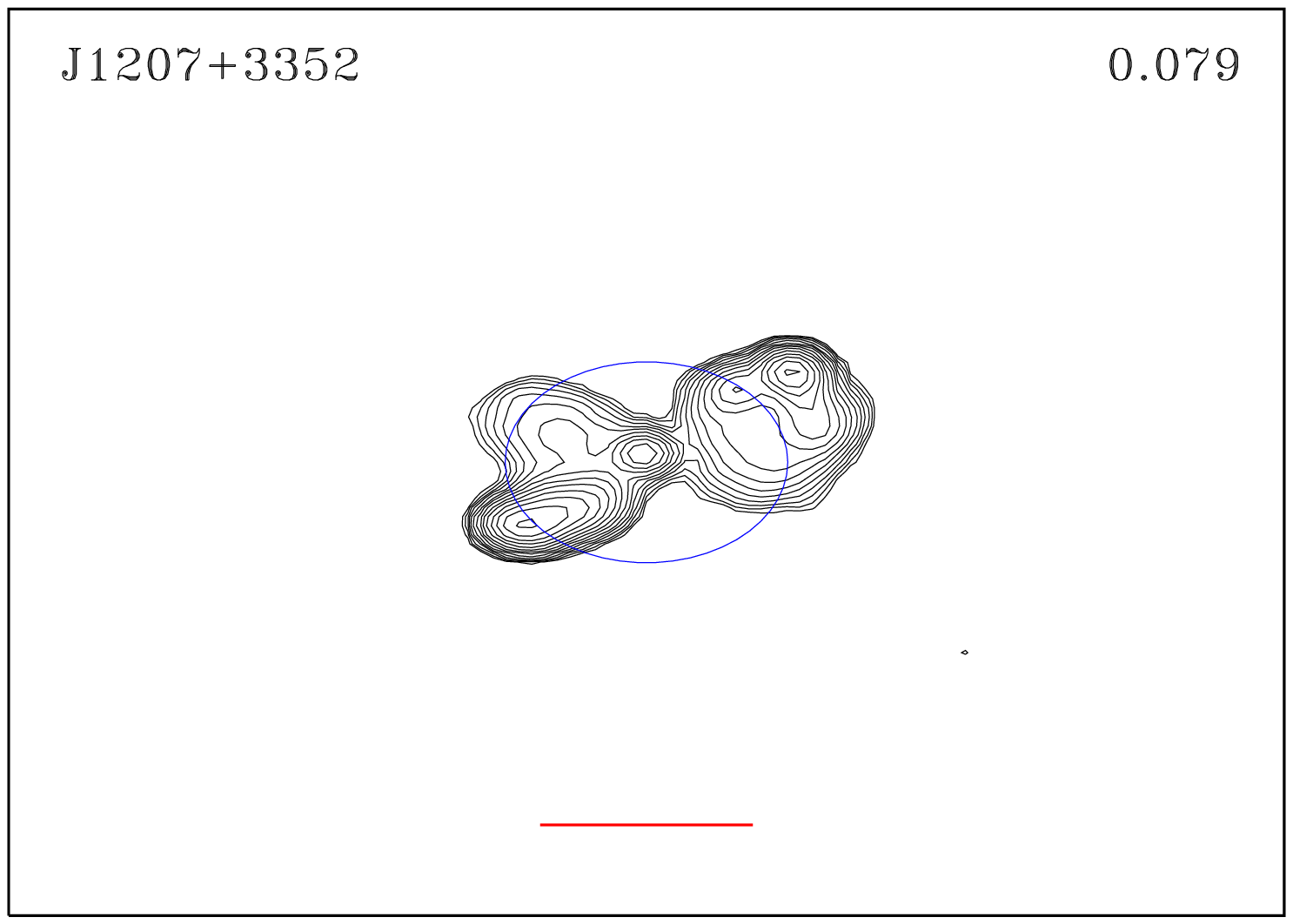}
\includegraphics[width=6.3cm,height=6.3cm]{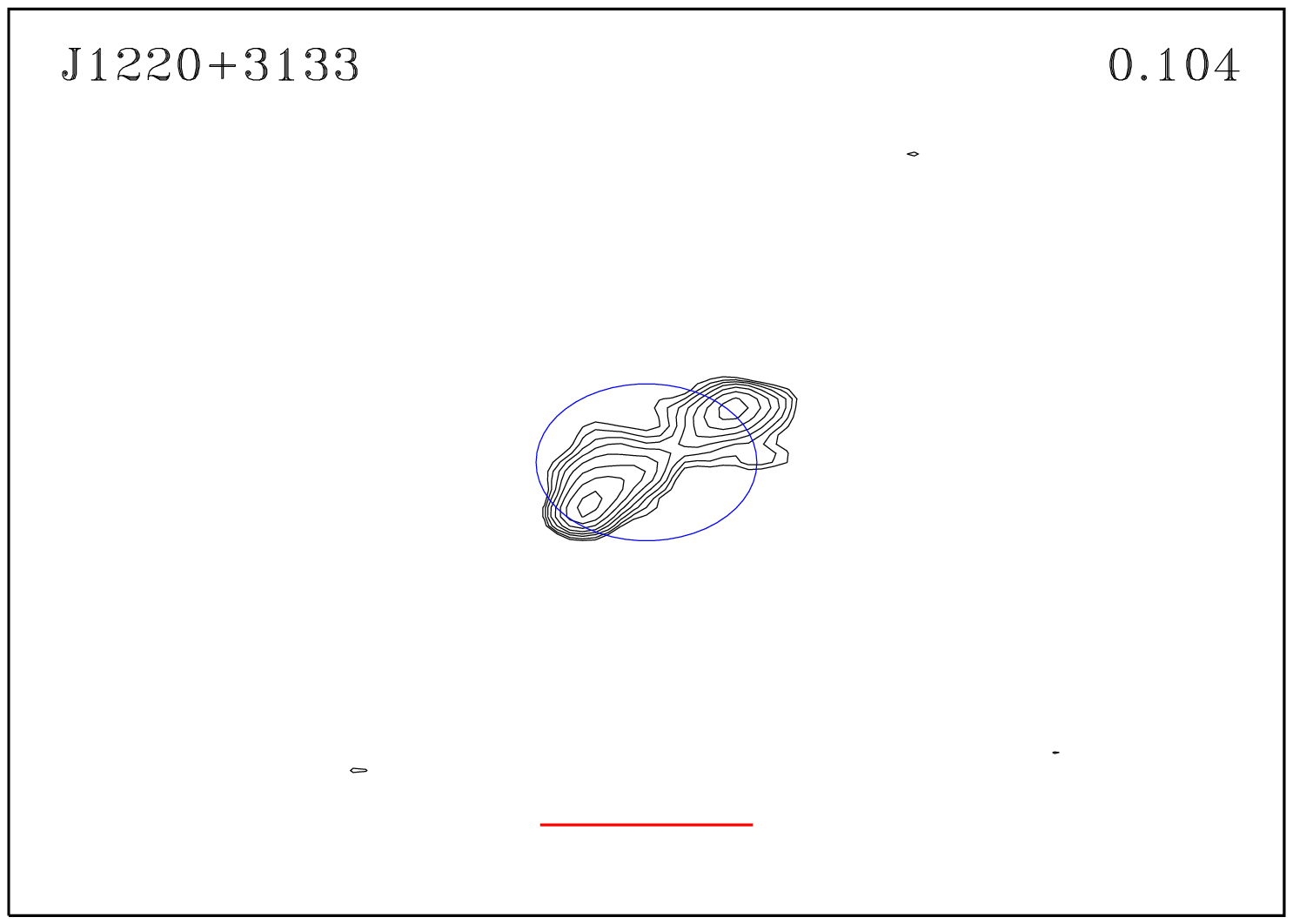}

\includegraphics[width=6.3cm,height=6.3cm]{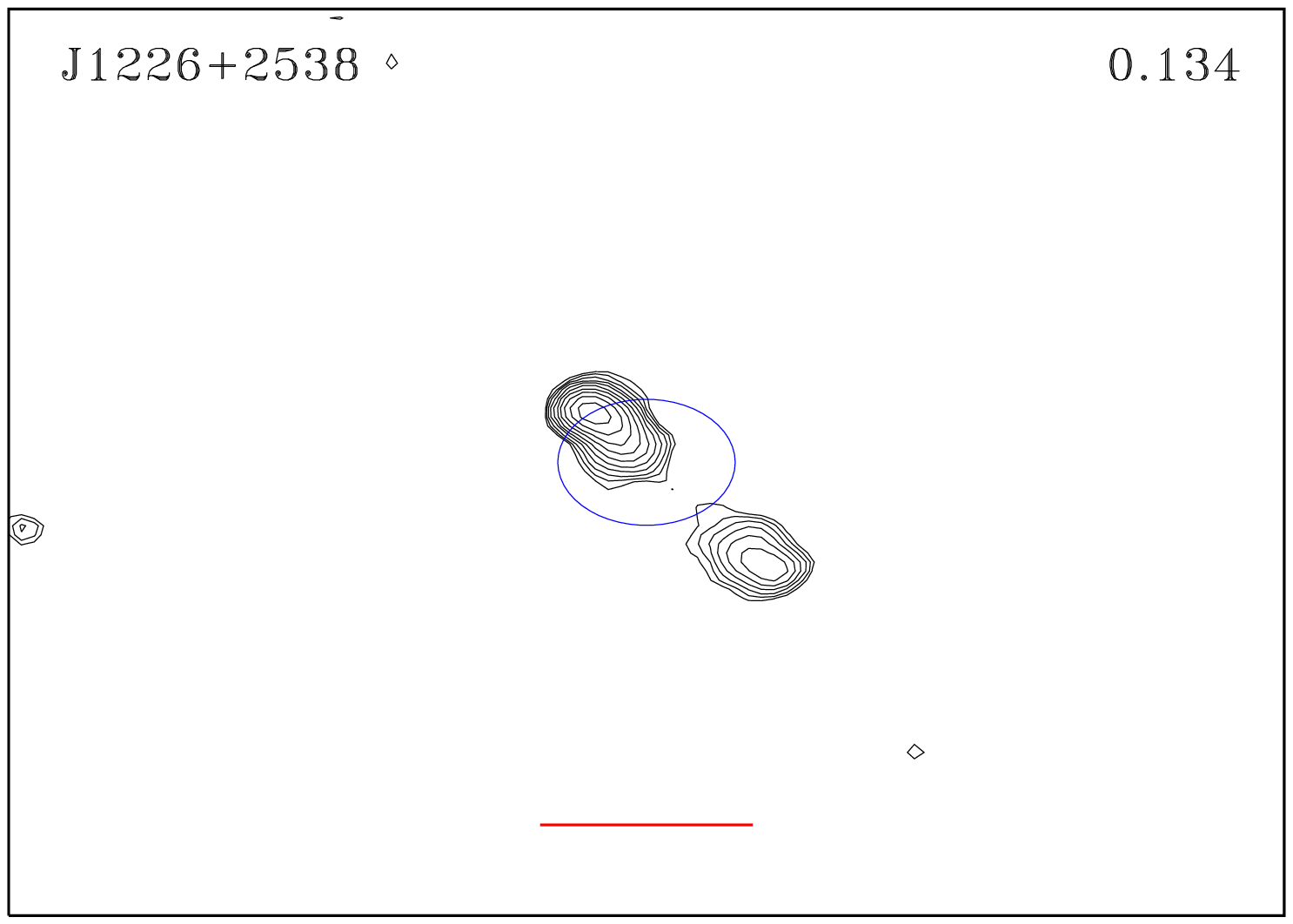}
\includegraphics[width=6.3cm,height=6.3cm]{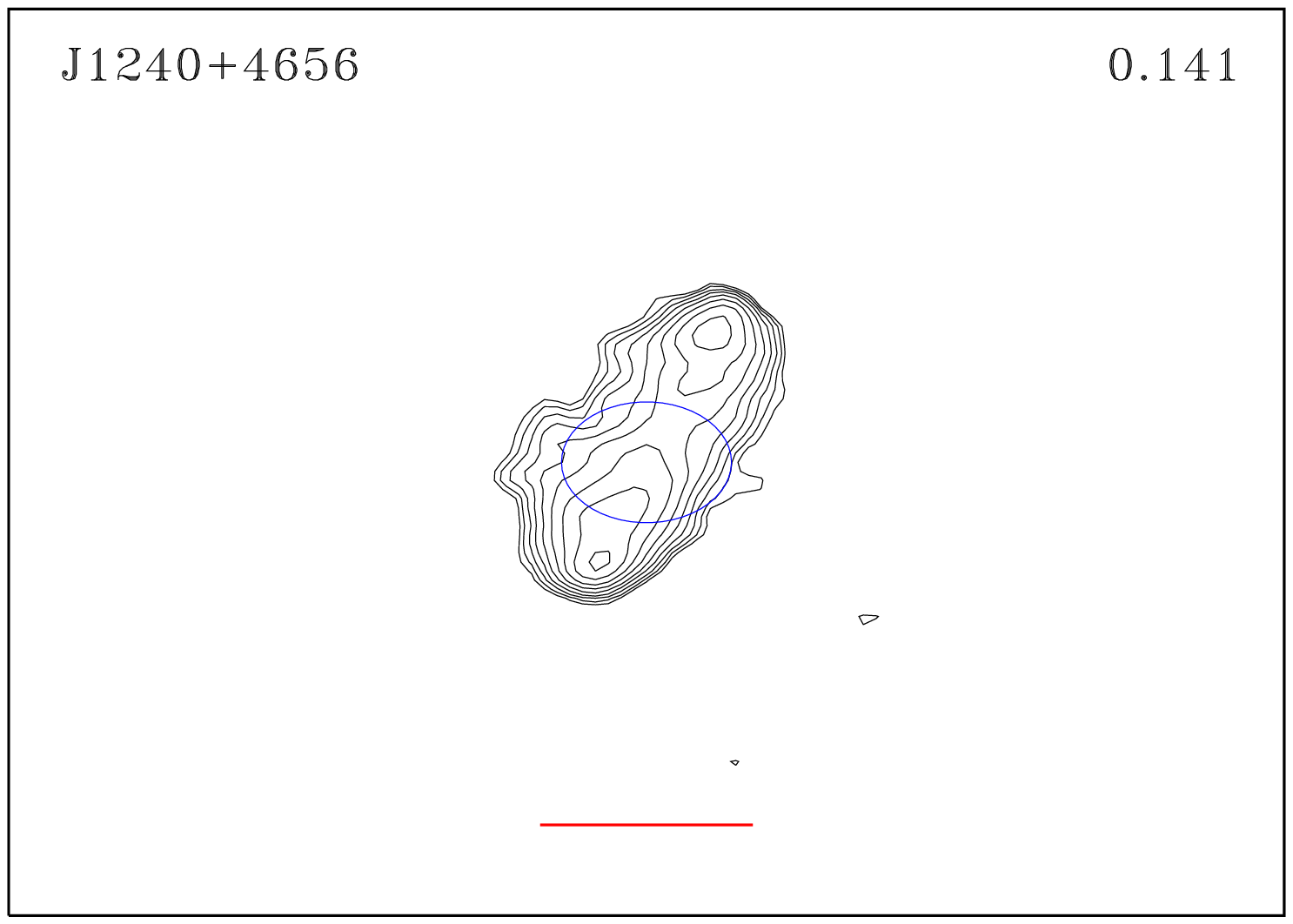}
\includegraphics[width=6.3cm,height=6.3cm]{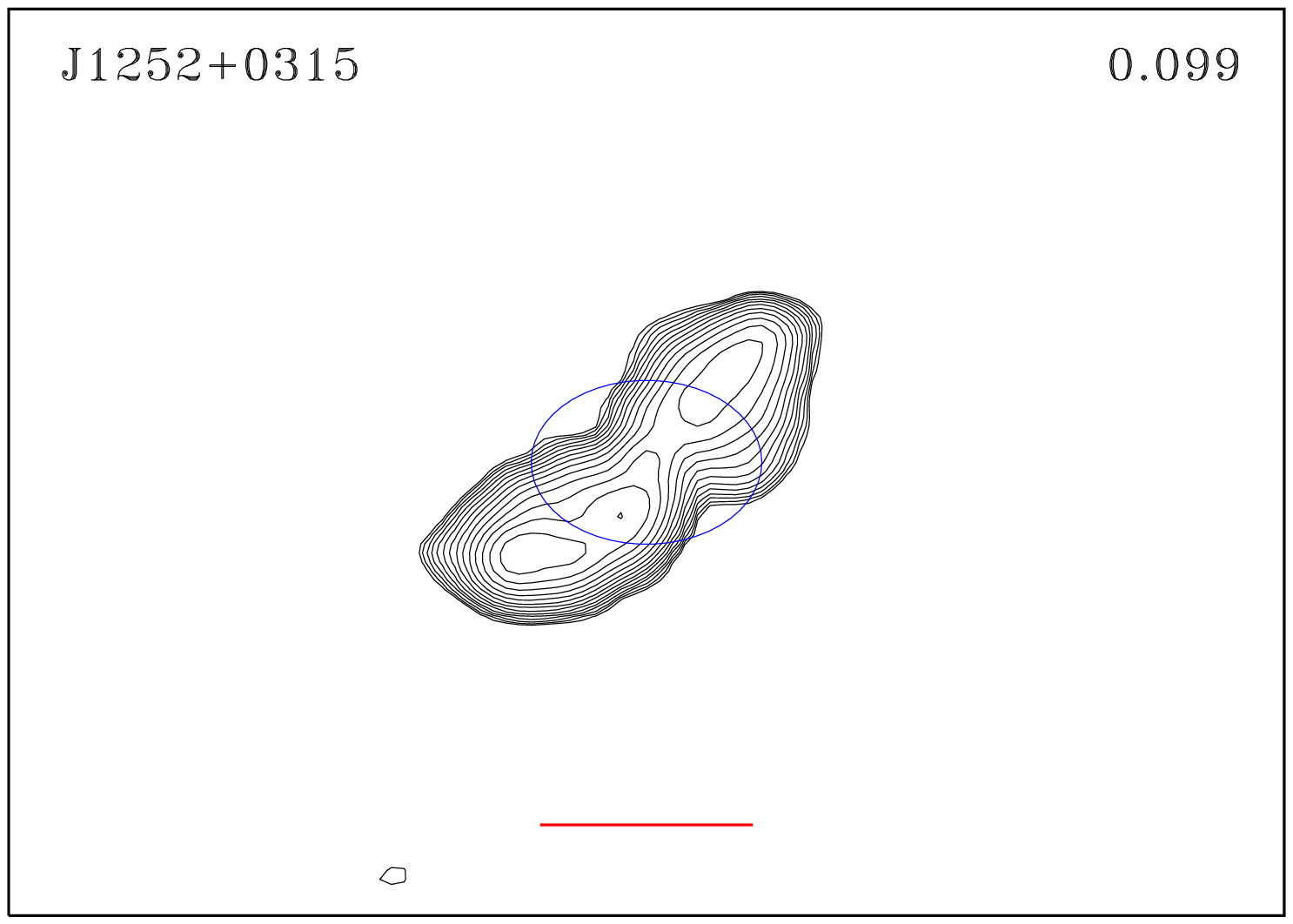}

\includegraphics[width=6.3cm,height=6.3cm]{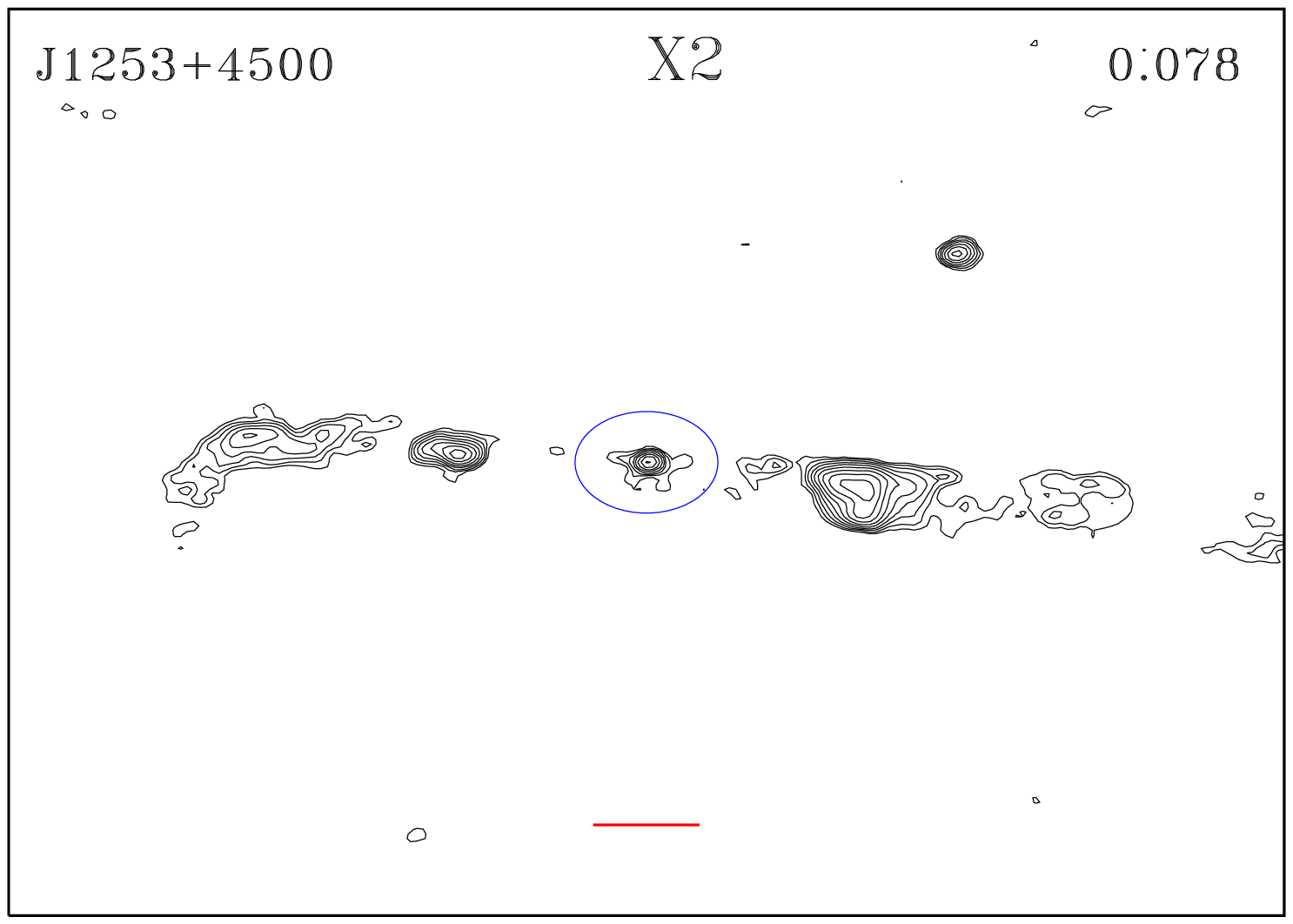}
\includegraphics[width=6.3cm,height=6.3cm]{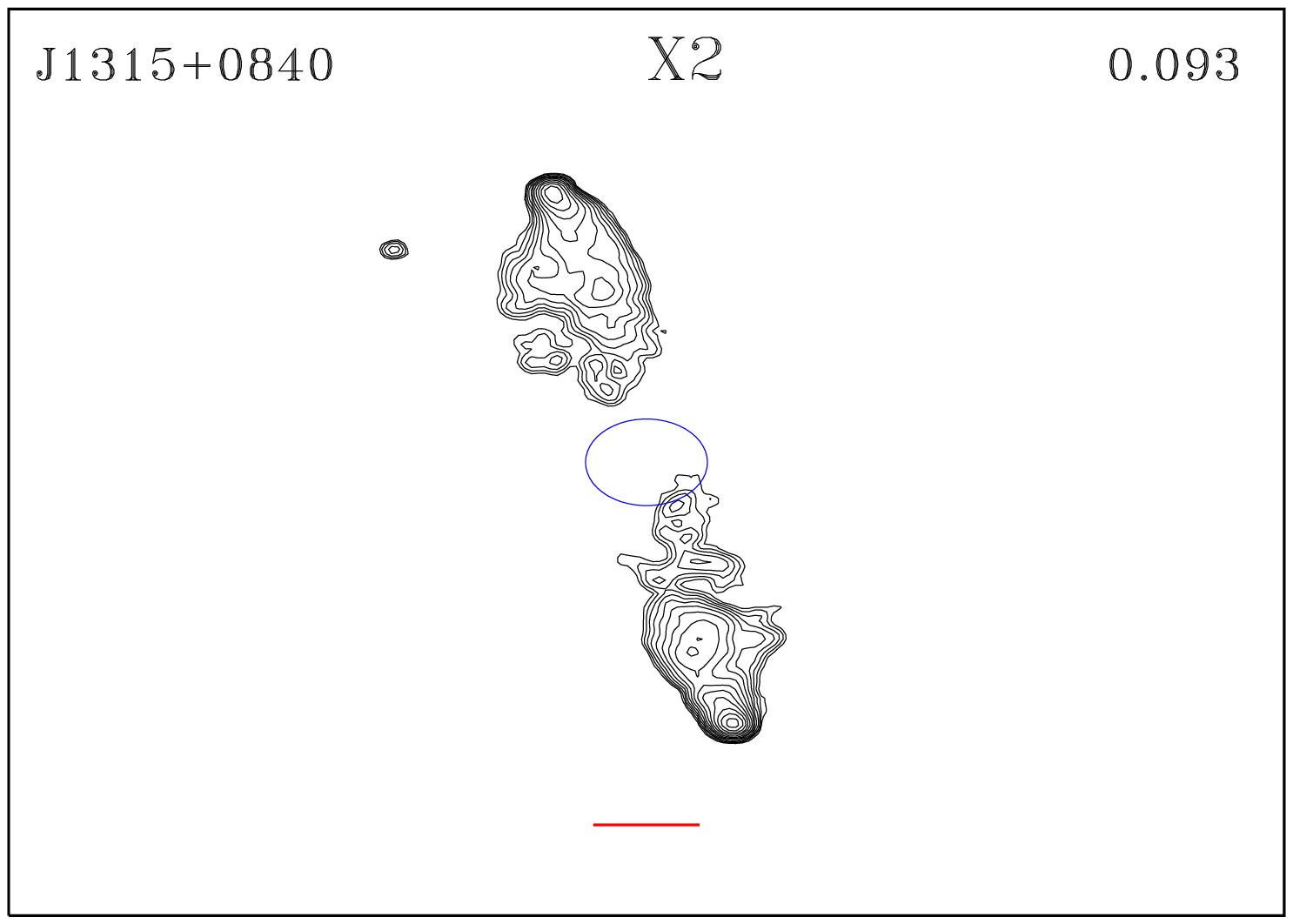}
\includegraphics[width=6.3cm,height=6.3cm]{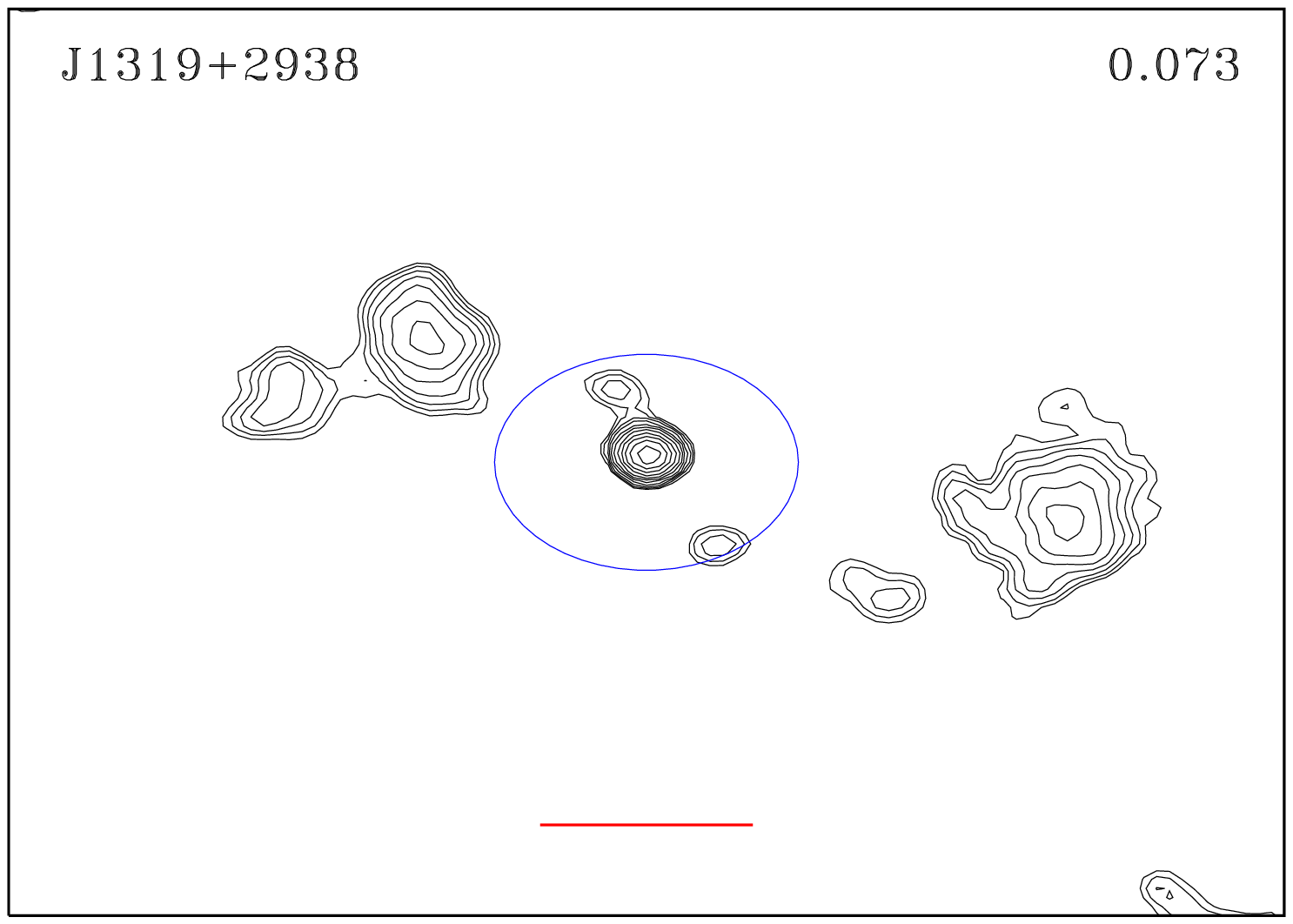}
\caption{(continued)}
\end{figure*}

\addtocounter{figure}{-1}
\begin{figure*}
\includegraphics[width=6.3cm,height=6.3cm]{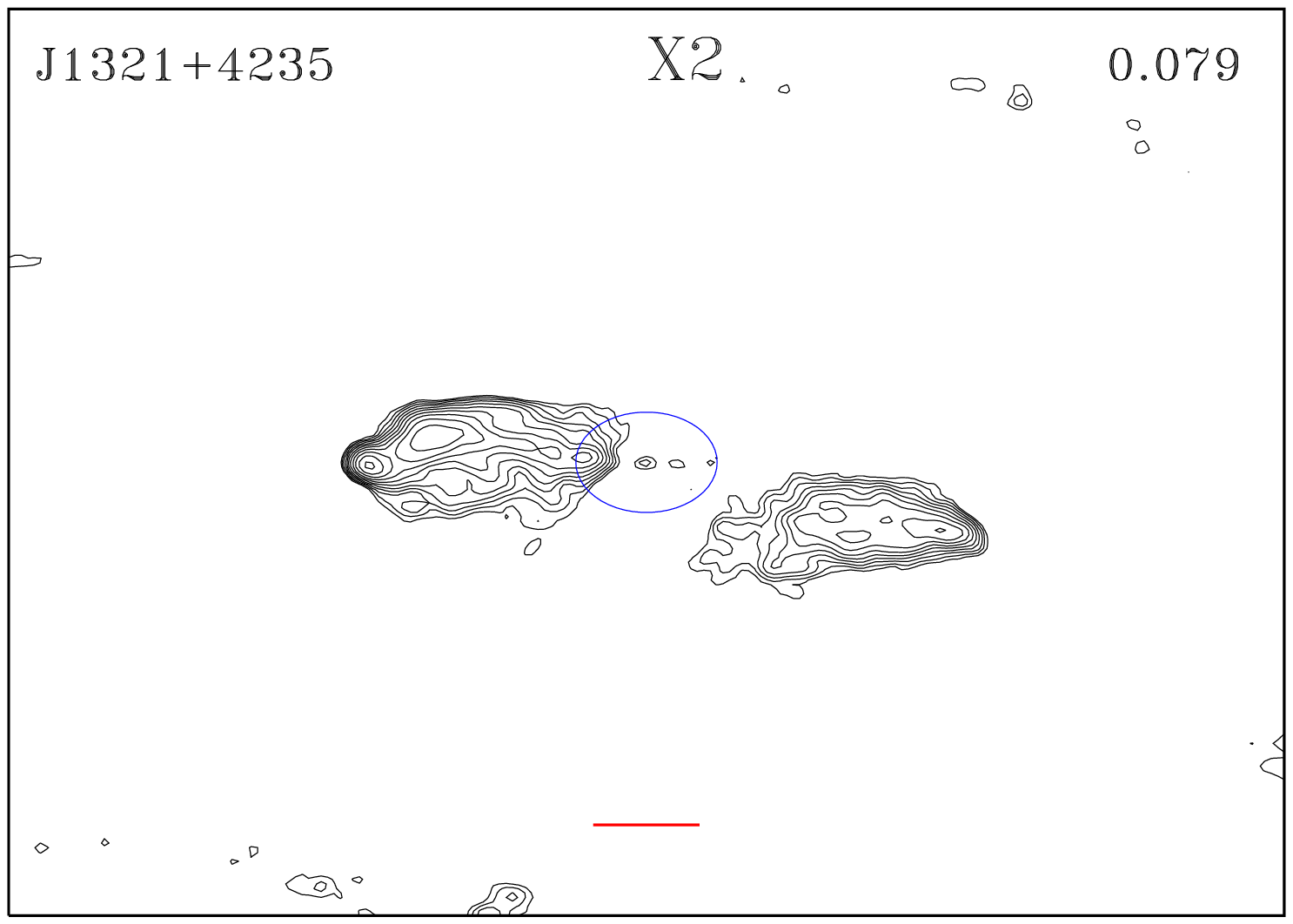}
\includegraphics[width=6.3cm,height=6.3cm]{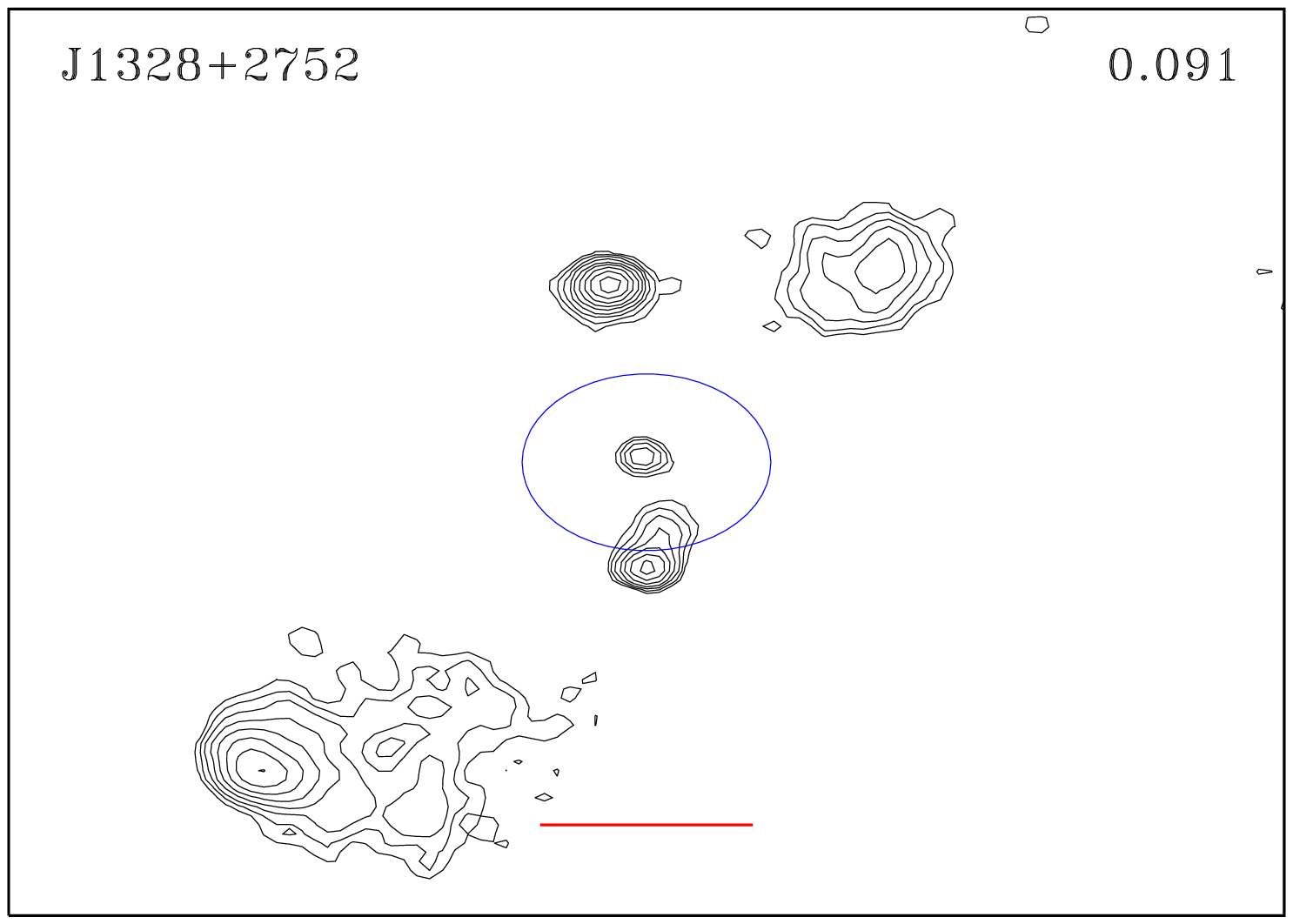}
\includegraphics[width=6.3cm,height=6.3cm]{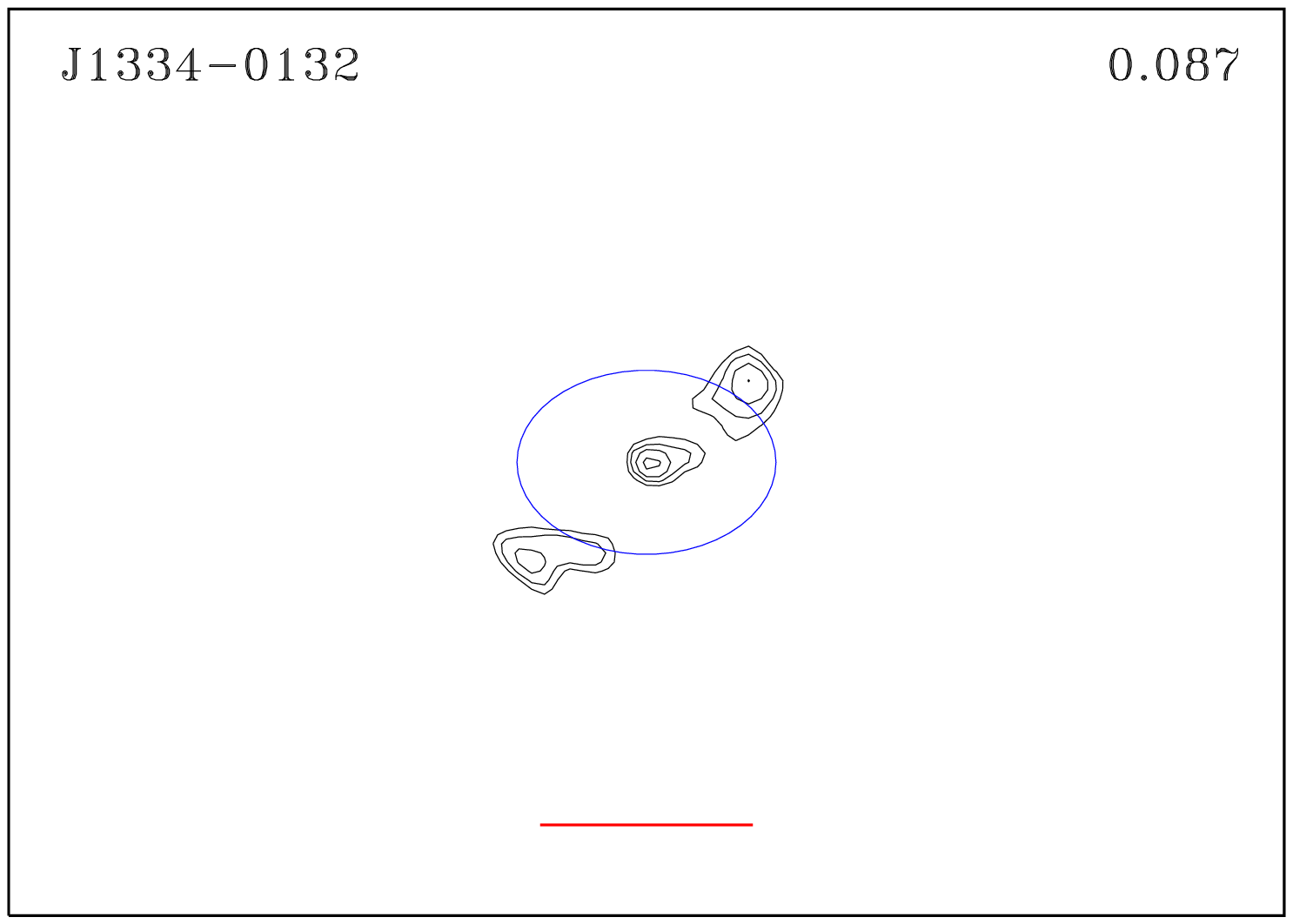}

\includegraphics[width=6.3cm,height=6.3cm]{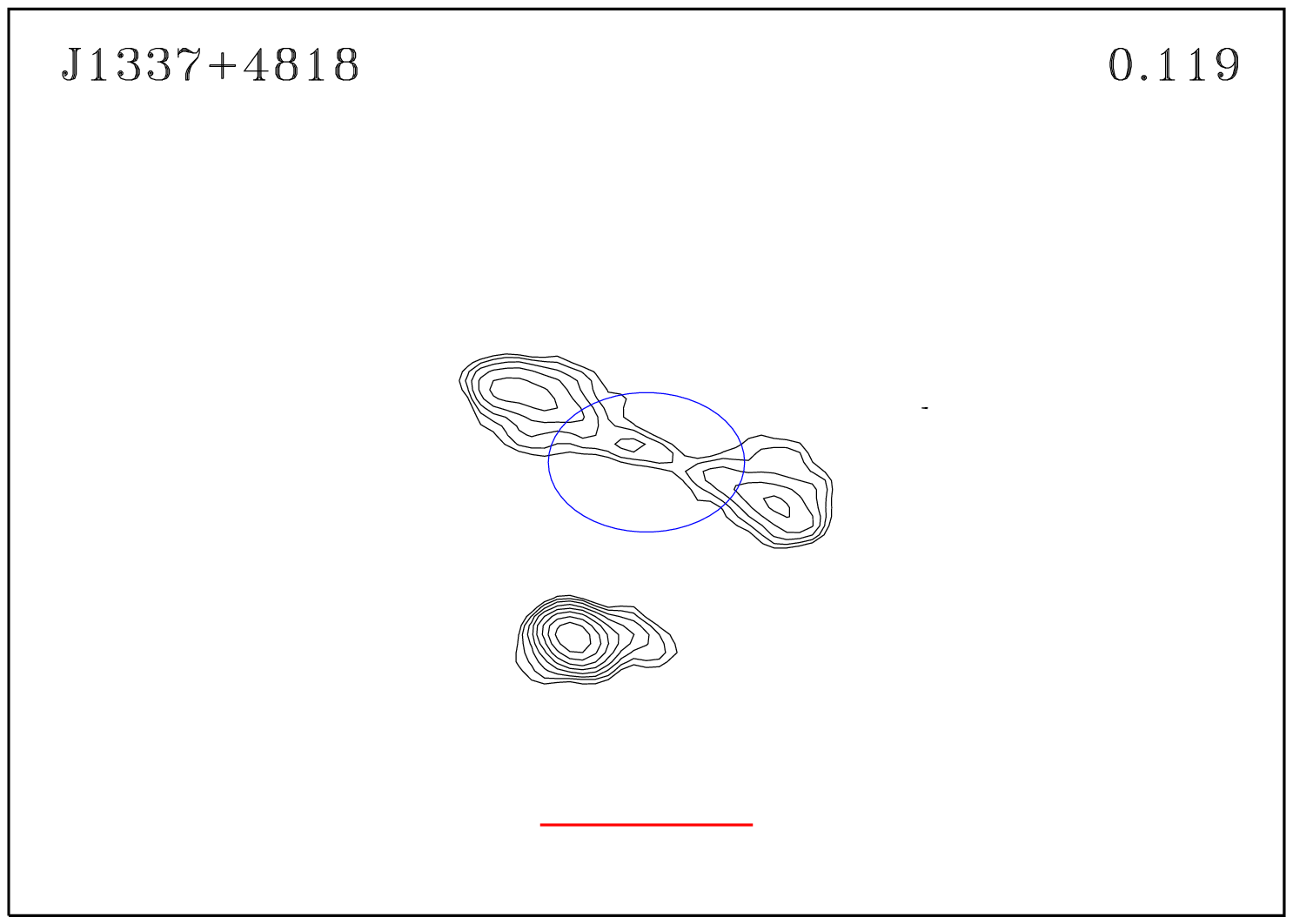}
\includegraphics[width=6.3cm,height=6.3cm]{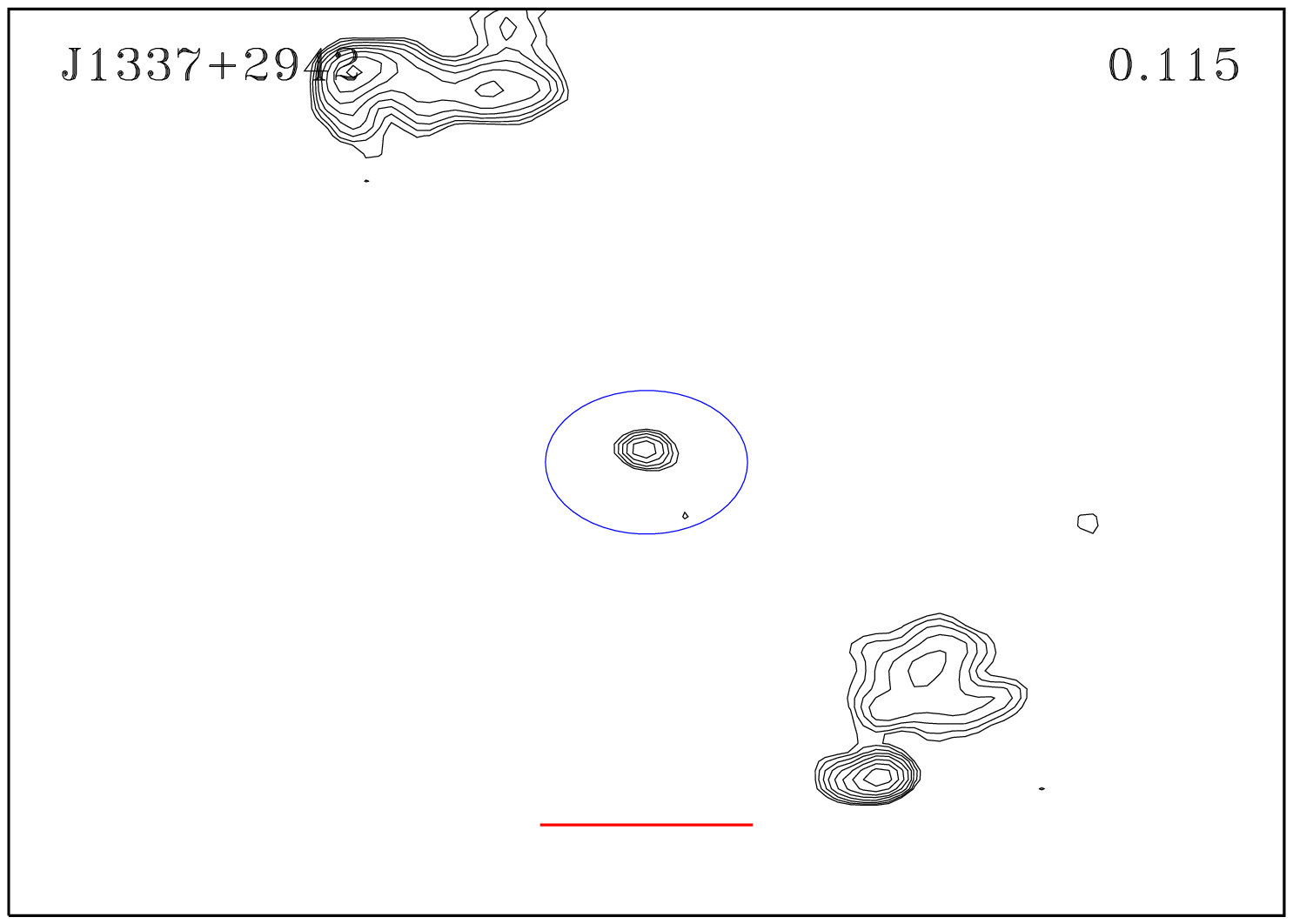}
\includegraphics[width=6.3cm,height=6.3cm]{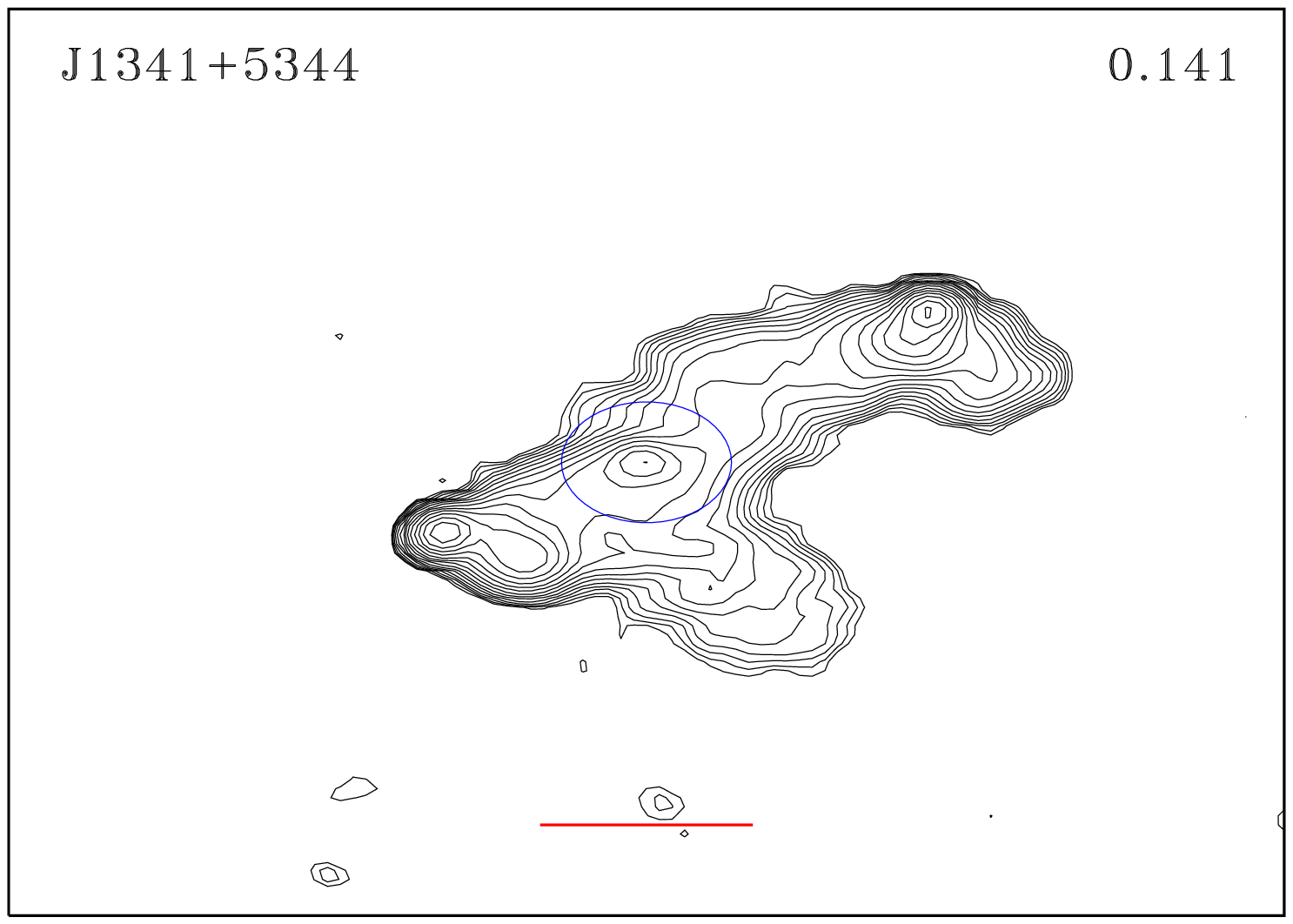}

\includegraphics[width=6.3cm,height=6.3cm]{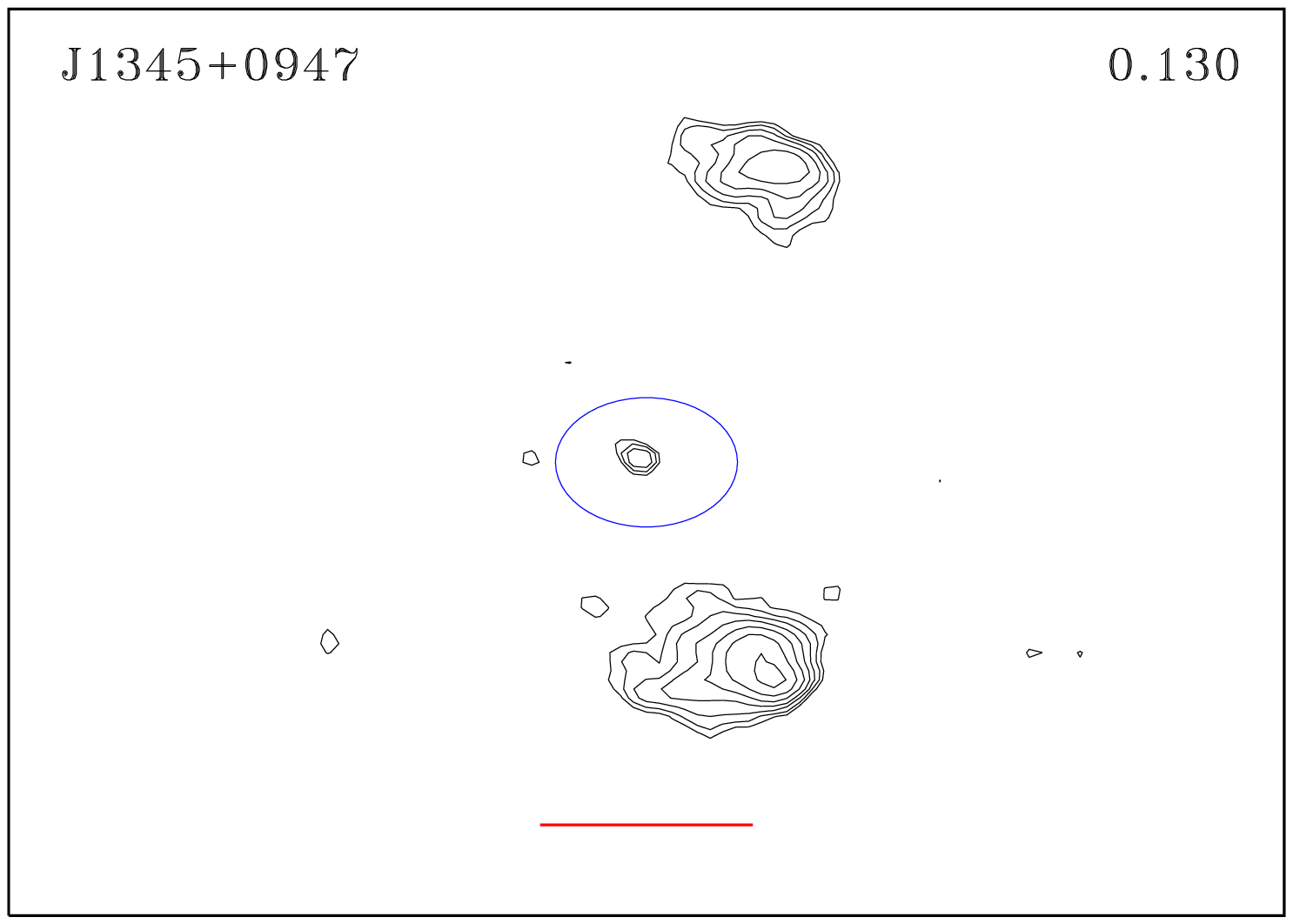}
\includegraphics[width=6.3cm,height=6.3cm]{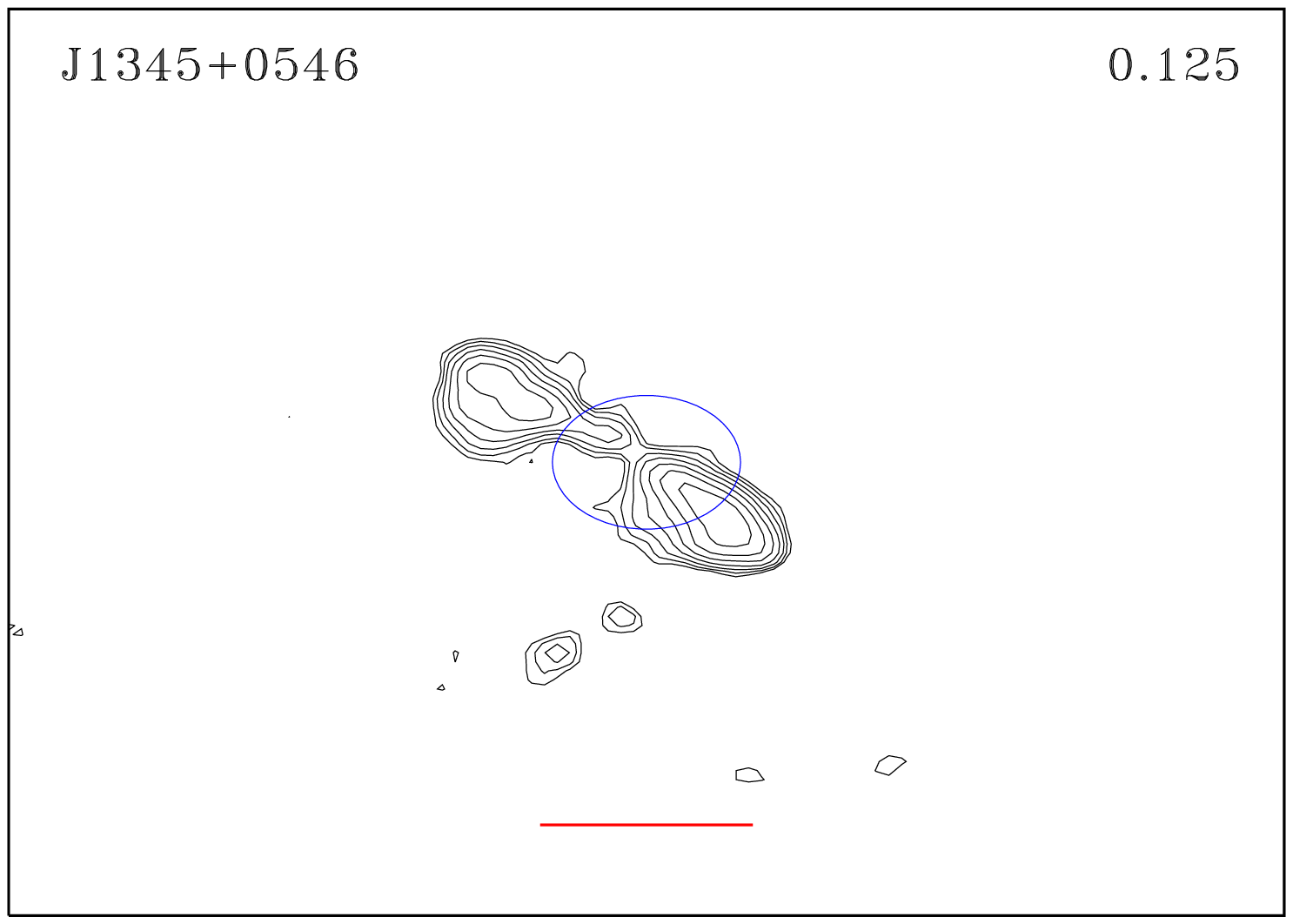}
\includegraphics[width=6.3cm,height=6.3cm]{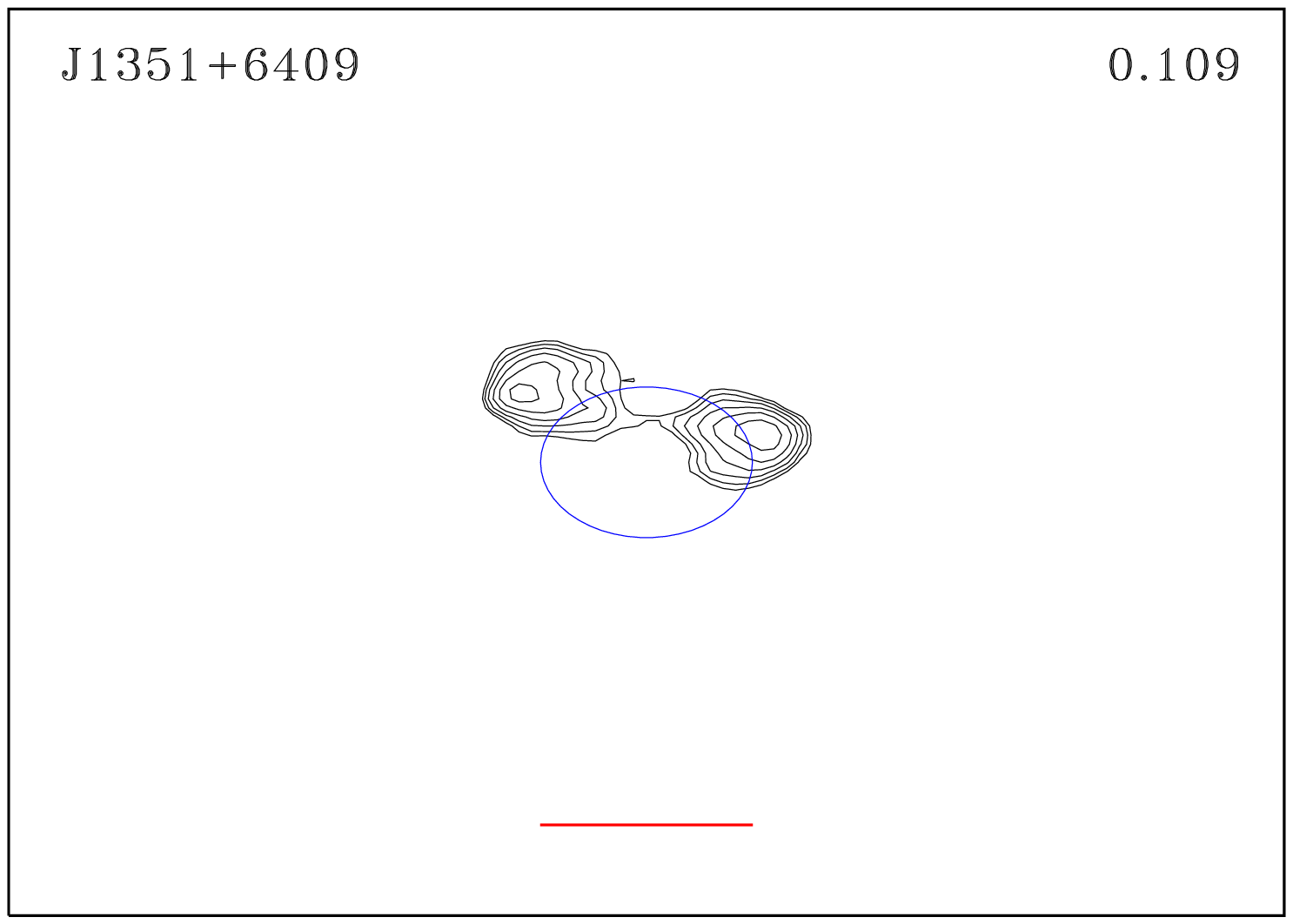}

\includegraphics[width=6.3cm,height=6.3cm]{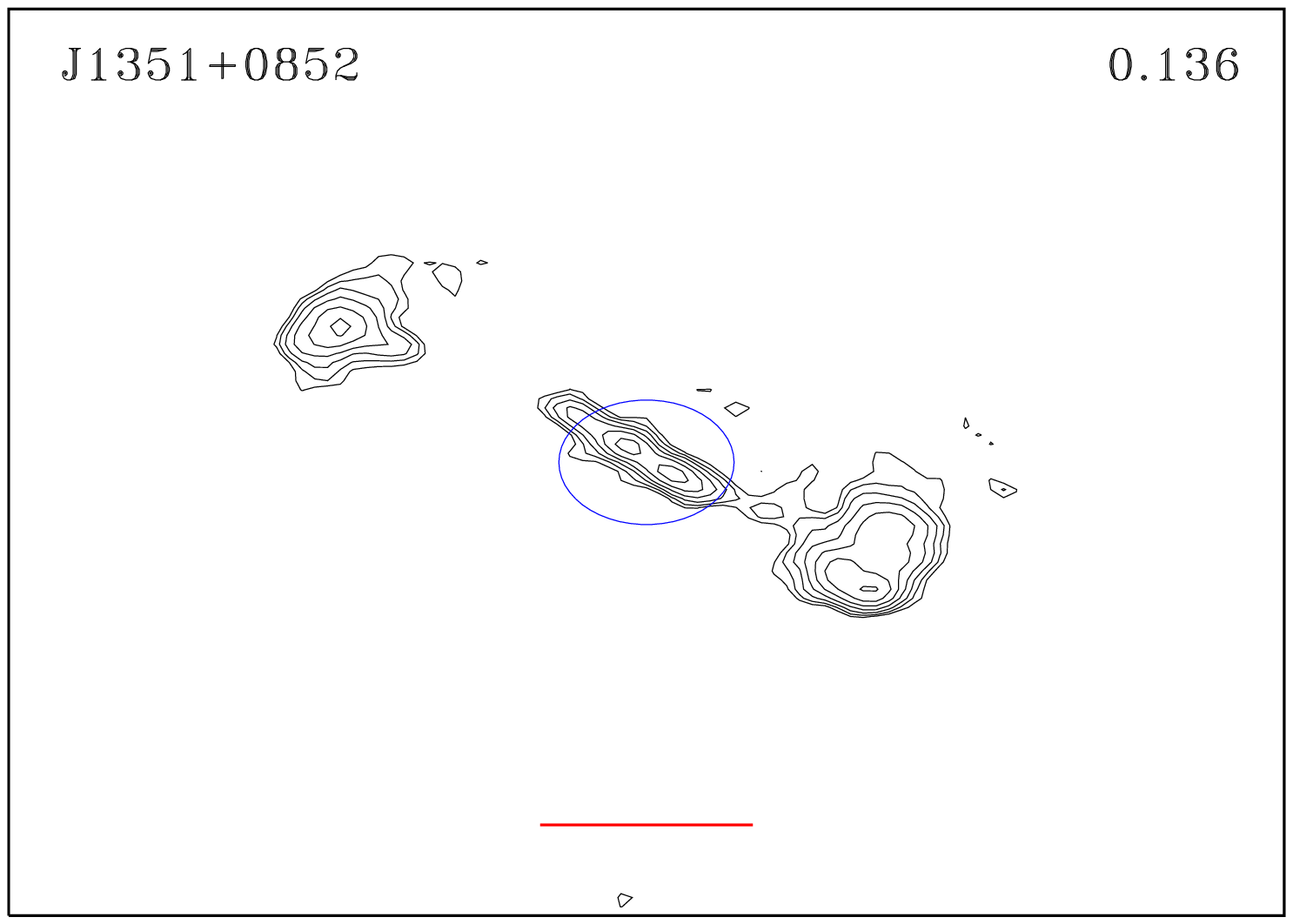}
\includegraphics[width=6.3cm,height=6.3cm]{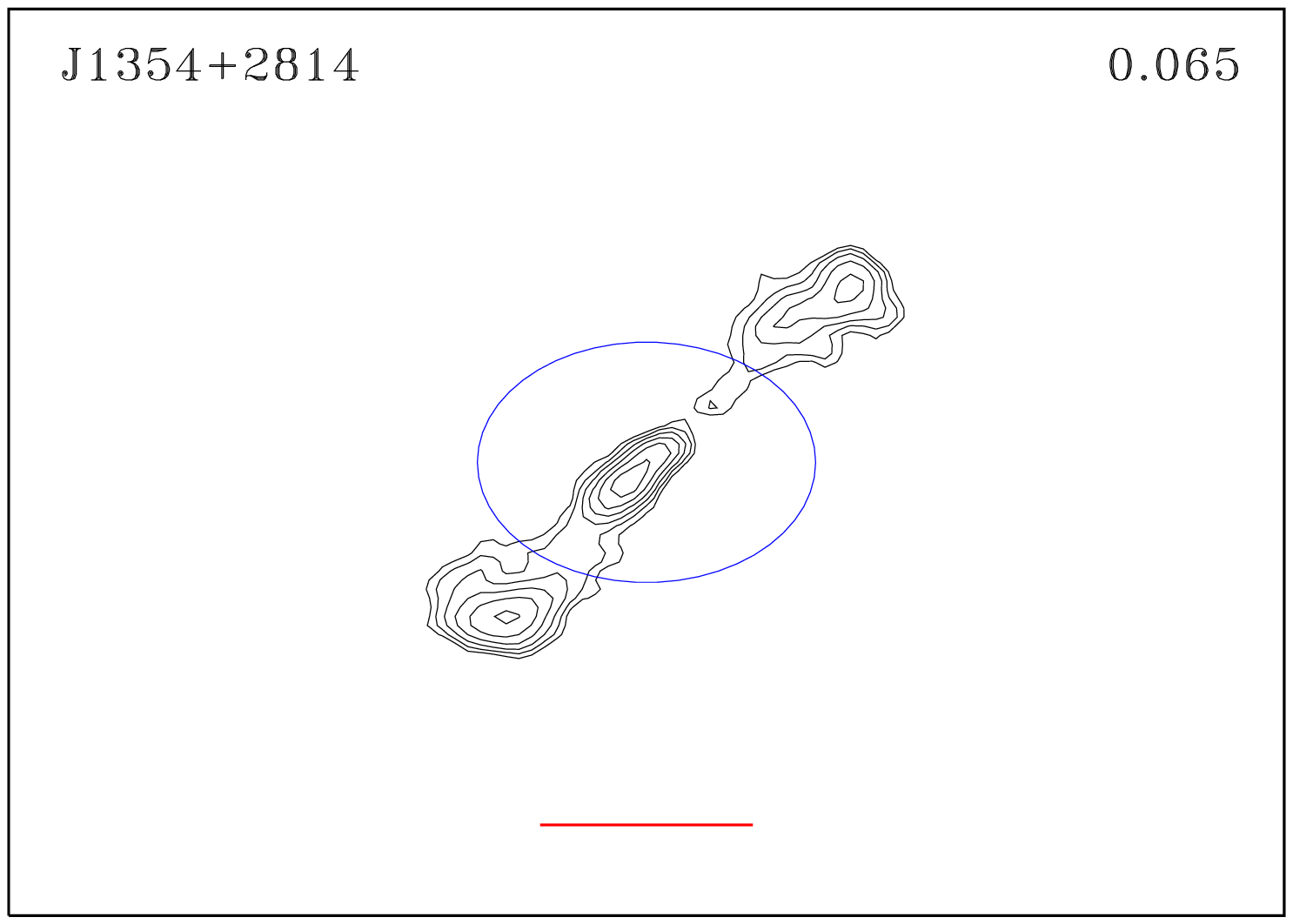}
\includegraphics[width=6.3cm,height=6.3cm]{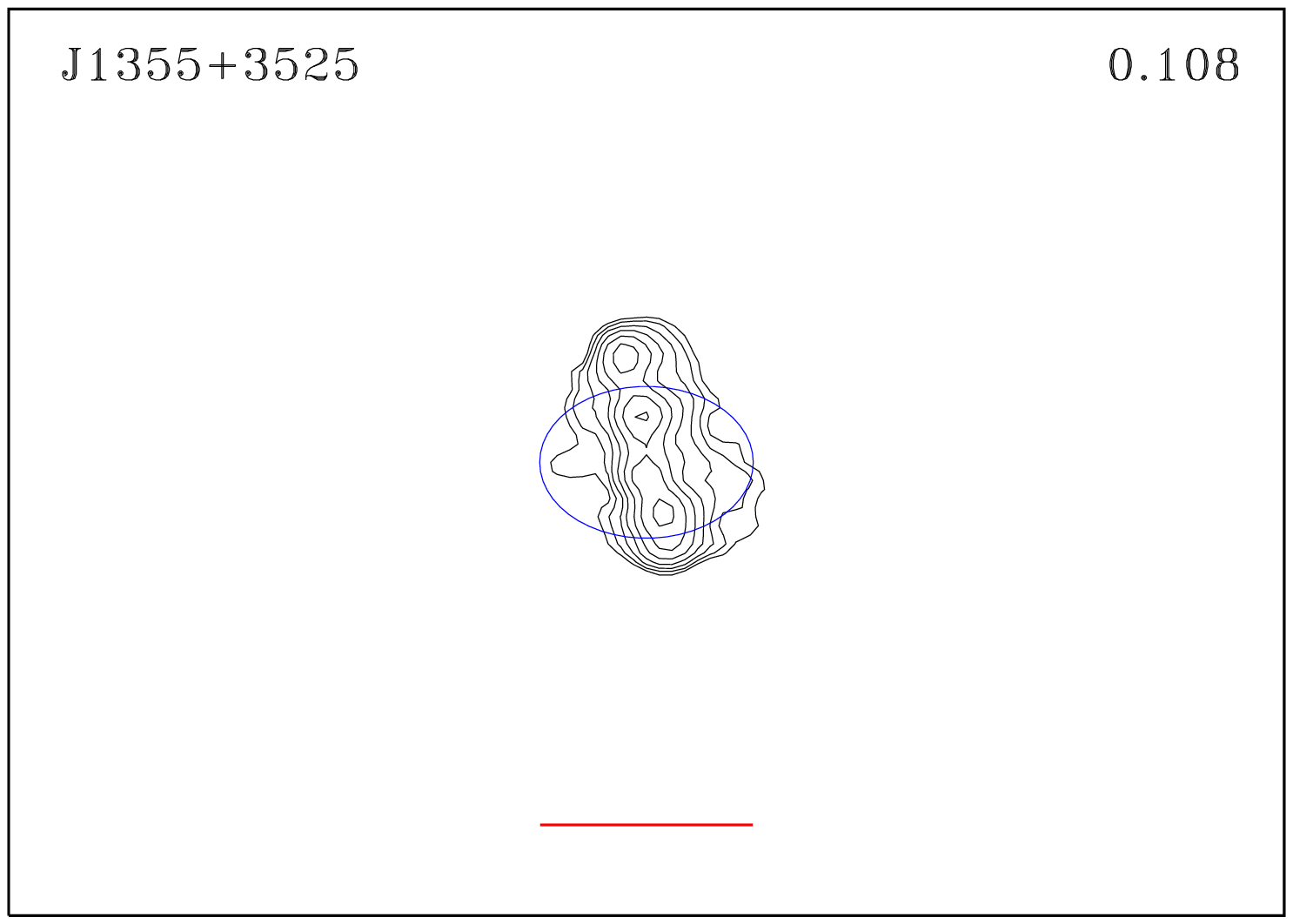}
\caption{(continued)}
\end{figure*}

\addtocounter{figure}{-1}
\begin{figure*}
\includegraphics[width=6.3cm,height=6.3cm]{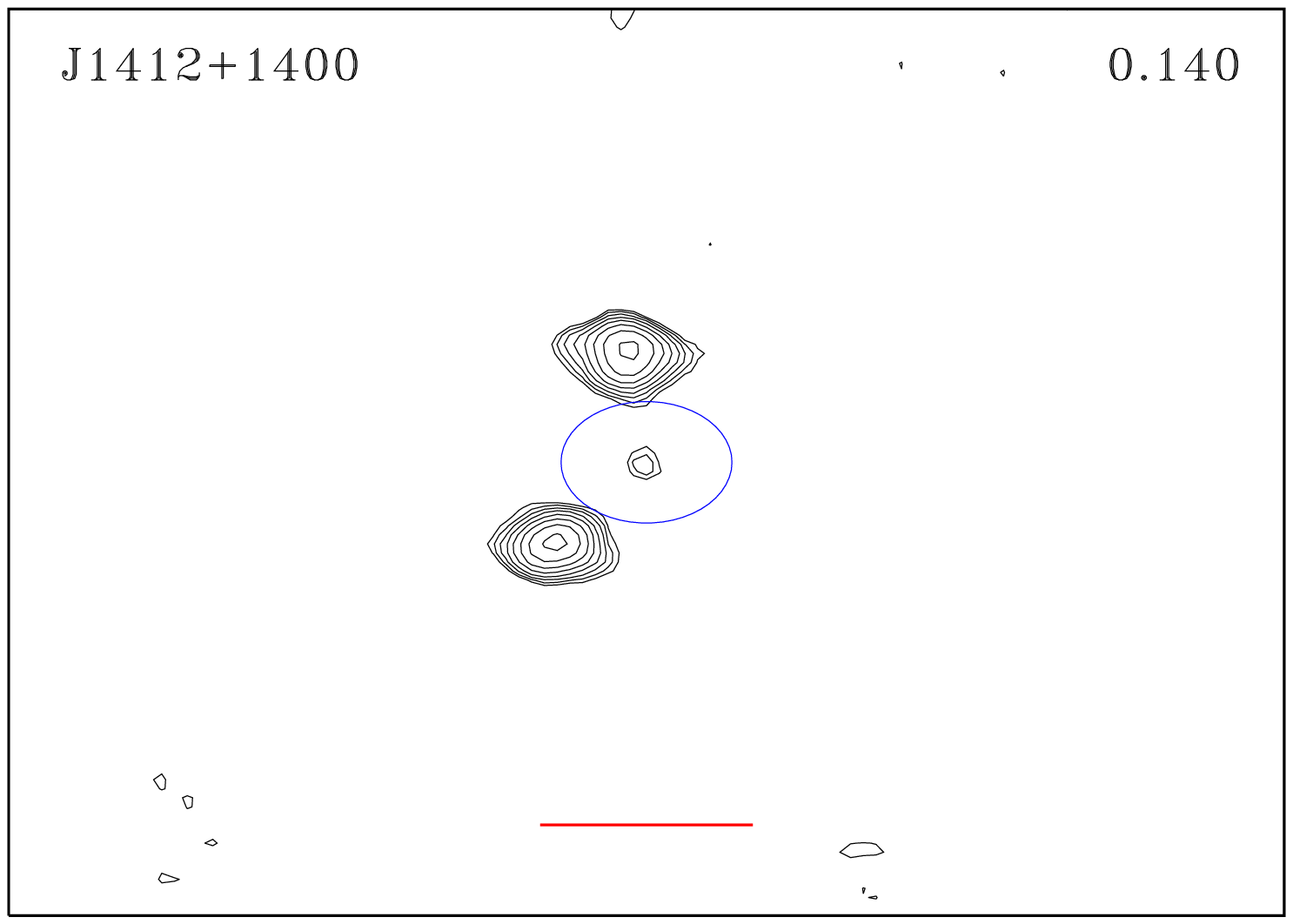}
\includegraphics[width=6.3cm,height=6.3cm]{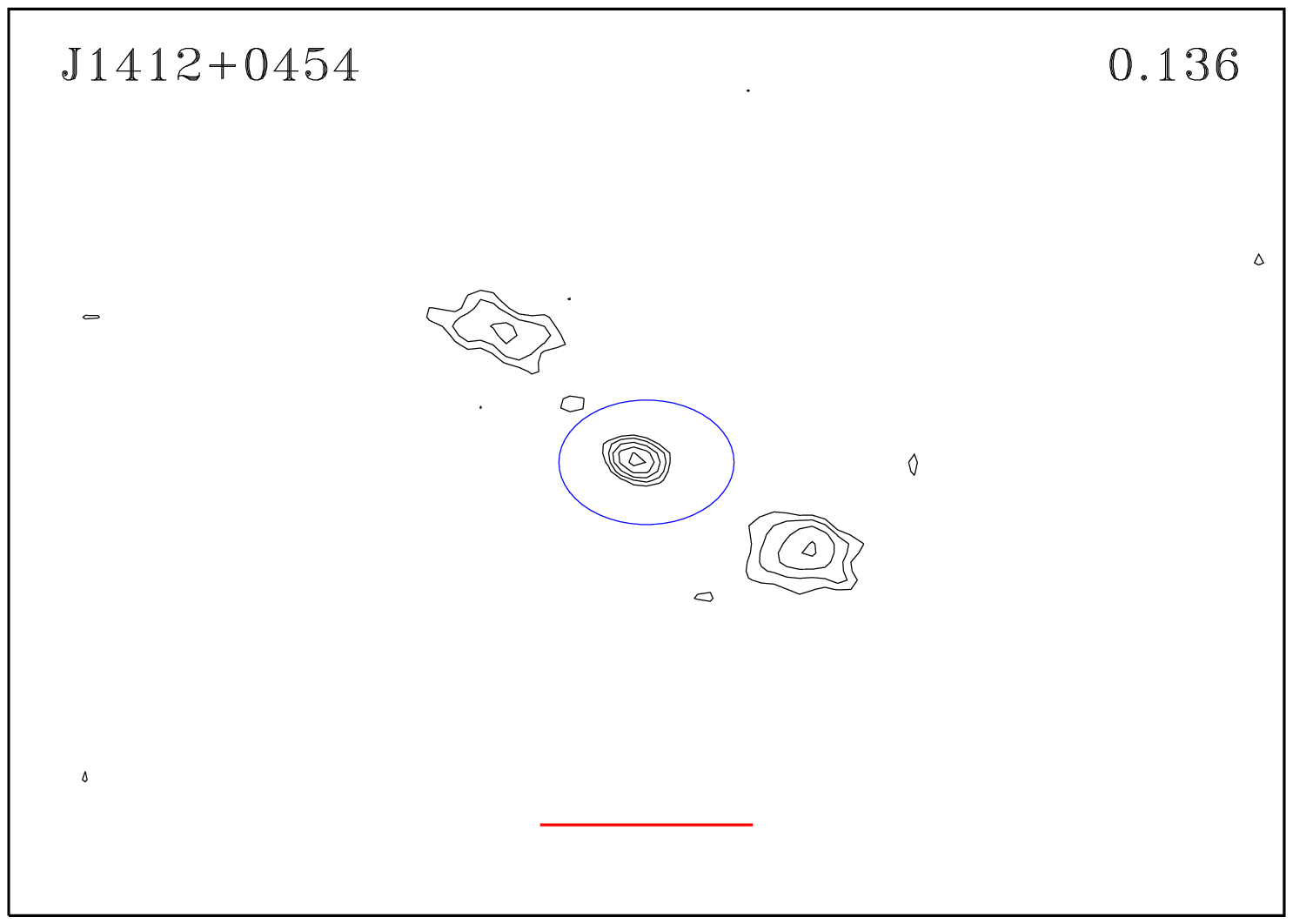}
\includegraphics[width=6.3cm,height=6.3cm]{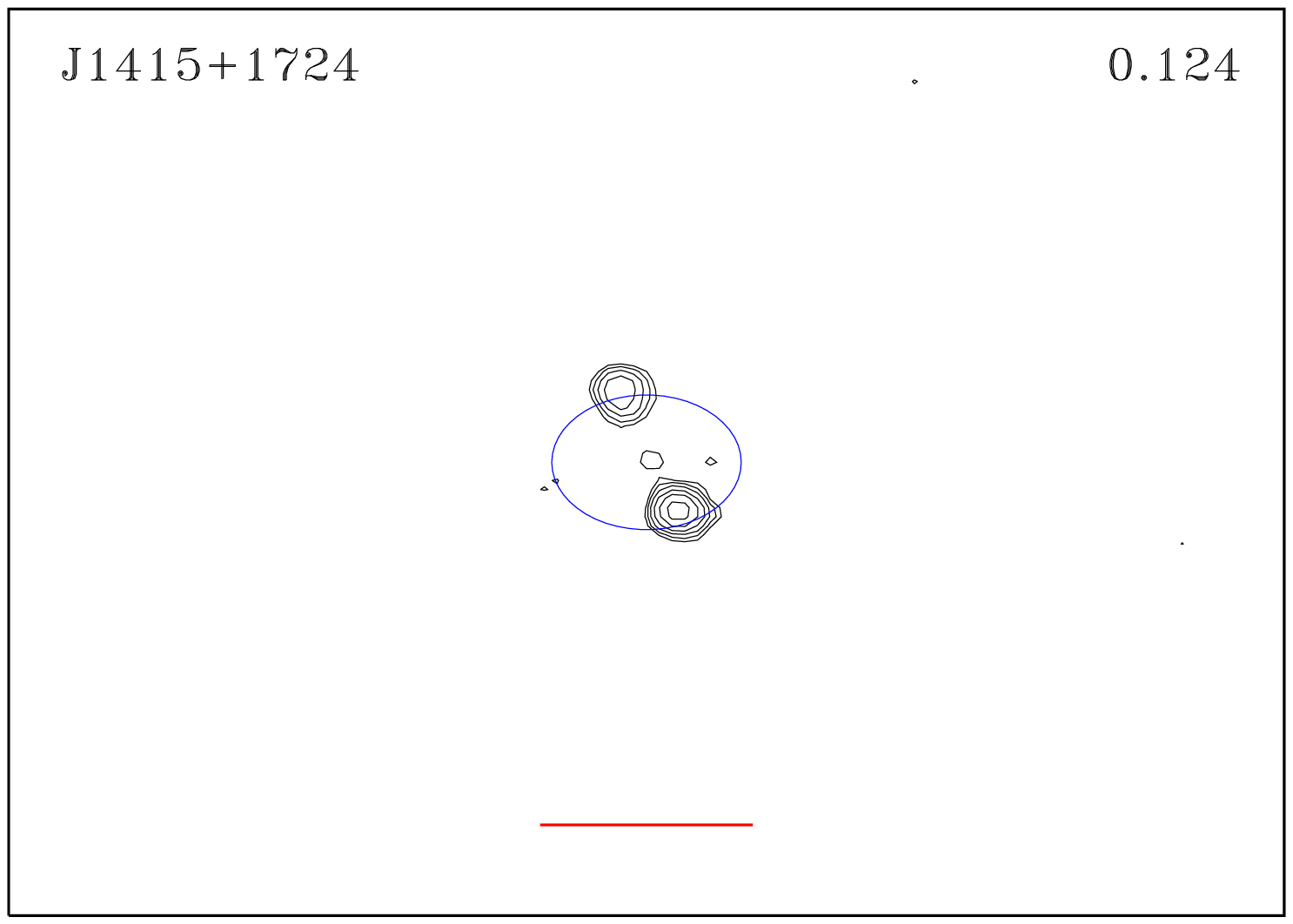}

\includegraphics[width=6.3cm,height=6.3cm]{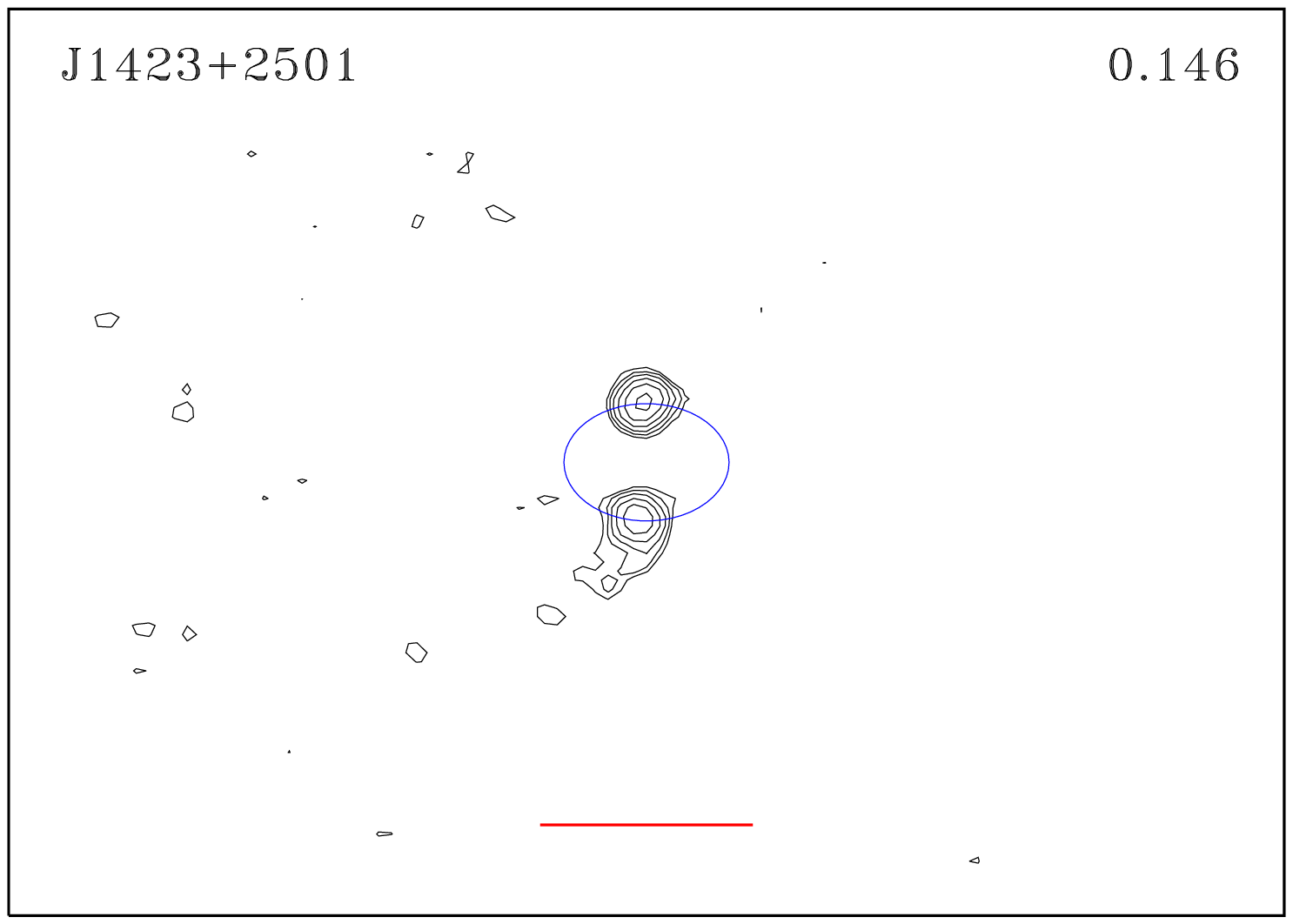}
\includegraphics[width=6.3cm,height=6.3cm]{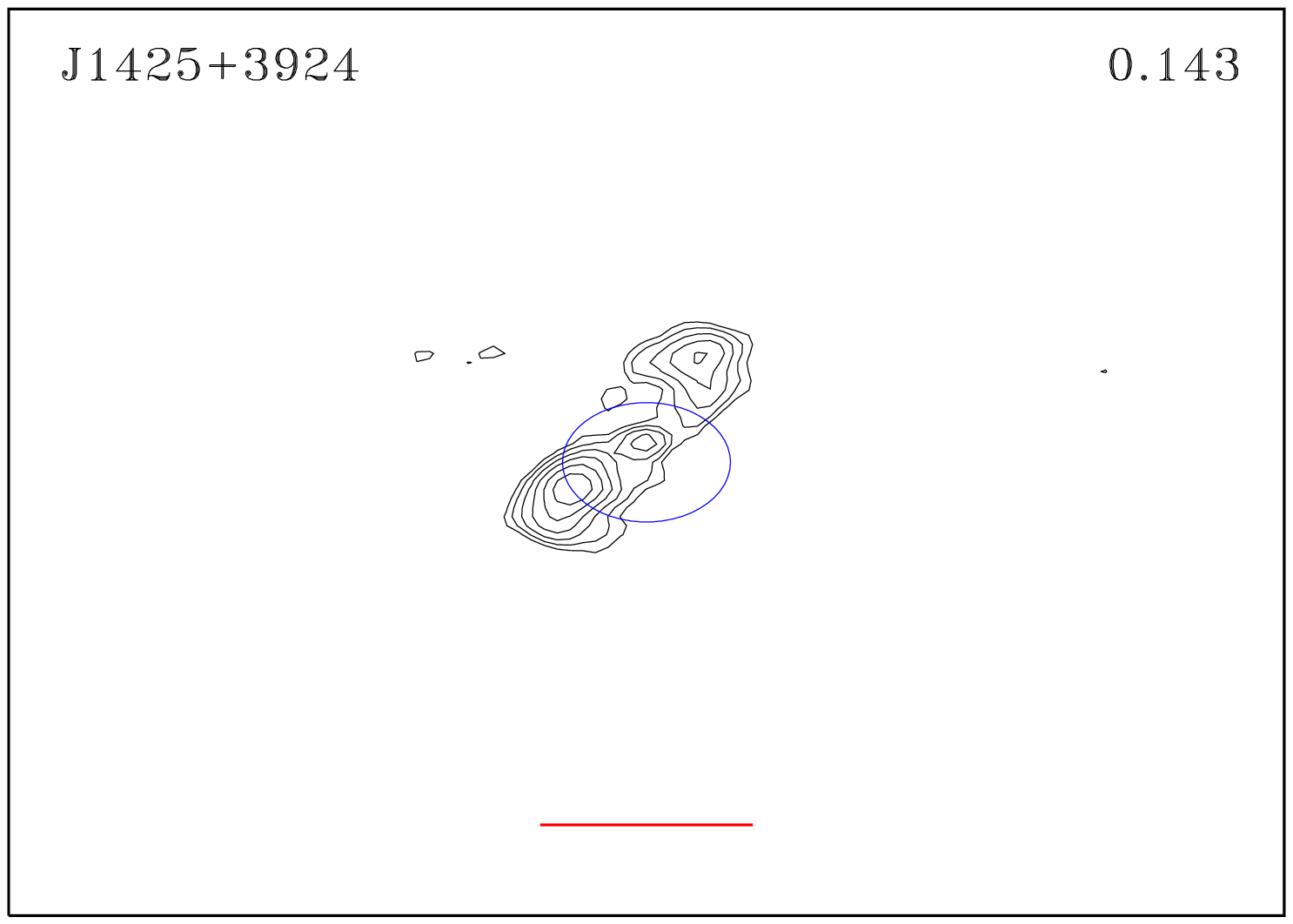}
\includegraphics[width=6.3cm,height=6.3cm]{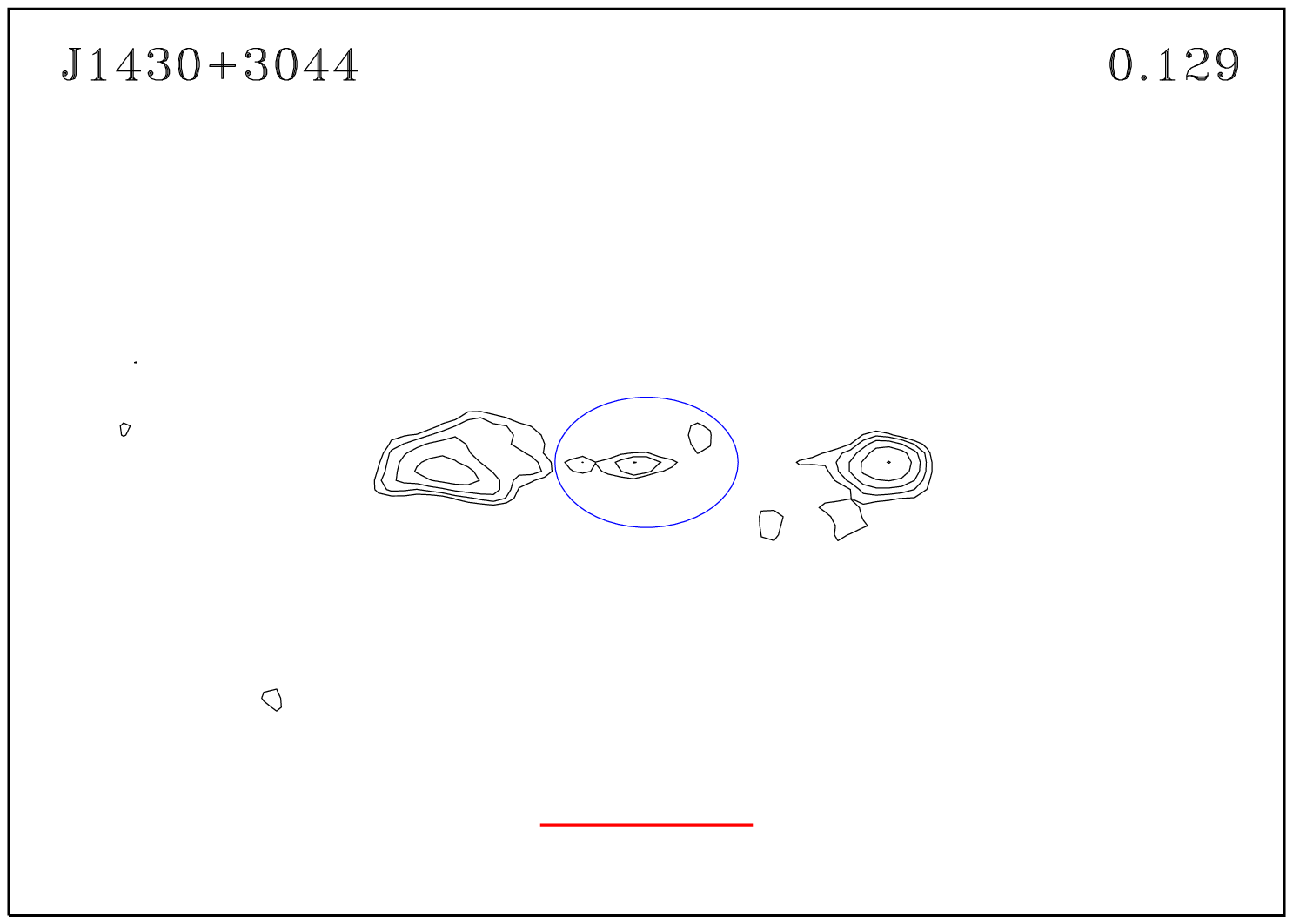}

\includegraphics[width=6.3cm,height=6.3cm]{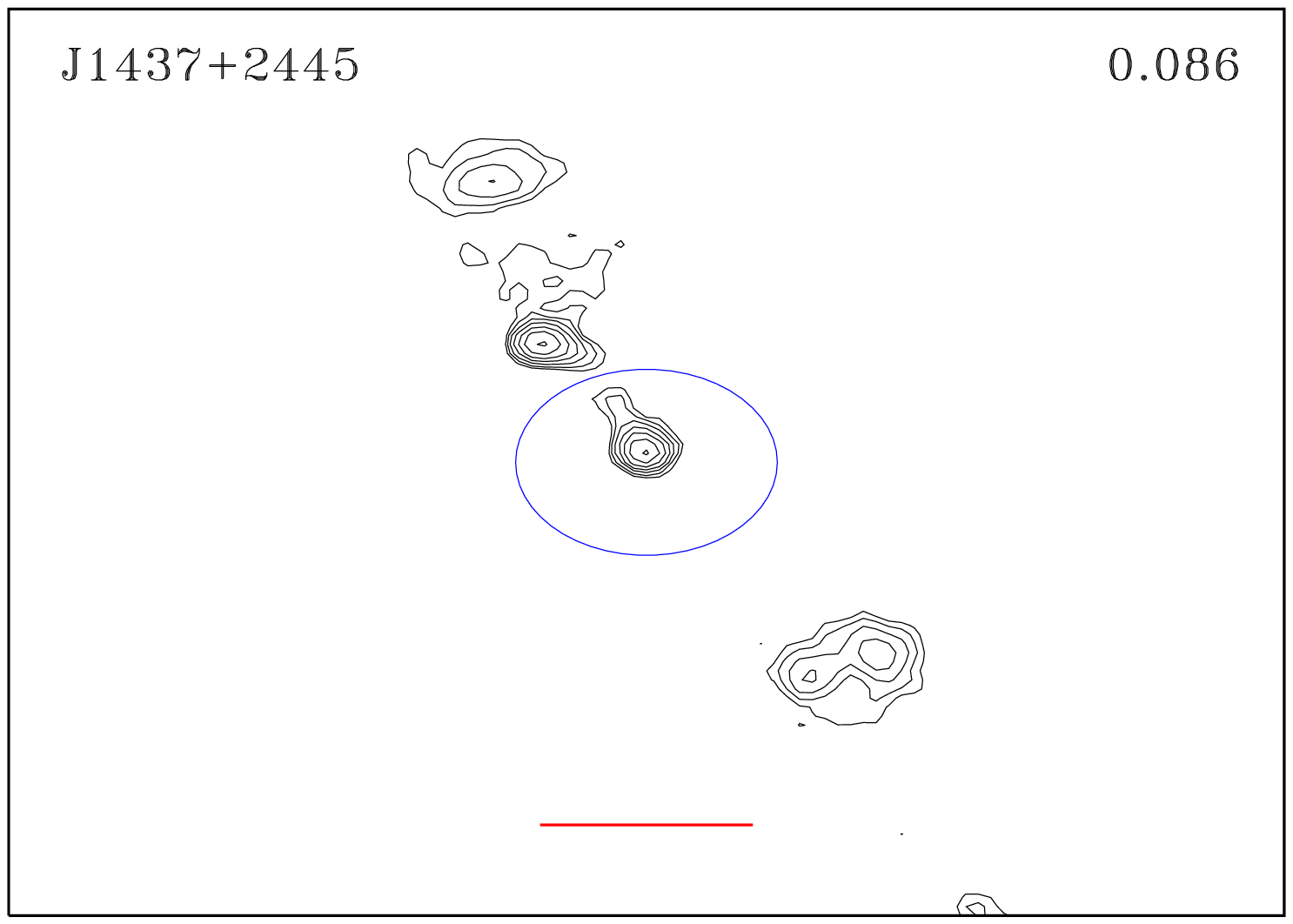}
\includegraphics[width=6.3cm,height=6.3cm]{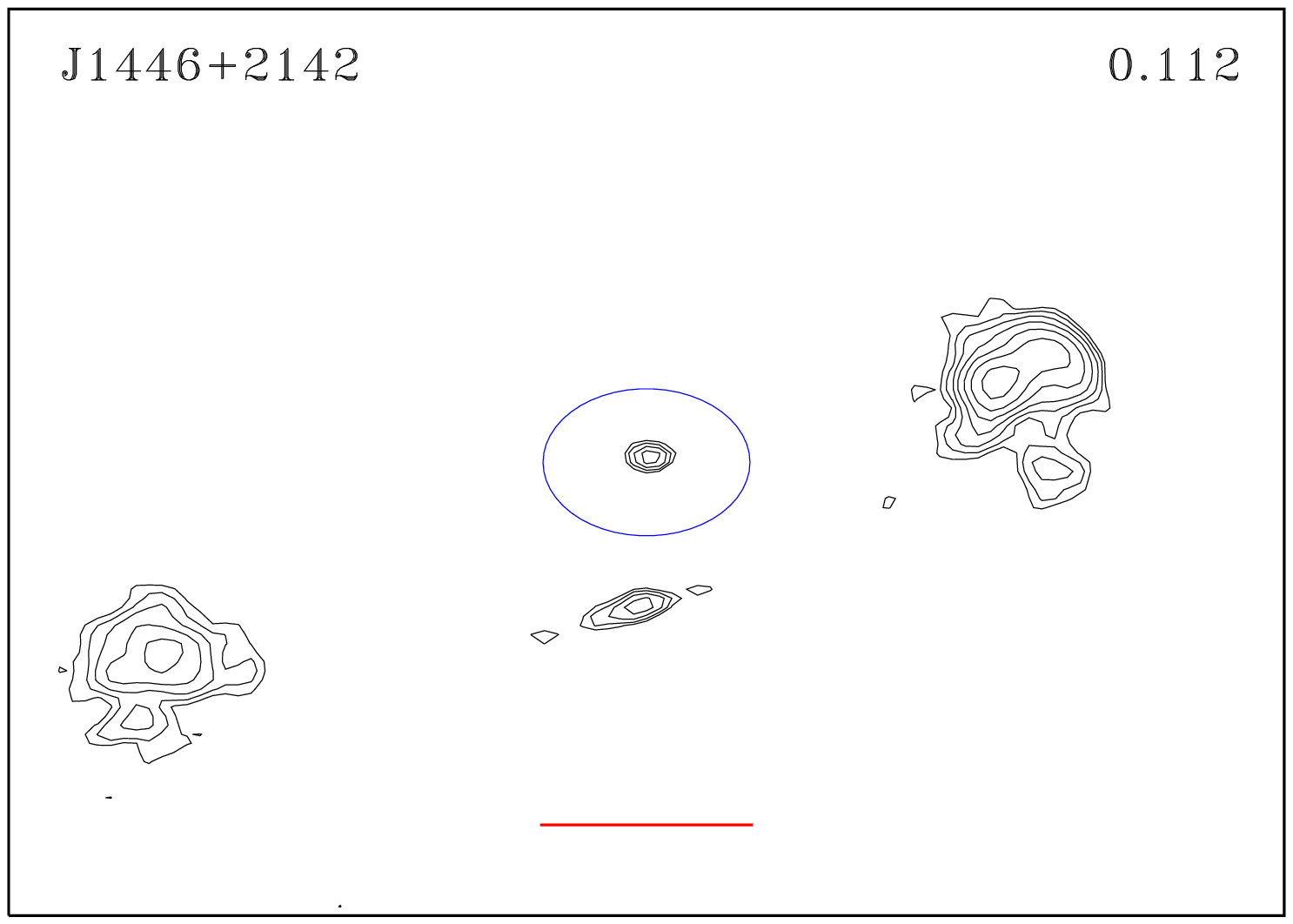}
\includegraphics[width=6.3cm,height=6.3cm]{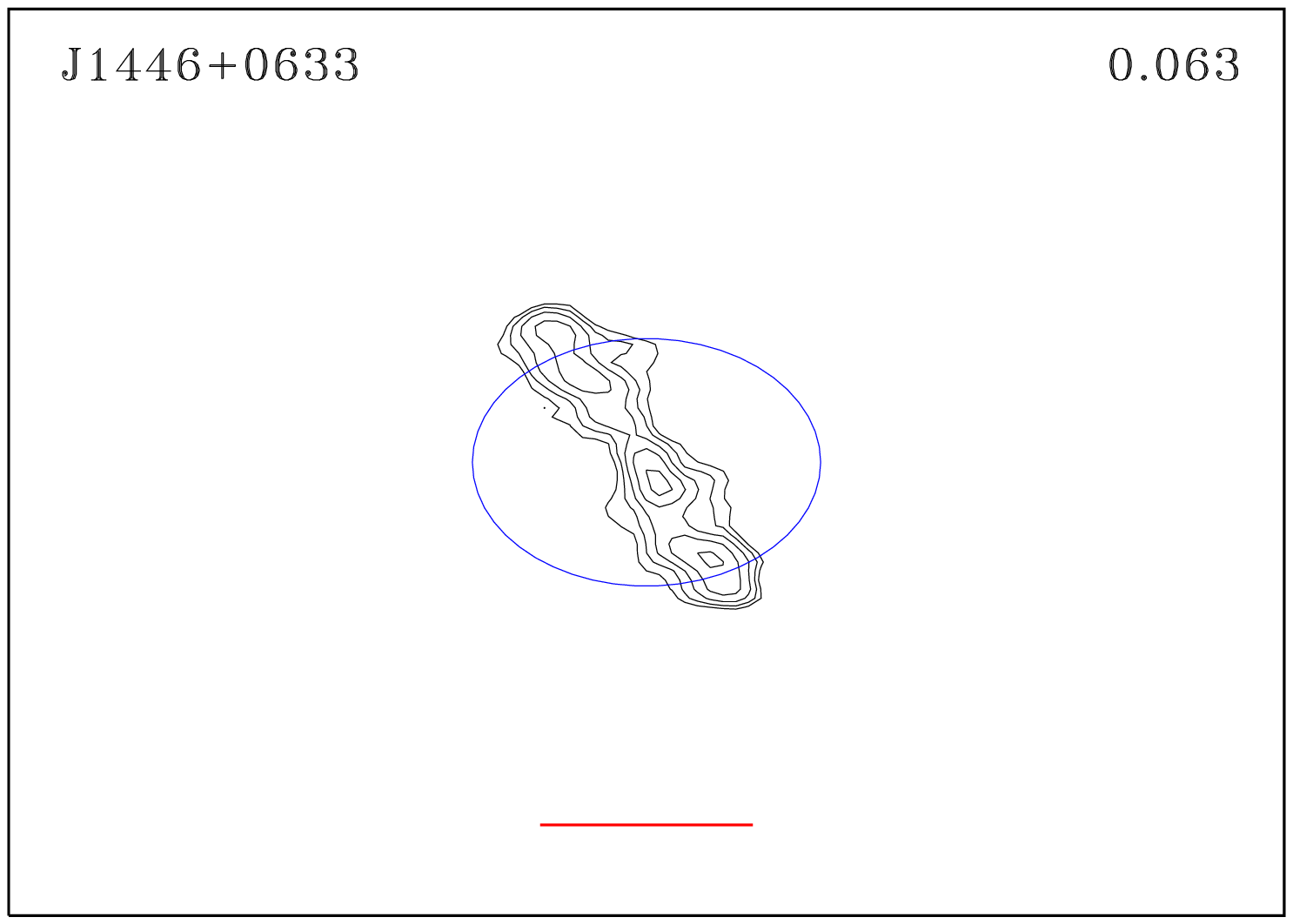}

\includegraphics[width=6.3cm,height=6.3cm]{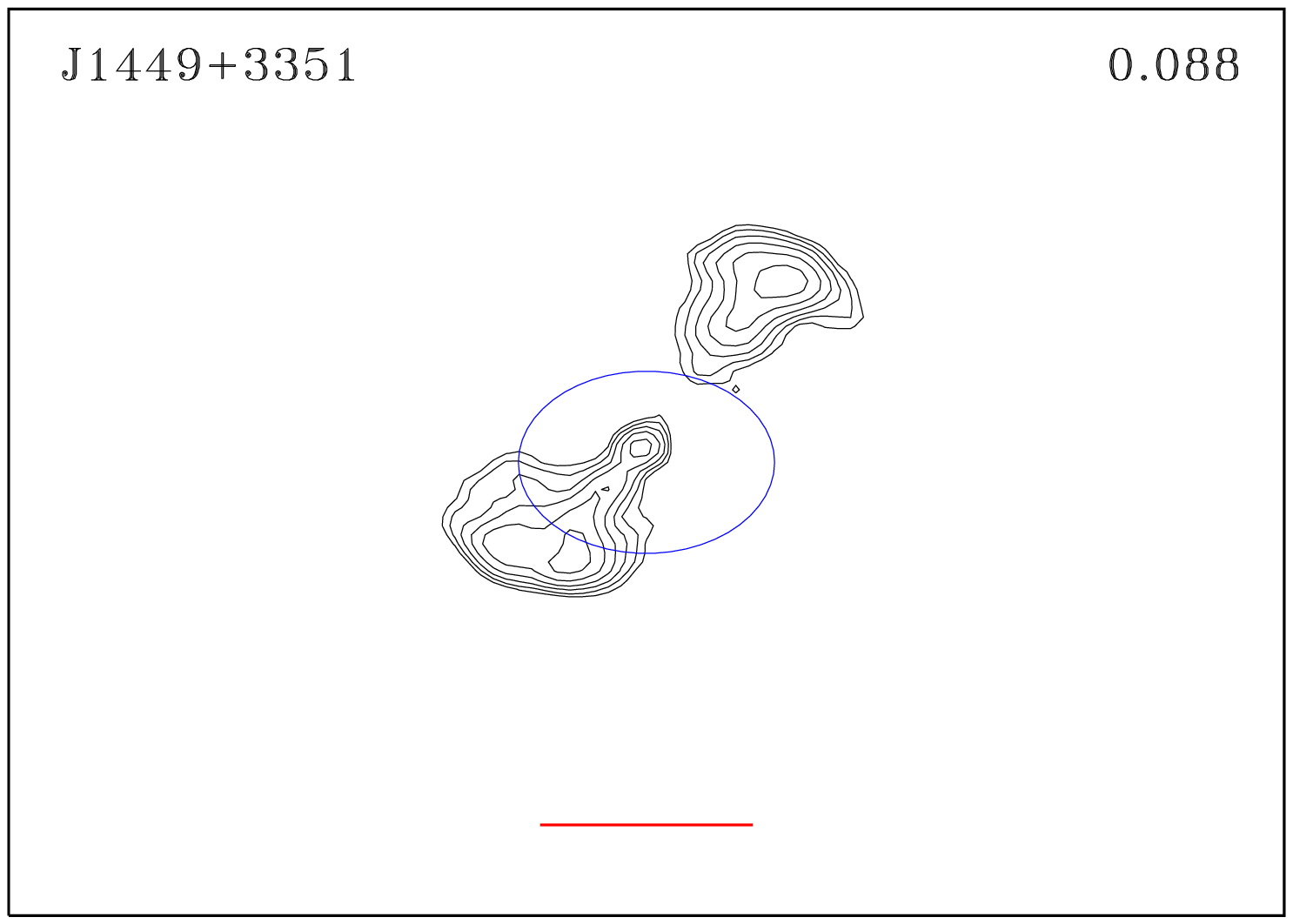}
\includegraphics[width=6.3cm,height=6.3cm]{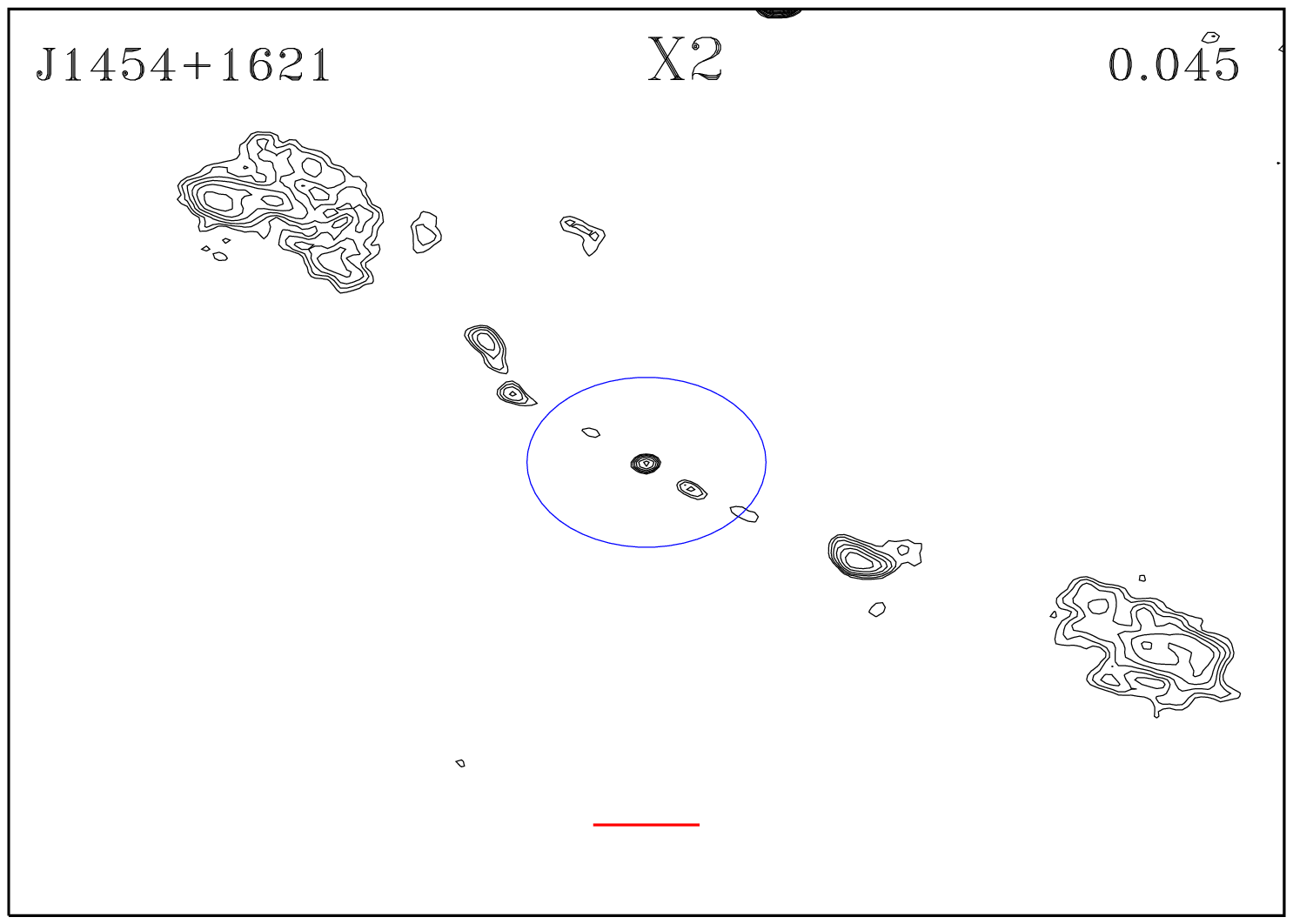}
\includegraphics[width=6.3cm,height=6.3cm]{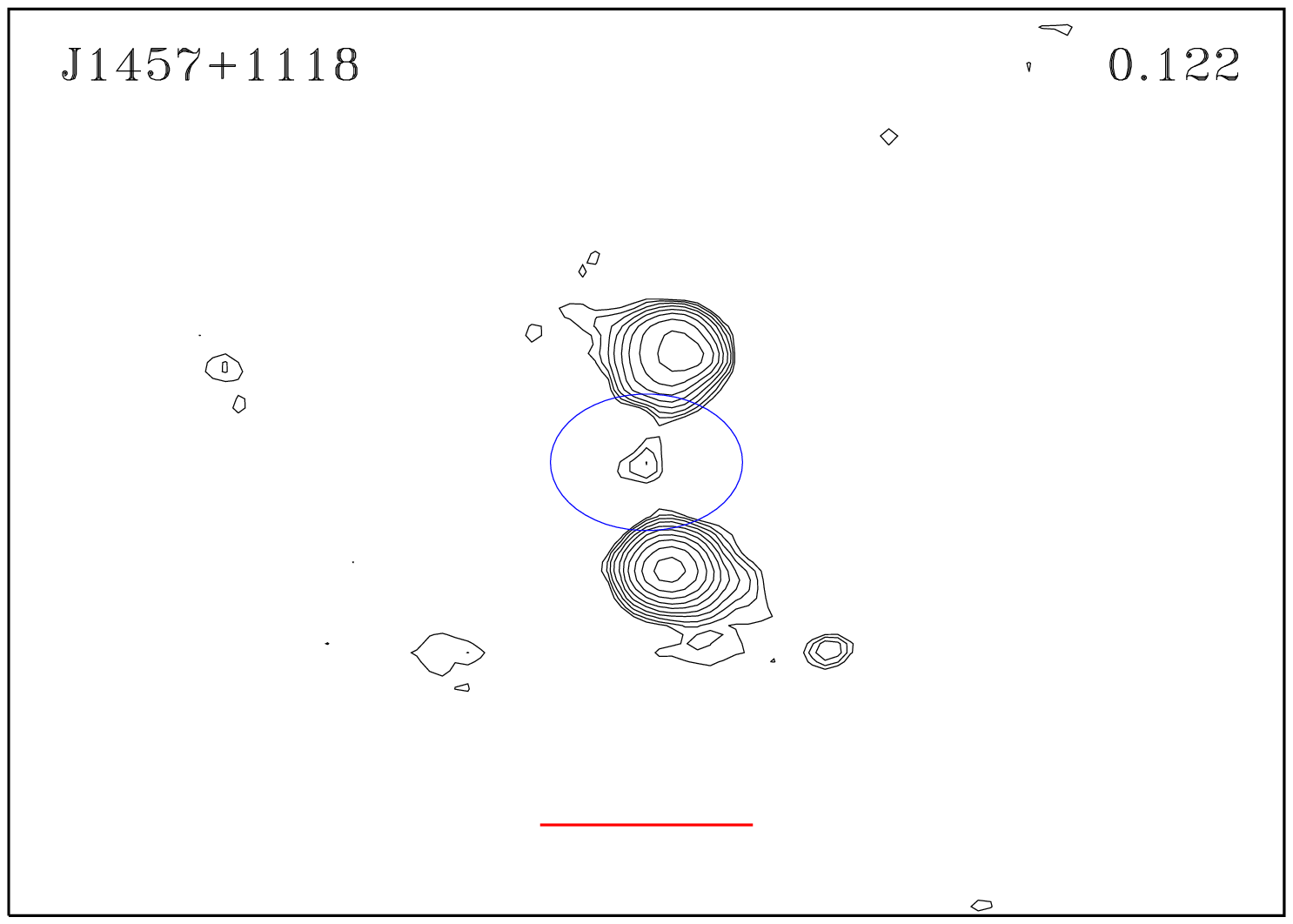}
\caption{(continued)}
\end{figure*}

\addtocounter{figure}{-1}
\begin{figure*}
\includegraphics[width=6.3cm,height=6.3cm]{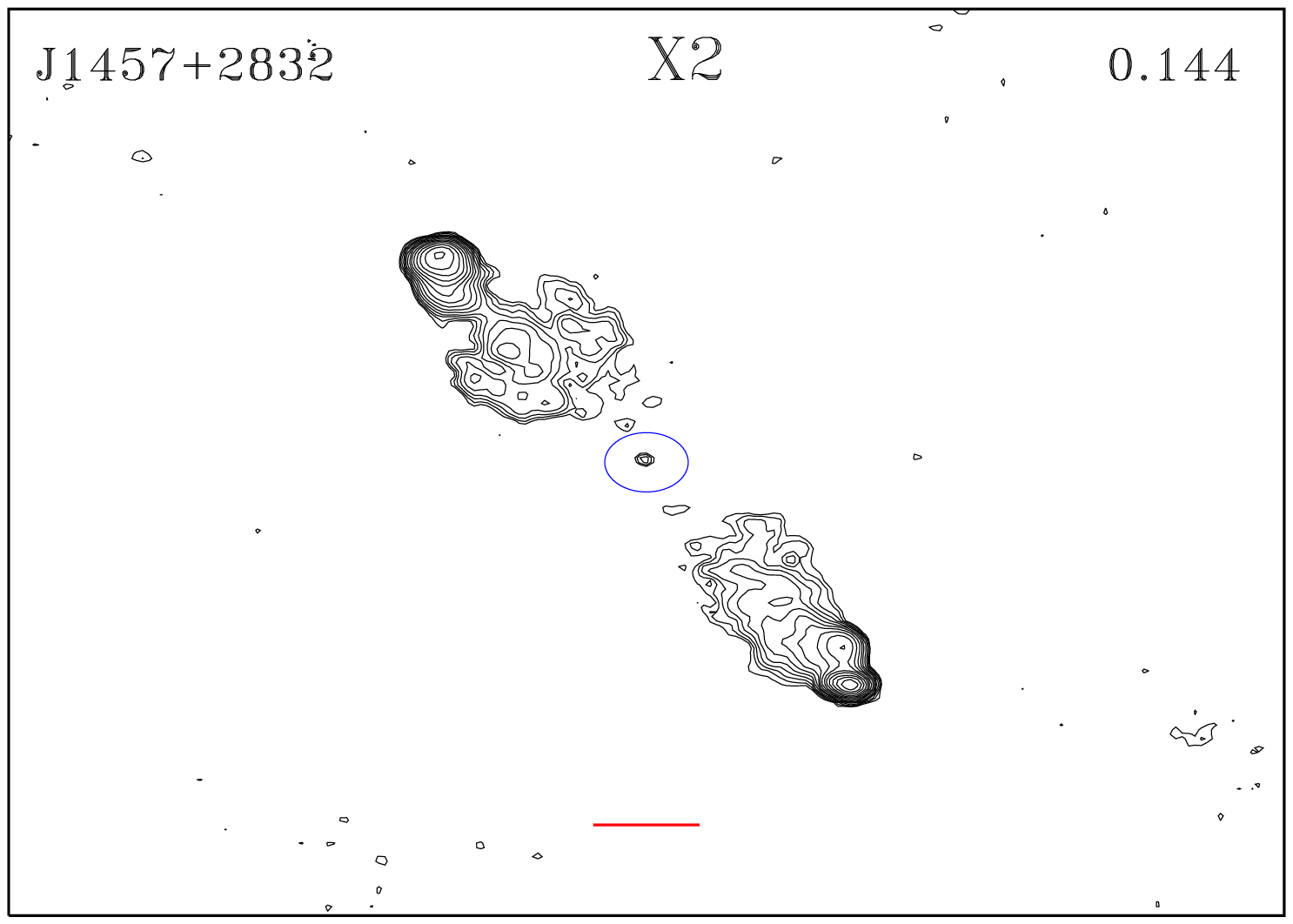}
\includegraphics[width=6.3cm,height=6.3cm]{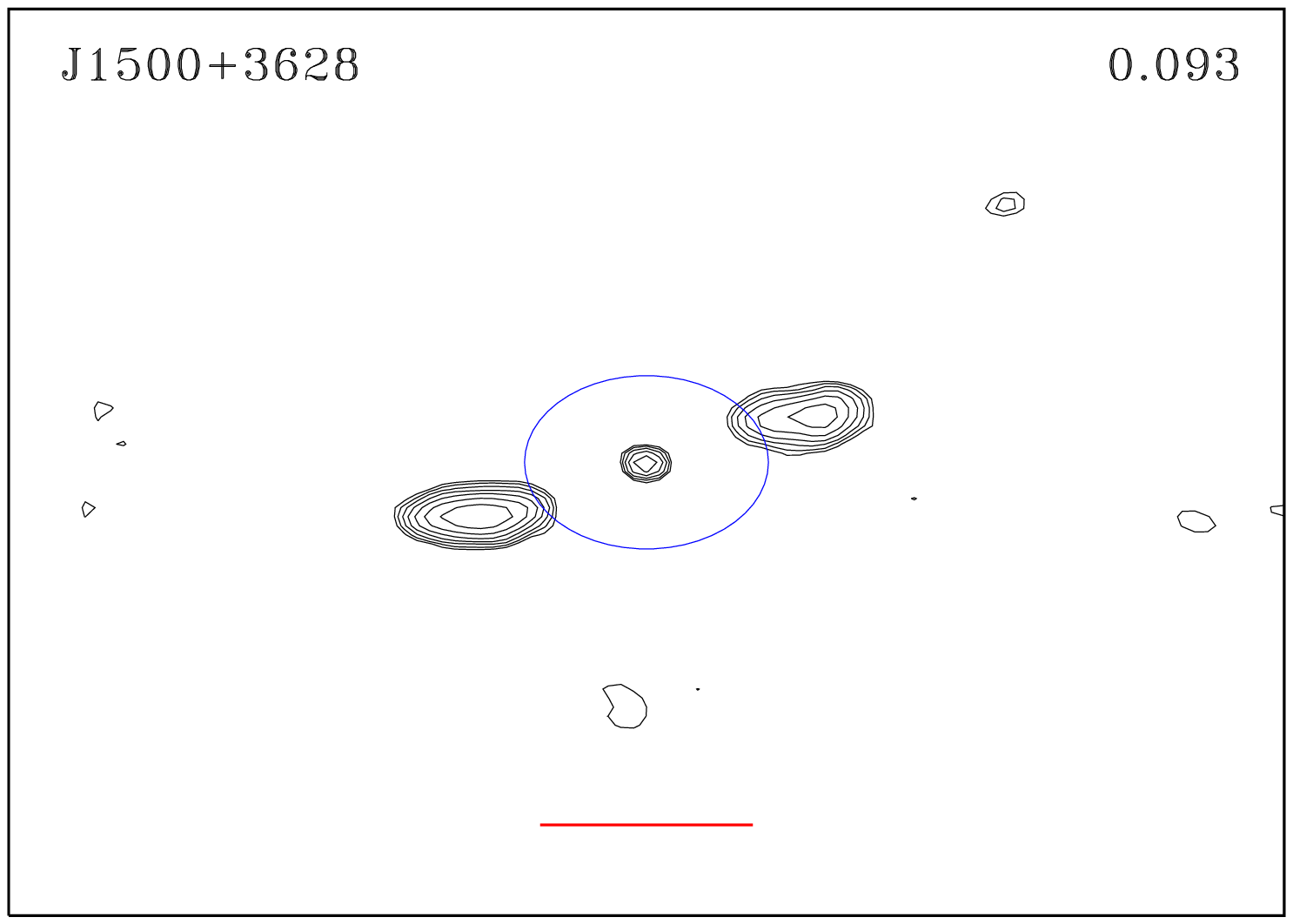}
\includegraphics[width=6.3cm,height=6.3cm]{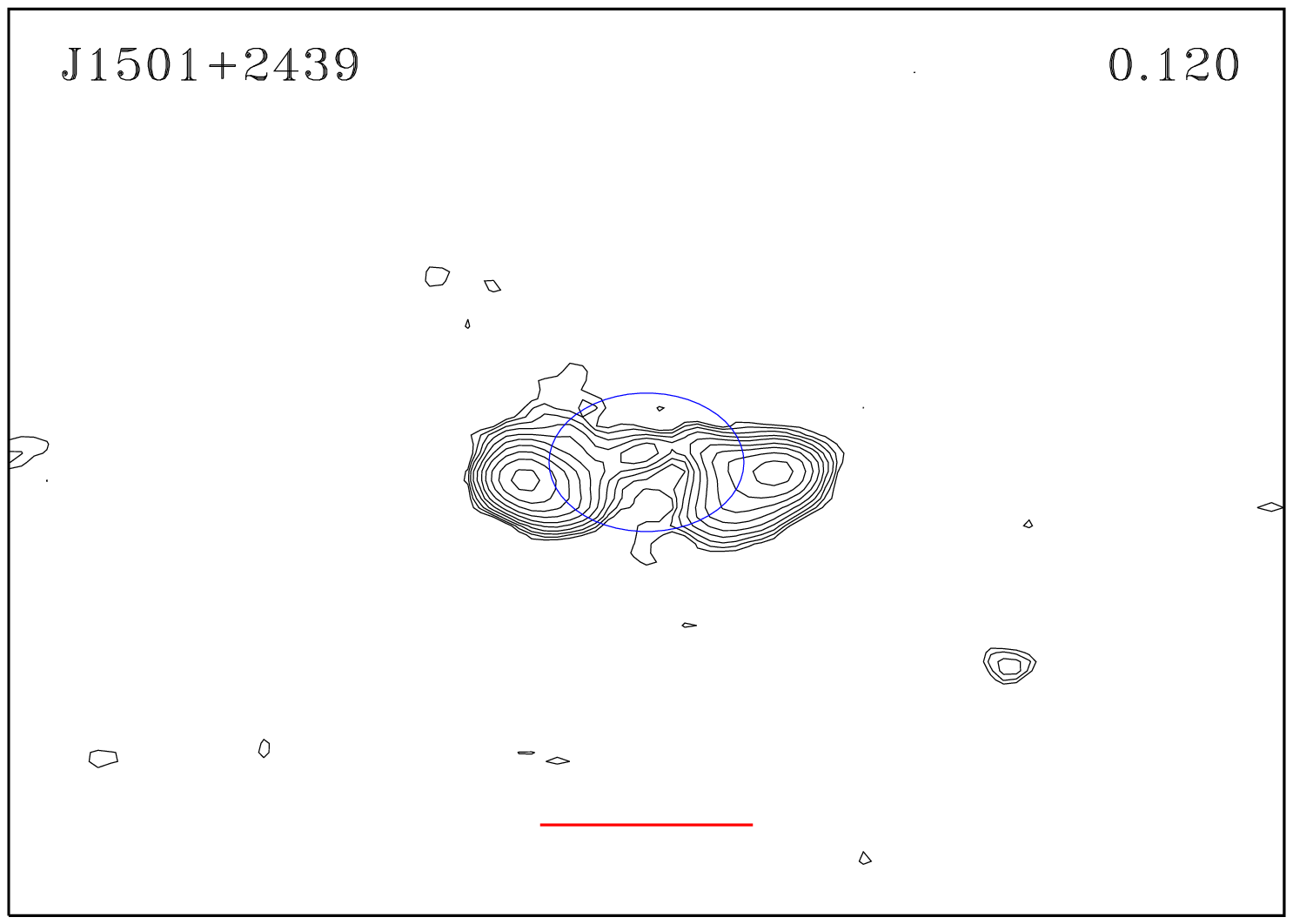}

\includegraphics[width=6.3cm,height=6.3cm]{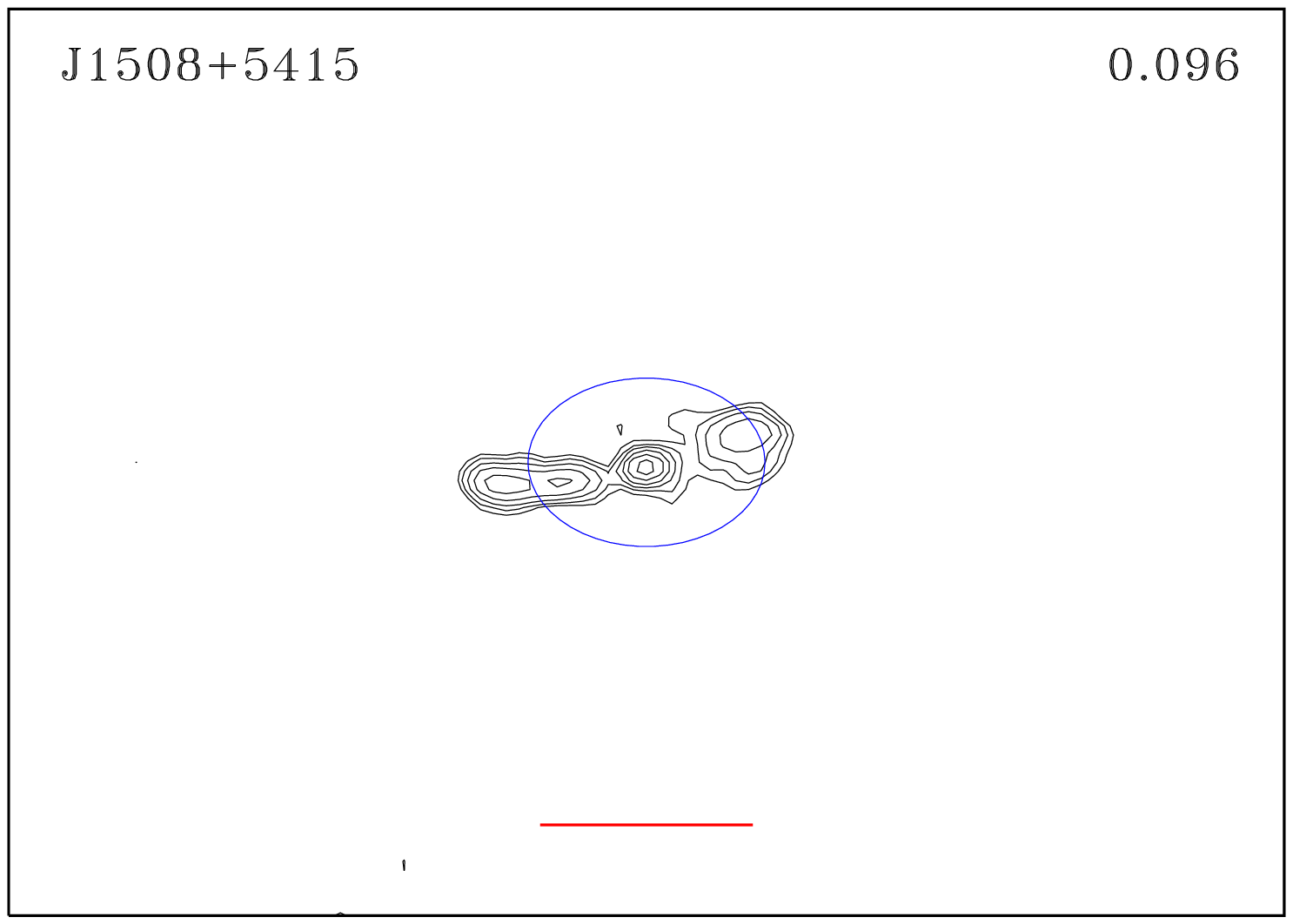}
\includegraphics[width=6.3cm,height=6.3cm]{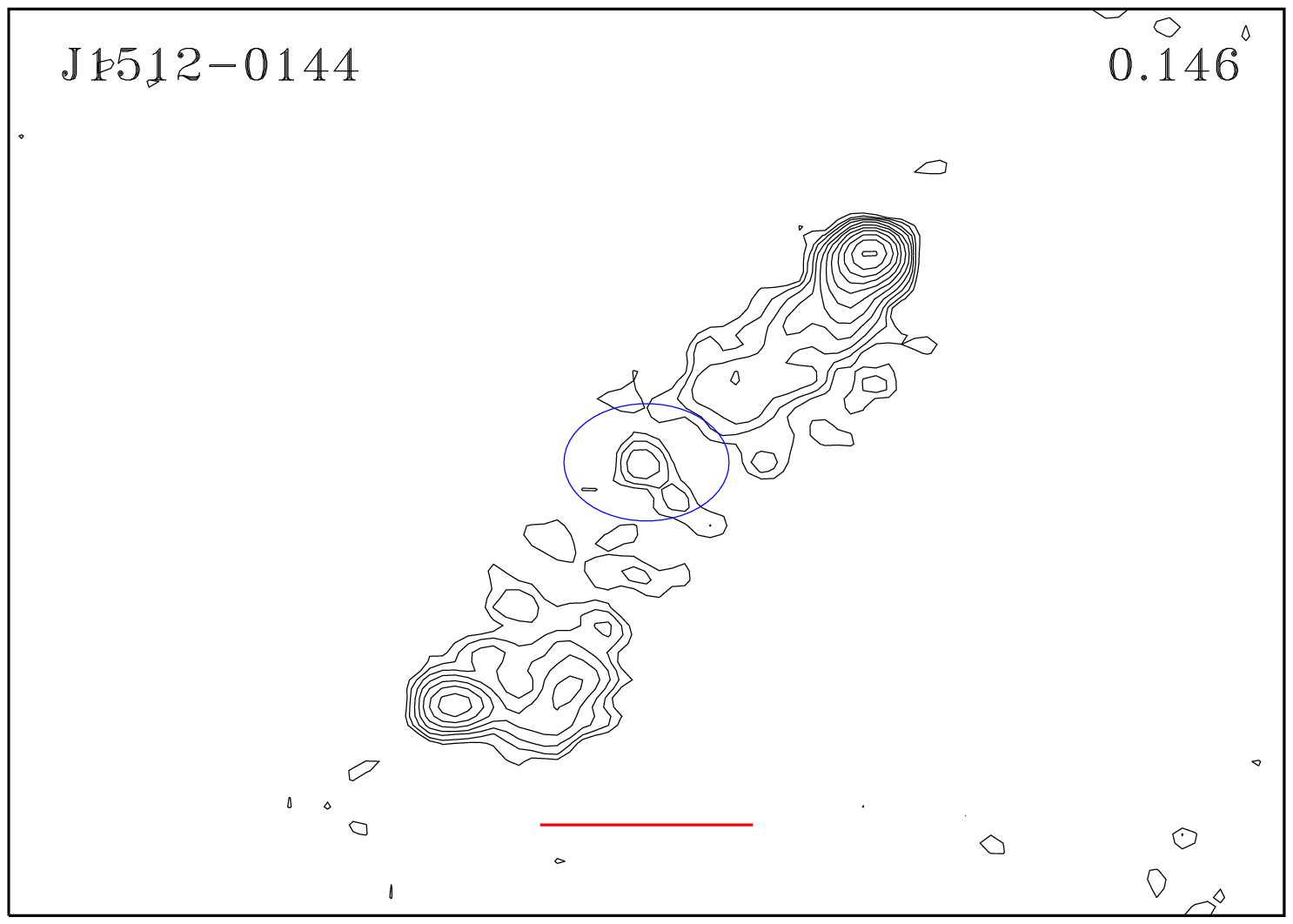}
\includegraphics[width=6.3cm,height=6.3cm]{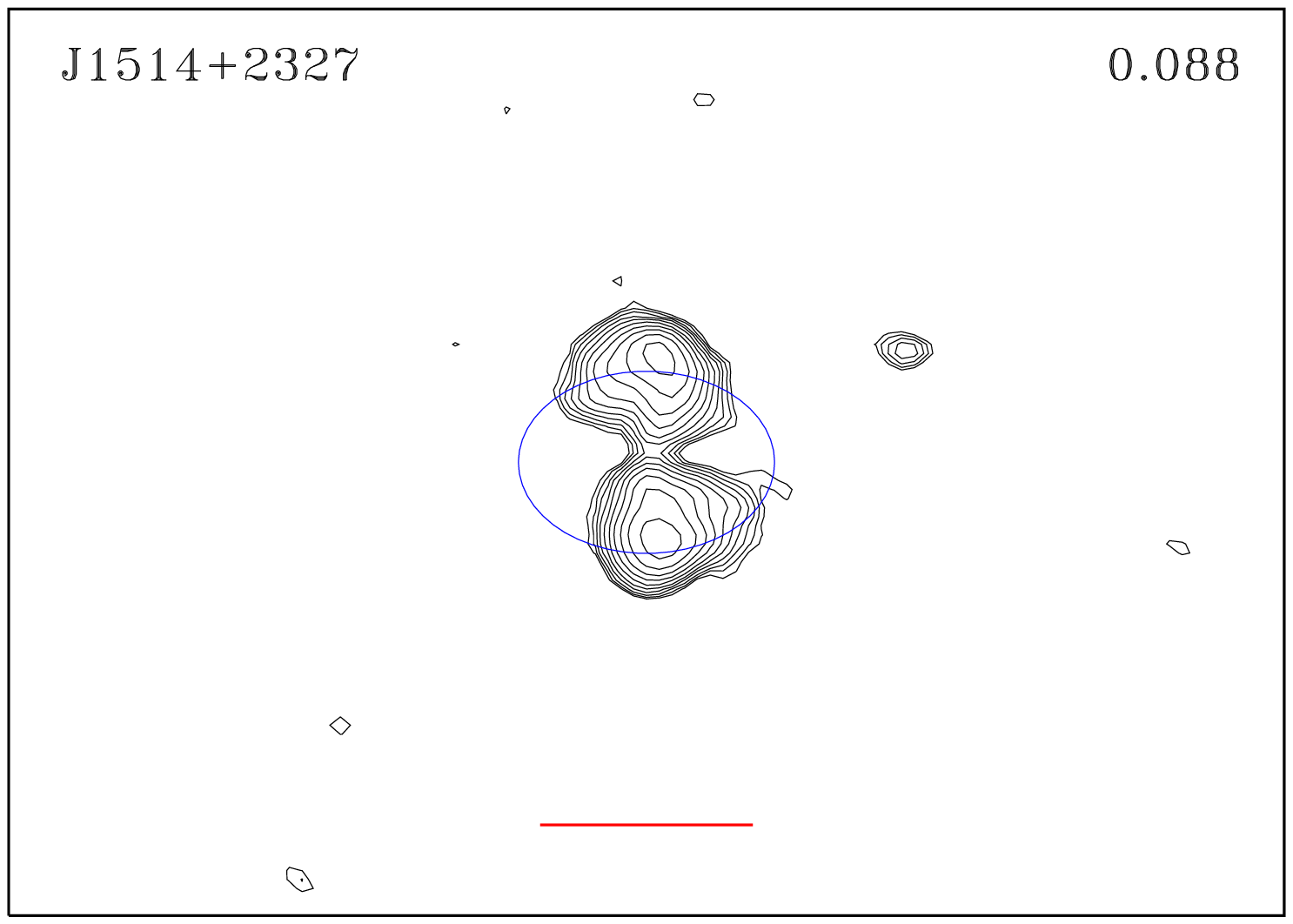}

\includegraphics[width=6.3cm,height=6.3cm]{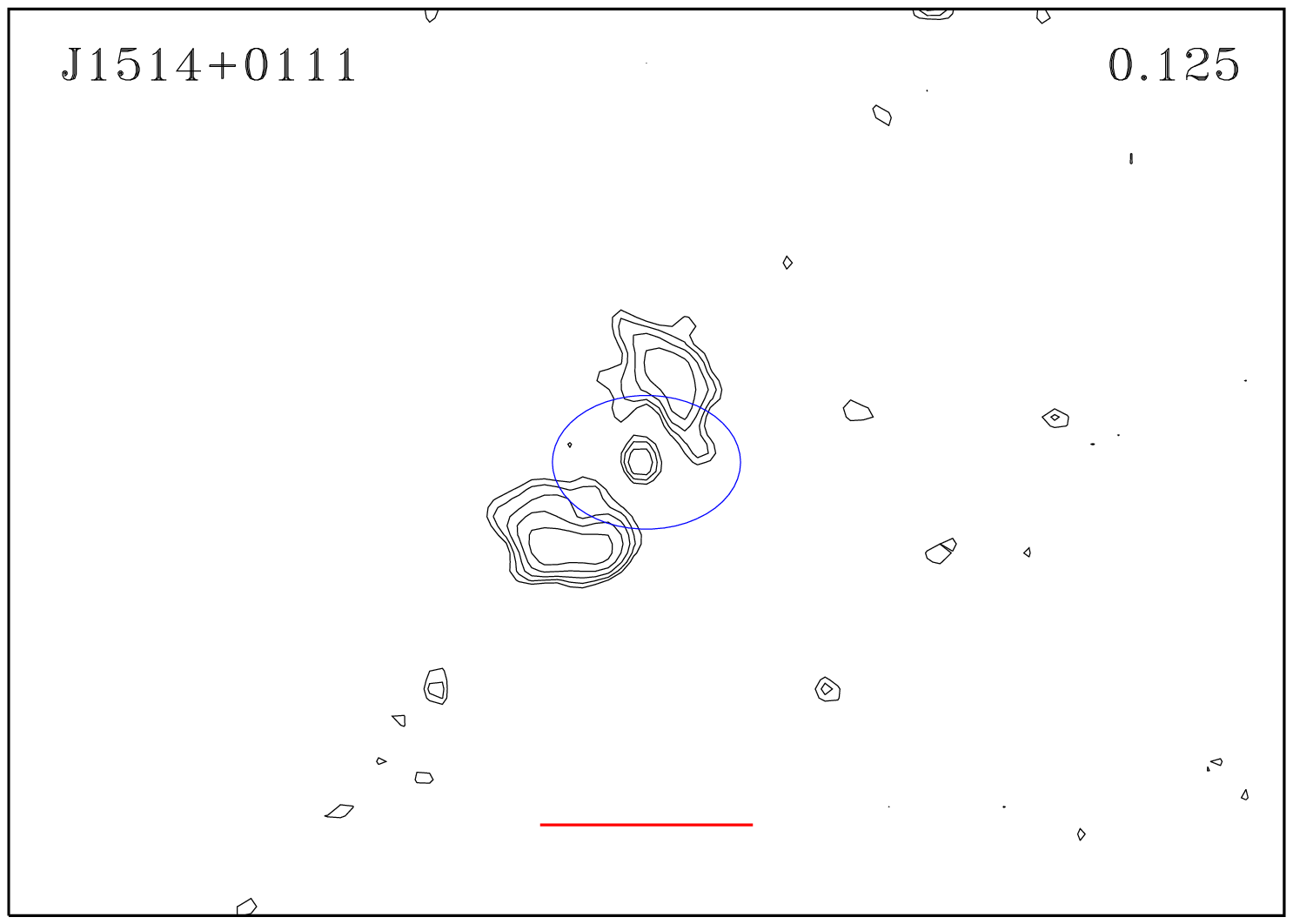}
\includegraphics[width=6.3cm,height=6.3cm]{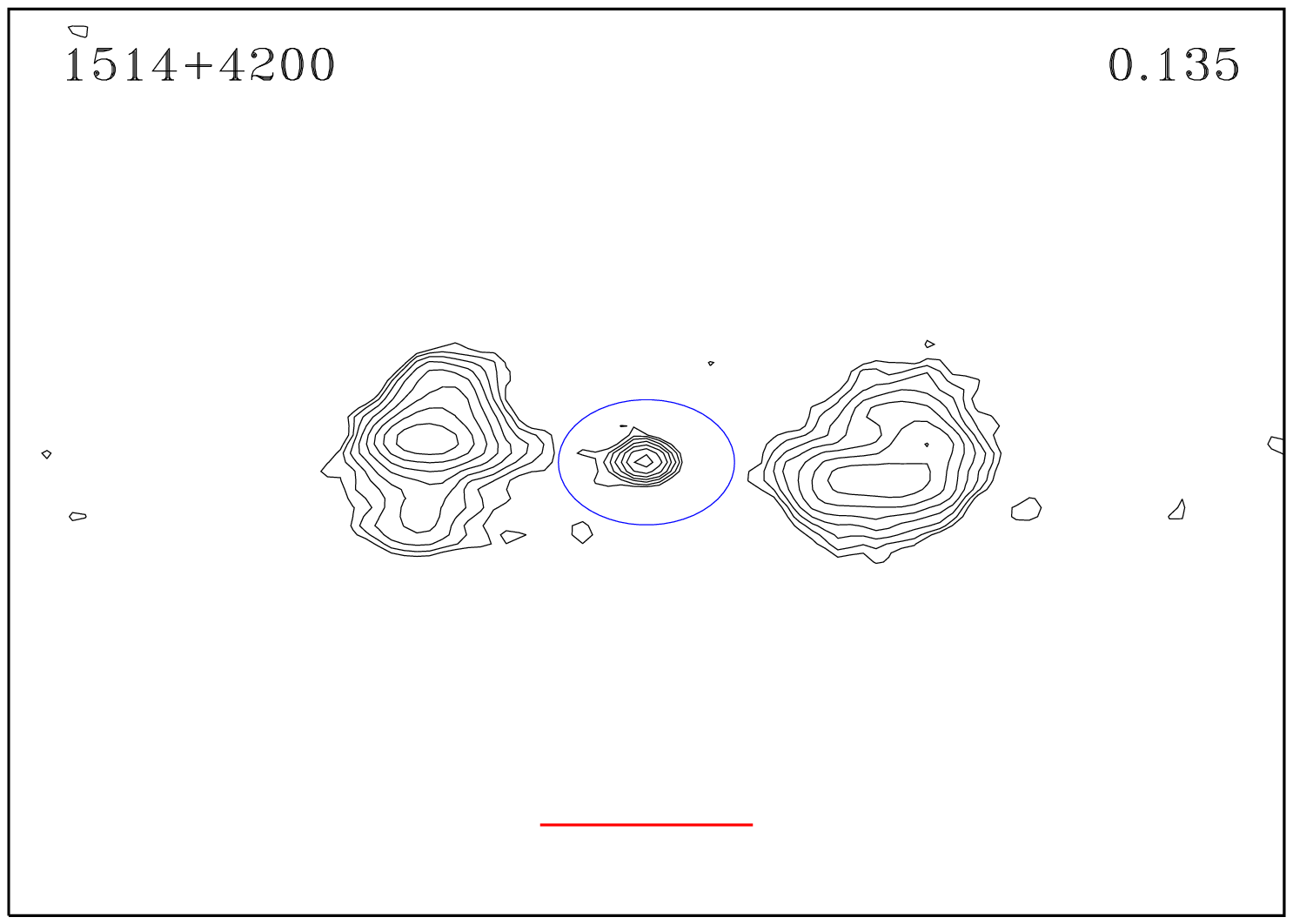}
\includegraphics[width=6.3cm,height=6.3cm]{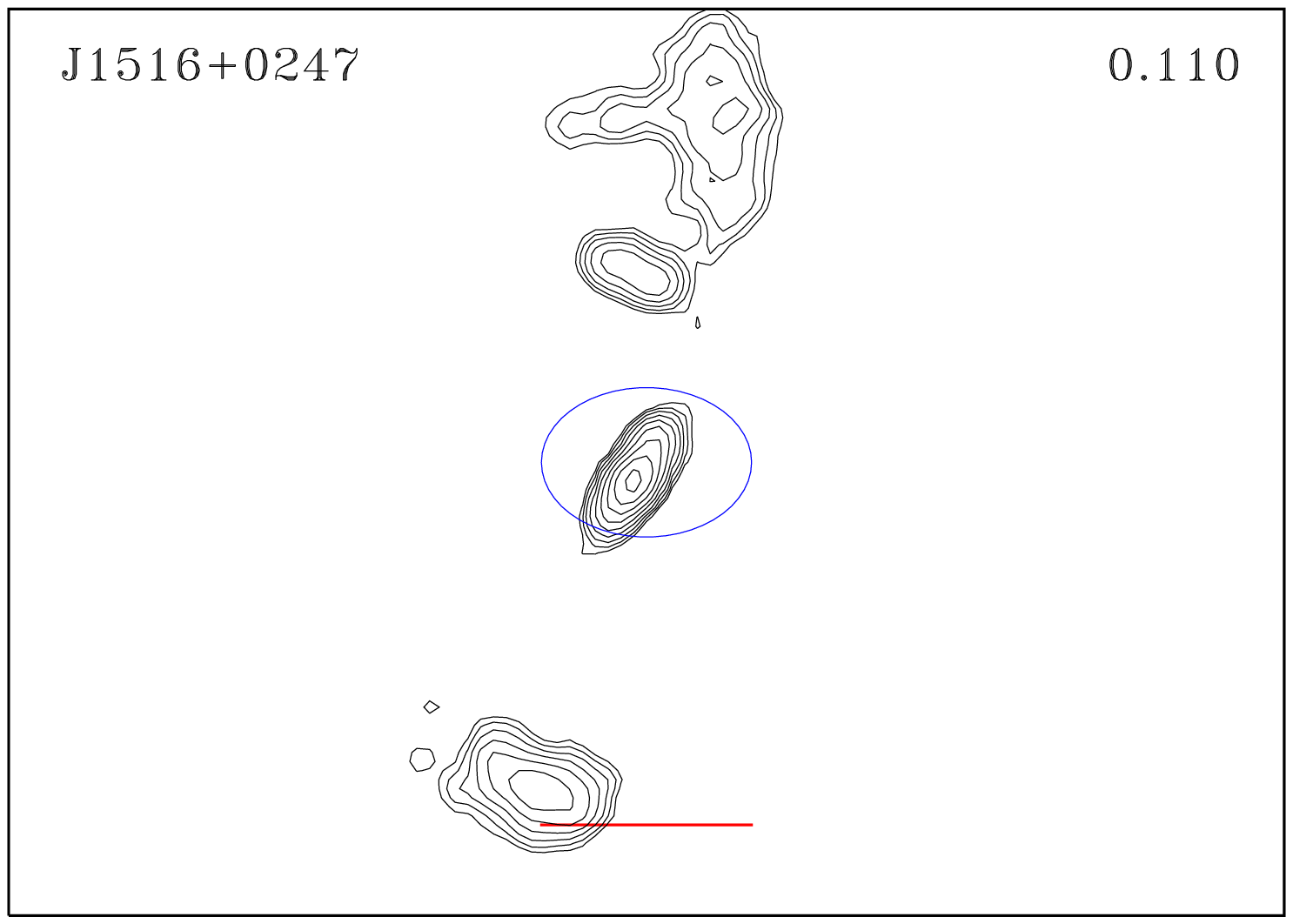}

\includegraphics[width=6.3cm,height=6.3cm]{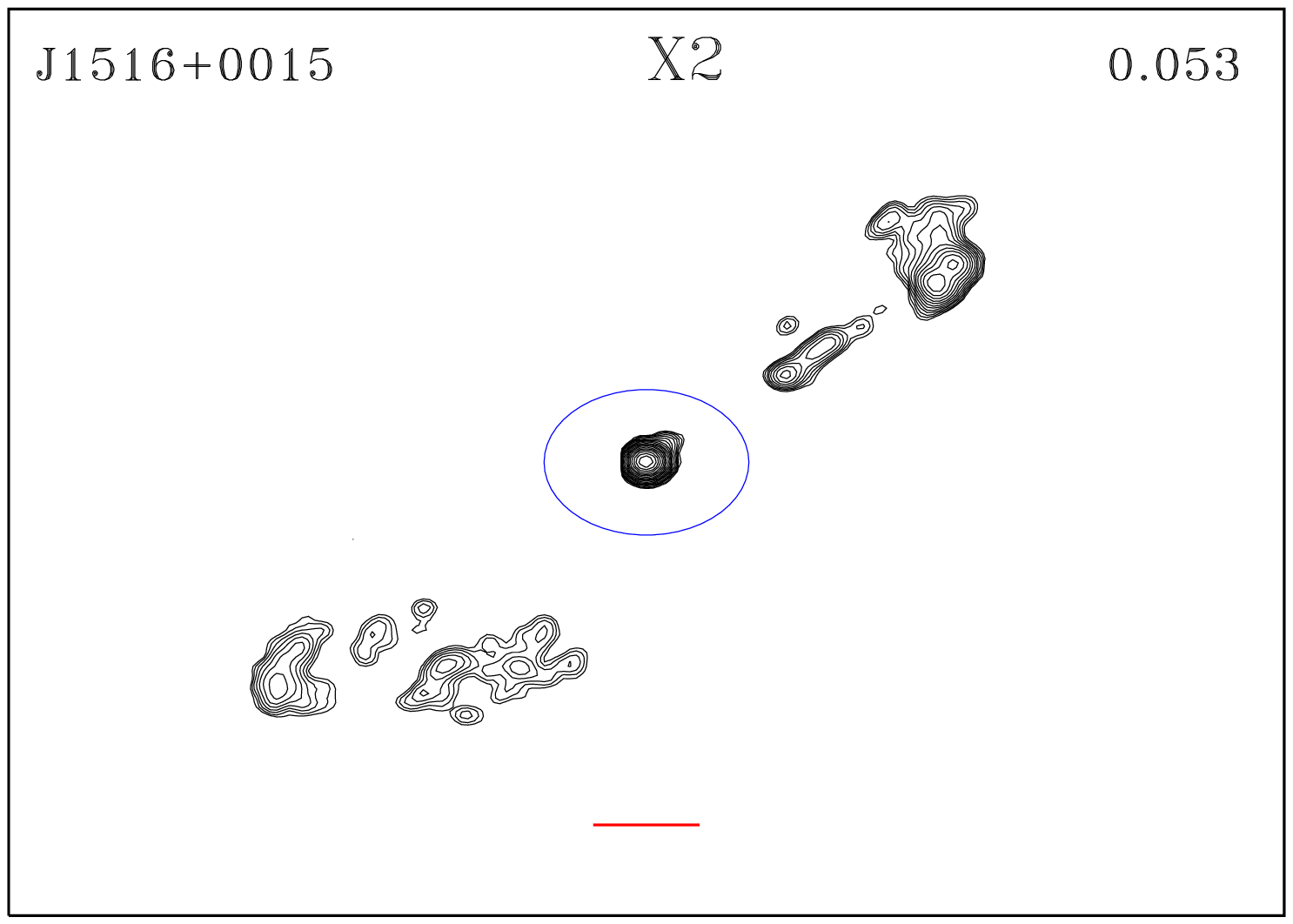}
\includegraphics[width=6.3cm,height=6.3cm]{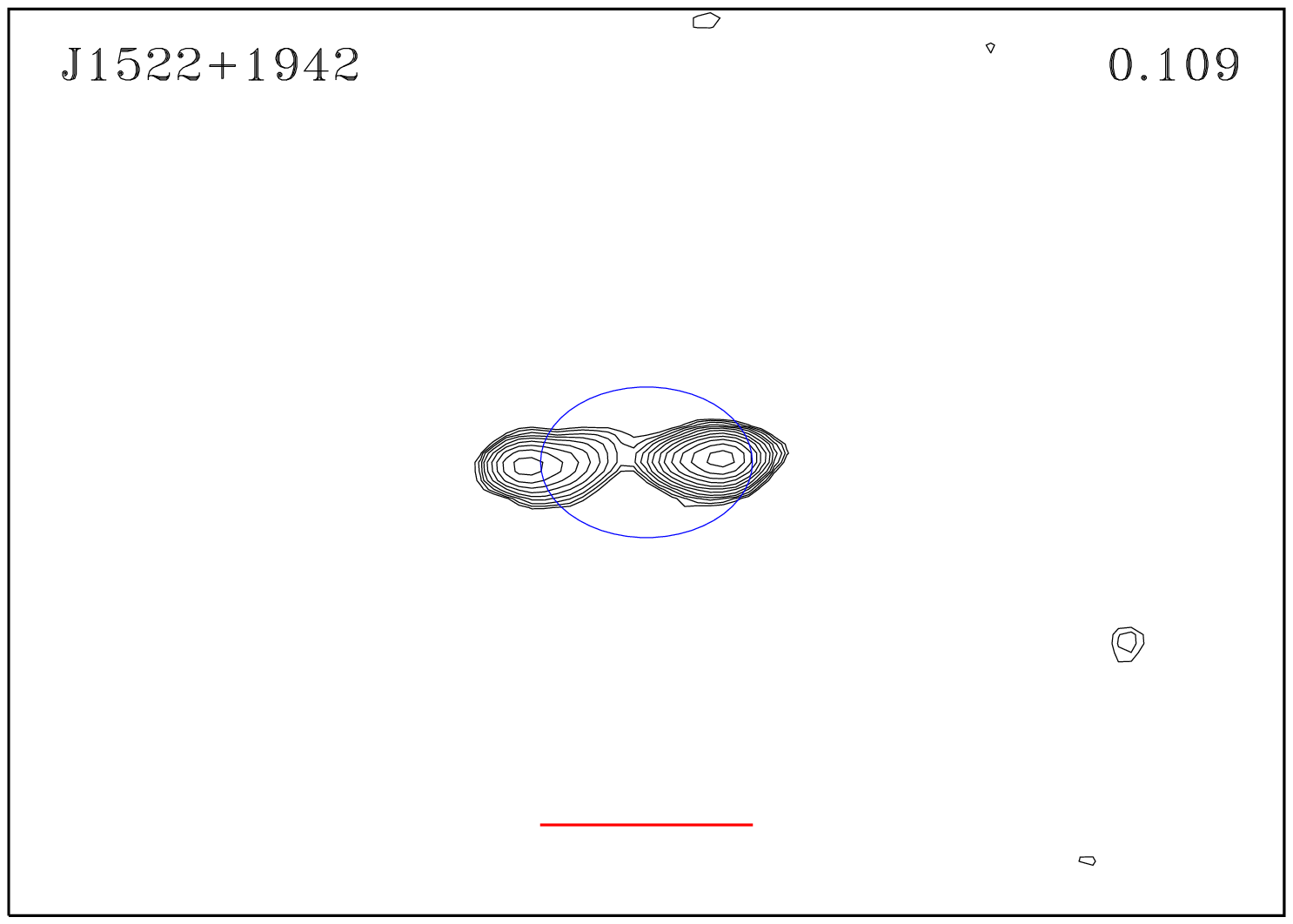}
\includegraphics[width=6.3cm,height=6.3cm]{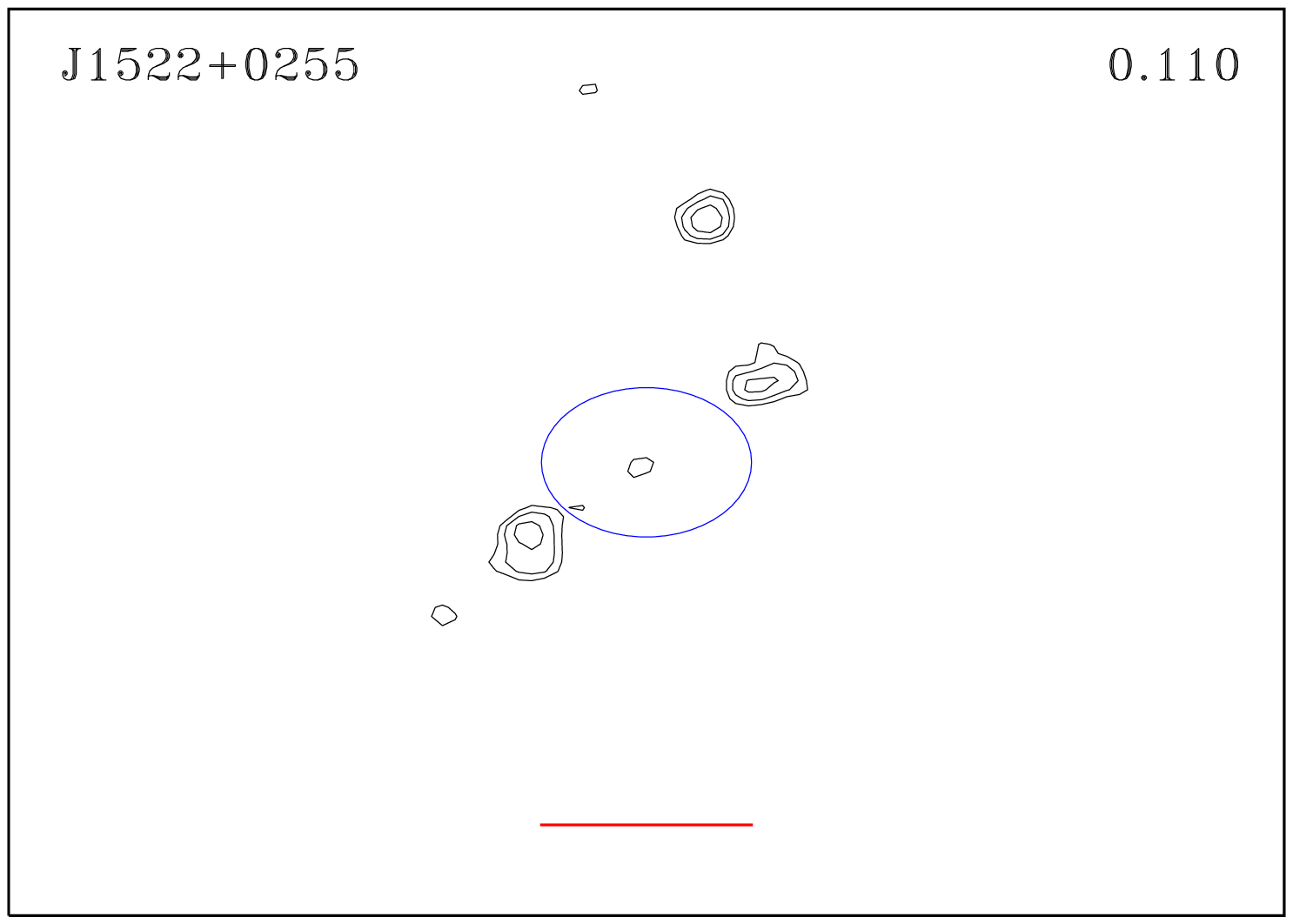}
\caption{(continued)}
\end{figure*}

\addtocounter{figure}{-1}
\begin{figure*}
\includegraphics[width=6.3cm,height=6.3cm]{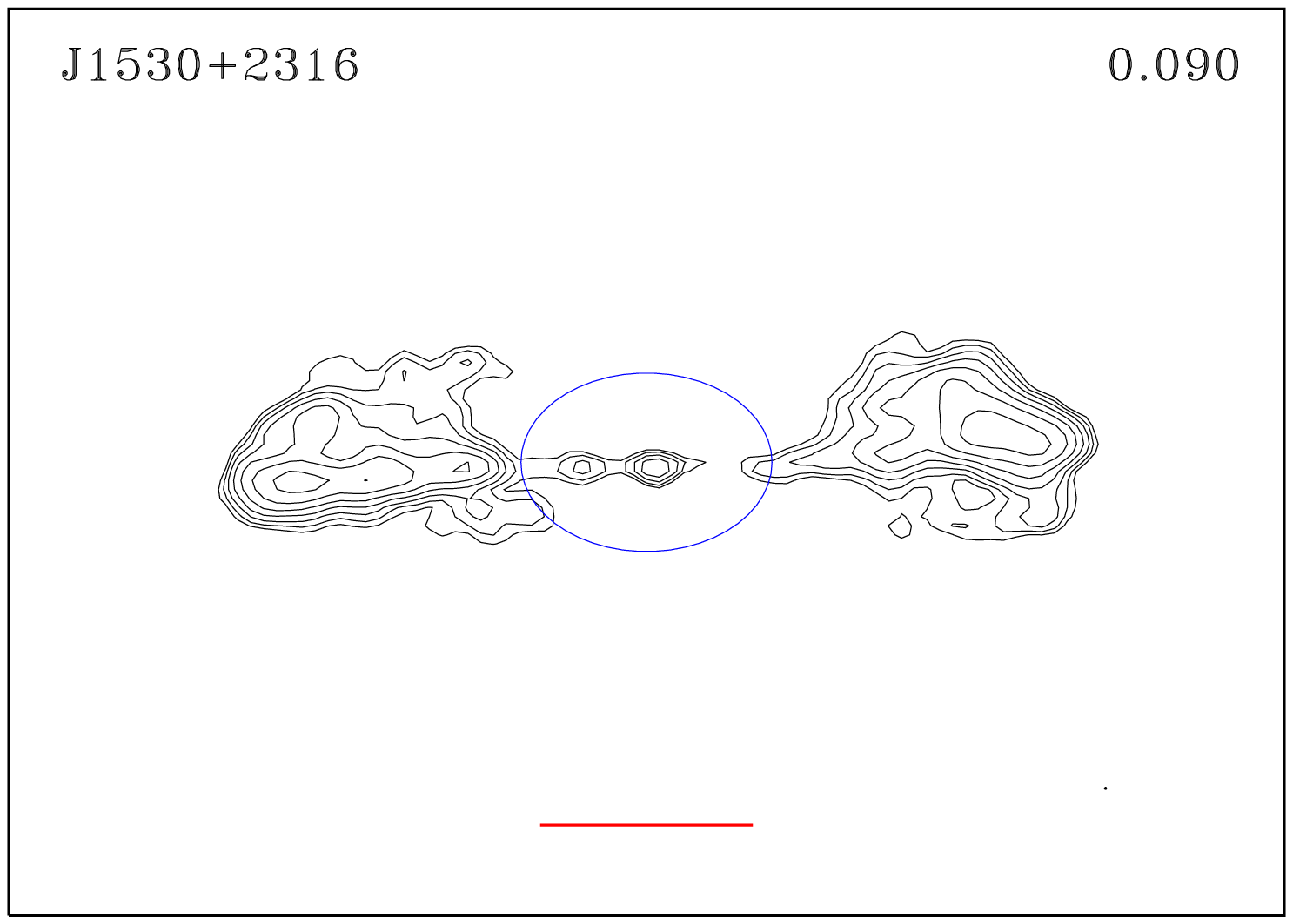}
\includegraphics[width=6.3cm,height=6.3cm]{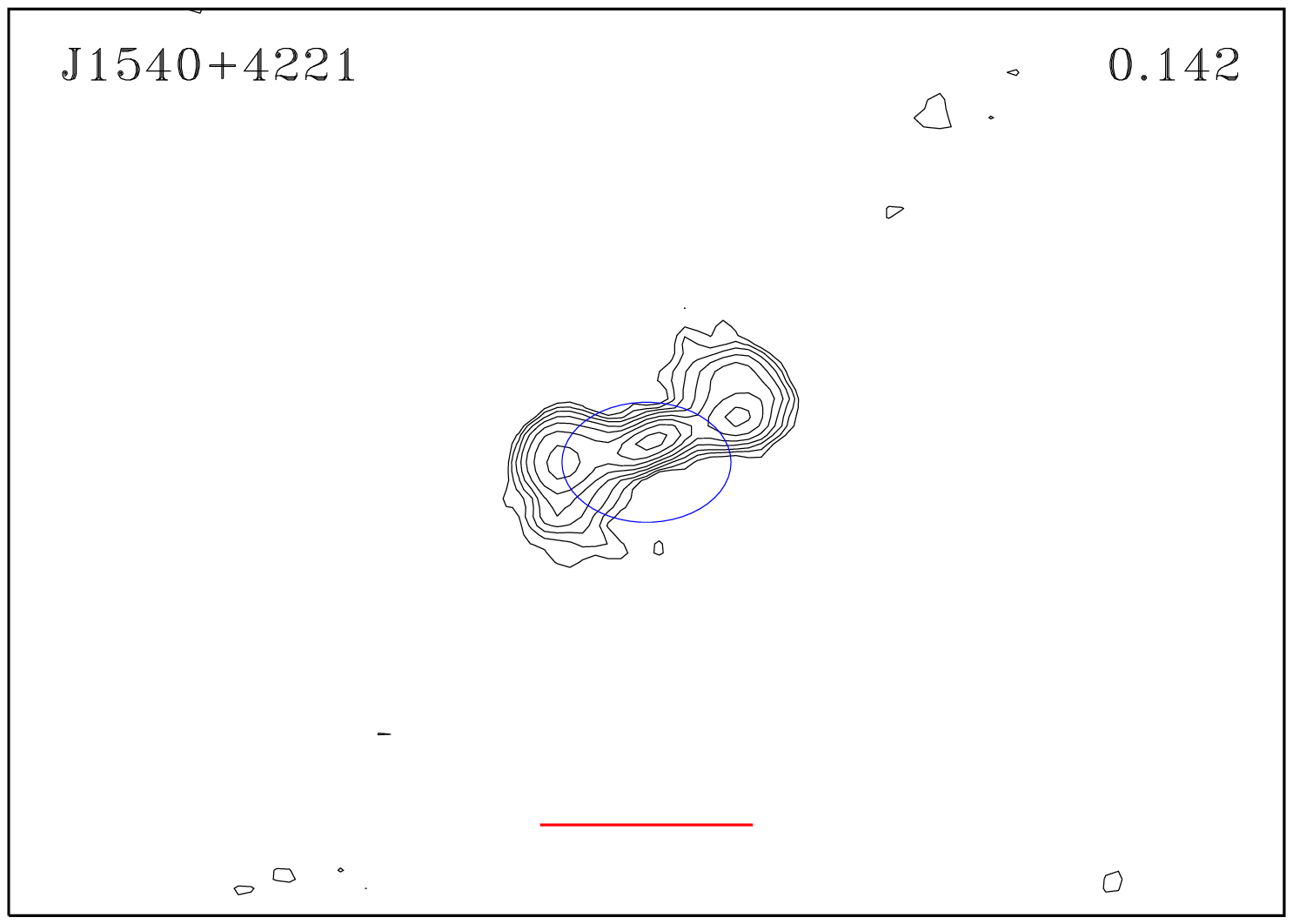}
\includegraphics[width=6.3cm,height=6.3cm]{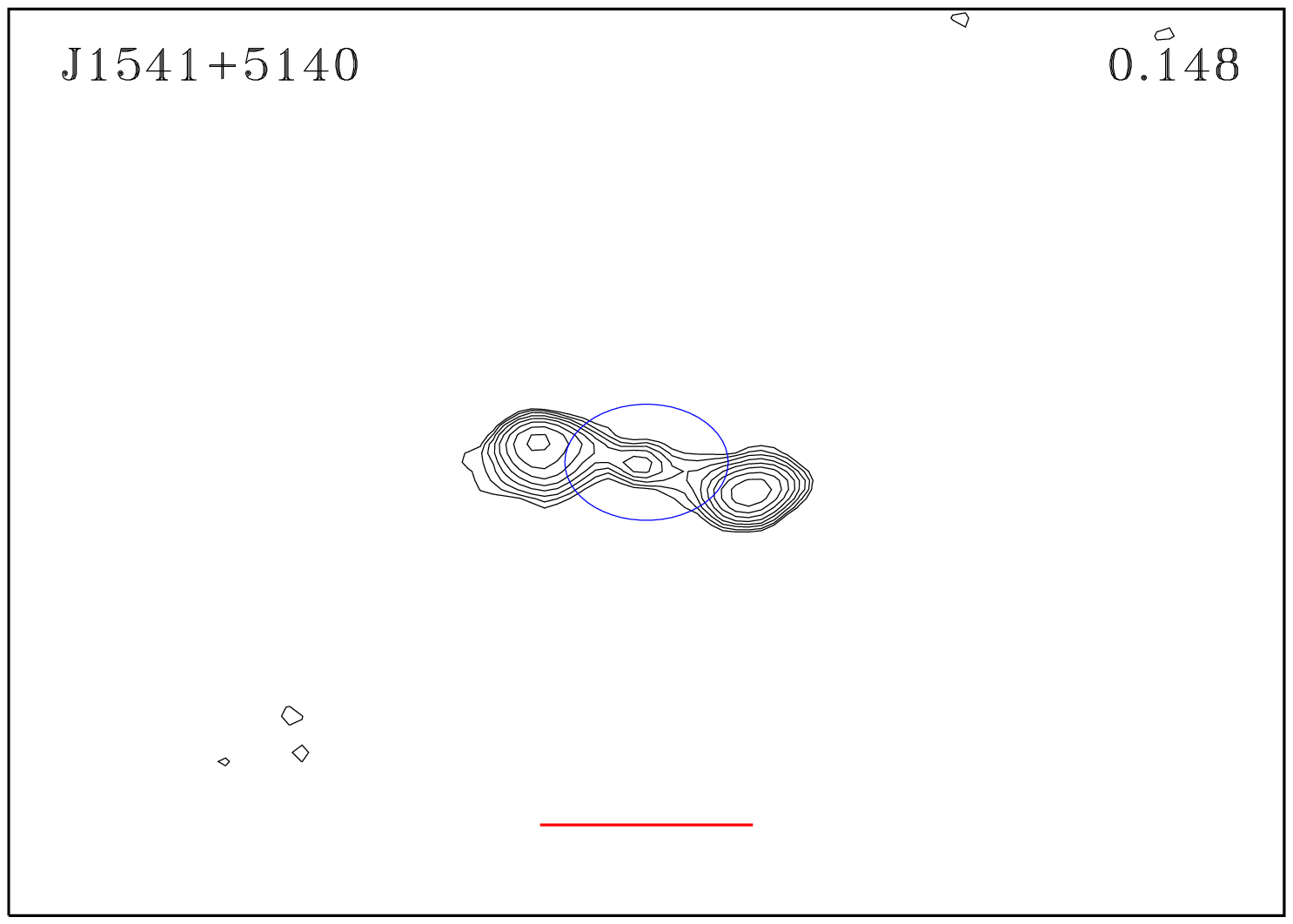}

\includegraphics[width=6.3cm,height=6.3cm]{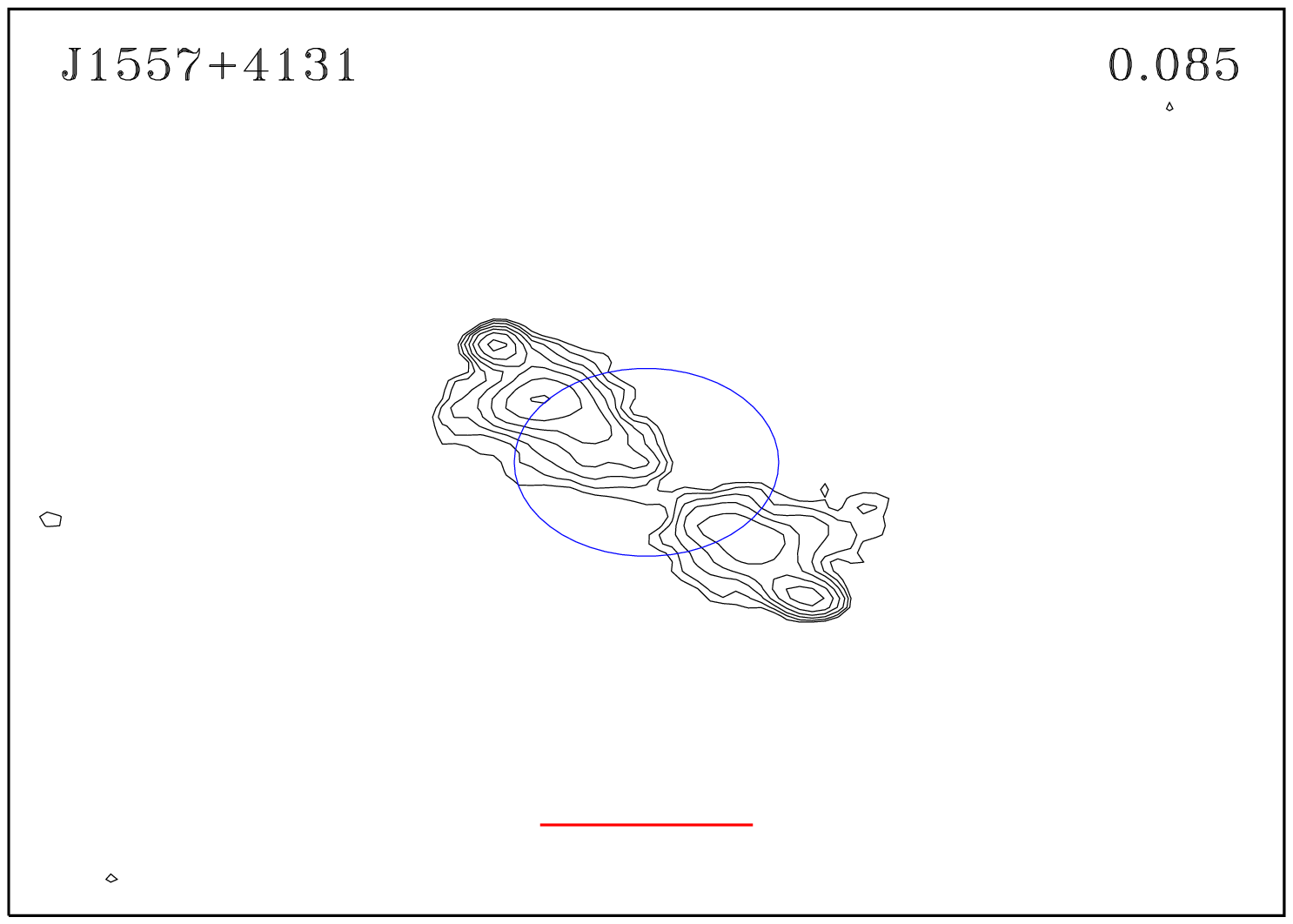}
\includegraphics[width=6.3cm,height=6.3cm]{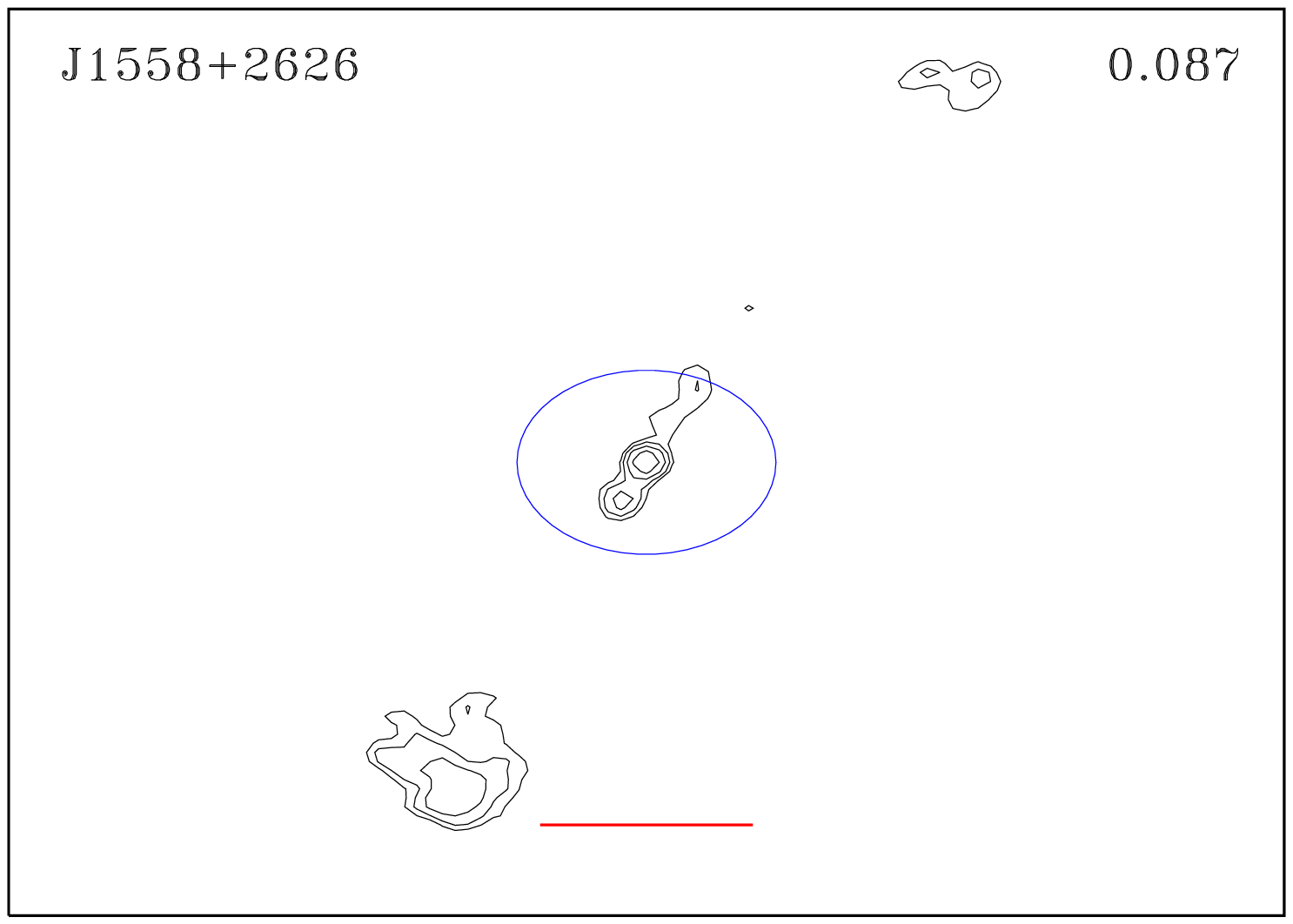}
\includegraphics[width=6.3cm,height=6.3cm]{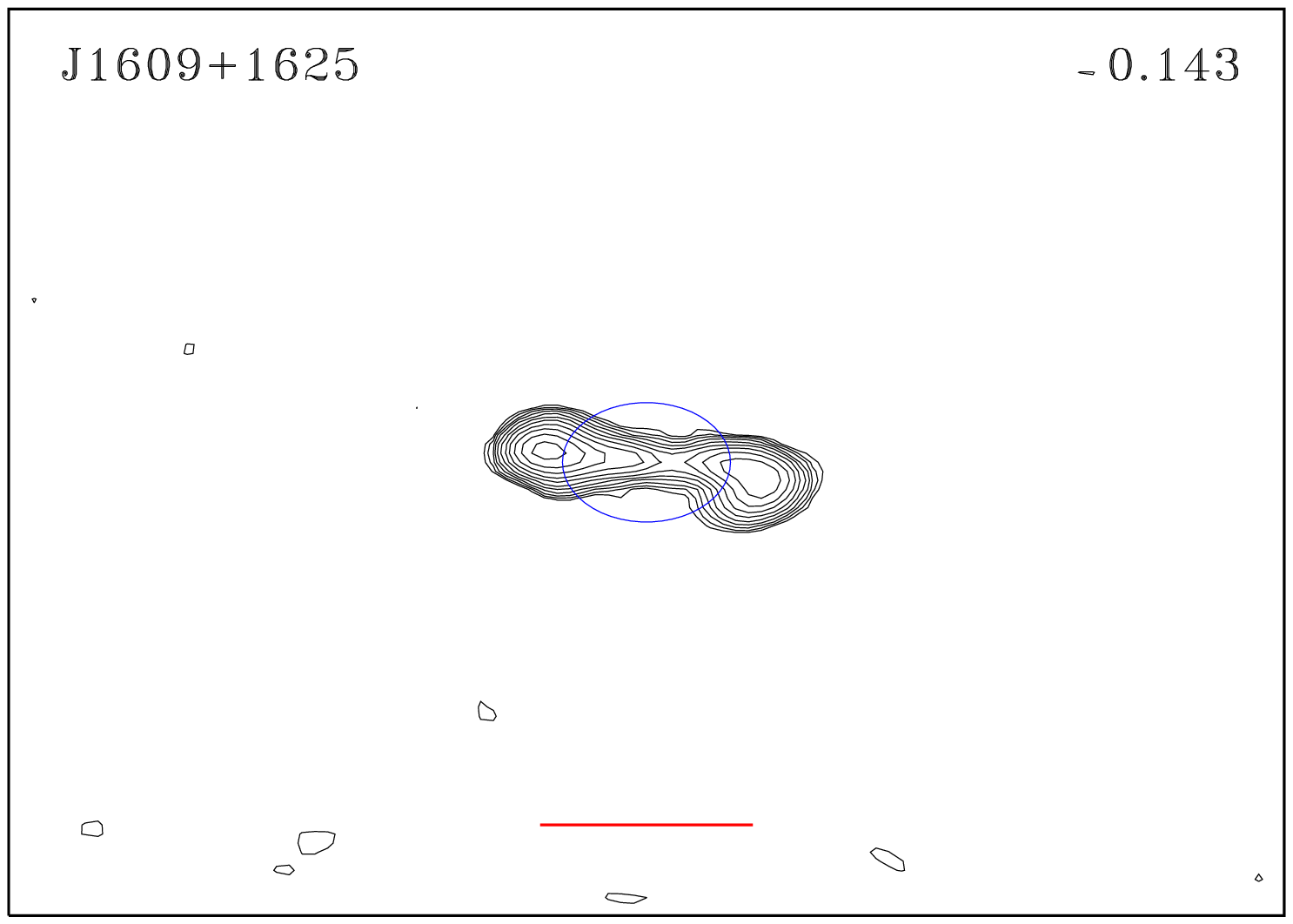}

\includegraphics[width=6.3cm,height=6.3cm]{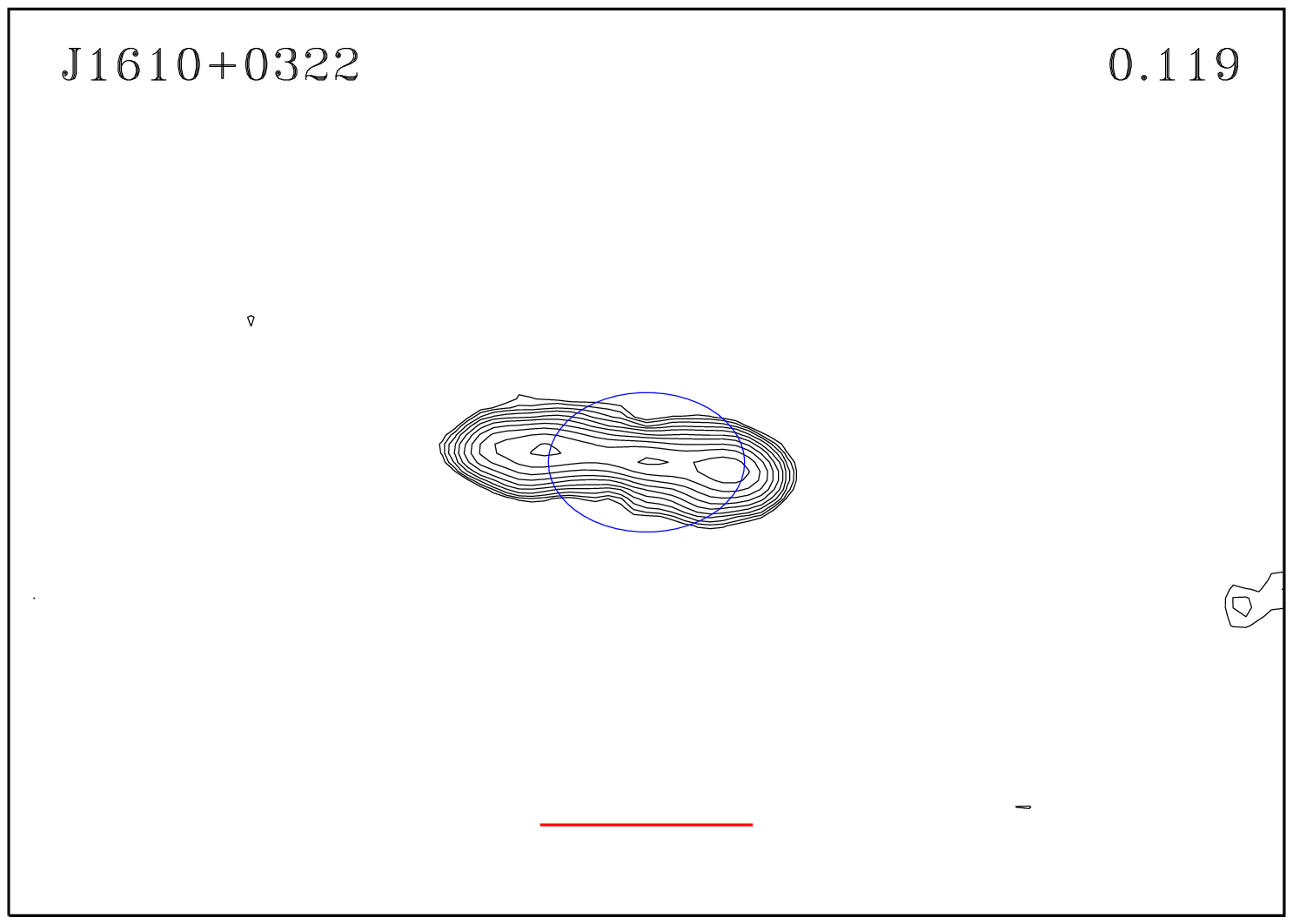}
\includegraphics[width=6.3cm,height=6.3cm]{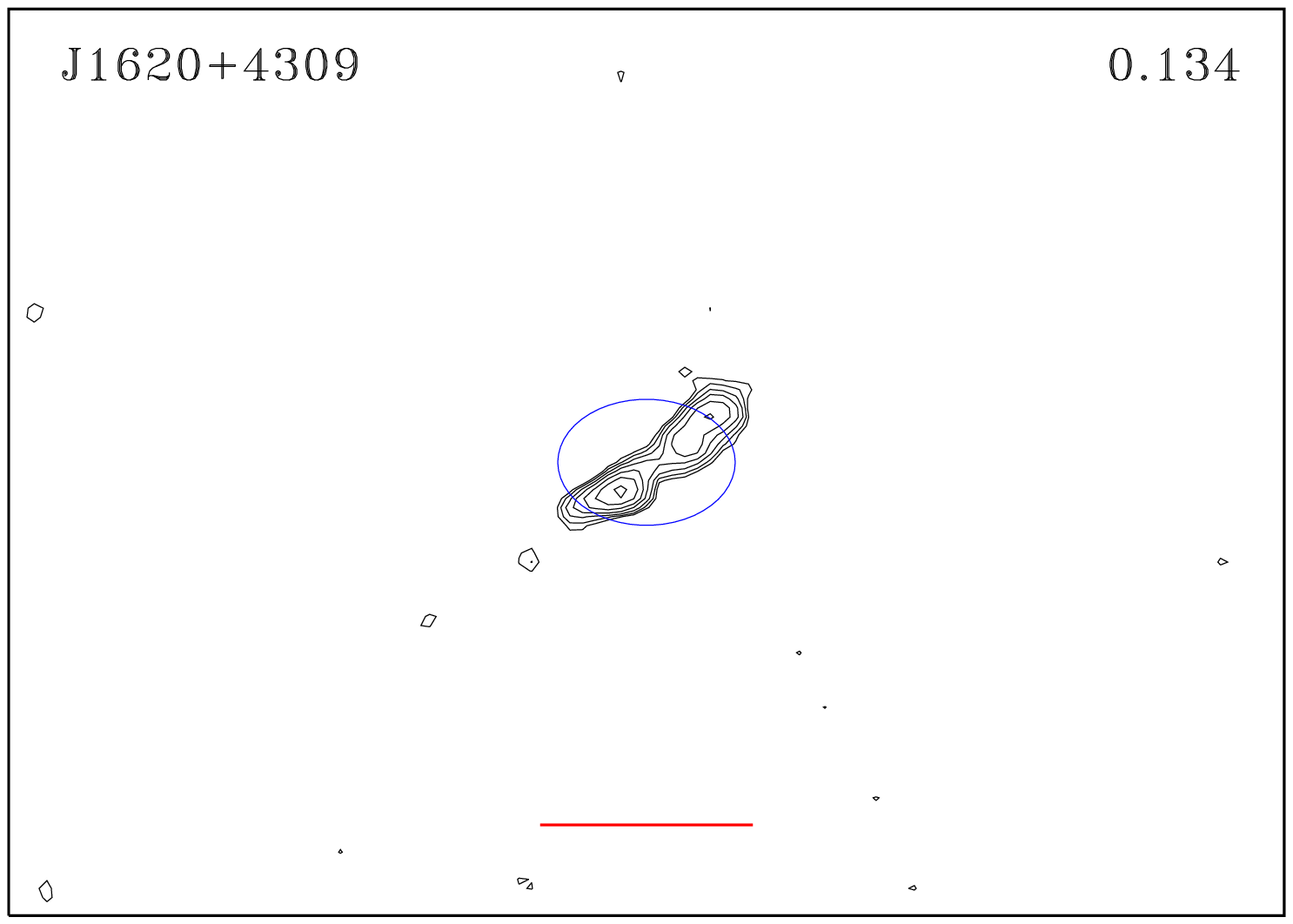}
\includegraphics[width=6.3cm,height=6.3cm]{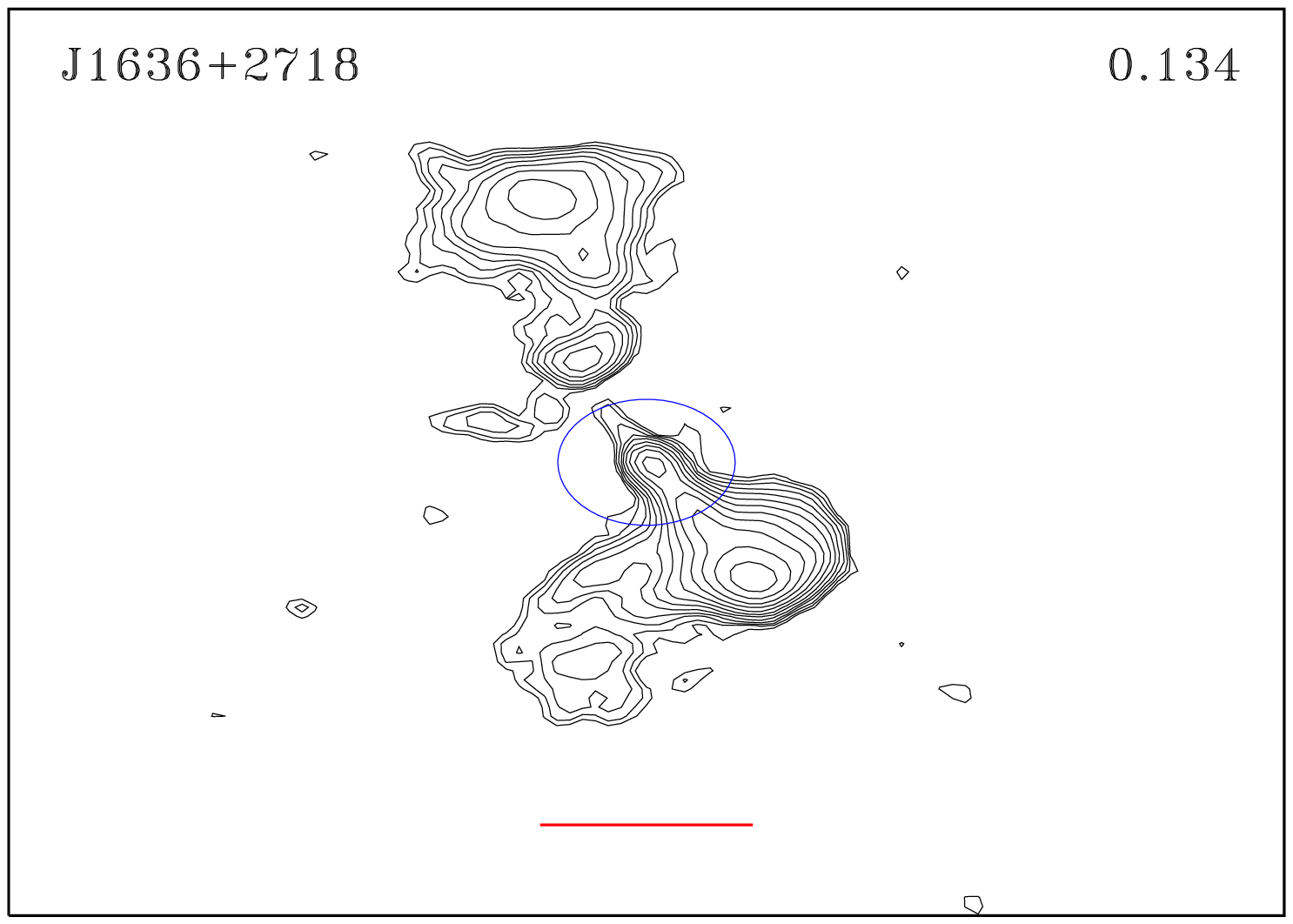}

\includegraphics[width=6.3cm,height=6.3cm]{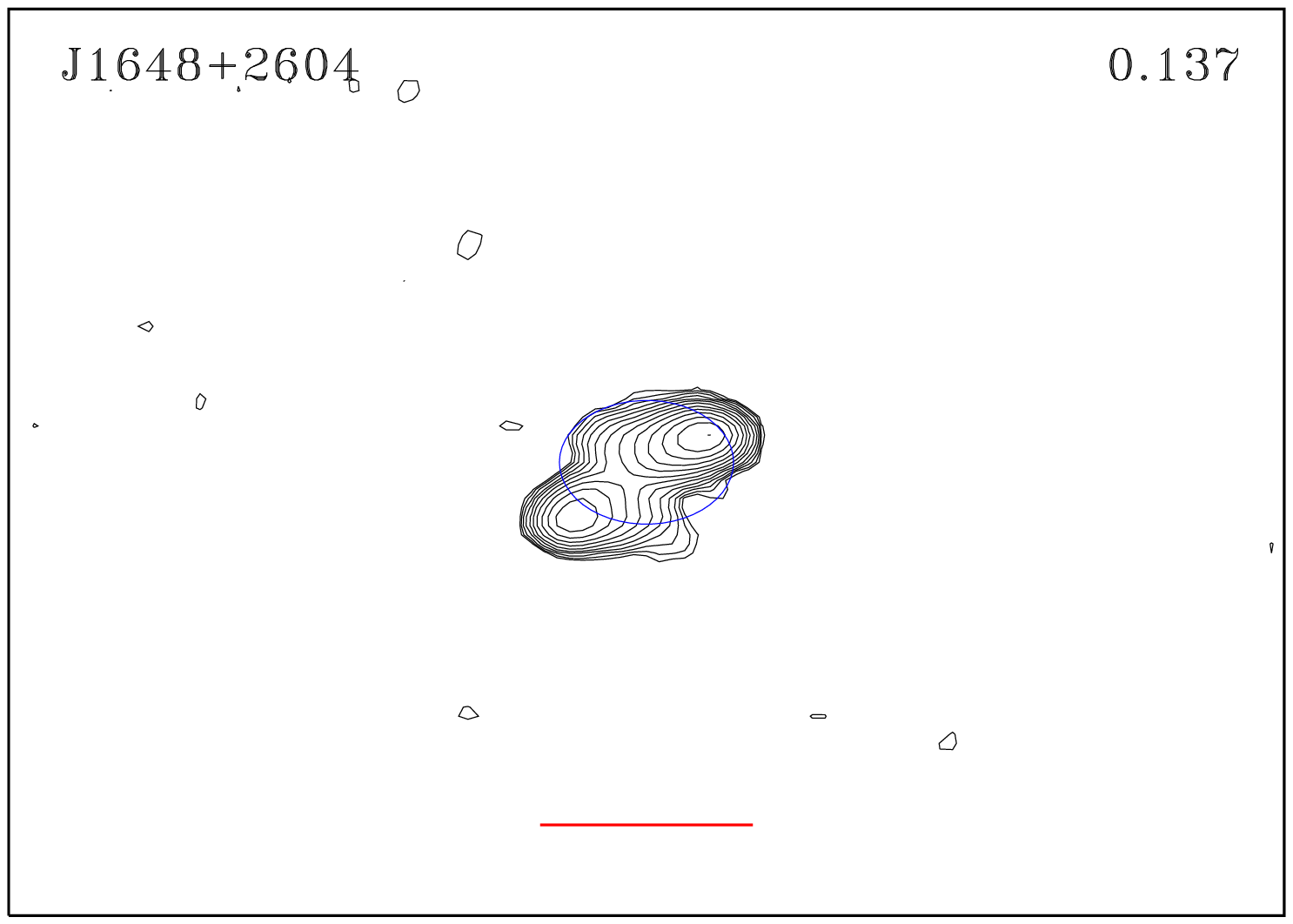}
\includegraphics[width=6.3cm,height=6.3cm]{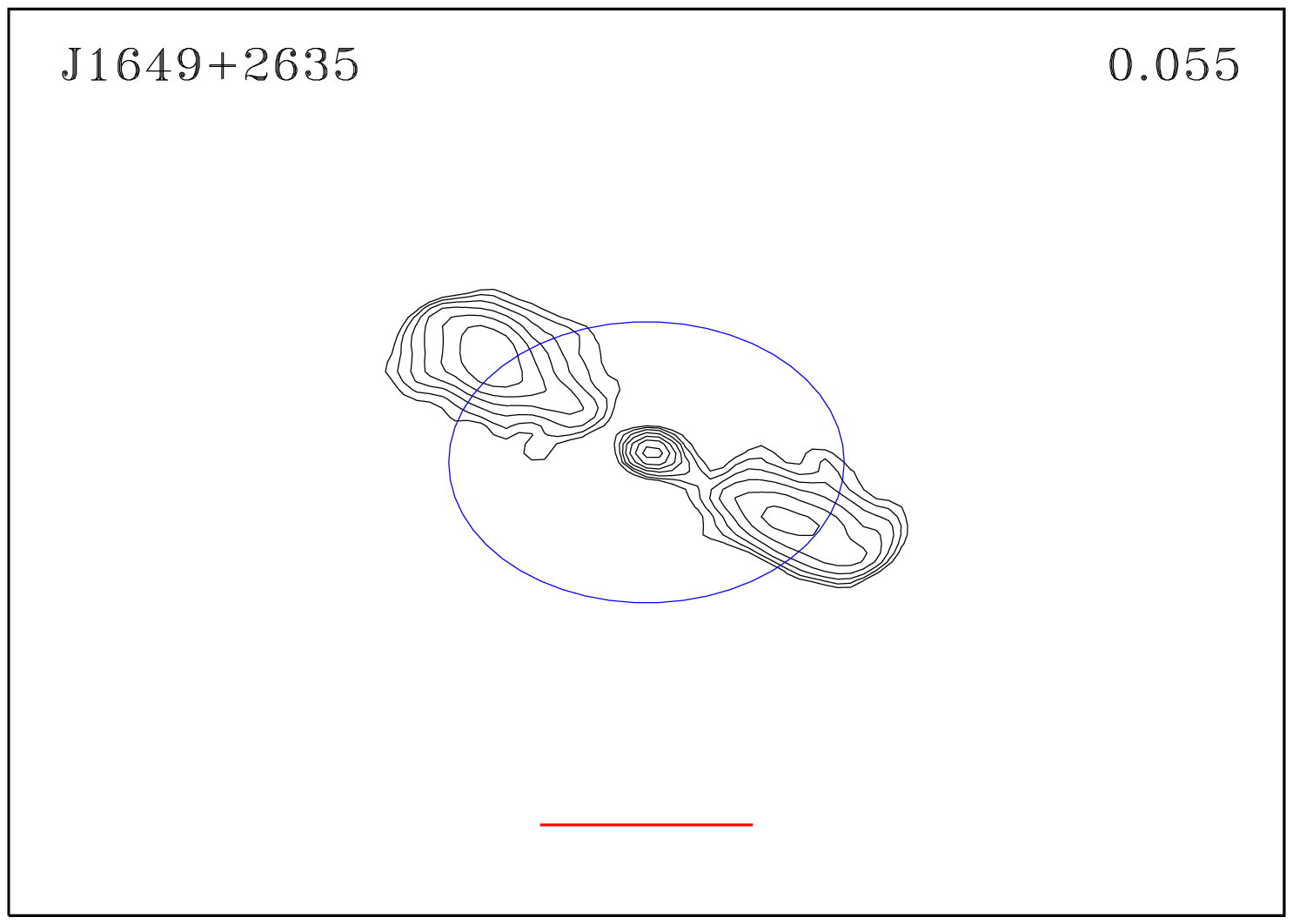}
\includegraphics[width=6.3cm,height=6.3cm]{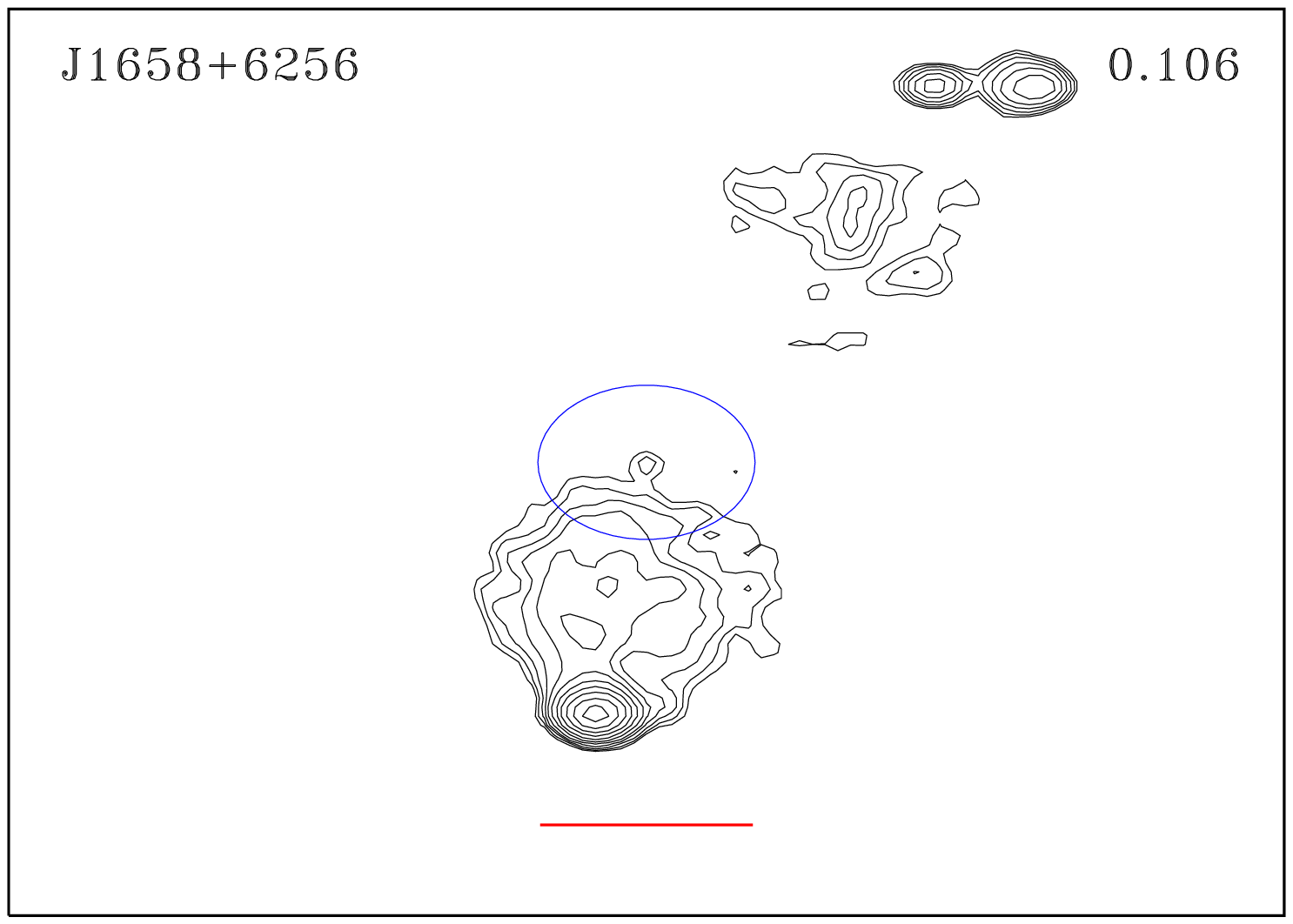}
\caption{(continued)}
\end{figure*}

\addtocounter{figure}{-1}
\begin{figure*}
\includegraphics[width=6.3cm,height=6.3cm]{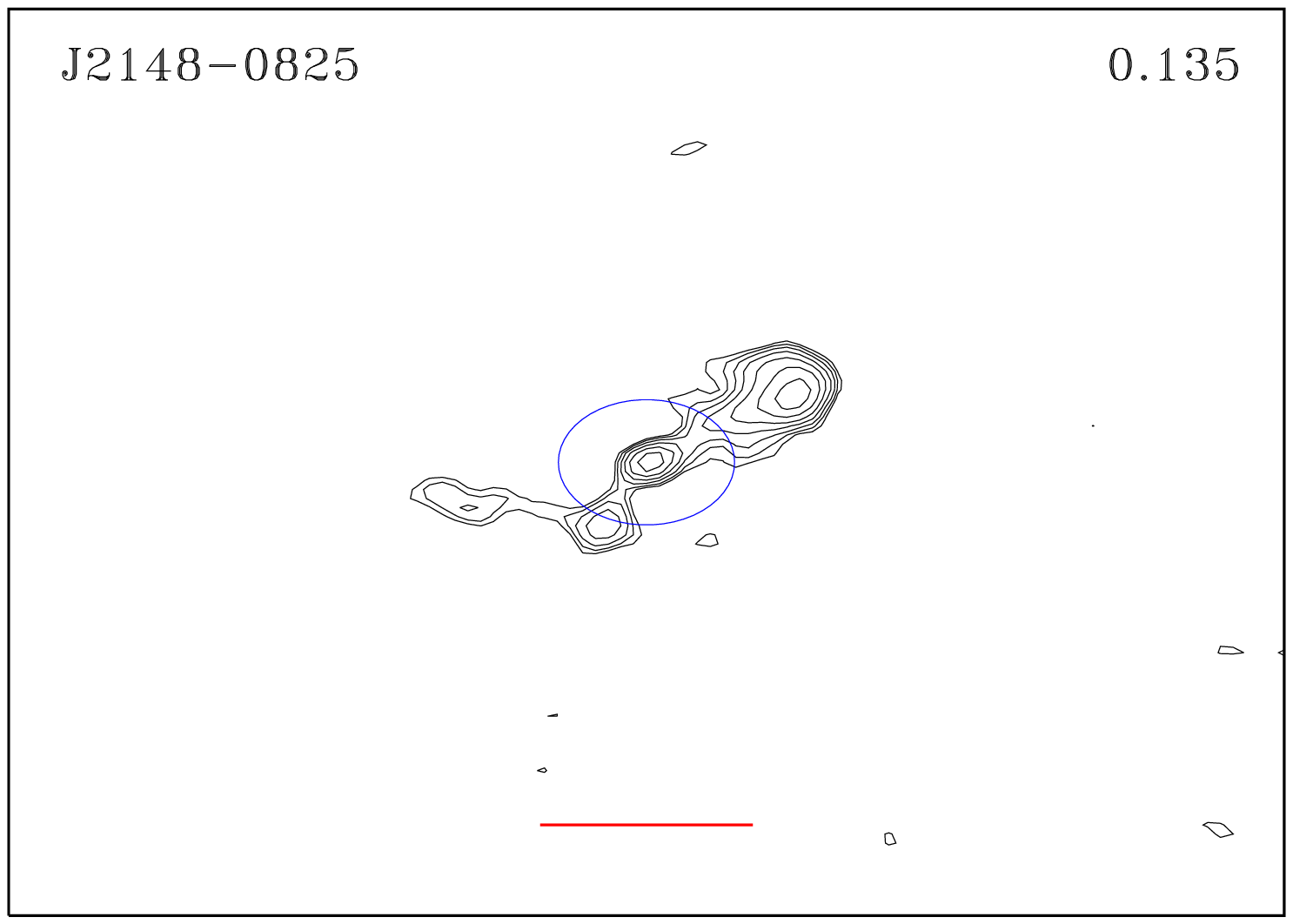}
\includegraphics[width=6.3cm,height=6.3cm]{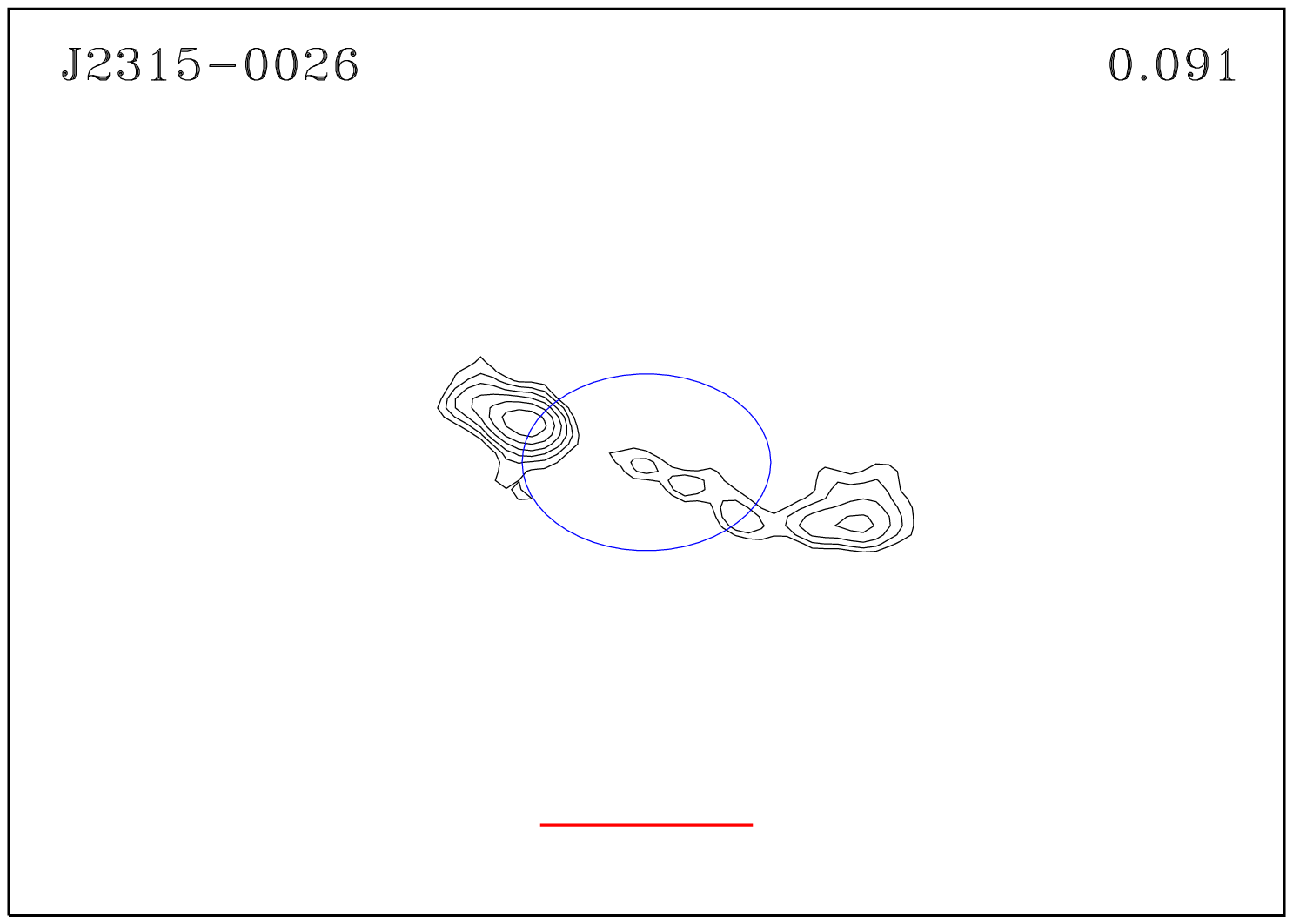}
\includegraphics[width=6.3cm,height=6.3cm]{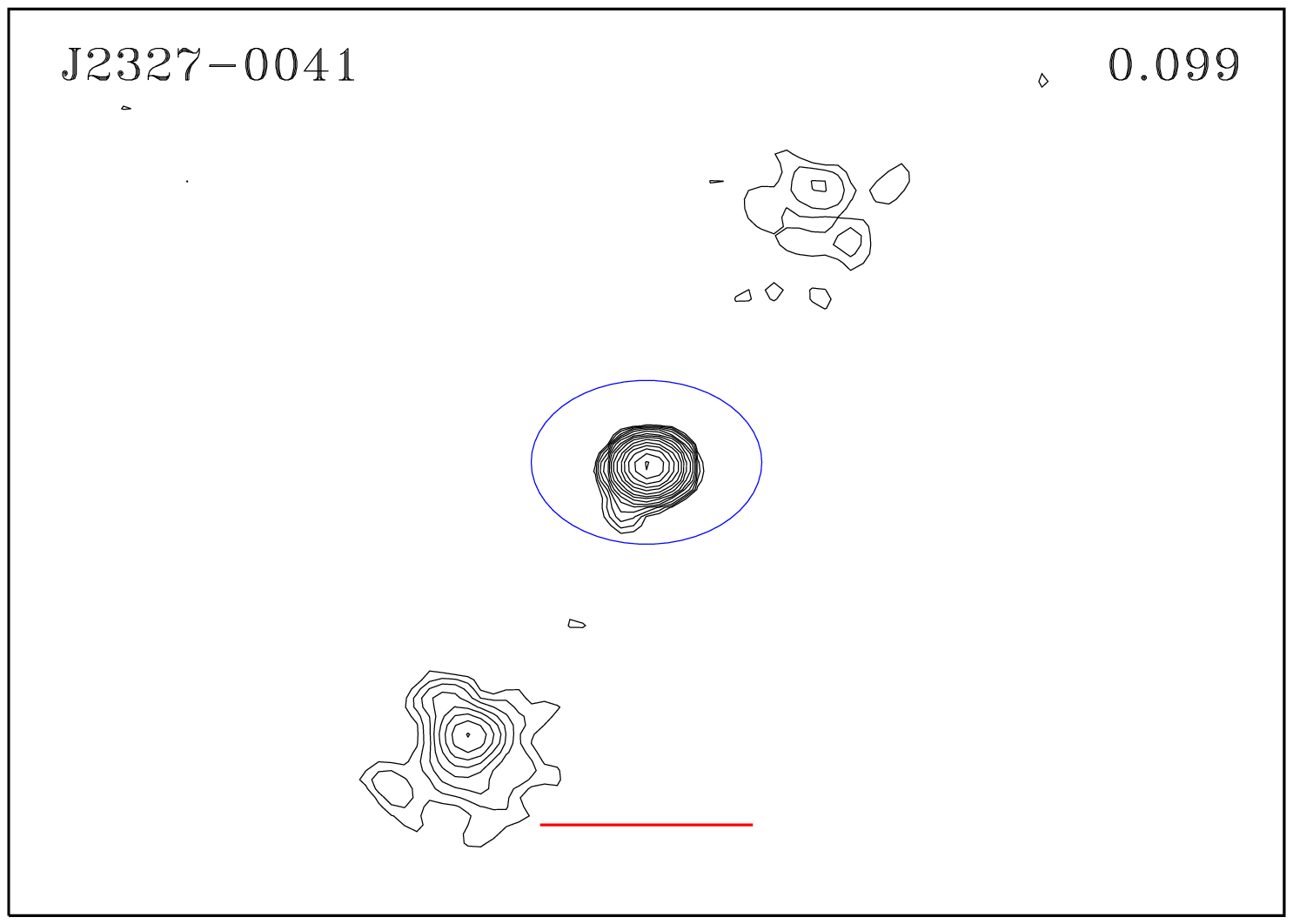}
\caption{(continued)}
\end{figure*}

\end{document}